\documentclass[showpacs,superscriptaddress,floatfix,prb,notitlepage,reprint]{revtex4-2}
\usepackage{graphicx,amsfonts,amssymb,amsmath,hyperref,hypcap,enumerate}
\usepackage{braket}
\usepackage{lipsum} 
\usepackage{amsthm}
\usepackage{multirow}
\usepackage{graphicx}
\usepackage{dcolumn}   %
\usepackage{bm}        %
\usepackage{amssymb}   %
\usepackage{amsmath}
\usepackage{braket}
\usepackage{epstopdf}
\usepackage{mathtools}
\usepackage{color}
\usepackage{subfigure}
\usepackage{diagbox}
\usepackage{longtable}
\usepackage{makecell}
\usepackage{bbold}

\newcommand{\ii}{\text{i}}
\newcommand{\AIIIz}{AIII$^{\dagger}$}
\newcommand{\AIIId}{AIII$^{\dagger}$}
\newcommand{\BDIz}{BDI$_0$}
\newcommand{\CIIz}{CII$_0$}
\newcommand{\DIIIz}{DIII$_0$}
\newcommand{\CIz}{CI$_0$}

\def\AIIp{AII + $\mathcal{S}_+$}
\def\Re{{\rm Re}}
\def\Im{{\rm Im}}
\newcommand{\iii}{{\bm i}}
\newcommand{\e}{{\bm e}}
\newcommand{\jjj}{{\bm j}}

\newcommand{\mS}{\mathcal{S}}
\newcommand{\mT}{\mathcal{T}}

\newcommand{\mC}{\mathcal{C}}
\newcommand{\mF}{\mathcal{F}}
\newcommand{\mK}{\mathcal{K}}

\def\sy{\sigma_y}

\def\ty{\tau_y}
\def\tz{\tau_z}
\def\tg{\tilde{\gamma}}

\DeclareMathAlphabet{\mathitb}{OT1}{cmr}{bx}{sl}
\usepackage[utf8]{inputenc}
\usepackage{enumitem}

\usepackage{etoolbox}
\makeatletter
\patchcmd{\frontmatter@abstract@produce}
  {\vskip200\p@\@plus1fil
   \penalty-200\relax
   \vskip-200\p@\@plus-1fil}
  {}
  {}
  {}
\makeatother

\begin{document}

\title{Universal hard-edge statistics of non-Hermitian random matrices}

\author{Zhenyu Xiao}
\email{wjkxzy@pku.edu.cn}
\affiliation{International Center for Quantum Materials, Peking University, Beijing 100871, China}

\author{Ryuichi Shindou}
\affiliation{International Center for Quantum Materials, Peking University, Beijing 100871, China}

\author{Kohei Kawabata}
\email{kawabata@issp.u-tokyo.ac.jp}
\affiliation{Institute for Solid State Physics, University of Tokyo, Kashiwa, Chiba 277-8581, Japan}

\date{\today}

\begin{abstract}
Random matrix theory is a powerful tool for understanding spectral correlations inherent in quantum chaotic systems.
Despite diverse applications of non-Hermitian random matrix theory, the role of symmetry remains to be fully established.
Here, we comprehensively investigate the impact of symmetry on the level statistics around the spectral origin---hard-edge statistics---and expand %
the classification of spectral statistics %
to encompass all the 38 symmetry classes of non-Hermitian random matrices.
Within this classification, we discern 28 symmetry classes characterized by distinct hard-edge statistics from the level statistics in the bulk of spectra, which are further categorized into two groups, namely the Altland-Zirnbauer$_0$ classification and beyond. 
We introduce and elucidate quantitative measures capturing the universal hard-edge statistics for all the symmetry classes.
Furthermore, through extensive numerical calculations, we study various open quantum systems in different symmetry classes, including quadratic and many-body Lindbladians, as well as non-Hermitian Hamiltonians.
We show that these systems manifest the same hard-edge statistics as random matrices and that their ensemble-average spectral distributions around the origin exhibit emergent symmetry conforming to the random-matrix behavior.
Our results establish a comprehensive understanding of non-Hermitian random matrix theory and are useful in detecting quantum chaos or its absence in open quantum systems.
\end{abstract}

\maketitle
\section{Introduction}

Symmetry plays a pivotal role in random matrix theory and is crucial for comprehending spectra of complex systems~\cite{Haake-textbook}.
Dyson's threefold way initially classifies Hermitian random matrices according to time-reversal symmetry (TRS)~\cite{Dyson-62}. 
The bulk of spectra is characterized by the threefold universal level statistics, termed the Wigner-Dyson statistics~\cite{Wigner-51, Wigner-58}. 
These universal level statistics appear in the following renowned random matrix ensembles: 
Gaussian unitary ensemble (lacking TRS), Gaussian orthogonal ensemble (possessing TRS with sign $+1$), and Gaussian symplectic ensemble (possessing TRS with sign $-1$). 
Extensive theoretical and experimental investigations have demonstrated that the spectral correlations of chaotic closed quantum systems align with these universal random-matrix statistics with corresponding symmetry~\cite{bohigas1984,  Friedrich1987, poilblanc1993, mirlin00, Serbyn2016}.
Subsequently, random matrix ensembles in the chiral and Bogoliubov-de Gennes (BdG) symmetry classes, which incorporate chiral symmetry~\cite{Verbaarschot-93, Verbaarschot-94} and particle-hole symmetry~\cite{altland1997}, were introduced. 
These symmetries dictate universal properties in the proximity of the symmetry-preserving point $E=0$ in the energy spectrum.
Together with TRS, chiral and particle-hole symmetries form the comprehensive classification of Hermitian random matrices, known as the Altland-Zirnbauer (AZ) tenfold classification~\cite{altland1997}.
Importantly, this well-established Hermitian random matrix theory has found broad applications across diverse fields of physics.
In condensed matter physics, 
it elucidates universal transport phenomena of disordered metals and superconductors~\cite{beenakker1997, beenakkerRMT15}
and also serves as 
a powerful tool to detect the Anderson transitions~\cite{Evers-review} and 
many-body-localization transitions~\cite{Huse-review, Abanin19}.
In nuclear physics, random matrices provide an effective model of chiral symmetry breaking~\cite{verbaarschot2000}.

While researchers historically focused 
on Hermitian operators, non-Hermitian operators are equally important in various physical systems.
Prime examples include chiral Dirac operators with chemical potential in quantum chromodynamics (QCD)~\cite{osborn2004, akemann2004, kanazawa2021} and scattering matrices in quantum transport phenomena~\cite{beenakker1997, beenakkerRMT15}.
Non-Hermitian matrices appear naturally in 
stochastic processes and network systems,  
having
important applications in biological systems 
~\cite{may1976, sommers88, sompolinsky88, ott1990, nelson98, amir2016, murugan2017, zhang2019a}.
Realistic
physical systems 
are inevitably coupled with
their surrounding environments, and non-Hermitian operators 
also play pivotal roles in the dynamics of such open physical systems~\cite{breuer2002theory}. 
For the Markovian environment, the dynamics of open quantum systems is described by the Lindblad master equation~\cite{GKS-76, Lindblad-76}, where the generator of the dynamics, known as the Lindbladian, is a non-Hermitian operator. 
The time evolution of classical synthetic 
materials such as optical and 
photonic systems with energy gain or 
loss~\cite{Makris08, guo2009, zeuner2015, tzortzakakis2020}, 
along with quantum systems under continuous measurement~\cite{lee2014, li2020, liang2022, Daley-review}, 
can also be characterized by non-Hermitian Hamiltonians. 
Thanks to advances in state-of-the-art 
experiments and theories, the past years have witnessed remarkable development in non-Hermitian physics~\cite{Bender-review, Konotop-review, Christodoulides-review, bergholtz2021}.

These advances necessitate a systematic understanding of symmetry inherent in open quantum systems, leading to the 38-fold classification~\cite{BL02, kawabata19} of non-Hermitian operators beyond the tenfold classification of Hermitian operators. 
This symmetry classification serves as a guiding principle in designing and comprehending non-Hermitian topological phases~\cite{Xu17, gong2018, Yao18, Yao18_2, kawabata2019a, kawabata19, Zhou19}, analyzing non-Hermitian Anderson~\cite{hatano1996localization, HatanoNelson97PRB, Efetov97, Xu16, longhi2019, Zeng20, Wang20, huang2020, Huang20SR, kawabata20, Luo21, Luo21TM, luo2021unifying, xiao2023} and many-body localization~\cite{hamazaki2019non, suthar2022, zhai2020}, as well as understanding spontaneous breaking of parity-time ($\mathcal{PT}$) 
symmetry~\cite{Bender98}.
Moreover, the chaotic behavior in open quantum systems has drawn substantial attention~\cite{grobe88, grobe89, Xu-19, Denisov-19, Can-19PRL, Can-19JPhysA, hamazaki20, akemann19, sa20, Wang-20, Xu-21, li2021a, garcia2023, costa2023, kawabata2023singular, 
roccati2023}, with symmetry playing a crucial role in its 
description. 
Researchers developed the symmetry classification of non-Hermitian generalizations of the Sachdev-Ye-Kitaev (SYK) model~\cite{Sachdev-Ye-93, kitaev15, Sachdev-15, Polchinski-Rosenhaus-16, Maldacena-Stanford-16, Rosenhaus-review, Sachdev-review}, a prototype of quantum chaotic model, revealing rich structures~\cite{garcia22}.
Substantial progress was also made in the symmetry classification of Lindbladians for both single-particle and many-body systems~\cite{lieu20a, altland2021, sa2023, kawabata2023}. 
Several prototypical models, such as Bogoliubov quasi-particles in dissipative superconductors~\cite{pikulin12, sanjose16, ghatak2018, okuma2020, cayao23} and non-Hermitian extensions~\cite{Esaki11, lieu18, Yao18} of the Su-Schrieffer-Heeger model~\cite{SSH1980}, exhibit particle-hole or sublattice symmetry and can display chaotic behavior in the presence of disorder. 

Non-Hermitian random matrix theory is useful for understanding the spectral correlations and dynamics of open quantum systems. 
Universal dynamical features of open quantum chaotic systems were investigated using non-Hermitian random matrices~\cite{Xu-19, Can-19PRL, Can-19JPhysA, Wang-20}.
It was conjectured that the statistics of complex energy levels in non-integrable open quantum systems follow those of non-Hermitian random matrices, which has been numerically verified in several models~\cite{grobe88, hamazaki20, akemann19, sa20, li2021a}. 
Level correlations of non-Hermitian random matrices 
in the bulk of complex spectra (i.e., away from special points, lines, and edges of complex spectra) 
belong to the threefold universality classes~\cite{ginibre1965statistical, grobe89, hamazaki20}. 
These universality classes depend only on 
an extension of TRS for non-Hermitian operators, 
time-reversal symmetry$^{\dag}$~\cite{kawabata19},
while the influence of other defining symmetries appears elusive. Given the prevalence of almost all 38 symmetry classes in realistic physical systems (e.g., see Refs.~\cite{garcia22, sa2023, kawabata2023}), characterizing the unique universality classes of level correlations in each symmetry class and exploring the impact of symmetry stands out as one of the crucial challenges in non-Hermitian random matrix theory. 
Despite its importance and 
some discrete works on level statistics beyond the threefold way, especially those on non-Hermitian but real random matrices~\cite{edelman1995, Sommers98, kanzieper2005, Forrester07, xiao22}, a comprehensive understanding of the universality classes within the 38-fold symmetry classification is still lacking.

In this work, we demonstrate that the level statistics around the spectral origin %
faithfully reveal the impact of all defining symmetries. 
In Hermitian random matrix theory, such level statistics around the spectral origin are known as the hard-edge statistics.
While the spectral origin of complex spectra may look like a hole or node, we find a rich analogy of the level statistics with the Hermitian case.
Thus, we call the level statistics around the spectral origin the hard-edge statistics even in non-Hermitian random matrix theory. %
We comprehensively explore possible universality classes of the hard-edge statistics of non-Hermitian random matrices in all 38 symmetry classes.
We identify 28 symmetry classes where the level statistics around the spectral origin differ from those in the bulk of the spectra or around the real or imaginary axis by analyzing their defining %
symmetries. %
This approach is akin to that in Hermitian random matrix theory, where the hard-edge statistics distinct from the bulk statistics appear in the chiral and BdG symmetry classes but not in the Wigner-Dyson symmetry classes~\cite{altland1997, beenakkerRMT15, Haake-textbook}. 
By contrast, 
non-Hermiticity enriches the symmetry classification and leads to more diverse universal spectral correlations. 
Specifically, 
we categorize these 28 symmetry classes into two groups and investigate all of them, summarized in Tables~\ref{tab: AZ0 classification} and~\ref{tab: AZ_real}.
Notably, similar to the Hermitian counterparts, only a few eigenvalues closest to the spectral origin contribute to the hard-edge statistics. 
Consequently, analyzing the hard-edge statistics and diagnosing symmetry classes requires studying ensembles of operators.

The first group, summarized in Table~\ref{tab: AZ0 classification}, comprises 
seven symmetry classes, 
where the spectral origin is the only high-symmetry 
point in the complex plane, and  
the hard-edge statistics exhibit spectral U(1) rotation symmetry. 
These seven symmetry classes, together with the threefold classes based on time-reversal symmetry$^{\dag}$, form the tenfold classification, which 
we dub Altland-Zirnbauer$_0$ (AZ$_0$) classification. 
We employ the density of complex eigenvalues and the distributions 
of the complex level ratios 
around the spectral origin to characterize the universality classes of the hard-edge statistics.
We numerically obtain seven universal ratio distributions and summarize their properties in Table~\ref{tab: AZ0 classification}.
In a similar spirit to the Wigner surmise,
we 
calculate these distributions analytically for small non-Hermitian random matrices.
We show that the asymptotic behavior observed in large matrices can be accurately estimated from that in small ones. 
Notably, distributions of the complex eigenvalue with the minimum modulus were investigated previously~\cite{splittorff04, akemann09, garcia22}.
By contrast, we introduce a new statistical quantity that captures the hard-edge statistics---ratio of the complex eigenvalue with the minimum modulus to that with the second minimum modulus, which we dub the complex level ratio.
We clarify the difference between the two approaches and discuss the potential advantages of the complex level ratio.

The second group, summarized in Table~\ref{tab: AZ_real}, comprises 
the remaining 21 symmetry classes 
which have no spectral U(1) rotation symmetry around the origin. 
We investigate the distributions of the eigenvalues 
$z_{\min}$ with the smallest modulus. 
We numerically demonstrate that in certain symmetry classes, the 
distributions exhibit delta-function peaks on the real and/or imaginary axes, in contrast to the preceding seven symmetry classes in the AZ$_0$ classification. 
We find that the probability of $z_{\min}$ being real or purely imaginary is characteristic and universal in each symmetry class, summarizing these universal values for all these 21 symmetry classes in Table~\ref{tab: AZ_real}. 
In particular, we focus on three representative symmetry classes
(i.e., classes BDI, CII, and \AIIp) 
and carefully examine their universal distributions of $z_{\min}$.
The obtained
distributions are qualitatively different when $z_{\min}$ is 
complex, real, or purely imaginary, and in certain classes, 
the distributions exhibit the point group symmetry D$_4$. 
We provide theoretical explanations for our numerical findings.  
 
To further demonstrate the universality of our newly found hard-edge statistics, we investigate diverse open quantum systems, including quadratic and many-body Lindbladians, as well as non-Hermitian Hamiltonians. For the seven symmetry classes in the AZ$_0$ 
classification, we construct corresponding physical models.
An ensemble of a physical model often has statistical symmetry 
(ensemble symmetry), while symmetry of a single realization 
of the model determines the relevant symmetry class.  
Notably, the ensemble symmetry of physical models is 
generally lower than the ensemble symmetry of random 
matrices in the same symmetry class. 
Nevertheless, the hard-edge statistics of non-Hermitian random matrices 
in the AZ$_0$ classification 
manifest themselves 
in these physical systems in the seven symmetry classes, 
signaling emergent spectral U(1) symmetry. For the 21 symmetry 
classes beyond the AZ$_0$ classification, we
construct models in the three representative symmetry classes.  
We show that they also exhibit the hard-edge statistics and emergent spectral D$_4$ symmetry consistent with non-Hermitian random matrices. 

The rest of this work is organized as follows.
In Sec.~\ref{sec: symmetry}, we begin with the symmetry 
classification of non-Hermitian random matrix theory.
We identify the symmetry classes exhibiting the unique hard-edge statistics.
In Secs.~\ref{sec: RM AZ0} and \ref{sec: RM AZ_real}, we investigate the hard-edge statistics of non-Hermitian random matrices in the Gaussian ensembles within and beyond the AZ$_0$ classification, respectively.
Sections~\ref{sec: quantum system} and \ref{sec: quantum system real} are respectively dedicated to open quantum physical models within and beyond the 
AZ$_0$ classification. 
We uncover that the hard-edge statistics of these physical models coincide with non-Hermitian random matrices, demonstrating that our characterization of the hard-edge statistics efficiently captures the chaotic behavior or integrability of open quantum systems.
Section~\ref{sec: conclusion} is devoted to conclusions and discussions.

\section{Symmetry classification of non-Hermitian random matrix theory}
\label{sec: symmetry}

We start with the AZ classification of Hermitian operators $H$. 
It
consists of two types of anti-unitary symmetry,
\begin{flalign}
&\text{time-reversal symmetry (TRS):}  \nonumber \\ 
& \qquad \mT_+ H^* \mT_+^{-1} = H \qquad (\mT_+ \mT_+^* = \pm 1)\, ; \label{eq: TRS_def}\\
&\text{particle-hole symmetry (PHS):}  \nonumber \\
& \qquad \mC_- H^{\rm T} \mC_-^{-1} = -H \qquad (\mC_- \mC_-^* = \pm 1)\, ;
\end{flalign}
and their
combined unitary symmetry 
\begin{align}
&\text{chiral symmetry (CS): } \nonumber \\ 
& \qquad \Gamma H^{\dagger} \Gamma^{-1} = -H \qquad (\Gamma^2 = 1) \, ,
\end{align}
where $\mT_{+}$, $\mC_{-}$, and $\Gamma$ are all unitary operators. %

Non-Hermiticity $H \neq H^{\dagger}$ ramifies the definitions of TRS, PHS, and CS, and enriches the symmetry classification~\cite{kawabata2019a}. 
Replacing $H$ in their definitions by $H^{\dagger}$ results in two types of additional anti-unitary symmetry,
\begin{flalign}
&\text{time-reversal }{\rm symmetry^{\dagger} \, (TRS^{\dagger}):}  \nonumber \\
& \qquad  \mC_+ H^{\rm T}  \mC_+^{-1} = H \qquad (\mC_+ \mC_+^* = \pm 1)\, ;\\
&\text{particle-hole } {\rm symmetry^{\dagger} \,   (PHS^{\dagger}):} \nonumber \\
& \qquad \mT_- H^{*} \mT_-^{-1} = -H  \qquad
(\mT_- \mT_-^* = \pm 1)\, ;
\end{flalign}
and additional unitary symmetry,
\begin{align}
& \text{sublattice symmetry (SLS): } \nonumber \\
& \qquad \mS H \mS^{-1} = -H \qquad (\mS^2 = 1) \, .\label{eq: SLS_def}
\end{align}
The combination of CS and SLS gives another unitary symmetry
\begin{align}
&\text{pseudo-Hermiticity (pH): } \nonumber \\  
& \qquad \eta H^{\dagger} \eta^{-1} = H \qquad (\eta^2 = 1) \,. \label{eq: pH_def} 
\end{align}
These four symmetries, together with the three original symmetries in the AZ classification, constitute the 38-fold symmetry classification~\cite{BL02, kawabata19} of non-Hermitian operators $H$.
Notably, if $H$ respects TRS (CS), $ \ii H$ respects PHS$^{\dagger}$ (pH), and vice versa. 
Hence, the spectral properties of $H$ and $\ii H$ are identical up to the multiplication by $\ii$.
Consequently, the symmetry classes of $H$ and $\ii H$ are counted only once in the 38-fold symmetry classification.

Symmetries impose constraints on spectra and eigenvectors of non-Hermitian operators.
As an example, let us consider a non-Hermitian Hamiltonian $H$ with PHS.
If ${\bm v}$ is a right eigenvector of $H$ with an eigenvalue $z$ (i.e., $H {\bm v} = z  {\bm v} $), ${\bm v}^{\rm T} \mC_-^{\dagger}$ is a left eigenvector of $H$ with an eigenvalue $-z$ [i.e., $({\bm v}^{\rm T} \mC_-^{\dagger})\,H = - z\,({\bm v}^{\rm T} \mC_-^{\dagger})$].
Thus, the particle-hole operation generally transforms an eigenvalue $z$ 
to another eigenvalue $-z$, and hence the complex spectrum is symmetric about the origin.
Notably, while PHS, in general, creates opposite-sign pairs $\left( z, -z \right)$ of complex eigenvalues, it imposes a special constraint on the eigenvalue invariant under the particle-hole operation, i.e., the zero eigenvalue $z=0$.
If a zero mode exists, its %
left eigenvector coincides with ${\bm v}^{\rm T} \mC_-^{\dagger}$. %
While exact zero modes typically do not appear in generic matrices, %
the level interaction around the spectral origin should be influenced by PHS. %
Similarly, SLS relates an eigenvalue $z$ to $-z$.
On the other hand, TRS, PHS$^{\dag}$, CS, and pH accompany complex or Hermitian conjugation:
while TRS and pH relate an eigenvalue $z$ to $z^*$, PHS$^{\dagger}$ and CS relate an eigenvalue $z$ to $-z^*$.
In contrast to these symmetries, TRS$^{\dagger}$ does not lead to such pairs of eigenvalues between different eigenvalues but imposes constraints on generic individual eigenvalues.

\begin{table*}[bt]
  \caption{Altland-Zirnbauer$^{\dagger}$ (AZ$^{\dagger}$) classification of non-Hermitian random matrices based on time-reversal symmetry (TRS), time-reversal symmetry$^{\dagger}$ (TRS$^{\dagger}$), and pseudo-Hermiticity (pH).
  If $H$ respects TRS or pH, $\ii H$ respects particle-hole symmetry$^{\dagger}$ (PHS$^{\dagger}$) or chiral symmetry (CS), respectively. 
  If $H$ belongs to the symmetry class in the column ``Class ($H$)'', $\ii H$ belongs to the corresponding class in ``Class ($\ii H$)''.
  In the nomenclature of symmetry classes, $\eta$ represents pH, and the subscript of $\eta_{\pm}$ denotes whether pH commutes ($+$) or anti-commutes ($-$) with TRS$^{\dagger}$. 
  In the columns ``TRS (PHS$^{\dag}$)" and ``TRS$^{\dag}$", the entry ``$ \pm 1$'' specifies the sign of the anti-unitary symmetry, and ``$0$'' means its absence.
  The influence of symmetries on the level statistics around the real axis is summarized~\cite{xiao22}.
  If there is a soft gap in the density of complex eigenvalues, $\rho_c(x+ \ii y)$, around the real axis $y=0$, its small-$|y|$ behavior is listed. 
  The column ``Real eigenvalues" indicates whether random matrices %
  have a finite number of real eigenvalues ($\surd$) or not ($\times$);
  for the five symmetry classes with ``$\surd$", 
  the number $N_{\rm real}$ of real eigenvalues after the ensemble average %
  is subextensive, $N_{\rm real} \propto \sqrt{N}$, with $N$ being the dimension of the matrix. %
  }
  \begin{tabular}{cc|ccc|cc}
  \hline \hline
\begin{tabular}{c} ~~~~Class~~~~  \\  ($H$) \end{tabular} &
\begin{tabular}{c} ~~~~Class~~~~  \\  ($\ii H$) \end{tabular}&
   \begin{tabular}{c} TRS \\ ~~(PHS$^{\dagger}$)~~ \end{tabular} &
   ~~TRS$^{\dagger}$~~
    &\begin{tabular}{c} pH \\ ~~(CS)~~ \end{tabular} & \begin{tabular}{c} Density $\rho_c(x + \ii y)$ around \\ the real axis  \end{tabular} & \begin{tabular}{c} Real eigenvalues  \end{tabular}  \\ \hline
  A & A & $0$ & $0$ & $0$ & - & $\times$ \\
  AI$^{\dagger}$ &AI$^{\dagger}$ & $0$ & $+1$ & $0$ & - & $\times$  \\
  AII$^{\dagger} $ & AII$^{\dagger}$ & $0$ & $-1$ & $0$ & - & $\times$   \\
  \hline
  A + $\eta$ & AIII & $0$ & $0$ & $1$ & $|y|$ &$\surd$  \\ 
  AI &D$^{\dagger}$ & $+1$ & $0$ & $0$ & $|y|$ &$\surd$  \\
  AII & C$^{\dagger}$ & $-1$ & $0$ & $0$ & $|y|^2$ &  $\times$  \\
  AI + $\eta_+$& BDI$^{\dagger}$ & $+1$ & $+1$ & $1$ & ~~$-|y|\ln |y|$~~ &$\surd$  \\
  AI + $\eta_-$& DIII$^{\dagger}$ & $+1$ & $-1$ & $1$  &$|y|$ &$\surd$  \\
  AII + $\eta_+$ &  CII$^{\dagger}$ & $-1$ & $-1$ &  $1$  &$|y|$ & $\surd$  \\
  AII + $\eta_-$ & CI$^{\dagger}$ & $-1$ & $+1$ & $1$ & $|y|$ & $\times$   \\
  \hline\hline
  \end{tabular}
  \label{tab: AZdagger classification}
\end{table*}

Importantly,
the influence of symmetry on the level statistics
is contingent upon the invariance of eigenvalues under the symmetry operation.
This is also the case for Hermitian random matrices, where TRS is relevant to generic real eigenvalues, but PHS and CS (or equivalently, SLS) are relevant only to the zero eigenvalue. 
As also discussed above, TRS$^{\dag}$ constrains generic complex eigenvalues and influences the spectral correlations in the bulk of complex spectra.
On the other hand, the other six types of symmetry affect the level statistics only around the symmetry-preserving lines or points.
For example, TRS transforms a complex eigenvalue $z$ to another complex eigenvalue $z^{*}$, as discussed previously.
As a result, TRS is respected only by eigenvalues satisfying $z = z^{*}$, i.e., real eigenvalues $z \in \mathbb{R}$, and hence changes the spectral statistics only around the real axis.
Similar to TRS, the other three types of symmetry, PHS$^{\dag}$, CS, and pH, are also relevant to the level statistics only around the real or imaginary axis.
By contrast, the remaining two types of symmetry, PHS and SLS, are not necessarily relevant even around the real or imaginary axis and influence the level statistics only around the spectral origin. 

Consistent with this general understanding, 
numerical investigations confirmed that symmetries other than TRS$^{\dagger}$ do not influence the level statistics within the bulk of spectra (i.e., generic eigenvalues away from the spectral origin, as well as real and imaginary axes)~\cite{hamazaki20}.
TRS$^{\dagger}$ alone gives rise to the threefold classification, comprising classes A, AI$^{\dagger}$, and AII$^{\dagger}$ (see the first three rows in Table~\ref{tab: AZdagger classification} or~\ref{tab: AZ0 classification}), and yields the three distinct universality classes of level statistics for generic complex eigenvalues~\cite{hamazaki20}. 
Specially, let $z_{\alpha}$ be a complex eigenvalue, $z_{\alpha; {\rm NN}}$ be its nearest-neighbor eigenvalue (i.e., $| z_{\alpha; {\rm NN}} - z_{\alpha} | = \min_{\beta} |z_{\alpha} - z_{\beta}|$), and $z_{\alpha;{\rm NNN}}$ be its next-nearest-neighbor eigenvalue.
The distributions of the complex level spacing~\cite{ginibre1965statistical, grobe88, grobe89, hamazaki20} $z_{\alpha; {\rm NN}} - z_{\alpha}$ and the complex level-spacing ratio $(z_{\alpha; {\rm NN}} - z_{\alpha})/(z_{\alpha; {\rm NNN}} - z_{\alpha})$~\cite{sa20} were used to characterize the spectral correlations and found to converge to universal level statistics for large random matrices.

Furthermore, TRS, TRS$^{\dagger}$, and pH give rise to the tenfold classification (see Table~\ref{tab: AZdagger classification}).
This tenfold classification is equivalent to the Altland-Zirnbauer$^{\dagger}$ (AZ$^{\dagger}$) classification~\cite{kawabata19} since TRS is equivalent to PHS$^{\dagger}$~\cite{kawabata2019a}.
In this classification, the three symmetry classes (i.e., classes A, AI$^{\dagger}$, and AII$^{\dagger}$) where real eigenvalues do not have higher symmetry than generic ones are characterized only by TRS$^{\dagger}$ and already included in the previous threefold classification. 
Consequently, $10-3 = 7$ distinct universality classes of level statistics on and around the real axis emerge.
It was found that these symmetries determine whether random matrices have real eigenvalues, 
as well as the behavior of the density of complex eigenvalues around the real axis (see a summary in Table~\ref{tab: AZdagger classification})~\cite{ginibre1965statistical, edelman1995, Efetov97, Sommers98, kanzieper2005, xiao22}.
The level spacings or level-spacing ratios of real eigenvalues (if exist) also exhibit universal and characteristic distributions in each symmetry class~\cite{xiao22}. 
The study of level statistics on or around the imaginary axis does not yield new universality classes since the real and imaginary axes are transformed into each other via the multiplication by $\ii$.

All the seven symmetries in Eqs.~(\ref{eq: TRS_def})-(\ref{eq: pH_def}) within the 38-fold classification influence the level statistics around the spectral origin. 
In the tenfold AZ$^{\dag}$ classification discussed above, 
the spectral origin does not possess higher symmetry compared to real eigenvalues.
In the remaining $38-10 = 28$ symmetry classes, by contrast, the spectral origin exhibits higher symmetry and hence is special. 
In this work, we show that the unique level statistics indeed appear around the spectral origin in these 28 symmetry classes.
We categorize these 28 symmetry classes into two groups according to whether they involve symmetries associated with complex or Hermitian conjugation (i.e., TRS, PHS$^{\dagger}$, CS, and pH).

\begin{figure}[tb]
    \includegraphics[width=1\linewidth]{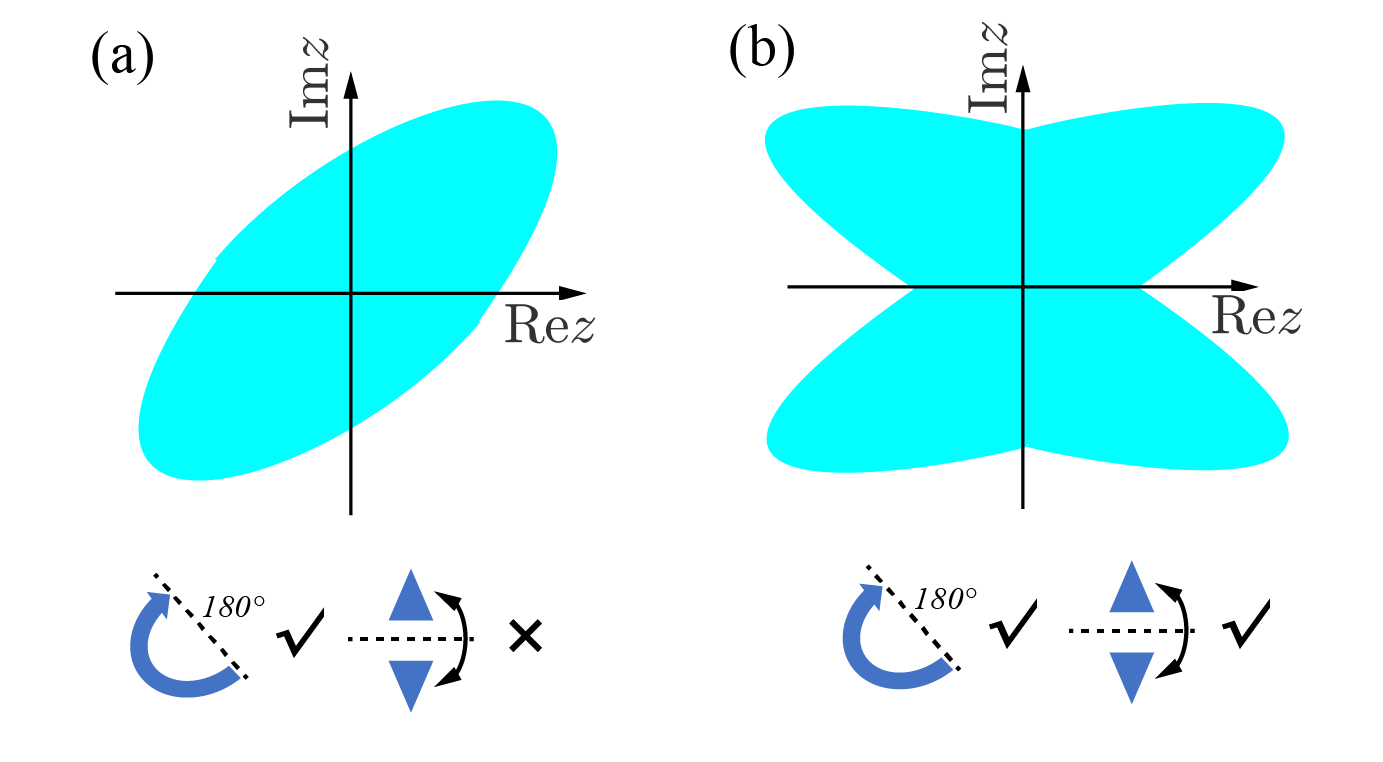}
    \caption{
    Symmetry of complex spectra (denoted by the colored region) for non-Hermitian random matrices. 
    (a)~Seven symmetry classes in the Altland-Zirnbauer$_0$ (AZ$_0$) classification.
    The complex spectrum is invariant under reflection about the spectral origin but not under reflection about the real or imaginary axis, resulting in the point group symmetry C$_2$. 
    (b)~Twenty-one symmetry classes beyond the AZ$_0$ classification.
    The complex spectrum is invariant under 
    reflections about both real and imaginary axes, 
    indicating the point group symmetry D$_2$.}
    \label{fig: D2}
\end{figure}

\begin{figure}[tb]
    \includegraphics[width=0.7\linewidth]{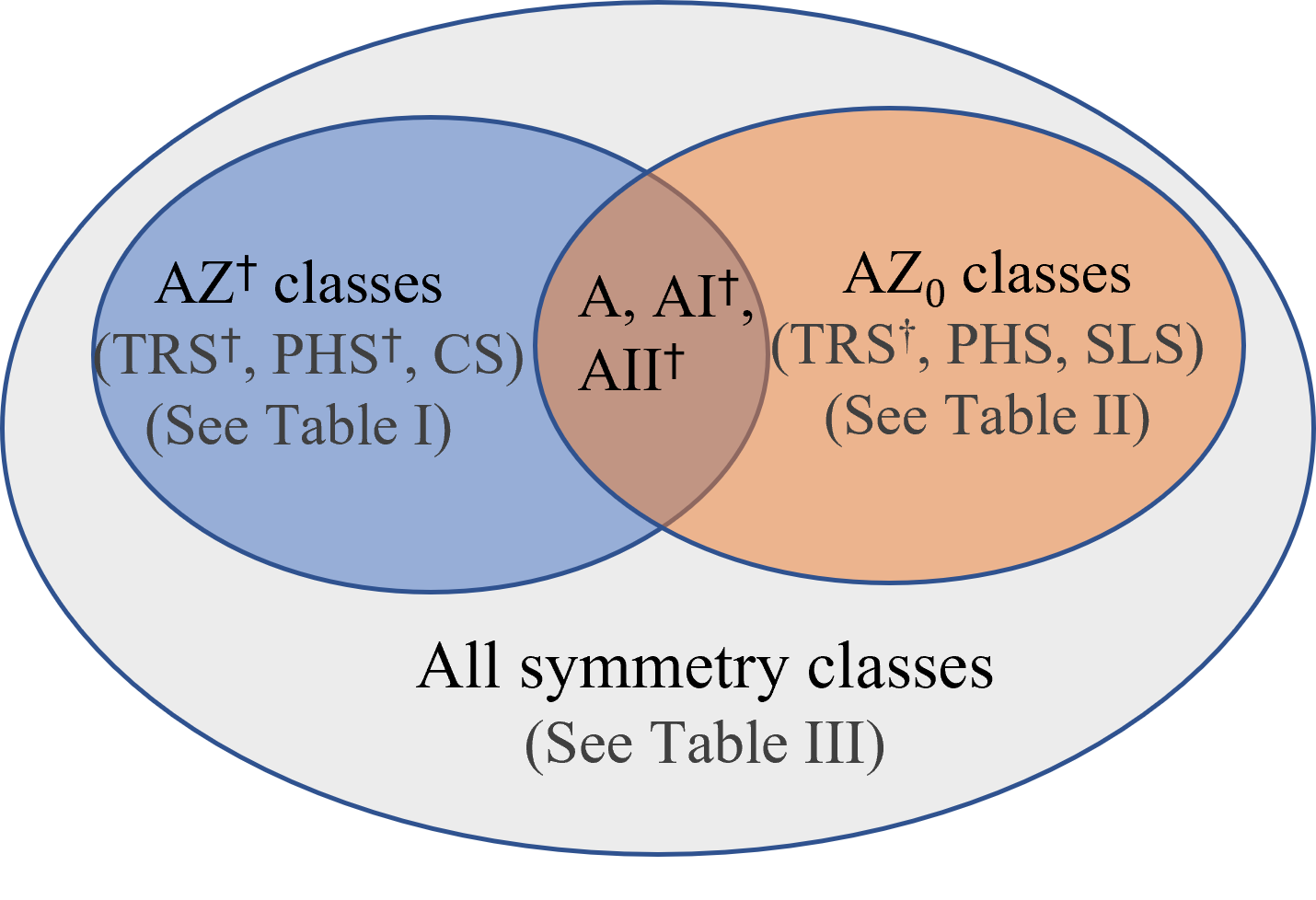}
    \caption{
    Symmetry classification of non-Hermitian random matrices. 
    The largest circle includes all the 38 symmetry classes based on all the seven symmetries [Eqs.~(\ref{eq: TRS_def})-(\ref{eq: pH_def})]. Time-reversal symmetry$^{\dag}$ (TRS$^{\dag}$), particle-hole symmetry$^{\dag}$ (PHS$^{\dag}$), and chiral symmetry (CS) form the 10-fold AZ$^{\dagger}$ classification (blue circle; see Table~\ref{tab: AZdagger classification} for details). 
    TRS$^{\dag}$, particle-hole symmetry (PHS), and sublattice symmetry (SLS) form the 10-fold AZ$_0$ classification (orange circle; see Table~\ref{tab: AZ0 classification} for details). 
    The intersection of the AZ$_0$ and AZ$^{\dagger}$ classification comprises classes A, AI$^{\dagger}$, and AII$^{\dagger}$, only characterized by TRS$^{\dag}$. 
    There are $38-10-10+3 = 21$ symmetry classes beyond the AZ$_0$ and AZ$^{\dagger}$ classification (see Table~\ref{tab: AZ_real} for details). } %
    \label{fig: illustrate classes}
\end{figure}

In the first group, symmetry classes involve TRS$^{\dagger}$, PHS, and SLS, but do not involve symmetries associated with complex or Hermitian conjugation.
TRS$^{\dagger}$, PHS, and SLS lead to the tenfold classification in Table~\ref{tab: AZ0 classification} 
different from the AZ$^{\dag}$ classification in Table~\ref{tab: AZdagger classification}.
Here, we call this additional tenfold classification Altland-Zirnbauer$_0$ (AZ$_0$) classification since it is concerned with the spectral origin. 
Besides the three symmetry classes (i.e., classes A, AI$^{\dagger}$, and AII$^{\dagger}$) that only involve TRS$^{\dagger}$ and are counted before, the seven symmetry classes (i.e., classes \AIIIz, \BDIz, \CIIz, D, C, \CIz, and \DIIIz) are in the first group.
Owing to PHS, SLS, or both of them, which maps an eigenvalue $z$ to another eigenvalue $-z$, the complex spectra of generic matrices in any of these symmetry classes
are invariant under the reflection about the spectral origin and hence respect the point group symmetry C$_2$ [see a schematic in Fig.~\ref{fig: D2}\,(a)].
Away from the origin, the spectral statistics depend solely on TRS$^{\dag}$ since PHS and SLS merely create opposite-sign pairs of eigenvalues, $\left( z, -z \right)$.
This is also the case for non-zero real or purely imaginary eigenvalues.
Around the origin, by contrast, even PHS and SLS change the spectral correlations and lead to unique spectral statistics, as shown below.
If a matrix $H$ belongs to one of these seven symmetry classes, $e^{\ii \phi} H$ also belongs to the same symmetry class for arbitrary $\phi \in \mathbb{R}$ (see Table~\ref{tab: AZ0 classification}).

The remaining $28 - 7 = 21$ symmetry classes are in the second group and involve at least one of TRS, PHS$^{\dagger}$, CS, and pH, as summarized in Table~\ref{tab: AZ_real}. 
These symmetries accompany complex or Hermitian conjugation and hence map an eigenvalue $z$ to $z^{*}$ or $-z^*$. 
In these classes, SLS, PHS, or both of them 
also exist.
Consequently, the real or imaginary axis exhibits higher symmetry than generic points, and the spectral origin exhibits even higher symmetry.
Owing to the combination of these symmetries, $H$ does not necessarily belong to the same symmetry class as $e^{\ii \phi} H$ for generic $\phi \in \mathbb{R}$, in contrast to the AZ$_0$ symmetry classification.
Rather, complex spectra of generic non-Hermitian matrices in these 21 symmetry classes are invariant under the reflections about the real and imaginary axes, and hence respect the point group symmetry D$_2$ [see a schematic in Fig.~\ref{fig: D2}\,(b)]. 
Figure~\ref{fig: illustrate classes} summarizes the aforementioned subgroups of symmetry classes and illustrates their relationship.

Below, we comprehensively study the spectral statistics around the origin---hard-edge statistics---of non-Hermitian random matrices in all the AZ$_0$ symmetry classes (Sec.~\ref{sec: RM AZ0}), 
as well as the other 21 symmetry classes with a particular focus on three typical classes (classes BDI, CI, and \AIIp) (Sec.~\ref{sec: RM AZ_real}).
Through these analyses, we expand %
the classification of the universal spectral statistics of non-Hermitian random matrices %
to encompass 38-fold symmetry classes.

\begin{table*}[th]
    \centering
    \caption{Altland-Zirnbauer$_0$ (AZ$_0$) classification of non-Hermitian random matrices based on time-reversal symmetry$^{\dagger}$ (TRS$^{\dagger}$), particle-hole symmetry (PHS), and sublattice symmetry (SLS). 
    The column ``Equivalent class" shows the equivalent symmetry classes in Ref.~\cite{kawabata19}.
    In the columns ``TRS$^{\dag}$" and ``PHS", the entry ``$ \pm 1$'' specifies the sign of the anti-unitary symmetry, and ``$0$'' means its absence. 
    The density $\rho(|z|)$ of the modulus of eigenvalues and the level-ratio distribution $p_r(r)$ [Eq.~(\ref{eq: ratio})] share the same small-argument behaviors, which are summarized in the columns ``$\rho(|z|)$ and $p_r(|z|)$''. 
    The mean values $\langle r \rangle$ and $\langle \cos \theta \rangle$ of random matrices are summarized, where $r e^{\ii \theta}$ is the level ratio around the spectral origin. 
    The statistics are obtained by diagonalizing $10^7$ realizations of $128\times 128$ random matrices. 
    The numbers in the parentheses denote standard deviations.
    }
    \begin{tabular}{cc|ccc|ccc}
    \hline \hline
    Class & Equivalent class & ~~TRS$^{\dagger}$~~ & ~~PHS~~ & ~~SLS~~ &~~$\rho(|z|)$ and $p_r(|z|)$~~&~~ ~~$\left\langle   r \right\rangle$ ~~ ~~& ~~ ~~ $\left\langle   \cos\theta \right\rangle$ ~~ ~~\\ \hline
    A & - & $0$ & $0$ & $0$   & - & - & - \\
    AI$^{\dagger}$ & - & $+1$ & $0$ & $0$ & - & - & -  \\
    AII$^{\dagger}$ & - & $-1$ & $0$ & $0$   & - & - & - \\
    \hline
    \AIIIz & A + $\mS$ & $0$ & $0$ & $1$ & $-|z|^3 \ln |z|$ & 0.6357(5) & 0.5391(7) \\ 
    \BDIz & ~D + $\mS_+$, AI$^{\dag}$ + $\mS_+$~ & $+1$ & $+1$ & $1$ & $|z|$  & 0.5778(6) & 0.5681(7) \\ 
    \CIIz & ~C + $\mS_+$, AII$^{\dag}$ + $\mS_+$~ &  $-1$ & $-1$ & $1$ & $|z|^3$ & 0.6623(5) & 0.5147(7) \\ 
     D & - & $0$ & $+1$ & $0$ & $|z|$  & 0.5411(6) & 0.5524(7) \\ 
     C & - & $0$ & $-1$ & $0$  & $|z|^3$ & 0.6746(5) & 0.5343(7) \\ 
    \CIz & C + $\mS_-$, AI$^{\dag}$ + $\mS_-$ & $+1$ & $-1$ & $1$ & $-|z|^3 \ln |z|$  & 0.6708(5) & 0.5589(7) \\ 
    \DIIIz & D + $\mS_-$, AII$^{\dag}$ + $\mS_-$ & $-1$ & $+1$ & $1$ & $-|z|^3 \ln |z|$  & 0.5950(6) & 0.5252(7) \\ 
    \hline \hline
    \end{tabular}
    \label{tab: AZ0 classification}
  \end{table*}

\begin{figure*}[t]
    \includegraphics[width=1\linewidth]{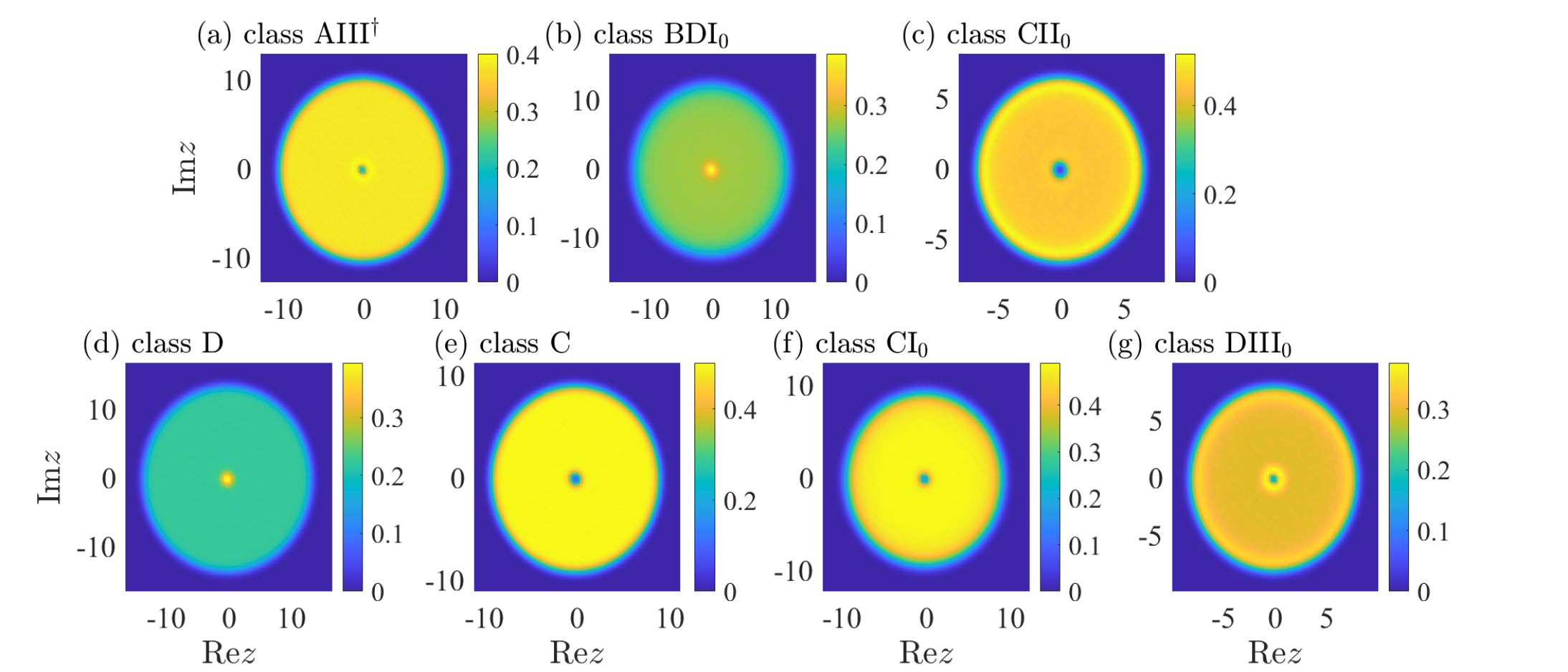}
    \caption{Density of complex eigenvalues of non-Hermitian random matrices in the Gaussian ensemble within the AZ$_0$ classification. 
    While the density is nearly constant away from the spectral origin, it deviates from this constant around the origin, manifesting distinct spectral correlations around the origin due to higher symmetry.
    The behavior of the density, as well as the concomitant spectral correlations,  around the origin varies among different symmetry classes. %
    In each symmetry class, the result is obtained by $10^7$ samples of $128 \times 128$ random matrices. 
    In classes (c)~\CIIz\ and (g)~\DIIIz, all eigenvalues are twofold degenerate, 
    and we only count the degenerate pairs once when calculating the density.}
    \label{figs: AZ0_RM_DoS_2D}
  \end{figure*}

\section{Non-Hermitian random matrices in the AZ$_0$ classification}
\label{sec: RM AZ0}

\begin{figure}[t]
    \includegraphics[width=1\linewidth]{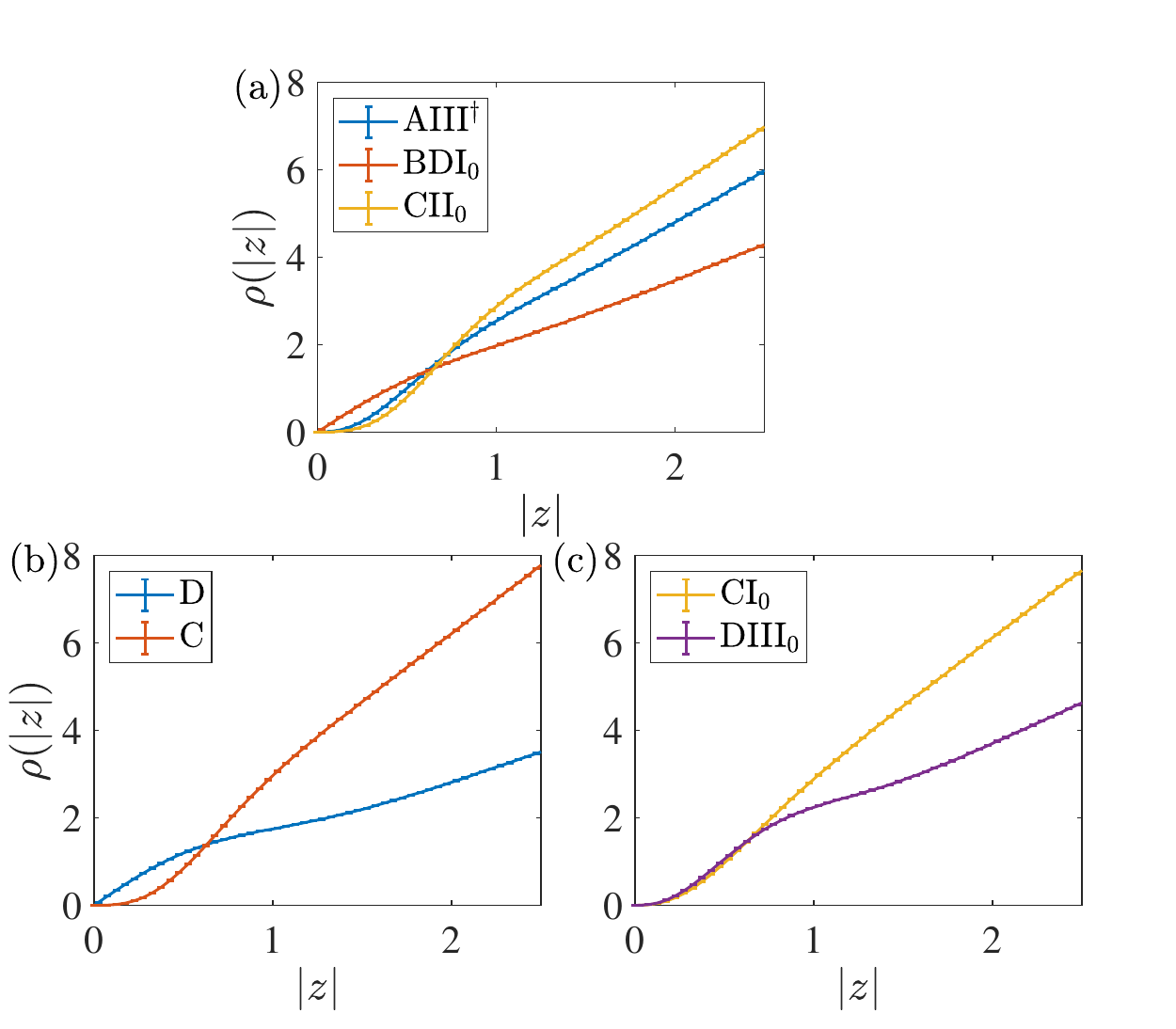}
    \caption{Density $\rho(|z|)$ of the modulus of complex eigenvalues of non-Hermitian random matrices in the Gaussian ensemble within the AZ$_0$ classification.
    (a)~%
    For $|z| \ll 1$, 
    $\rho(|z|) \approx -11.3 |z|^3 \ln |z|$ (class \AIIIz), 
    $\rho(|z|) \approx 2.64 |z|$ (class \BDIz), and
    $\rho(|z|) \approx 8.80 |z|^3$ (class \CIIz). %
    (b)~%
    For $|z| \ll 1$, 
    $\rho(|z|) \approx 2.80 |z|$ (class D) and 
    $\rho(|z|) \approx 10.6 |z|^3$ (class C). %
    (c)~%
    For $|z| \ll 1$,
    $\rho(|z|) \approx -9.40 |z|^3 \ln |z|$ (class \CIz) and 
    $\rho(|z|) \approx -11.2 |z|^3 \ln |z|$ (class \DIIIz)~\cite{footnote1}. %
    }
    \label{figs: AZ0_RM_DoS_modulus}
\end{figure} 

\begin{figure*}[bht]
    \centering
    \includegraphics[width=0.8\linewidth]{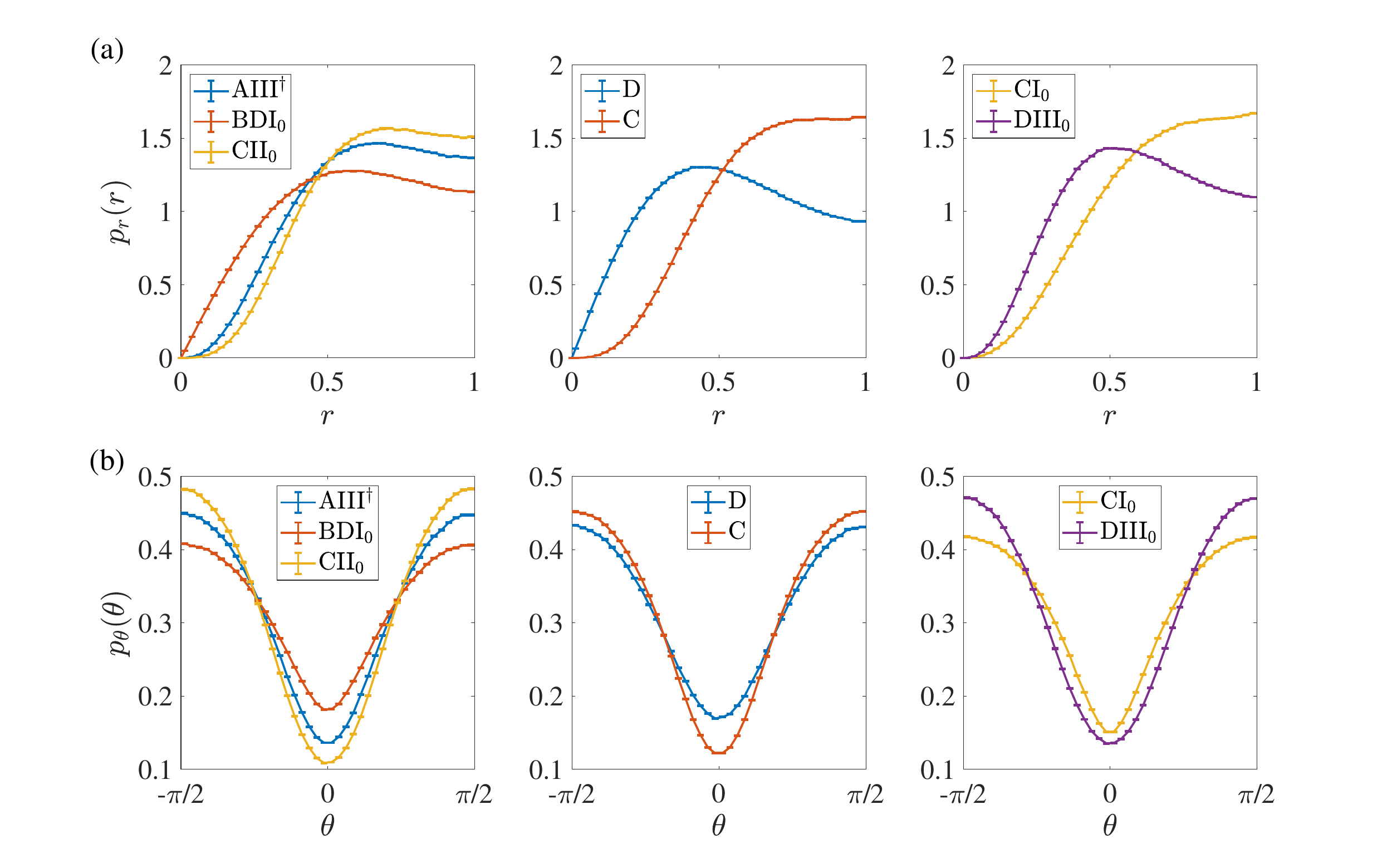}
  \caption{(a)~Radial distributions $p_r(r)$ and (b)~angular distributions $p_{\theta}(\theta)$ of the complex level ratio $r e^{\ii \theta}$ [Eq.~(\ref{eq: ratio})] around the spectral origin for the seven symmetry classes in the AZ$_0$ classification. 
   The error bars represent one standard deviation evaluated by the bootstrap method~\cite{press07}.}
  \label{figs: pdf_r_theta_DoS_AZ0}
\end{figure*} 

\begin{figure}[ht]
    \includegraphics[width=1\linewidth]{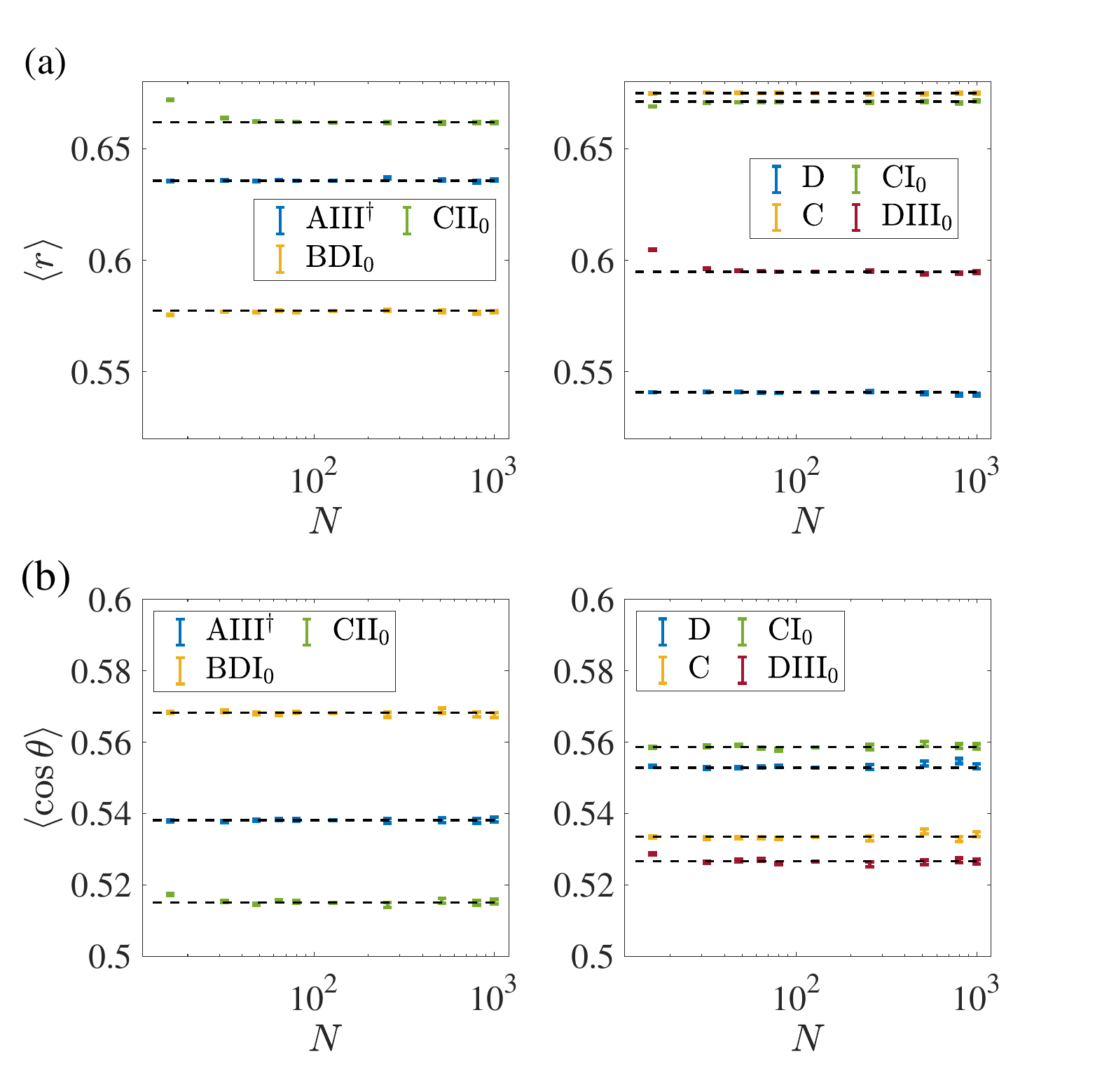}
    \caption{Mean values (a) $\left\langle r \right\rangle$ and (b) $\left\langle \cos\theta \right\rangle$ [Eq.~(\ref{eq: ratio})] as functions of the matrix size $N$. 
    For each symmetry class and each matrix size, 
    the number of diagoanlized samples is $N_{\rm sample} = 10^7,10^6,2\times 10^5$ for $N = 128$, $N < 128$, $N > 128$, respectively. 
    The error bars represent one standard deviation. 
    The dashed lines denote the mean values obtained by matrices with $N = 128$. 
    }
    \label{figs: mean_r_cos_AZ0}
\end{figure} 

We consider non-Hermitian random matrices in the seven symmetry classes within the AZ$_0$ classification (Table~\ref{tab: AZ0 classification}).
Specifically, we take the Gaussian ensemble with the probability density function 
\begin{equation}
    p(H) \propto e^{-\beta\,{\rm Tr}\,(H^{\dagger}H)}
    \label{eq: Gaussian ensemble}
\end{equation}
with a positive constant $\beta > 0$. 
We follow the terminology in the AZ classification: 
classes \AIIIz, \BDIz, and \CIIz\ are categorized as chiral symmetry classes, and  
classes D, C, \CIz, and \DIIIz\ are categorized as BdG classes.
The explicit forms of non-Hermitian random matrices in each symmetry class are enumerated in Appendix~\ref{app: form RM}.

Similar to Hermitian random matrix theory~\cite{beenakkerRMT15}, the total number $\nu d_E$ of topologically protected exact zero modes can manifest themselves and influence the level statistics around the origin in some symmetry classes.
Here, $d_E$ refers to the degrees of degeneracy of generic eigenvalues: $d_E = 2$ 
in the presence of TRS$^{\dag}$ with the negative sign $-1$
(i.e., classes \AIIIz, \CIIz, and \DIIIz)
and $d_E = 1$ otherwise.
Specifically, a non-Hermitian random matrix $H$ in class D ($H^{\rm T} = - H$) has $\nu d_E = 1$ zero mode if the matrix dimension of $H$ is odd. 
A non-Hermitian random matrix $H$ in class \DIIIz\ ($H^{\rm T} = - H$ and $\sigma_y H^{\rm T} \sigma_y = H$) has $\nu d_E = 2$ zero modes if the matrix dimension of $H$ is $4n + 2$ ($n \in \mathbb{Z}$). 
For non-Hermitian random matrices in classes \AIIIz, \BDIz, and \CIIz, on the other hand, $\nu d_E$ equals the difference between the numbers of positive and negative eigenvalues of the symmetry operator $\mS$.
While random matrices having such zero modes can be 
physically relevant, 
for example,
to non-Hermitian superconductors with vortices~\cite{brezin1999, ivanov2002}, 
dissipative SYK-type models~\cite{garcia2023topo},
and QCD with topological charges~\cite{osborn2004}, we focus on random matrices with $\nu d_E = 0$ in this work.

We numerically diagonalize $10^7$ samples of $N \times N$ ($N = 128$) non-Hermitian random matrices in each symmetry class.
Let $z_1$ be the eigenvalue with the smallest modulus. 
We normalize eigenvalues such that $\langle |z_1| \rangle = 1$, where $\langle \cdot \rangle$ denotes the average over the disorder ensemble. 
After the ensemble average, the density of complex eigenvalues does not depend on the phases of eigenvalues but only on their modulus (Fig.~\ref{figs: AZ0_RM_DoS_2D}).
It is rotationally invariant with respect to the origin and exhibits
spectral U(1) rotation symmetry.
This spectral U(1) symmetry arises because the Gaussian ensemble includes $e^{\ii \phi} H$ with the same probability for arbitrary $\phi \in \mathbb{R}$.
The eigenvalues are distributed almost uniformly in a circle except around the origin and boundary.
In classes \BDIz\ and D, the density of eigenvalues around the origin is non-zero and higher than that in the bulk, revealing the absence of level repulsion at the spectral origin.
On the other hand, in classes \AIIIz, \CIIz, C, \CIz, and \DIIIz, the density vanishes at the origin. 

Notably, the small-$|z|$ behavior of the density of the modulus of eigenvalues $z_i$,
\begin{equation}
\rho \left( |z| \right) \equiv \sum_i \left\langle \delta \left( |z| - |z_i| \right) \right\rangle,
\end{equation} 
characterizes the different strength of %
the %
level repulsion in different symmetry classes (Fig.~\ref{figs: AZ0_RM_DoS_modulus}).
Among the three chiral symmetry classes, the descending order of $\rho(|z|)$ for $|z| \ll 1$ is classes \BDIz, \AIIIz, and~\CIIz\ [see Fig.~\ref{figs: AZ0_RM_DoS_modulus} (a)].
Among the four BdG symmetry classes, for $|z| \ll 1$, $\rho(z)$ in class D is the largest, that in class C is the smallest, and those in classes \CIz\ and \DIIIz\ are close to each other [see Figs.~\ref{figs: AZ0_RM_DoS_modulus} (b) and (c)].
Note that the density $\rho(|z|)$ of the modulus differs from the density of eigenvalues in the complex plane by a Jacobian, and hence we have $\rho(|z| = 0) = 0$ even in 
the absence of the level repulsion for 
classes D and \BDIz.
Fitting $\rho(|z|)$, we find the asymptotic behavior of $\rho(|z|)$ for $|z| \ll 1$ (also summarized in Table~\ref{tab: AZ0 classification}),
\begin{equation} \label{eq: rho_z_small}
    \rho(|z|)  \propto \begin{cases}
        |z| & (\text{classes D and BDI}_0)  \, ; \\
        -|z|^3 \ln |z| & (\text{classes AIII}^{\dagger} \text{, CI}_0 \text{, and DIII}_0) \, ; \\
        |z|^3 & (\text{classes CII}_0 \text{ and C}) \, .
    \end{cases} 
\end{equation} 
The coefficients of this proportional relationship are given in the caption of Fig.~\ref{figs: AZ0_RM_DoS_modulus}. %
As %
noted above, 
the linear decay of $\rho(|z|)$ for classes D and \BDIz~does not originate from the level repulsion but solely from the integral measure along the angular direction.
The higher density of eigenvalues, given as $\rho(|z|)/\pi|z|$, around the spectral origin compared to that in the bulk is a characteristic feature of these two symmetry classes. %
By contrast, the cubic decay (with a potential logarithmic correction) for the other symmetry classes indicates the level repulsion around the spectral origin.
In a similar spirit to the Wigner surmise,
the qualitative behavior of level statistics for large random matrices can be estimated from those of small matrices.
In Appendix~\ref{app: analytic}, we analytically calculate $\rho(|z|)$ of small random matrices, exhibiting the identical small-$|z|$ behavior observed in large ones. 

To further characterize the level correlations around the spectral origin, we introduce a new quantitative measure of the hard-edge statistics for non-Hermitian operators. 
Owing to PHS, SLS, or both of them, complex eigenvalues must appear in the opposite-sign pairs $\left( +z, -z \right)$.
Let $\pm z_1, \pm z_2, \pm z_3, \ldots$ be eigenvalues (or Kramers pairs of eigenvalues) of $H$ in the ascending order of the modulus 
(i.e., $\left| z_1 \right| \leq \left| z_2 \right| \leq \left| z_3 \right| \leq \cdots$). 
We introduce the complex level ratio by
\begin{equation}  \label{eq: ratio}
    r e^{\ii \theta} \equiv \pm \frac{z_1}{z_2} \qquad (r, \theta \in \mathbb{R}). 
\end{equation}
We characterize the level-ratio statistics by the distribution 
$p_{r}(r)$ of the module $r$ of the ratio 
and the distribution $p_{\theta}(\theta)$ of its angle (phase) $\theta$,  
\begin{align}
    p_{r}(r) &\equiv \left\langle \delta \left( r - \left| \frac{z_1}{z_2} \right| \right) \right\rangle, \\ 
    p_{\theta}(\theta) &\equiv 
\left\langle \delta \left(\theta - {\rm arg} \left( \frac{z_1}{z_2} \right) \right) \right\rangle,
\end{align}
where we have $r \in [0,1]$ and $\theta \in [-\pi/2,\pi/2)$ by definition.  
When the levels are uncorrelated, $z_1$ is uniformly distributed inside 
the circle $|z_1| \leq |z_2|$, resulting in 
\begin{equation} \label{eq: 2D Poisson}
p_r(r) = 2r,\quad p_{\theta}(\theta) = \frac{1}{\pi}.
\end{equation}
In Sec.~\ref{subsec: integrable model}, we show that 
such level-ratio statistics indeed appear in an integrable open quantum system.

The complex level ratio is the ratio between two complex eigenvalues, whereas 
a complex spacing ratio previously used 
for studies of 
spectral correlations in the bulk spectra~\cite{Oganesyan07,atas13,sa20} 
is a ratio between two spacings of eigenvalues. 
For Hermitian operators, one can arrange real eigenvalues 
$\lambda_i$'s in ascending order and introduce the spacing ratio as 
$ \min(s_i,s_{i-1})/ \max(s_i,s_{i-1})$ with $s_i \equiv \lambda_{i+1} - \lambda_{i}$~\cite{Oganesyan07,atas13}. 
For non-Hermitian operators, %
one can 
select 
a complex eigenvalue $z_{\alpha}$, find %
its nearest-neighboring eigenvalue $z_{\alpha; {\rm NN}}$ and next-nearest-neighboring eigenvalue $z_{\alpha; {\rm NNN}}$ in the complex plane, and introduce the complex spacing ratio as $(z_{\alpha; {\rm NN}} - z_{\alpha})/(z_{\alpha; {\rm NNN}} - z_{\alpha})$~\cite{sa20}.  %
The complex level ratio %
and the complex spacing ratio share the same aim---to utilize quantities independent of the normalization of the random-matrix ensembles.
Since the former involves two eigenvalues while the latter involves three, %
the two measures quantify the different types of spectral correlations.

For the random-matrix ensemble, 
both the radial distribution $p_r(r)$ 
and the angular (phase) distribution $p_{\theta}(\theta)$ 
take non-trivial forms owing to the level correlations between the level pairs $(+z_1, -z_1)$ and $(+z_2, -z_2)$. 
From the numerical diagonalization of $128 \times 128$ random matrices, we obtain the radial distribution $p_r(r)$ and angular distribution $p_{\theta}(\theta)$ of the complex level ratio $r e^{\ii \theta}$ (Fig.~\ref{figs: pdf_r_theta_DoS_AZ0}). 
In each symmetry class, the distributions $p_r(r)$ and $p_{\theta}(\theta)$ show characteristic behavior, significantly distinct from that of uncorrelated levels. 
The small-$r$ behavior of $p_r(r)$ is similar to the small-$|z|$ behavior of the density $\rho(|z|)$ of the modulus of eigenvalues in the same symmetry class. 
For example, among the three chiral symmetry classes, the descending order of $p_r(r)$ for $r \ll 1$ is the same as that of $\rho(|z|)$ for $|z| \ll 1$ [compare Fig.~\ref{figs: AZ0_RM_DoS_modulus}\,(a) with Fig.~\ref{figs: pdf_r_theta_DoS_AZ0}\,(a)].
Fitting $p_r(r)$ with $r \ll 1$, we find that $p_r(r)$ shows the same power-law decay as $\rho(|z|)$ in Eq.~(\ref{eq: rho_z_small}) and Table~\ref{tab: AZ0 classification}.
The density $\rho(|z|)$ should be mostly contributed by $z_1$ for $|z_1| \ll 1$, leading to $r \propto |z_1|$.
Consequently, the small-argument behaviors of both $\rho(|z|)$ and $p_r(r)$ are mainly determined by the level correlations between the eigenvalue pair 
$\left( +z_1, -z_1 \right)$ closest to the spectral origin,
which underlies the similarity between $\rho(|z|)$ and $p_r(r)$.
Moreover, the angular distributions $p_{\theta}(\theta)$ are unique in different symmetry classes and characterize the different universality classes
[Fig.~\ref{figs: pdf_r_theta_DoS_AZ0}\,(b)]. 
First, we have $p_{\theta}(\theta) = p_{\theta}(-\theta)$ since the matrix ensembles are invariant under complex conjugation, 
transforming
$\theta$ to $-\theta$.
Second, $p_{\theta}(\theta)$ reaches its minimum and maximum at $\theta = 0$ and $\pi/2$, respectively.
The pairs
$\left( +z_1, -z_1 \right)$ and $\left( +z_2, -z_2 \right)$
tend to be oriented perpendicularly to each other rather than in parallel, revealing the presence of level correlations. 
We also obtain the analytic formulas of $p_r^{N = 4}(r)$ and $p_{\theta}^{N = 4}(\theta)$ for $4\times 4$ non-Hermitian random matrices in classes \AIIIz\ and D, which show behavior consistent with large ones (see Appendix~\ref{app: analytic}).

The mean values $\langle r \rangle$ and $\langle \cos\theta \rangle$ of the complex level ratio $re^{\ii \theta}$ serve as useful quantitative indicators 
of the hard-edge statistics among the different universality classes.
These values in each symmetry class are summarized in Table~\ref{tab: AZ0 classification}. 
Notably, the distribution of $r$ 
in each symmetry class should converge to a universal form for sufficiently large matrix sizes. 
To clarify this convergence, we utilize $\left\langle r \right\rangle$ and $\left\langle \cos\theta \right\rangle$ as functions of the matrix size $N$ 
and find that they 
already
reach the convergence for $N \geq 128$ (see Fig.~\ref{figs: mean_r_cos_AZ0}).
A similar rapid convergence of the distribution of $\left| z_1 \right|$ was also reported in Ref.~\cite{garcia22}.
In Sec~\ref{sec: quantum system}, we also compare the random-matrix behavior of the hard-edge statistics with the distributions of physical models, which further provides evidence of the universality.

In the literature,
the distributions of the eigenvalue $z_1$ with the smallest modulus 
were mainly used
to characterize the hard-edge statistics 
(see Ref.~\cite{sun20} for the Hermitian case and Ref.~\cite{garcia22} for the non-Hermitian case, as well as the references therein).
Our newly proposed distributions of the complex level ratios in Eq.~(\ref{eq: ratio}) differ from those of $z_1$ in several aspects.
The distribution of $r e^{\ii \theta}$ does not only depend on the modulus $r$ but also on the phase $\theta$ and hence is more informative than that of $z_1$.
Studying $p_r(r)$ and $p_{\theta}(\theta)$ does not require normalization while studying the distribution of $z_1$ requires normalization, for example, by the mean value of $|z_1|$.
Furthermore,
the distribution of 
$re^{\ii \theta}$
can be more effective in distinguishing between integrable and non-integrable models,
as demonstrated in Sec.~\ref{subsec: integrable model}.

\section{Non-Hermitian random matrices beyond the AZ$_0$ and  AZ$^{\dagger}$ classification}
\label{sec: RM AZ_real}

We consider non-Hermitian random matrices in the 21 symmetry classes beyond the AZ$_0$ and AZ$^{\dagger}$ classification.
Similar to the previous analysis, we take the Gaussian ensemble in Eq.~(\ref{eq: Gaussian ensemble}).
The relevant symmetries of these classes are summarized in Table~\ref{tab: AZ_real}.
We first study classes BDI, CI, and \AIIp~as representative classes (see Sec.~\ref{sec: RM AZ_real - BDI, CI, AIIp}), and later discuss all the other classes in a similar manner (see Sec.~\ref{subsec: AZ all general} and Appendix~\ref{app: sec alternative}).
In Appendix~\ref{app: sec alternative}, we provide additional characterization of the hard-edge statistics for these general symmetry classes.

\subsection{Classes BDI, CI, and \AIIp}
    \label{sec: RM AZ_real - BDI, CI, AIIp}

\begin{figure*}[th]
  \centering
  \includegraphics[width=1\linewidth]{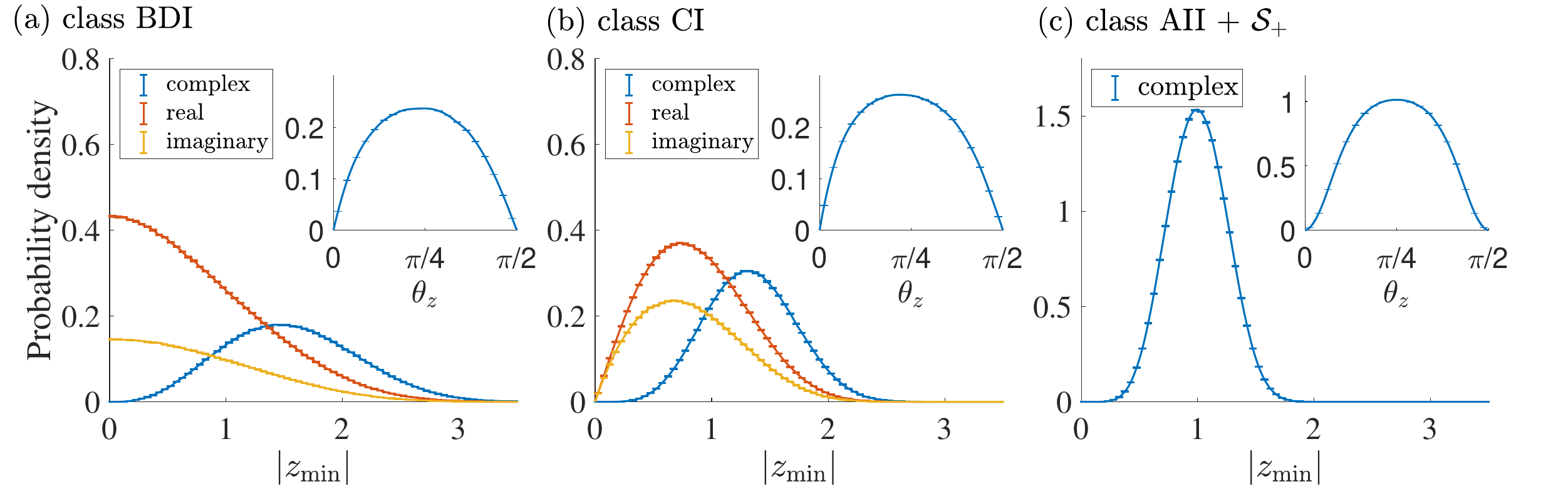}
  \includegraphics[width=1\linewidth]{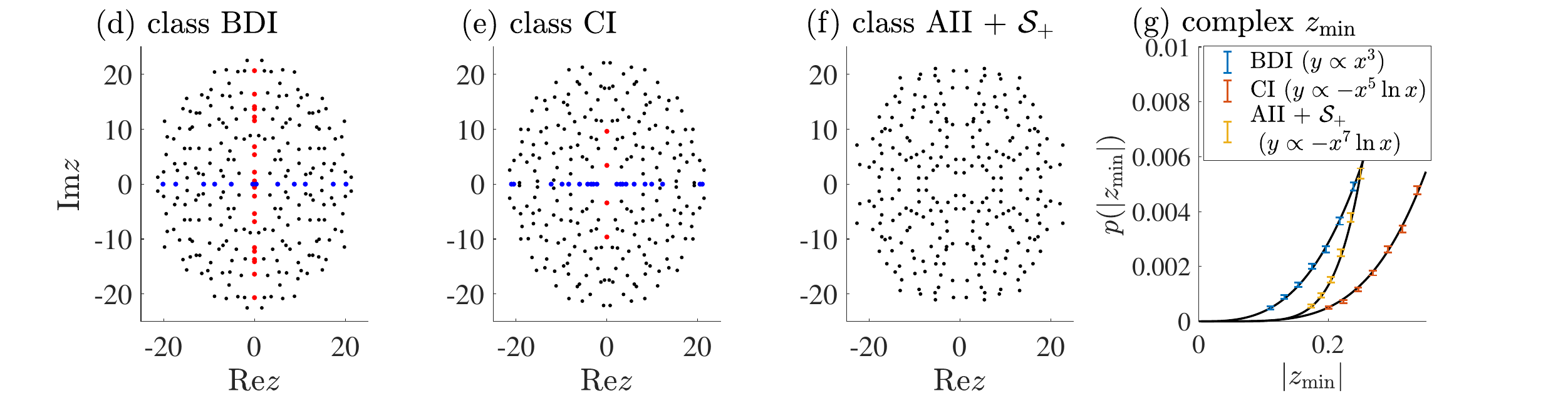}
\caption{(a)-(c)~Distributions of the eigenvalue $z_{\min}$ with the smallest modulus of non-Hermitian random matrices in classes BDI, CI, and \AIIp.
The blue, orange, and yellow curves represent the radial distributions of $z_{\min}$ when it is complex, real, and purely imaginary, respectively.
Inset:~angular distribution of complex $z_{\min}$.
The distributions are obtained by diagonalizing $10^7$ samples of $256 \times 256$ random matrices in each symmetry class.
(d)-(f)~Eigenvalues of a single realization of a $256 \times 256 $ random matrix in each symmetry class. 
Complex, real, and purely imaginary eigenvalues are denoted by the black, blue, and red points, respectively.
(g)~Small-$|z_{\min}|$ behavior of the radial distributions of complex $z_{\min}$ in different symmetry classes, with the solid curves being fitting curves.}
    \label{figs: pdf_z_min_AZ_real}
\end{figure*}

Non-Hermitian random matrices in class BDI (CI) respect TRS with sign $+1$ and CS commuting (anti-commuting) with TRS.
On the other hand, non-Hermitian random matrices in class \AIIp\ respect TRS with sign $-1$ and SLS commuting with TRS.
See Table~\ref{tab: AZ_real}, as well as Appendix~\ref{app: form RM} for explicit forms of non-Hermitian random matrices in each symmetry class. 
We numerically diagonalize
$10^7$ samples of $N \times N$ ($N = 256$) non-Hermitian random matrices in each of the three symmetry classes.
In these symmetry classes, the combination of symmetries leads to quartets $\left( z, -z, z^{*}, - z^{*} \right)$ of complex eigenvalues, and hence the spectrum exhibits %
D$_2$ symmetry [see Fig.~\ref{fig: D2}\,(b)].
In all the three symmetry classes, the eigenvalues are distributed almost uniformly in a circle in the complex plane, except around its origin, its circumference, and the real and imaginary axes.
This observation is consistent with the symmetries of these random matrices since the spectral origin, as well as the real and imaginary axes, possesses higher symmetries than generic points in the bulk of the spectrum.
In classes BDI and CI, a finite number $N_{\rm real/imag}$ of eigenvalues are real or purely imaginary.
Here, $N_{\rm {real/imag}}$ scales as $\sqrt{N}$ after the ensemble averaging (not shown here), exhibiting subextensive behavior similar to that of random matrices in the AZ$^{\dag}$ classification (Table~\ref{tab: AZdagger classification}). %
In class AII + $\mathcal{S}_+$, by contrast, all eigenvalues are complex [Figs.~\ref{figs: pdf_z_min_AZ_real}\,(d)-(f)]. 
We find that the presence of real or purely imaginary eigenvalues is determined by the relevant symmetries on the real and imaginary axes.

\begin{figure}[th]
  \includegraphics[width=1\linewidth]{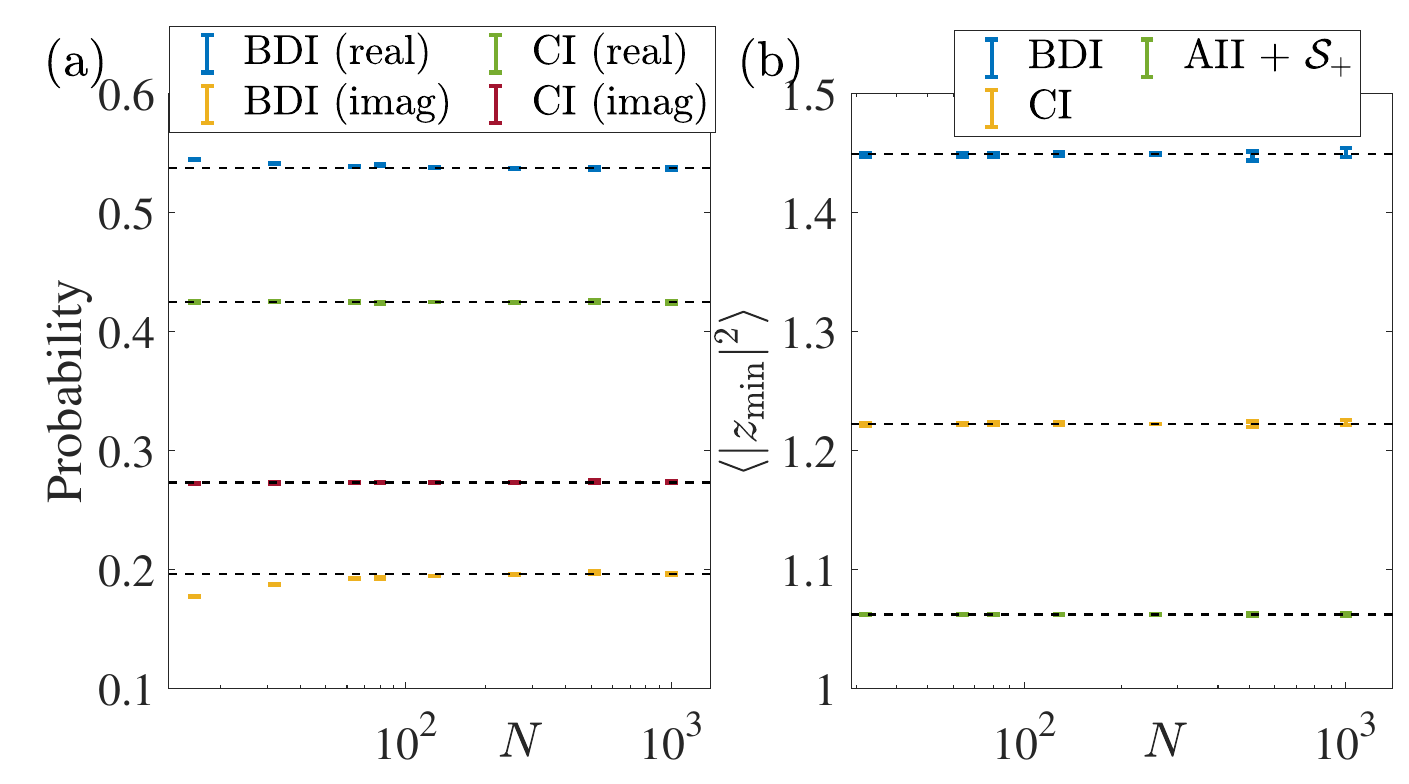}
\caption{(a)~Probability of the eigenvalue $z_{\rm min}$ with the smallest modulus being real or purely imaginary as functions of the matrix size $N$ in different symmetry classes. 
In class AII + $\mathcal{S}_+$, $z_{\rm min}$ is always complex.  
(b)~$\langle |z_{\rm min}|^2  \rangle$ as functions of the matrix size $N$ in different symmetry classes. 
The statistics are obtained by diagonalization of $N_{\rm sample}$ matrices for each class and size with $N_{\rm sample} = 10^7,10^6,2\times 10^5$ for $N = 256$, $N<256$, $N > 256$, respectively. 
The error bars represent one standard deviation. 
The dashed lines denote the mean values obtained by matrices with $N = 256$.}
\label{figs: p_real_AZ_real}
\end{figure} 

\begin{table*}[bt]
	\caption{Twenty-one symmetry classes of non-Hermitian random matrices beyond the AZ$^{\dagger}$ and AZ$_0$ classification. 
   If $H$ belongs to the symmetry class in the column ``Class ($H$)'', $\ii H$ belongs to the corresponding class in ``Class ($\ii H$)''.
  If $H$ and $\ii H$ belong to the same symmetry class, the class is in bold.
  In the nomenclature of symmetry classes, $\mS$ represents sublattice symmetry (SLS) and $\eta$ represents pseudo-Hermiticity (pH). 
  The subscript of $\mS_{\pm}$ or $\eta_{\pm}$ denotes whether SLS or pH commutes ($+$) or anti-commutes ($-$) with time-reveal symmetry (TRS), particle-hole symmetry (PHS), or chiral symmetry (CS).
  The first and second subscripts of $\mS_{\pm\pm}$ denote the commutation ($+$) or anti-commutation ($-$) relations between SLS and TRS, and between SLS and PHS, respectively.
  In the columns ``TRS", ``TRS$^{\dag}$", ``PHS", and ``PHS$^{\dag}$", the entry ``$ \pm 1$'' specifies the sign of the anti-unitary symmetry, and ``$0$'' means its absence. 
  The column ``$[\Gamma,\mathcal{S}]_{\pm}$=0'' specifies whether CS $(\Gamma)$ and SLS $(\mathcal{S})$ commute ($+$) or anti-commute ($-$) with each other if both of them exist. 
  The symmetry classes are categorized into different groups by the number ($N = 0,2,4$) of anti-unitary symmetries. 
  $z_{\rm min}$ denotes the eigenvalue with the smallest modulus. 
  The level statistics around the spectral origin are characterized by $\langle |z_{\rm min}|^2 \rangle$ and the probabilities of $z_{\min}$ being real or purely imaginary.
  If these probabilities equal each other within the range of statistical errors, they are in bold.
  The statistics are obtained by diagonalizing $10^7$ samples of $256 \times 256$ random matrices for classes BDI, CI, and \AIIp\ and diagonalizing $10^6$ samples of $256 \times 256$ random matrices for the other symmetry classes.
  }
  \resizebox*{1\textwidth}{!}{
	\begin{tabular}{c c  | c c c c | c c c c| c c c}
		\hline \hline
	\begin{tabular}{c} ~~~~Class~~~~  \\  ($H$) 
\end{tabular}		
  &
    \begin{tabular}{c} ~~~~Class~~~~  \\  ($\ii H$) 
\end{tabular}
		&
		~TRS~  &
		~PHS~  &
		~TRS$^{\dagger}$~ &
		~PHS$^{\dagger}$~ &
    ~CS~  &
		~SLS~&
    ~pH~&
    ~$[%
    \Gamma,\mathcal{S}]_{\pm}$=0~ &
    ~$\langle |z_{\rm min}|^2 \rangle$~ & 
    ~Pr($z_{\min} \in \mathbb{R}$)~ & 
    ~Pr($\ii z_{\min} \in \mathbb{R}$)~
	  \\ 
\hline
$N = 0$ & & & & & & & & & &\\ \hline
\textbf{AIII} + $\bm{\mathcal{S}_{+}}$ &  \textbf{AIII} + $\bm{\mathcal{S}_{+}}$  & $0$   &  $0$  & $0$   &  $0$  & $1$ & $1$ &  $1$ & $+$  & $1.1680(9)$ & $\textbf{0.2279(5)}$ & $\textbf{0.2280(5)}$ \\  
\textbf{AIII} + $\bm{\mathcal{S}_{-}}$ &  \textbf{AIII} + $\bm{\mathcal{S}_{-}}$  & $0$  &  $0$  &  $0$  &   $0$ & $1$  &  $1$ & $1$ & $-$  & $1.2707(12)$ & $\textbf{0.3336(5)}$ & $\textbf{0.3334(5)}$ \\ \hline  
$N = 2$ & & & & & & & & & & \\ \hline 
BDI &  D + $\eta_{+}$ & $+1$  & $+1$  &   $0$ &   $0$ & $1$ &      $0$   & $0$ &     & $1.4488(6)$ & $0.5373(2)$ & $0.1960(2)$  \\  
CI   &  C + $\eta_{-}$          & $+1$  & $-1$ &   $0$ &  $0$  & $1$ &     $0$  & $0$   &     & $1.2223(4)$ &$0.4247(2)$ & $0.2732(2)$ \\  
DIII   &  D + $\eta_{-}$        & $-1$ & $+1$  &   $0$ &   $0$ & $1$ &       $0$  & $0$ &     & $1.3390(13)$ & $ 0 $ & $0.5419(5)$ \\  
CII  &  C + $\eta_{+}$     & $-1$ & $-1$ &  $0$  & $0$   & $1$ &       $0$ & $0$ &     & $1.0926(7)$ & $ 0 $ & $0.2531(5)$ \\  
\textbf{AI} + $\bm{\mathcal{S}_{+}}$ &  \textbf{AI} + $\bm{\mathcal{S}_{+}}$  & $+1$  &  $0$  &  $0$  & $+1$  &      $0$   & $1$ & $0$ &     & $1.3094(14)$ & $\textbf{0.3885(5)}$ & $\textbf{0.3894(5)}$ \\  
AI + $\mathcal{S}_{-}$ &  AII + $\mathcal{S}_{-}$    &  $+1$  &   $0$ &   $0$ & $-1$ &       $0$  & $1$  & $0$ &     & $1.2055(10)$ & $0.5788(5)$ & $ 0 $ \\  
\textbf{AII} + $\bm{\mathcal{S}_{+}}$ & \textbf{AII} + $\bm{\mathcal{S}_{+}}$  & $-1$ &  $0$  &  $0$  & $-1$ &       $0$  & $1$ & $0$ &     & $1.0623(2)$ & $ \textbf{0} $ & $ \textbf{0} $ \\  
\hline
$N = 4$ & & & & & & & & & & \\ \hline 
\textbf{BDI} + $\bm{\mathcal{S}_{++}}$ &  \textbf{BDI} + $\bm{\mathcal{S}_{++}}$ & $+1$  & $+1$  & $+1$ & $+1$  & $1$ & $1$ & $1$ & $+$   & $1.4387(17)$ &  $\textbf{0.3502(5)}$ & $\textbf{0.3504(5)}$\\  
BDI + $\mathcal{S}_{--}$ & DIII + $\mathcal{S}_{--}$ & $+1$  & $+1$  & $-1$   & $-1$ & $1$ & $1$ & $1$ & $+$  & $1.2097(10)$ & $0.5293(5)$ & $0.0823(3)$ \\  
\textbf{DIII} + $\bm{\mathcal{S}_{++}}$ & \textbf{DIII}  + $\bm{\mathcal{S}_{++}}$ & $-1$ & $+1$  & $+1$ & $-1$ & $1$ & $1$ & $1$ & $+$  & $1.1071(7)$ & $ \textbf{0} $ & $ \textbf{0} $ \\  
\textbf{CI} + $\bm{\mathcal{S}_{++}}$ & \textbf{CI} + $\bm{\mathcal{S}_{++}}$ & $+1$  & $-1$ & $-1$   & $+1$  & $1$ & $1$ & $1$ & $+$   & $1.2134(10)$ &  $\textbf{0.3655(5)}$ & $\textbf{0.3661(5)}$\\  
CI + $\mathcal{S}_{--}$ & CII + $\mathcal{S}_{--}$ &  $+1$  & $-1$ & $+1$ & $-1$ & $1$ & $1$ & $1$ & $+$  & $1.1692(9)$ & $0.4193(5)$ & $ 0 $ \\  
\textbf{CII} + $\bm{\mathcal{S}_{++}}$ & \textbf{CII} + $\bm{\mathcal{S}_{++}}$ & $-1$ & $-1$ & $-1$ & $-1$ & $1$ &  $1$ & $1$  & $+$   & $1.0648(6)$ & $\textbf{0.1022(4)}$ & $\textbf{0.1025(4)}$ \\  
\textbf{BDI} + $\bm{\mathcal{S}_{+-}}$  &\textbf{BDI} + $\bm{\mathcal{S}_{+-}}$ & $+1$  & $+1$  & $-1$ & $+1$  & $1$ & $1$ & $1$ & $-$  & $1.3162(14)$ &  $\textbf{0.3878(5)}$ & $\textbf{0.3881(5)}$\\  
BDI + $\mathcal{S}_{-+}$ & DIII + $\mathcal{S}_{-+}$ & $+1$  & $+1$  & $+1$ & $-1$ & $1$ & $1$ & $1$ & $-$  & $1.4015(16)$ & $0.6190(5)$ & $ 0 $ \\  
\textbf{DIII} + $\bm{\mathcal{S}_{+-}}$ &  \textbf{DIII} + $\bm{\mathcal{S}_{+-}}$ & $-1$ & $+1$  & $-1$ & $-1$ & $1$ & $1$ & $1$ & $-$  & $1.2328(11)$ & $\textbf{0.2797(5)}$ & $\textbf{0.2802(5)}$ \\  
\textbf{CI} + $\bm{\mathcal{S}_{+-}}$ & \textbf{CI} + $\bm{\mathcal{S}_{+-}}$ & $+1$  & $-1$ & $+1$ & $+1$  & $1$ & $1$ & $1$ & $-$  & $1.2929(13)$ &  $\textbf{0.3684(5)}$ & $\textbf{0.3692(5)}$ \\  
CI + $\mathcal{S}_{-+}$ & CII +  $\mathcal{S}_{-+}$ & $+1$  & $-1$ & $-1$  & $-1$ & $1$ & $1$  & $1$ & $-$  & $1.1840(9)$ & $0.4316(5)$ & $0.2140(5)$ \\  
\textbf{CII} + $\bm{\mathcal{S}_{+-}}$ & \textbf{CII} + $\bm{\mathcal{S}_{+-}}$ & $-1$ & $-1$ & $+1$ & $-1$ & $1$ & $1$ & $1$ & $-$  & $1.0795(6)$ & $ \textbf{0} $ & $ \textbf{0} $ \\  
\hline \hline
	\end{tabular}
}
	\label{tab: AZ_real}
\end{table*}

To characterize the level statistics around the spectral origin, we first consider the distribution $p_{\min}(z_{\rm min})$ of the eigenvalue $z_{\rm min}$
closest to the origin. 
To make $z_{\min}$ unique, we consider the distribution for 
$\Re\,z_{\min} \geq 0$ and 
$\Im\,z_{\min} \geq 0$. 
We normalize $z_{\min}$ such that $\left\langle|z_{\min}|\right\rangle = 1$. 
Because of the possible presence of real and purely imaginary eigenvalues, the probability density $p_{\min}(z_{\min} = x + \ii y)$ of 
$z_{\min}$
is decomposed as
\begin{align}
    & p_{\rm min}(z_{\min} = x + \ii y) \nonumber \\
   &  =  p_{\rm min;r}(x) \delta(y ) + p_{\rm min;i}(y) \delta(x) + p_{\rm min;c}(x,y) \, ,
\end{align}
where $\delta(\cdot)$ is the Dirac delta function, and $p_{\rm min;r}(x) $, $ p_{\rm min;i}(y)$, and $p_{\rm min;c}(x,y) $ represent the distributions of $z_{\min}$ when it is real, purely imaginary, and complex, respectively.
The distribution $p_{\rm min}(z_{\min})$ is normalized by 
\begin{equation}
\int_0^{\infty} dx  \int_0^{\infty} dy~p_{\rm min}(x,y) = 1, 
\end{equation}
and 
\begin{align}
    \mathrm{Pr}\,(z_{\min} \in \mathbb{R}) &\equiv \int_0^{\infty} dx~p_{\rm min; r} (x) \leq 1, \label{eq: Pr-R} \\
    \mathrm{Pr}\,(\ii z_{\min} \in \mathbb{R}) &\equiv \int_0^{\infty} dy~p_{\rm min; i} (y) \leq 1 \label{eq: Pr-I}
\end{align}
represent the probability of $z_{\min}$ being real and purely imaginary, respectively.
Even if $z_{\min}$ is real (purely imaginary), it can have a tiny imaginary (real) part in numerical diagonalization due to the machine inaccuracy.
In our numerical investigation,
$z_{\min}$ is considered as real (purely imaginary) if it satisfies $|\Im\,z_{\min}| < C$ ($|\Re\,z_{\min}|< C$) with a cutoff $C = 10^{-12}$. %
We provide both $ \mathrm{Pr}\,(z_{\min} \in \mathbb{R})$ and $\mathrm{Pr}\,(\ii z_{\min} \in \mathbb{R})$ in Table~\ref{tab: AZ_real}. 
For $N \geq 256$, the probability of $z_{\min}$ being real or purely imaginary converges to a universal value characteristic to each symmetry class [Fig.~\ref{figs: p_real_AZ_real}\,(a)].
Additionally, $\langle |z_{\min}|^2  \rangle$ also converges to a universal value in each symmetry class for large $N$ [Fig.~\ref{figs: p_real_AZ_real}\,(b)], which verifies
the convergence of the statistics of $z_{\min}$ for $N \geq 256$.

We consider the radial distribution $p^{(r)}_{\rm min;c}(|z_{\rm min}|) $ for complex $z_{\min} \in \mathbb{C}$, given as 
\begin{align}
    &p^{(r)}_{\rm min;c}\,(|z_{\rm min}|) = \int_0^{\infty} dx  \int_0^{\infty} dy~p_{\rm min;c}\,(x,y) \nonumber \\
    &\qquad\qquad\qquad\qquad\qquad \times \delta \left( |z_{\rm min}| - \sqrt{x^2 + y^2} \right).
\end{align}
Notably, $ p^{(r)}_{\rm min;c}(|z_{\rm min}|)$ shows distinct behavior in each symmetry class [Figs.~\ref{figs: pdf_z_min_AZ_real}\,(a)-(c)].
Fitting $ p^{(r)}_{\rm min;c}(|z_{\rm min}|)$ with small $|z_{\min}|$ [see Fig.~\ref{figs: pdf_z_min_AZ_real}\,(g)], we find %
\begin{equation} \label{eq: small z_min}
  p^{(r)}_{\rm min;c}(|z_{\rm min}|) \propto \begin{cases}
    |z_{\min} |^3 & \text{(class BDI)} \, ; \\
    -|z_{\min} |^5\ln |z_{\min}| & \text{(class CI)} \, ; \\
    -|z_{\min} |^7\ln |z_{\min}| & \text{(class AII} + \mathcal{S}_+) \, ,\\
  \end{cases}
\end{equation}
for $|z_{\min}| \ll 1$. 
The joint probability distribution of complex eigenvalues for non-Hermitian random matrices in class \AIIp\ with
arbitrary $N$
was obtained previously~\cite{akemann2005}.
In Appendix~\ref{app: analytic}, we verify that $4 \times 4$ random matrices capture the same small-$|z_{\min}|$ behavior as that of large random matrices.

The distribution of complex $z_{\min}$ 
is qualitatively different from 
the distribution $p_{\rm min;r/i}(|z_{\min}|)$ of real or purely imaginary $z_{\min}$ (if exists) even in the same symmetry class. 
In fact, in class BDI, we have 
[Fig.~\ref{figs: pdf_z_min_AZ_real}\,(a)] 
\begin{equation}
    p_{\rm min;r}(0) > 0,\quad p_{\rm min;i}(0) > 0.
\end{equation}
In class CI, we have [Figs.~\ref{figs: pdf_z_min_AZ_real}\,(b)]
\begin{equation}
    p_{\rm min;r/i}(|z_{\rm min}|) \propto |z_{\rm min}| \quad \left( |z_{\min}| \ll 1 \right).
\end{equation}
They are significantly distinct from the small-$|z_{\min}|$ behavior of $p^{(r)}_{{\rm min};c}(|z_{\min}|)$.  
We provide an intuitive explanation 
for the 
stronger level repulsion for 
complex $z_{\rm min}$.
As discussed before, all of $\pm z_{\min}$ and $\pm z_{\min}^{*}$ are eigenvalues.
For complex $z_{\min}$, these four eigenvalues differ from each other and share the smallest modulus.
By contrast, for real $z_{\min}$ ($z_{\min} = z_{\min}^*$) or purely imaginary $z_{\min}$ ($z_{\min} = -z_{\min}^*$), only the two different eigenvalues share the smallest modulus. 
For complex $z_{\rm min}$ 
with $|z_{\min}| \ll 1$, there are at least three eigenvalues (i.e., $-z_{\min}$ and $\pm z_{\min}^*$) whose distance to $z_{\min}$ is smaller than $2|z_{\min}|$
and hence much smaller than unity. 
In contrast, for real or purely imaginary $z_{\rm min}$
with $|z_{\min}| \ll 1$,
only at least one eigenvalue %
satisfies
this requirement. 
This difference leads to the %
distinct
strength of level repulsion.
In addition, we analytically calculate the hard-edge statistics 
for $4 \times 4$ non-Hermitian random matrices in class BDI in Appendix~\ref{app: analytic}, which also show distinct small-$|z_{\min}|$ behaviors for complex, real, and purely imaginary $z_{\min}$, consistent with those of large random matrices.

Furthermore, the angular distribution 
\begin{align}
    &p^{(\theta)}_{\rm min;c}\,(\theta_z) =  \int_0^{\infty} dx  \int_0^{\infty} dy~p_{\rm min;c}\,(x,y) \nonumber \\
    &\qquad\qquad\qquad\qquad\qquad \times \delta \left( \theta_z - \arctan \left( y/x \right) \right)
\end{align}
for complex $z_{\min} = |z_{\min}| e^{\ii \theta_z}$ [$\theta_z \in (0,\pi/2)$] shows characteristic behavior in each symmetry class 
[see the insets of Figs.~\ref{figs: pdf_z_min_AZ_real}\,(a)-(c)].
In all the three symmetry classes, the distributions $p^{(\theta)}_{\rm min;c}(\theta_z)$ vanish for $\theta_z \rightarrow 0$ or $\pi/2$ and get maximal for $\theta_z \simeq \pi/4$. 
This behavior is due to the level repulsion between $z_{\min}$ and $z_{\min}^*$ or $-z_{\min}^*$. 
Specifically, we have for $|\theta_z| \ll 1$
\begin{equation} \label{eq: small theta_z 1}
  p^{(\theta)}_{\rm min;c}(\theta_z) \propto \begin{cases}
    |\theta_z| & \text{(classes BDI and CI)} \, ; \\
    |\theta_z|^2 & \text{(class AII} + \mathcal{S}_+) \, .\\
  \end{cases}
\end{equation}
For $|\pi/2 - \theta_{z}| \ll 1$, we also have 
\begin{equation} \label{eq: small theta_z 2}
  p^{(\theta)}_{\rm min;c}(\pi/2 - \theta_{z}) \propto \begin{cases}
    |\pi/2 - \theta_{z}| & \text{(classes BDI and CI)} \, ; \\
    |\pi/2 - \theta_{z}|^2 & \text{(class AII} + \mathcal{S}_+) \, .\\
  \end{cases}
\end{equation}
Notably, these linear or quadratic decays of $p^{(\theta)}_{\rm min;c}$ are reminiscent
of the soft gap of the density $\rho_c(x+\ii y)$ of complex eigenvalues around the real axis in the presence of TRS or pH.
In fact, in the presence of TRS with sign $+1$ or pH, we have $\rho_c(x+\ii y) \propto |y|$ for $|y| \ll 1$, and in the presence of TRS with sign $-1$, we have $\rho_c(x+\ii y) \propto|y|^2$ for $|y| \ll 1$, as summarized in Table~\ref{tab: AZdagger classification}~\cite{Efetov97, kanzieper2005, xiao22}.
In both classes BDI and CI, non-zero real eigenvalues respect TRS with sign $+1$ while purely imaginary eigenvalues respect CS, equivalent to pH. 
The exponents in the power-law decay of $p_{\rm min;c}^{(\theta)}$ align with those of $\rho_c(x + \ii y)$. 
The difference between classes BDI and CI lies in the signs of PHS, not influencing symmetries of non-zero real or purely imaginary eigenvalues. 

In class \AIIp, non-zero real eigenvalues respect TRS with sign $-1$ while non-zero purely imaginary eigenvalues respect PHS$^{\dagger}$ with sign $-1$, equivalent to TRS with sign $-1$~\cite{kawabata2019a}.
The exponent in the power-law decay of $p_{\rm min;c}^{(\theta)}$ also aligns with that of $\rho_c(x,y)$.
It is also notable that $p^{(\theta)}_{\rm min;c}(\theta_z) $ in class \AIIp\ exhibits higher symmetry than those in classes BDI and CI. 
Because of the equivalence between TRS and PHS$^{\dagger}$, if $H$ belongs to class \AIIp, $\ii H$ also belongs to class \AIIp,
both of which occur with the same probability in the Gaussian ensemble.
Consequently, the probability of $z$ being the eigenvalue with the smallest modulus is the same as that of $\ii z$.
In combination with the spectral D$_2$ symmetry due to TRS and SLS, the distribution of $z_{\rm min}$ should thus exhibit the point group symmetry D$_4$ after the ensemble average.
This spectral D$_4$ symmetry requires $p^{(\theta)}_{\rm min;c}(\theta_z) =  p^{(\theta)}_{\rm min;c}(\pi/2 - \theta_z)$, consistent with the numerical results in the inset of Fig.~\ref{figs: pdf_z_min_AZ_real}\,(c).
On the other hand, such symmetry is absent in classes BDI and CI, and numerical results show that $p^{(\theta)}_{\rm min;c}(\theta_z)$ equals $p^{(\theta)}_{\rm min;c}(\pi/2 - \theta_z)$ only approximately.

\subsection{General classes}
\label{subsec: AZ all general}
Beyond the AZ$^{\dagger}$ and AZ$_0$  classification, we now investigate the remaining 18 symmetry classes.
As indicators of the universal classes of the level statistics around the spectral origin, we use $\langle |z_{\rm min}|^2 \rangle$ and the probability of $z_{\rm min}$ being real or purely imaginary, denoted by Pr($z_{\min} \in \mathbb{R}$)
in Eq.~(\ref{eq: Pr-R})
and Pr($\ii z_{\min} \in \mathbb{R}$)
in Eq.~(\ref{eq: Pr-I}), respectively.
Since these indicators already reach the convergence for $256\times 256$ non-Hermitian random matrices in the three representative classes (see Fig.~\ref{figs: p_real_AZ_real}), we obtain these values in all the remaining 18 classes through the exact diagonalization of random matrices with the same size, as summarized in Table~\ref{tab: AZ_real}.
We expect these values to be universal and characterize the universality classes of the complex spectral statistics. 
For the three representative classes, we shortly verify the universality also by comparison with physical models (see Table~\ref{tab: model AZ_real} in Sec.~\ref{sec: quantum system real} for details); the exhaustive verification for physical models in the other symmetry classes is left for future study.

In each symmetry class, the distributions of $z_{\min}$ exhibit %
characteristic 
behaviors.
Within a subset of symmetry classes, the probability of $z_{\min}$ being real or purely imaginary is non-zero, similar to classes BDI and CI. %
We find that whether Pr($z_{\min} \in \mathbb{R}$) is zero depends only on symmetries respected by the real axis (i.e., TRS, TRS$^{\dagger}$, and pH), regardless of the other symmetries. 
These symmetries, comprising the AZ$^{\dagger}$ classification, also determine whether non-Hermitian random matrices in the AZ$^{\dagger}$ classification have real eigenvalues (see Sec.~\ref{sec: symmetry} and a summary in Table~\ref{tab: AZdagger classification}).
For example, in the AZ$^{\dagger}$ classification, a random matrix in class 
AI + $\eta_{-}$
has TRS with sign $+1$ and TRS$^{\dagger}$ with sign $-1$, leading to a subextensive number of real eigenvalues (see Table~\ref{tab: AZdagger classification}).
Correspondingly, classes BDI + $\mathcal{S}_{--}$, CI + $\mathcal{S}_{++}$, BDI + $\mathcal{S}_{+-}$, and CI + $\mathcal{S}_{-+}$ have the same signs of TRS and TRS$^{\dagger}$, and non-Hermitian random matrices in these symmetry classes exhibit Pr$(z_{\min} \in \mathbb{R})\neq 0$ (see Table~\ref{tab: AZ_real}). 
Similarly, whether Pr($\ii z_{\min} \in \mathbb{R}$) is zero depends only on the symmetries respected by pure imaginary eigenvalues (i.e., PHS$^{\dagger}$, TRS$^{\dagger}$, and CS).

Another notable feature is the 
spectral D$_4$ symmetry of the level statistics in some symmetry classes  
where $H$ and $\ii H$ belong to the same symmetry class. 
Such symmetry classes are shown in bold font in Table~\ref{tab: AZ_real}.  
When Pr$(z_{\min} \in \mathbb{R})$ and Pr$(\ii z_{\min} \in \mathbb{R})$ overlap within the range of statistical errors, they are also shown in bold font in Table~\ref{tab: AZ_real}. 
These two bold-font items consistently appear in the same rows in Table~\ref{tab: AZ_real} due to the spectral D$_4$ symmetry of the distributions of $z_{\min}$, analogous to our discussion for class \AIIp.

In the remaining 18 symmetry classes, we also find the distinct small-$|z_{\min}|$ behavior between the distribution of complex $z_{\min}$ and that of real or purely imaginary $z_{\min}$.
Moreover, the angular distribution $p^{(\theta)}_{\rm min;c}\,(\theta_z)$ vanishes as $|\theta_z| \rightarrow 0$ or $|\pi/2 - \theta_z| \rightarrow 0$ (see Appendix~\ref{app: hard-edge supp} 
for details). 
Thereby, our discussions on the level repulsion in the three representative classes can also be applied to the 
remaining 18
symmetry classes.
A more detailed numerical study of the asymptotic behavior of 
the distributions, in a similar manner to Eqs.~(\ref{eq: small z_min})-(\ref{eq: small theta_z 2}), and their theoretical explanations are left for future study. %

\section{Open quantum systems in the AZ$_0$ classification}
\label{sec: quantum system}

In quantum chaotic systems that respect PHS or SLS 
and are isolated from the environment, 
the spectral statistics around zero energy $E = 0$ can be described by Hermitian random matrices in the same symmetry class (see, for example, Ref.~\cite{beenakkerRMT15} and references therein). Here, we calculate the density of complex eigenvalues and the distribution of the complex level ratios in Eq.~(\ref{eq: ratio}) around the spectral origin of open quantum systems that respect PHS or SLS.
We find that they are consistent with the statistics of non-Hermitian random matrices studied in Sec.~\ref{sec: RM AZ0}. As a comparison, we also study the level statistics of an integrable 
non-Hermitian many-body Hamiltonian with PHS.

\subsection{Lindblad equation}
\label{subsec: Lindblad}
When a quantum system is coupled to a Markovian environment, the time evolution of its reduced density matrix $\rho$ is described by the Lindblad equation~\cite{GKS-76, Lindblad-76, breuer2002theory}
\begin{align} 
   \frac{d \rho}{dt} &= \mathcal{L}(\rho) \nonumber \\ &\equiv  -\ii [H,\rho] +  \sum_{\mu} \left( L_{\mu} \rho L_{\mu}^{\dagger} - \frac{1}{2} \left\{  L_{\mu}^{\dagger} L_{\mu}, \rho \right\}   \right) \, .
\end{align}
Here, $H$ is %
the 
Hamiltonian of the system that governs the unitary dynamics of $\rho$, and $L_{\mu}$'s represent dissipators originating from the coupling with the environment. The Lindbladian $\mathcal{L}$ is a superoperator that acts on the density matrix. Through the vectorization $\rho = \rho_{ij}\ket{i}\bra{j} \rightarrow \ket{\rho} = \rho_{ij}\ket{i}\ket{j}$, the density matrix can be mapped to a state in the double Hilbert space. 
Correspondingly, $\mathcal{L}$ becomes a non-Hermitian operator 
in the double Hilbert space, given as  
\begin{equation} \label{eq: Lindbladian}
\begin{aligned}
  &\mathcal{L} = -\ii \left(H \otimes 1 - 1 \otimes H^* \right) \\ &+ \sum_{\mu} \left[ L_{\mu} \otimes L_{\mu}^* - \frac{1}{2} \left(L_{\mu}^{\dagger}L_{\mu} \otimes 1 \right)  - \frac{1}{2} \left(1 \otimes L_{\mu}^{\rm T}L_{\mu}^* \right) \right] \, .
\end{aligned}
\end{equation}
For a generic system with $N$ complex fermions, the dimension of the double Hilbert space is $(2^N)^2 = 4^N$.
Therefore, $\mathcal{L}$ is a $4^N \times 4^N$ operator, and 
numerical evaluation of its spectrum for larger $N$ costs 
computation time. 

Notably, quadratic Lindbladians, comprising quadratic Hamiltonians
and linear dissipators, effectively reduce to single-particle non-Hermitian Hamiltonians~\cite{prosen2008a, prosen2010}, which facilitates numerical investigations of relatively larger systems. 
A non-interacting fermionic Hamiltonian and linear dissipators are generally given by
\begin{equation} 
  H = \frac{1}{2} \sum_{i,j = 1}^{2N} \Psi_i^{\dagger} H_{ij} \Psi_j, \quad L_{\mu} = \sum_{i = 1}^{2N} l_{\mu \ii } \Psi_{i} \, ,
\end{equation}
in the Nambu basis ${ \Psi} = ( c_1,\cdots, c_N, c_1^{\dagger},\cdots, c_N^{\dagger} )^{\rm T}$. 
By construction, the $2N \times 2N$ Hermitian 
matrix $H$ respects PHS $H = -\tau_x H^{\rm T} \tau_x$, where $\tau_x$ is the Pauli matrix acting on the $(c,c^{\dagger})$ space. 
We define a $2N \times 2N$ matrix by
\begin{equation} 
  N_{ij} \equiv (l^{\rm T}l^*)_{ij} =  \sum_{\mu } l_{\mu i} l_{\mu j}^* \, ,
\end{equation}
and its particle-hole-symmetric combination 
\begin{equation} 
  \Gamma \equiv  \frac{1}{2} (N + \tau_x N^{\rm T} \tau_x) \, .
\end{equation}
Using $H$ and $\Gamma$, we can construct a 
$2N \times 2N $ single-particle non-Hermitian Hamiltonian
\begin{equation} 
  K \equiv H - \ii \Gamma \, .
\end{equation} 
From complex eigenvalues $\lambda_i$'s of this effective single-particle non-Hermitian Hamiltonian $K$, the $4^N$ eigenvalues of $\mathcal{L}$ are given as 
\begin{equation} 
  \Lambda_{\iii} = -2\ii \sum_{i = 1}^{2N} n_i \lambda_i \, ,
\end{equation}
with $n_i = 0$ or $1$. 
Notably, $K$ always satisfies PHS$^{\dagger}$, 
\begin{equation}
\tau_x K^{*}\tau_x = - K.
    \label{eq: K-PHSdag}
\end{equation}

In this work, 
we consider $H$ with the conserved particle number, given by 
\begin{equation} 
  H = \frac{1}{2} \begin{pmatrix} 
    h & 0 \\
    0 & -h^{\rm T}
  \end{pmatrix} \, .
\end{equation}
The coupling with the environment is assumed to  
manifest itself in single-particle loss; 
the dissipators $L_{\mu}$ consist only of annihilation 
operators, 
\begin{equation}
L_{\mu} = \sum_{i = 1}^{N} l_{\mu i 
} c_{i}.
\end{equation}
In such a case, the effective single-particle non-Hermitian Hamiltonian $K$ is block-diagonalized as 
\begin{equation} 
  K =  \frac{1}{2} \begin{pmatrix} 
    h - \ii l^{\rm T}l^*  & 0 \\
    0 & -h^{\rm T} -  \ii l^{\dagger}l
  \end{pmatrix} \, .
\end{equation} 
While 
PHS$^{\dagger}$ in Eq.~(\ref{eq: K-PHSdag}) 
relates 
the two reduced blocks $h - \ii l^{\rm T}l^*$ and 
$-h^{\rm T} - \ii l^{\dag}l  = - (h - \ii l^{\rm T}l^*)^{ *}$, %
the reduced block $h - \ii l^{\rm T}l^*$ itself does not necessarily respect PHS$^{\dagger}$.
The symmetry of quadratic Lindbladians is determined solely 
by $h - \ii l^{\rm T}l^*$ without intrinsic PHS$^{\dag}$.
Since $h$ is Hermitian and $ l^{\rm T}l^* $ is positive semidefinite, complex eigenvalues $z_i$'s of $h - \ii l^{\rm T}l^*$ must satisfy the constraint 
\begin{equation}
    \Im\,z_i \leq 0.
        \label{eq: positivity}
\end{equation}
It was argued that quadratic Lindbladians cannot respect TRS, PHS, pH, or SLS since these symmetries violate this %
non-positivity
constraint~\cite{lieu20a}.
As discussed previously, TRS or pH (PHS or SLS) requires complex eigenvalues to appear in $\left( z, z^{*} \right)$ [$\left( z, -z \right)$] pairs, which is incompatible with Eq.~(\ref{eq: positivity}).
However, if we shift the Lindbladian by its trace, the traceless part 
is free from 
such constraints~\cite{Prosen-12, kawasaki2022, sa2023, kawabata2023}.
In principle, the diagonal block of the traceless part, given as 
\begin{equation} \label{eq: h_eff}
 h_{\rm eff} \equiv  h - \ii l^{\rm T}l^* - \frac{1}{N} {\rm Tr} K  
\end{equation}
can respect any of the seven symmetries in Eqs.~(\ref{eq: TRS_def})-(\ref{eq: pH_def}). 
Hereafter, we investigate the symmetry classes and the hard-edge statistics of the shifted quadratic Lindbladian, represented by the effective non-Hermitian Hamiltonian $h_{\rm eff}$.

\begin{table}[bt]
  \centering
  \caption{Mean values $\langle r \rangle$ and $\langle \cos \theta \rangle$ of non-Hermitian random matrices (RM) and Hamiltonians %
  in the AZ$_0$ classification (Secs.~\ref{subsec: quadratic L} and \ref{subsec: model BdG}), where %
  $r e^{\ii \theta}$ denotes the complex level ratio [Eq.~(\ref{eq: ratio})] around the spectral origin.}
  \begin{tabular}{c|cccc}
  \hline \hline
  Class & \makecell[c]{$\langle r \rangle$ \\ (RM) } & \makecell[c]{$\langle r \rangle$ \\(Hamiltonian) } &  \makecell[c]{$\langle \cos \theta \rangle$ \\ (RM)\\ } &  \makecell[c]{$\langle \cos \theta \rangle$ \\ (Hamiltonian)\\ }\\ \hline
  \AIIIz  & 0.6357(1) & 0.6357(5) &  0.5381(1) & 0.5391(7) \\ 
  \BDIz  & 0.5773(1) & 0.5778(6) &  0.5682(1) & 0.5681(7) \\ 
  \CIIz  & 0.6618(1) & 0.6623(5) &  0.5150(1) & 0.5147(7) \\ 
   D   & 0.5408(1) & 0.5411(6) &  0.5528(1) & 0.5524(7) \\ 
   C   & 0.6749(1) & 0.6746(5) &  0.5334(1) & 0.5343(7) \\ 
  \CIz  & 0.6711(1) & 0.6708(5) &  0.5585(1) & 0.5589(7) \\ 
  \DIIIz  & 0.5949(1) & 0.5950(6) &  0.5265(1) & 0.5252(7) \\ 
  \hline
  \end{tabular}
  \label{tab: model AZ0}
\end{table}

\subsection{Quadratic Lindbladians}
\label{subsec: quadratic L}

We study three prototypical models of quadratic Lindbladians in the AZ$_0$ classification. 
We consider the following non-interacting Hermitian Hamiltonians on the three-dimensional cubic lattice $\iii \in \mathbb{Z}^3$ in the chiral symmetry classes (i.e., classes AIII, BDI, and CII):
\begin{equation}    
  H =  \sum_{\left\langle \iii,\jjj \right\rangle} c_{\iii}^{\dagger} t_{\iii,\jjj}  c_{\jjj} + \text{H.c.} \, , 
    \label{eq: H_chiral_class}
\end{equation} 
where the bracket $\left\langle \iii,\jjj \right\rangle$ represents the nearest neighboring sites. 
Each site contains only one orbital but can accompany the spin degree of freedom. 
We assume the periodic boundary conditions and the system size of $L^3$. 
For any hopping amplitude $t_{\iii, \jjj }$ and $L \in 2 \mathbb{Z}$, the Hamiltonian respects SLS 
\begin{equation}
    \mS H \mS = - H
\end{equation}
with $\mS_{\iii,\jjj} \equiv \delta_{\iii,\jjj} (-1)^{(\iii_x + \iii_y + \iii_z)}$.
Depending on the choices of $t_{\iii,\jjj}$, $H$ realizes models in all three chiral symmetry classes, as follows. 

Let us first consider %
cases without the spin degree of freedom:
\begin{itemize}
    \item[(i)] {\it Class AIII}.---Choose $t_{\iii, \jjj} = e^{\ii \theta_{\iii, \jjj}}$, where $\theta_{\iii, \jjj} \in \mathbb{R}$ is distributed uniformly and independently in $[0,2\pi)$. 
    The Hamiltonian realizes the random flux model~\cite{lee1981, furusaki1999} and is denoted by $H_{\rm AIII}$. 
    The Hamiltonian $H_{\rm AIII}$ only respects SLS and hence belongs to class AIII.

    \item[(ii)] {\it Class BDI}.---Choose $t_{\iii, \jjj}$ to be real-valued and distributed $t_{\iii, \jjj} \in \mathbb{R}$ uniformly and independently in $[t-W/2,t+ W/2]$.
    The Hamiltonian realizes the random hopping model~\cite{lee1981, furusaki1999} and is denoted by $H_{\rm BDI}$.
    In addition to SLS, $H_{\rm BDI}$ respects TRS 
    with sign $+1$ %
    \begin{equation}
    H_{\rm BDI}^* = H_{\rm BDI} 
    \end{equation}
    and hence belongs to class BDI.
    
\end{itemize}
Consider next the %
case with 
the spin degree of freedom:
\begin{itemize}
    \item[(iii)] {\it Class CII}.---Choose $2\times 2$ hopping matrices $t_{\iii, \jjj}$ that are distributed uniformly and independently on the Harr measure of SU(2) matrices. 
    The Hamiltonian realizes the SU(2) model~\cite{asada2002} and is denoted by $H_{\rm CII}$. 
    In addition to SLS, $H_{\rm CII}$ respects TRS with sign $-1$ %
    \begin{equation}
    \sigma_y H_{\rm CII}^* \sigma_y = H_{\rm CII} 
    \end{equation}
    and hence belongs to class CII.
    
\end{itemize}
These models were originally proposed  
for solid-state materials but
also found realizations in synthetic materials such as cold atoms and photonic systems~\cite{kondov2011, meier2018, stutzer2018, liu2020}.  
The coupling between cold atoms can be designed to respect CS~\cite{meier2018}, and the hopping $t_{\iii, \jjj}$ can be controlled by synthetic gauge fields~\cite{livi2016, ozawa2019}.

We introduce the single-particle loss by dissipators that consist of a linear combination of annihilation operators~\cite{diehl2011, gong2018, song2019a, kawabata2019a}. 
For each Hamiltonian $H$, we choose the 
following linear dissipators such that  
the relevant symmetries 
are respected:
\begin{itemize}
    \item[(i)] {\it Class \AIIIz}.---For $H_{\rm AIII}$, we introduce the dissipators as $L_{\iii} =  \sqrt{{\gamma}} (c_{\iii} + \ii c_{\iii + \e_z})$ ($ \iii \in \mathbb{Z}^3 $).  

    \item[(ii)] {\it Class \BDIz}.---For $H_{\rm BDI}$, we introduce the dissipators as $L_{\iii} =  \sqrt{{\gamma}} (c_{\iii} + c_{\iii + \e_z})$ ($ \iii \in \mathbb{Z}^3 $).

    \item[(iii)] {\it Class \CIIz}.---For $H_{\rm CII}$, we introduce the dissipators as $L_{\iii  \sigma }=  \sqrt{{\gamma}} (c_{\iii \sigma} + c_{(\iii + \e_z) \sigma})$ ($ \iii \in \mathbb{Z}^3 , \sigma = \uparrow, \downarrow$).
\end{itemize}
In all the cases, $\gamma \geq 0$ denotes the dissipative coupling strength between the system and environment. 
The dissipative coupling is assumed to be identical among different sites and maintain translation invariance. 
The effective non-Hermitian Hamiltonians [Eq.~(\ref{eq: h_eff})] associated with these dissipators are denoted by $h_{\rm eff}^{\rm AIII^{\dagger}}$, $h_{\rm eff}^{\rm BDI_0}$, and $h_{\rm eff}^{\rm CII_0}$, respectively. 
Even in the presence of dissipation, they still respect SLS and TRS$^{\dagger}$ with the same symmetry operators and hence belong to Classes \AIIIz, \BDIz, and \CIIz. 
The explicit forms of the Hamiltonians and their symmetries can be found in Appendix~\ref{app: model detail}. %

\begin{figure}[th]
  \includegraphics[width=1\linewidth]{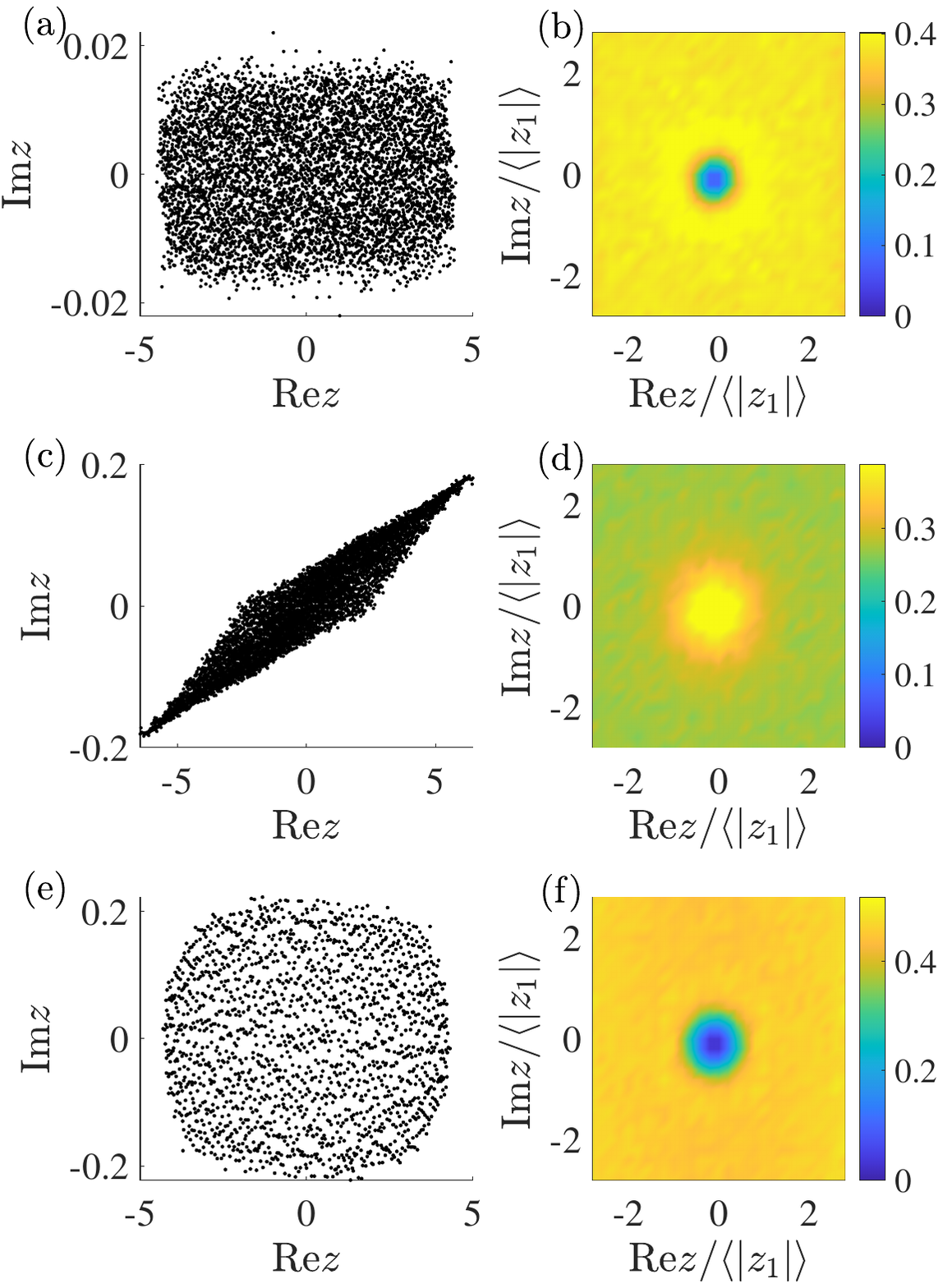}
  \caption{Complex eigenvalues of single realizations of the effective non-Hermitian Hamiltonians for  
  the quadratic Lindbladians in classes (a) \AIIId, (c) \BDIz, and (e) \CIIz~(see Sec.~\ref{subsec: quadratic L}). 
  (b), (d), (f)~Ensemble average of the density of complex eigenvalues around the spectral origin of the same Hamiltonians as the left subfigures.
  }
  \label{figs: z_dis chiral_AZ0}
\end{figure}

We investigate these effective non-Hermitian Hamiltonians numerically (see the parameters in Appendix~\ref{app: model detail}).
For each class, we diagonalize $2 \times 10^5$ samples in the disorder ensemble.
We study the statistics of single-particle eigenvalues of the effective non-Hermitian Hamiltonians, rather than
many-body 
eigenvalues of the Lindbladians. 
Figures~\ref{figs: z_dis chiral_AZ0}\,(a), (c), and (e) show the distributions of complex eigenvalues for a single realization of each Hamiltonian.
The full complex spectra deviate from the circular law of 
non-Hermitian random matrices~\cite{ginibre1965statistical, Girko-85} and are generally non-universal. 
The range of each spectrum along the imaginary-axis direction is 
much smaller than that along the real-axis direction because 
of the small dissipation strength $\gamma$.

\begin{figure*}[th]
  \centering
  \includegraphics[width=0.8\linewidth]{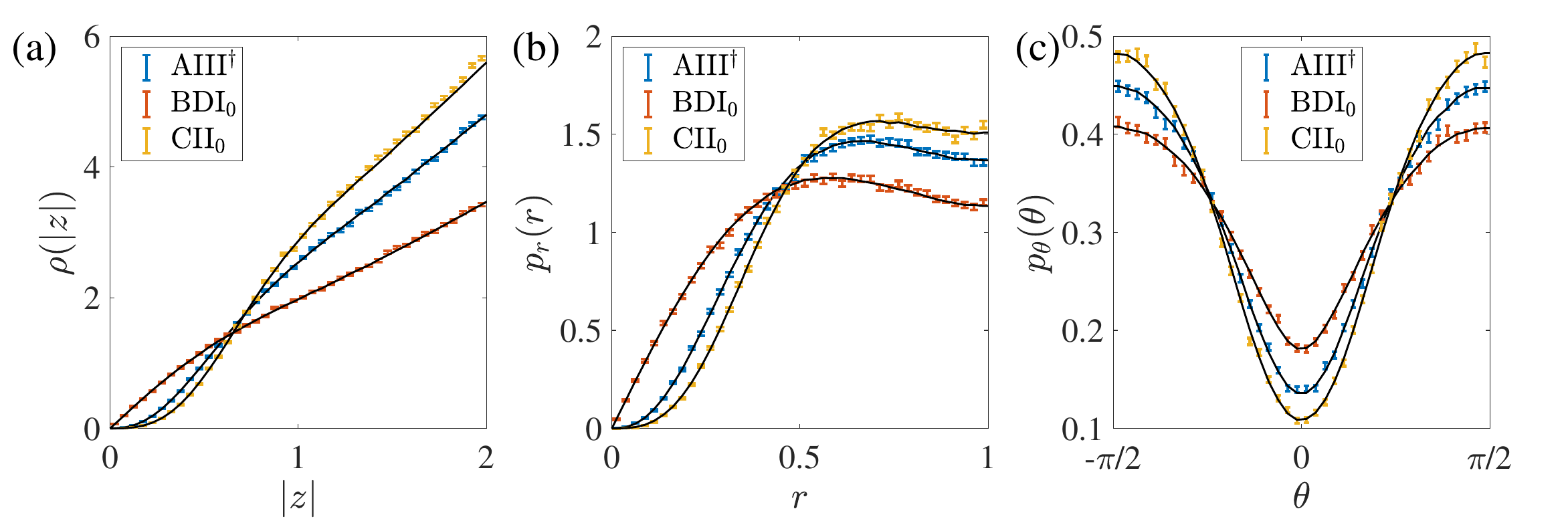}
  \caption{Hard-edge statistics of the effective non-Hermitian Hamiltonians 
  of the quadratic Lindbladians in classes \AIIId, \BDIz, and \CIIz~(see Sec.~\ref{subsec: quadratic L}). 
  (a)~Density $\rho(|z|)$ of the modulus of complex eigenvalues.
  (b)~Radial distributions $p_r(r)$ and (c)~angular distributions $p_{\theta}(\theta)$ of the complex level ratios $r e^{\ii \theta}$ [Eq.~(\ref{eq: ratio})] around the spectral origin. 
  The colored points with the error bars refer to the Hamiltonians,
  and the black solid lines are obtained 
  from
  non-Hermitian random matrices in the same symmetry classes. 
  }
  \label{figs: rho_ratio chiral_AZ0}
\end{figure*}

We investigate the spectral properties around the origin 
and demonstrate that they conform to the universal hard-edge statistics 
determined solely by symmetry. 
We normalize the eigenvalues of the Hamiltonians such that $\langle |z_1| \rangle = 1$, where $z_1$ denotes the complex eigenvalue with the smallest modulus.
After the ensemble average, the density of complex eigenvalues around the spectral origin does not depend on the phases but only on the modulus, and hence exhibits spectral U(1) rotation symmetry [Figs.~\ref{figs: z_dis chiral_AZ0} (b), (d), and (f)].
For the Hamiltonians in different symmetry classes, $\rho(|z|)$'s show distinct behavior consistent with the characteristic forms of non-Hermitian random matrices in the corresponding symmetry classes [Fig.~\ref{figs: rho_ratio chiral_AZ0}\,(a)].
The complex level ratios $r e^{\ii \theta} \equiv z_1/z_2$ in Eq.~(\ref{eq: ratio}) around the spectral origin also obey the same distributions of non-Hermitian random matrices in the same symmetry class [Figs.~\ref{figs: rho_ratio chiral_AZ0}\,(b) and (c)].
To further validate this consistency, in Table~\ref{tab: model AZ0}, we compare $\langle r \rangle$ and $\langle \cos \theta  \rangle$ of the effective non-Hermitian Hamiltonians (and also those studied in Sec.~\ref{subsec: model BdG}) and non-Hermitian random matrices. 
In each symmetry class, the corresponding statistical quantities 
overlap with each other within two standard deviations. 
Our results demonstrate the universality of 
the hard-edge statistics of non-Hermitian random matrices and 
reveal the chaotic nature of the quadratic Lindbladians. 

As also discussed 
above, the complex spectra around the origin exhibit U(1) rotation symmetry even for the physical models.
Notably, this spectral U(1) symmetry of the ensemble-average density 
does not arise from the original symmetry of the disorder ensemble of the Hamiltonians but from the chaotic behavior.
In fact, unlike the Gaussian ensemble of random matrices, $e^{\ii \phi} h_{\rm eff}^{\rm X}$ ($\rm X = AIII^{\dagger}, BDI_0, CII_0$) does not belong to the same disorder ensemble of $h_{\rm eff}^{\rm X}$ for generic $\phi \in \mathbb{R}$.
Nevertheless, its hard-edge statistics can still be described by random matrices,
resulting in the emergence of the spectral U(1) symmetry.
Notably, the spectra away from the origin, especially spectral boundaries [see Fig.~\ref{figs: z_dis chiral_AZ0}\,(a), (c), and (e)], do not respect the U(1) symmetry.
Thus, in such systems, an energy threshold $E_c \geq 0$ should exist, akin to the Thouless energy~\cite{JTEdwards1972}, such that the spectra are well described by the random matrices for $|z| \leq E_c$.
In the study of closed quantum systems, the Thouless energy was shown to satisfy $E_c \propto L^{-2}$~\cite{altshuler1986repulsion}, exhibiting the same order as the mean level spacing of two-dimensional systems.
In our open quantum systems studied here, $E_c$ is at least larger than $2 \langle |z_1| \rangle$, while an accurate estimate 
of $E_c$ is left for future study.
It should also be noted that
the emergent spectral U(1) symmetry, as well as the universality of the hard-edge statistics, is exact only in the 
infinite-size limit $L \rightarrow \infty$. 

The eigenstates around the spectral origin of the effective non-Hermitian Hamiltonians $h_{\rm eff}$ correspond to the decay modes in the original Lindbladians. 
Before the spectral shift in Eq.~(\ref{eq: h_eff}), these eigenvalues have non-zero imaginary parts
$-2\gamma$.
In our models, we have $\gamma = 0.1$ for classes AIII$^{\dagger}$ and BDI$_0$, and $\gamma = 0.4$ for class CII$_0$.
Owing to the weak dissipation, $2\gamma$ is much smaller than the bandwidth $\approx 10$ (i.e., range of the spectrum 
along the real axis; see Fig.~\ref{figs: z_dis chiral_AZ0}), and the decay time of these eigenstates 
is
estimated as
$1/(2\gamma)$ 
(the reduced Planck constant is set to $\hbar = 1$). 
Thus, %
this decay should only have a small effect on
the short-time dynamics in the time regime $t \lesssim 1/( 2\gamma)$.
It is also notable that physically relevant initial states 
can
have a large 
overlap with the eigenstates around the spectral origin of the shifted Lindbladians~\cite{kawasaki2022}.
In such a case, the short-time dynamics should be predominantly determined by the level statistics and the properties of eigenstates around the origin.

\subsection{Non-Hermitian superconductors}
\label{subsec: model BdG}

\begin{figure}[h]
  \includegraphics[width=1\linewidth]{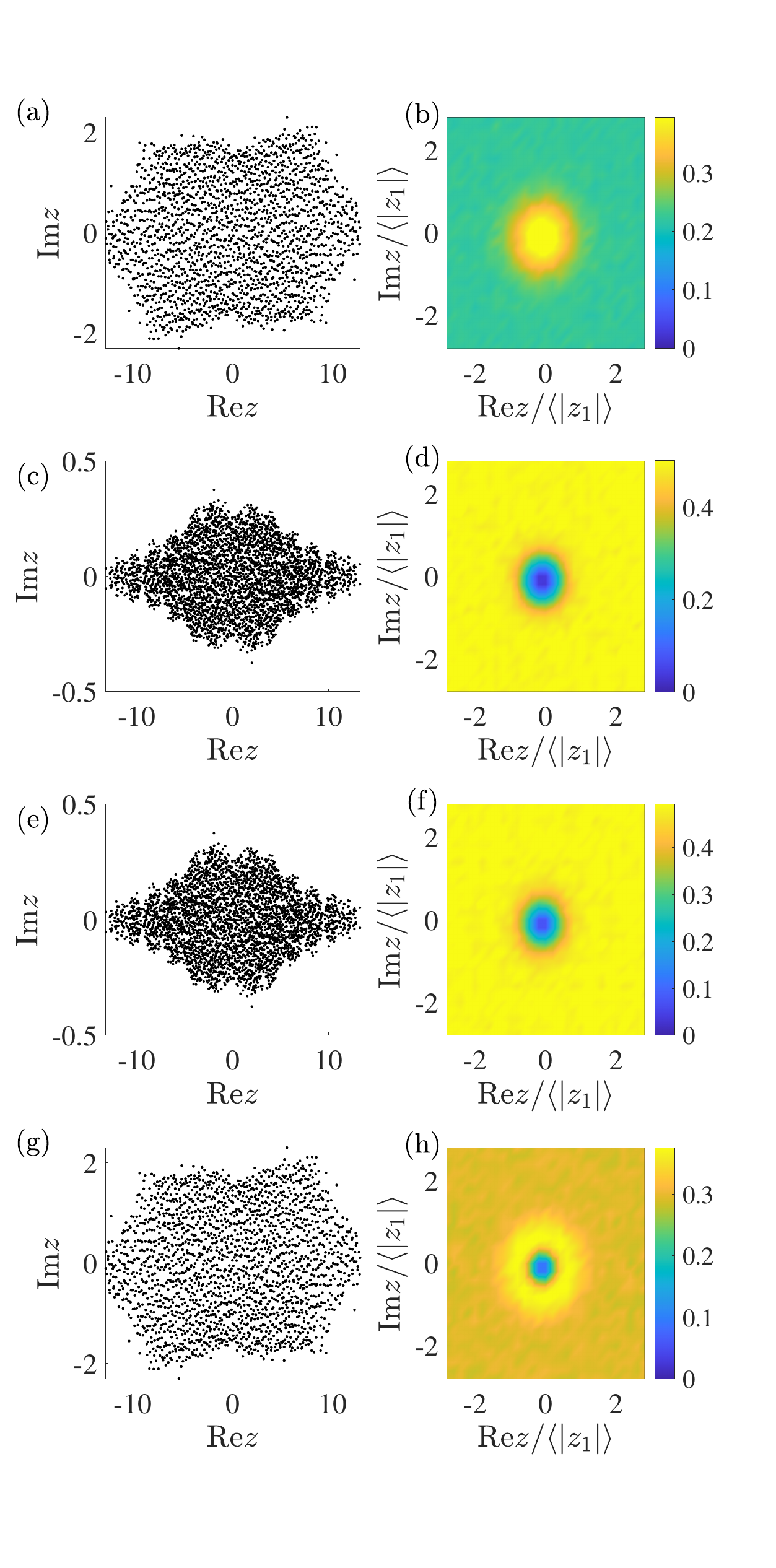}
  \caption{Complex eigenvalues of single realizations of the Hamiltonians 
  of non-Hermitian superconductors in classes (a)~D, (c)~C, (e)~\CIz, and (g)~\DIIIz (see Sec.~\ref{subsec: model BdG}). 
  (b), (d), (f), (h)~Ensemble average of the density of complex eigenvalues around the spectral origin of the same Hamiltonians as the left subfigures.  %
  }
  \label{figs: z_dis BdG_AZ0}
\end{figure}

\begin{figure*}[th]
  \centering
  \includegraphics[width=0.8\linewidth]{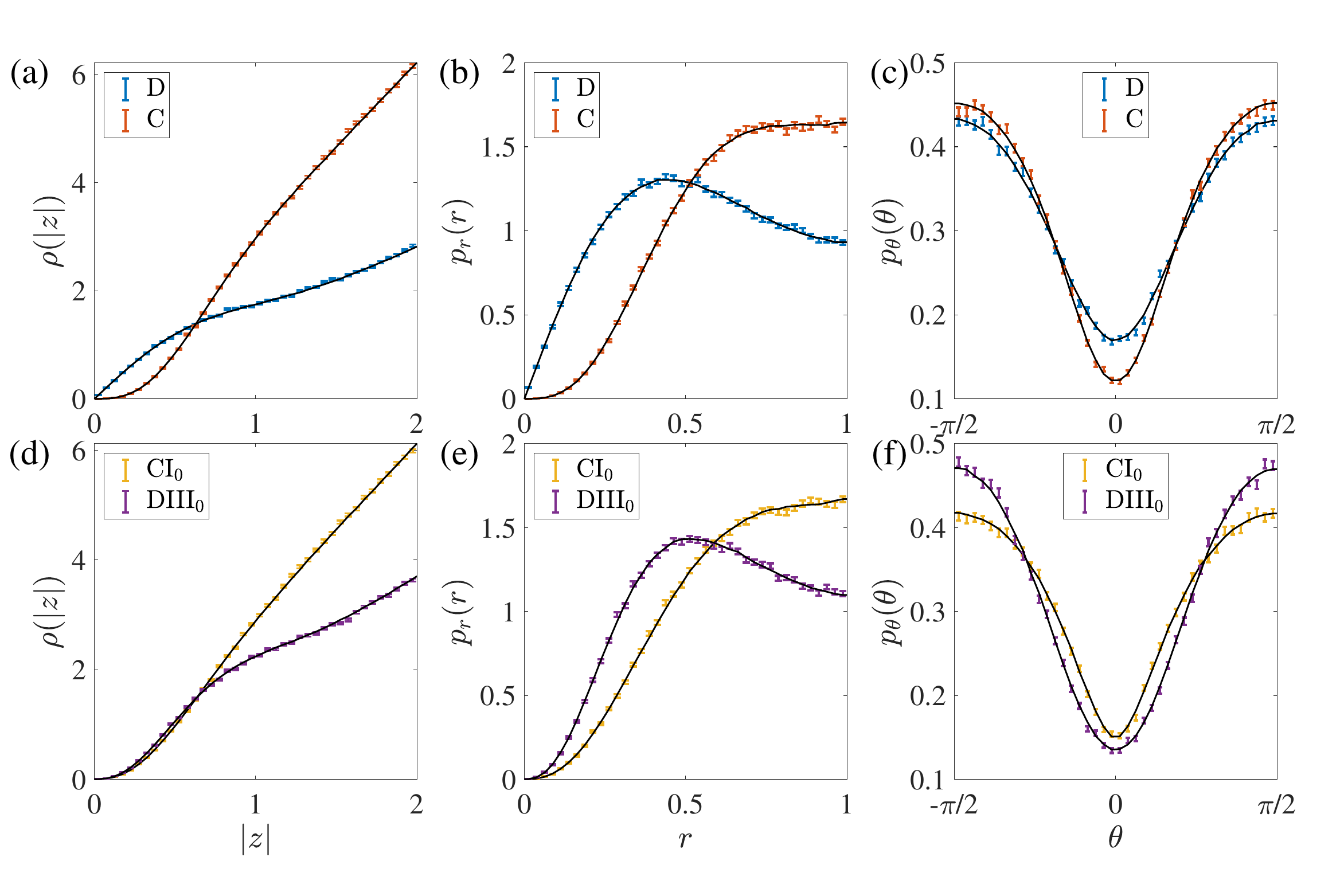}
  \caption{Hard-edge statistics of the Hamiltonians 
  of non-Hermitian superconductors in classes D, C, \CIz, and \DIIIz~(see Sec.~\ref{subsec: model BdG}). 
  (a)~Density $\rho(|z|)$ of the modulus of eigenvalues. 
  (b)~Radial distributions $p_r(r)$ and (c) angular distributions $p_{\theta}(\theta)$ of the complex level ratios $r e^{\ii \theta}$ [Eq.~(\ref{eq: ratio})] around the spectral origin. 
  The colored points with the error bars refer to the Hamiltonians, and the black solid curves are obtained from
  non-Hermitian random matrices in the corresponding symmetry classes.}
  \label{figs: rho_ratio BdG_AZ0}
\end{figure*} 

Superconductors with energy gain and loss are one of the most 
intensively studied systems in the field of non-Hermitian 
physics~\cite{ghatak2018, kawabata19}.
Superconductors attached to leads or reservoirs are also effectively 
described by non-Hermitian Hamiltonians~\cite{pikulin12, sanjose16, cayao23}.
In the following, we study four prototypical models of non-Hermitian superconductors that belong to classes D, C, CI$_{0}$, and DIII$_0$ in the AZ$_0$ classification
and demonstrate its dissipative 
quantum chaos described by non-Hermitian random matrices with symmetry.

(i)~For symmetry class D, we introduce an energy gain and loss into 
a two-dimensional $p$-wave superconductor~\cite{pientka13}, 
\begin{align}  \label{eq: H_D} 
    &H_{\rm D}  = 2 \sum_{\iii} \left[ \left( \epsilon^r_{\iii} + \ii \epsilon^i_{\iii} -\mu\right)  c_{\iii}^{\dagger} c_{\iii} + \left( te^{\ii \phi} c_{\iii+\e_x}^{\dagger} c_{\iii} \right. \right.  \nonumber \\
    & \left. \left. + t c_{\iii+\e_y}^{\dagger} c_{\iii} + \ii \Delta c_{\iii  + \e_x } c_{\iii}   +  \Delta c_{\iii  + \e_y} c_{\iii} + {\rm H.c.}\right)\right] \, .  
\end{align}
Here, $\epsilon^r_{\iii}$ and $\epsilon^i_{\iii}$ are %
random chemical potential and random energy gain or loss that are distributed uniformly and independently in $
[-W_1/2,W_1/2]$ and $[-W_2/2,W_2/2]$, respectively. 
Additionally, $te^{\ii \phi}$ and $t$ represent the hopping amplitudes, and $\Delta$ and $\ii \Delta$ are the $p$-wave superconducting paring potentials.
In the Nambu basis $\Psi_{\iii} = (c_{\iii}, c_{\iii}^{\dagger})^{\rm T}$, this Hamiltonian $H_{\rm D}$ satisfies 
\begin{equation}
    \tau_x H_{\rm D}^{\rm T} \tau_x = - H_{\rm D} \, ,
\end{equation}
where $\tau_x$ is the Pauli matrix acting on the particle-hole space. 
Hence, $H_{\rm D}$ belongs to class D (see Appendix~\ref{app: model detail} for further details). %

(ii)~For symmetry class C, 
we introduce a random imbalanced $s$-wave pairing potential into 
a three-dimensional superconductor with conserved spin,
\begin{align}   
\label{eq: AZ0 model C}
    H_{\rm C} & = \sum_{\iii} \left\{  \Psi_{\iii}^{\dagger} \left[ (  \epsilon_{\iii}-\mu) \tau_z + \ii \Delta^s_{\iii} \tau_y  \right] \Psi_{\iii}  \right. \nonumber \\
  &\left. + \left[ \Psi_{\iii+\e_z}^{\dagger} ( t \cos \phi \tau_z + \ii t \sin \phi \tau_0  + \Delta \tau_y) \Psi_{\iii} \right. \right. \nonumber \\
    & \left. \left. + \sum_{\mu = x,y} \Psi_{\iii+\e_{\mu}}^{\dagger} ( t  \tau_z  +  \Delta \tau_x   ) \Psi_{\iii}+ {\rm H.c.}\right] \right\} \, 
\end{align}
with the Nambu basis $\Psi_{\iii} = (c_{\uparrow},c_{\downarrow}^{\dagger})$. 
Here, $\epsilon_{\iii}$ is a random potential distributed uniformly and 
independently in $[-W_1/2,W_1/2]$.
$t$ and $\phi$ characterize
the hopping amplitude %
and phase, respectively, %
$\Delta \in \mathbb{R}$ represents a Hermitian pairing potential, and $\Delta^s_{\iii} \in \mathbb{R}$ represents the random imbalanced $s$-wave paring potential~\cite{li2018, kornich2022, kornich2022a, kornich23, zhang23a} distributed uniformly in $[-W_2/2,W_2/2]$. 
This Hamiltonian satisfies 
\begin{equation}
    \tau_y H_{\rm C}^{\rm T} \tau_y = - H_{\rm C},
\end{equation}
and hence belongs to class C.

(iii)~For symmetry class CI$_0$, we consider a non-Hermitian superconductor similar to $H_{\rm C}$ in Eq.~(\ref{eq: AZ0 model C}) but with a different pairing potential. 
We denote its Hamiltonian as $H_{\rm CI_0}$; for further details, refer to Appendix~\ref{app: model detail}.

(iv) For symmetry class DIII$_0$, we study a non-Hermitian extension of a three-dimensional time-reversal-invariant superconductor with the spin-orbit coupling~\cite{schnyder08a, schnyder11}.
We denote its Hamiltonian as $H_{\rm DIII_0}$; 
additional information is provided in Appendix~\ref{app: model detail}.

We study single-particle eigenvalues of the BdG Hamiltonians of these four non-Hermitian superconductors numerically with the parameters 
provided in Appendix~\ref{app: model detail}. %
For each Hamiltonian, we diagonalize $2 \times 10^5$ disorder realizations.
Figures~\ref{figs: z_dis BdG_AZ0}\,(a), (c), (e), and (g) show the spectral distributions of single realizations.
The ensemble average of the density of complex eigenvalues around the origin exhibits emergent spectral U(1) rotation symmetry [Figs.~\ref{figs: z_dis BdG_AZ0} (b), (d), (f), and (h)].
Importantly, both the density of the modulus (after the normalization such that $\langle |z_1| \rangle = 1$) and the distributions of the complex level ratio $z_1/z_2$ in Eq.~(\ref{eq: ratio}) match well with those of non-Hermitian random matrices in the same symmetry class, as shown in Fig.~\ref{figs: rho_ratio BdG_AZ0} and Table~\ref{tab: model AZ0}.
These results further demonstrate that the hard-edge statistics of non-Hermitian random matrices 
well describe the universal level statistics
of open quantum chaotic systems.

The random-matrix behavior of the hard-edge statistics in these models suggests
the delocalization of the eigenstates around the spectral origin.
Disorder can drive quadratic non-Hermitian Hamiltonians into the Anderson-localized phases~\cite{hatano1996localization, HatanoNelson97PRB, Efetov97, Xu16, longhi2019, Zeng20, Wang20, huang2020, Huang20SR, kawabata20, Luo21, Luo21TM, luo2021unifying, xiao2023}, 
where eigenvalues should be uncorrelated.
For the quadratic Lindbladians 
$h_{\rm eff}^{\rm X}$ ($\rm X = AIII^{\dagger}, BDI_0, CII_0$) in Sec.~\ref{subsec: quadratic L}, disorder manifests not in on-site potential but in random hopping, resulting in 
SLS and the concomitant 
absence of localization~\cite{Evers-review}.
By contrast, the non-Hermitian superconductors %
studied in this subsection
include on-site disorder. 
Hence, we study their weak disorder regimes and anticipate that they have yet to undergo the Anderson localization, supported by their hard-edge statistics. 
Notably, the finite-size scaling of the hard-edge statistics can capture %
possible Anderson transitions for stronger disorder. 

Additionally, symmetry plays a fundamental role in spectral statistics and Anderson localization. 
In the models studied in this work, the spectral origin exhibits higher symmetry than generic points in the bulk of spectra.
Thus, it is %
plausible
that %
eigenstates around the spectral origin are extended while those away from the origin 
are localized, or vice versa. 
A notable Hermitian example of the former case 
is found in two-dimensional disordered superconductors 
in class D~\cite{Evers-review, Mildenberger07}, 
where %
the spectral statistics exhibit
the random-matrix behavior around the origin but 
become uncorrelated away from the origin. 
We still consider %
such %
models 
with extended states only around the origin %
as chaotic, %
because the extended states may %
contribute significantly to 
chaotic behavior
in quantum transport phenomena.

\subsection{Integrable model: 
non-Hermitian SYK$_2$ model}
\label{subsec: integrable model}

In the preceding sections, we 
classify the universal 
hard-edge
statistics in various open quantum 
chaotic models. 
In this section, on the other hand, we study an open quantum integrable model  
and demonstrate that the hard-edge statistics provide a 
diagnosis for the integrability of non-Hermitian many-body operators.
We study a non-Hermitian extension of the two-body SYK 
model~\cite{Sachdev-Ye-93, kitaev15, Sachdev-15, Polchinski-Rosenhaus-16, Maldacena-Stanford-16, Rosenhaus-review, Sachdev-review}, with the non-Hermitian Hamiltonian~\cite{liu2021, garcia22prl, zhang2021, jian2021, altland2021}
\begin{equation} 
  H_{\rm SYK}^{q = 2} = \sum_{i,j = 1}^N \left( J_{ij} + \ii M_{ij} \right)  \gamma_i \gamma_j \, ,
  \label{eq: SYK q=2}
\end{equation} 
where $\gamma_i$'s are Majorana operators satisfying $ \gamma_i = \gamma_i^{\dagger}$ and $ \{ \gamma_i , \gamma_j \} = \delta_{i,j}$, 
and $N$ is the number of Majorana fermions. 
The coefficients satisfy $J_{ij} = -J_{ji}$ and $M_{ij} = -M_{ji}$, and are independent real Gaussian random variables with zero mean and variance $\sigma^2 = 1/12N$.
The single-particle Hamiltonian $M_{ij} + \ii J_{ij}$ is a non-Hermitian random matrix in class D within the Gaussian ensemble. 
By contrast, all $2^{N/2}$ many-body eigenvalues $z_{\iii}$ are 
derived from the single-body eigenvalues $\lambda_{j}$'s,
\begin{equation} \label{eq: z SYK}
 z_{ \iii } = \sum_{ j = 1}^N \left( n_j - \frac{1}{2} \right) \lambda_j \, ,
\end{equation}
with the single-body eigenvalues $\pm \lambda_1$, $\ldots$, $\pm \lambda_{N/2}$ of $J_{ij} + \ii M_{ij}$, and $n_j = 0$ or 1. 
Consequently, while the single-particle random matrix $M_{ij} + \ii J_{ij}$ is non-integrable, $H_{\rm SYK}^{q = 2}$ is considered integrable as a non-Hermitian many-body operator.

It was found that $H_{\rm SYK}^{q = 2}$ as a many-body Hamiltonian respects PHS with sign $+1$ for $N \in 8 \mathbb{Z}$~\cite{garcia22}.
Additionally, its level statistics in the bulk of the spectrum (i.e., away from the origin or the edge of the spectrum) were found to be accurately described by the Poisson statistics of complex numbers (or equivalently, two-dimensional Poisson statistics).
In the presence of integrability, 
many-body PHS does not necessarily imply that the model 
belongs to symmetry class D. 
Generally, the integrability implies a block-diagonalized structure of many-body Hamiltonians, and their relevant symmetry classes should be determined by symmetries of the reduced blocks. Conversely, the hard-edge statistics can also be used as a diagnostic tool for detecting 
the integrability of many-body Hamiltonians with symmetries. 

\begin{figure*}[htb]
  \centering
  \includegraphics[width=1\linewidth]{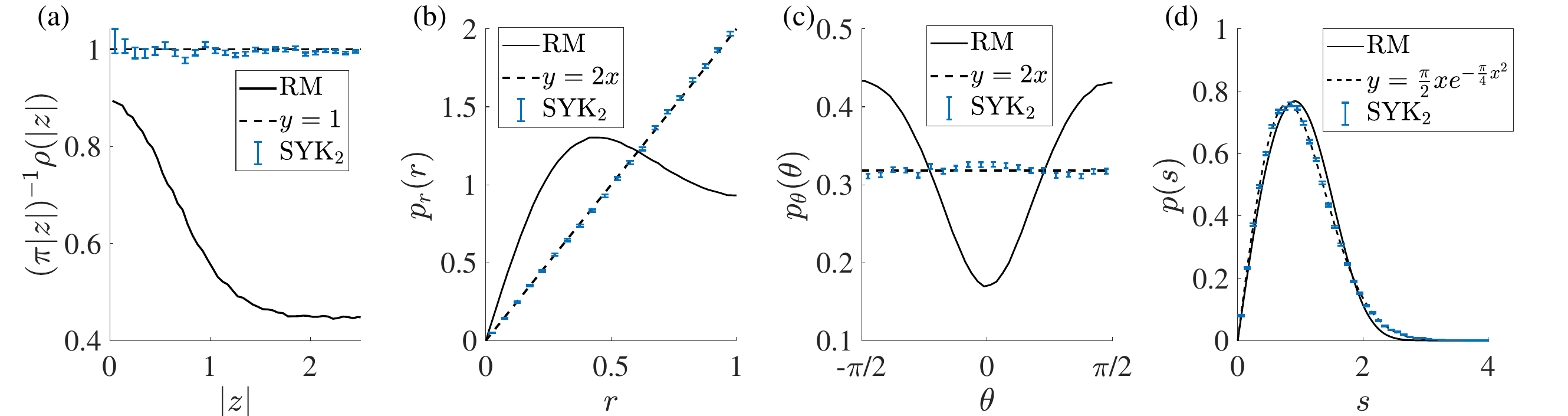}
  \caption{Hard-edge statistics of the two-body non-Hermitian SYK model $ H_{\rm SYK}^{q = 2}$ in Eq.~(\ref{eq: SYK q=2}). 
  (a)~Density of eigenvalues in the complex plane, given by $\rho(|z|)/\pi |z|$. %
  (b)~Radial distributions $p_r(r)$ and (c) angular distributions $p_{\theta}(\theta)$ of the complex level ratios $r e^{\ii \theta}$ [Eq.~(\ref{eq: ratio})] around the spectral origin. 
  (d)~Radial distribution of the eigenvalue $z_{\min}$ with the smallest modulus.
  The points with the error bars refer to the non-Hermitian SYK model with $q = 2$, and the solid curves are obtained from %
  non-Hermitian random matrices (RM) in class D.}
  \label{figs: AZ0 model SYK D}
\end{figure*}

To demonstrate this in the non-Hermitian SYK$_2$ model, we numerically diagonalize $2 \times 10^5$ realizations of $H_{\rm SYK}^{q = 2}$ with $N = 24$,
and %
compare the statistics of its many-body eigenvalues $z_i$'s [Eq.~(\ref{eq: z SYK})] %
around the spectral origin with the hard-edge statistics from non-Hermitian random matrices in class D (Fig.~\ref{figs: AZ0 model SYK D}).
The radial distribution $p_r(r)$ and the angular distribution of the complex level ratio $re^{\ii \theta} \equiv z_1/z_2$ are consistent with the complex Poisson statistics in Eq.~(\ref{eq: 2D Poisson}) [see Figs.~\ref{figs: AZ0 model SYK D}\,(b) and (c)]~\cite{akemann19, huang2020, Luo21}, 
which indicates the absence of level correlations.
The density $\rho(|z|)$ of the modulus of eigenvalues (after the normalization by $\langle |z_1| \rangle = 1$) satisfies [Fig.~\ref{figs: AZ0 model SYK D}\,(a)]
\begin{equation}
\rho(|z|) \simeq \pi |z| \quad \left( |z|  \lessapprox 2 \right) \, ,
\end{equation}
implying that the 
density of complex eigenvalues, given by $\rho(|z|)/\pi|z|$, is %
not higher but
almost constant around the spectral origin.
All these distributions are significantly different from the random-matrix behavior in class D, providing clear evidence of the integrability of the model. 
We also study the statistics of the eigenvalue $z_{\min} \equiv z_1$ with the smallest modulus~\cite{sun20, garcia22}.
Its radial distribution satisfies [Fig.~\ref{figs: AZ0 model SYK D}\,(d)] 
\begin{equation}
p_{\rm min} \left( \left| z_{\min} \right| \right) = \frac{\pi}{2} \left| z_{\min} \right| e^{-\pi \left| z_{\min} \right|^2/4}.
\end{equation}
The distribution conforms to the complex Poisson statistics and deviates from the 
random-matrix statistics in class D, indicating that $p_{\rm min}(|z_{\min}|)$ also 
serves as a tool to detect quantum chaos or its absence.  
However, $p_{\rm min} \left( \left| z_{\min} \right| \right)$ of non-Hermitian
random matrices in class D is close to the complex Poisson statistics 
[Fig.~\ref{figs: AZ0 model SYK D}\,(d)]. Thus, 
a practical use of $p_{\rm min} \left( |z_{\min}| \right)$ for detecting quantum chaos 
is less efficient than the level ratio statistics, 
especially for many-body models, where
numerical calculations 
are constrained by limited system 
size and finite-size effects are substantial.

\section{Open quantum systems beyond the AZ$_0$ and AZ$^{\dagger}$ classification}
\label{sec: quantum system real}

In this section, we study the hard-edge statistics of open quantum systems beyond the AZ$_0$ classification.
In Sec.~\ref{subsec: AZ_real free model}, we study the single-particle spectra of non-interacting systems. 
In Sec.~\ref{subsec: SYK Lindbladian}, by contrast, we consider the many-body spectra of interacting Lindbladians. %
Specifically, we focus on the distributions of the complex eigenvalue $z_{\rm min}$ closest to the spectral origin in the three representative symmetry classes, i.e., classes BDI, CI, and \AIIp.
We show that the numerically obtained distributions of the physical models match well with the distributions of non-Hermitian random matrices in the corresponding symmetry classes.  
Notably, in classes BDI and CI, the distributions of $z_{\min}$ show delta-function peaks on the real and imaginary axes, consistent with the random-matrix behavior.

\subsection{Non-interacting systems}
\label{subsec: AZ_real free model}
\begin{figure*}[htb]
  \centering
  \includegraphics[width=0.8\linewidth]{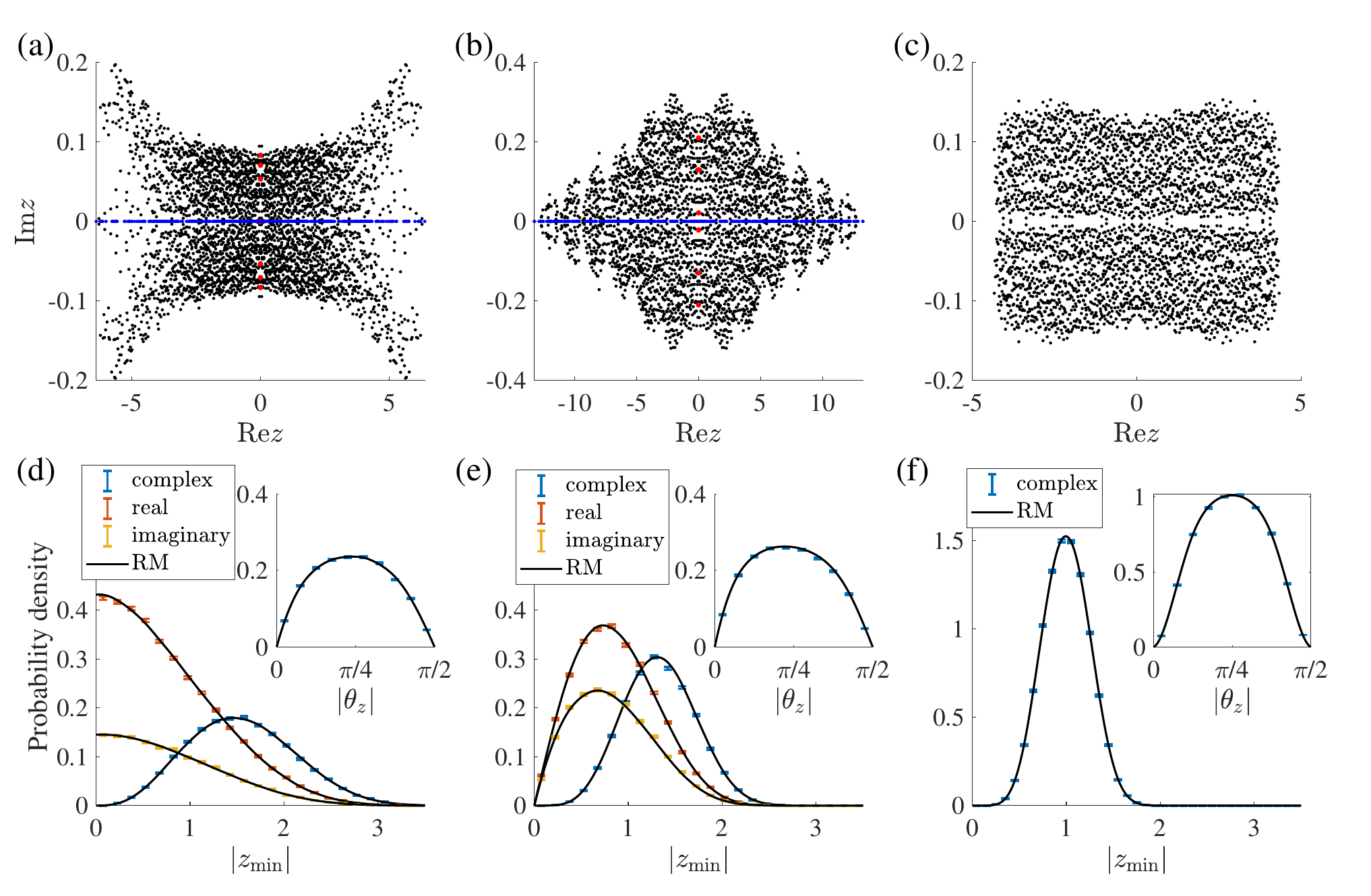}
  \caption{(a)-(c)~Complex eigenvalues of single realizations of the non-Hermitian Hamiltonians
(see Sec.~\ref{subsec: AZ_real free model}).
  Complex, real, and purely imaginary eigenvalues are denoted by the black, blue, and red points, respectively.
  (d)-(f)~Distributions of the complex eigenvalue $z_{\min}$ with the smallest modulus. 
  The blue, orange, and yellow points with the error bars represent the radial distributions of $z_{\min}$ when it is complex, real, and purely imaginary, respectively. 
  Inset:~angular distribution of complex $z_{\min}$. 
  The solid curves correspond to non-Hermitian random matrices (RM).
  (a), (d)~Effective non-Hermitian Hamiltonian $h_{\rm eff}^{\rm BDI}$
  for class BDI. 
  (b), (e)~Hamiltonian $H_{\rm CI}$ %
  of the non-Hermitian superconductor for class CI. 
  (c), (f)~Effective non-Hermitian Hamiltonian $h_{\rm eff}^{{\rm AII} + \mathcal{S}_+}$
  for class \AIIp. 
  }
  \label{figs: AZ_real model}
\end{figure*}

We introduce three non-interacting models in classes BDI, CI, and \AIIp, respectively. 
For class BDI, we add the following dissipators 
\begin{equation} \label{eq: L BDI}
L_{\iii} =  \sqrt{{\gamma}} (c_{\iii} + \ii c_{\iii + 2 \e_z})
\quad \left( \iii \in \mathbb{Z}^3 \right),   
\end{equation}
to the random-hopping model $H_{\rm BDI}$ [i.e., Eq.~(\ref{eq: H_chiral_class}) with $t_{\iii, \jjj} \in \mathbb{R}$]. 
For class \AIIp, we add the following dissipators 
\begin{equation} \label{eq: L AIIp}
L_{\iii  \sigma }=  \sqrt{{\gamma}} 
(c_{\iii \sigma} 
+ \ii c_{(\iii + \e_z) \sigma}
) \quad \left( \iii \in \mathbb{Z}^3 , \sigma = \uparrow, \downarrow \right)
\end{equation}
to the SU(2) model $H_{\rm CII}$ [i.e., Eq.~(\ref{eq: H_chiral_class}) with $t_{\iii, \jjj} \in \rm SU(2)$].
The effective non-Hermitian Hamiltonians of these quadratic Lindbladians in Eq.~(\ref{eq: h_eff}) %
are denoted by $h_{\rm eff}^{\rm BDI}$ and $h_{\rm eff}^{{\rm AII} + \mathcal{S}_+}$, respectively. 
Here, $h_{\rm eff}^{\rm BDI}$ satisfies TRS with sign $+1$ and CS; 
$h_{\rm eff}^{{\rm AII} + \mathcal{S}_+}$ satisfies TRS with sign $-1$ and SLS (see Appendix~\ref{app: model detail} for details). %

As a non-Hermitian model in class CI, we investigate a three-dimensional non-Hermitian superconductor similar to Eq.~(\ref{eq: AZ0 model C}) but with a different form of imbalanced pairing potentials.
We refer to its Hamiltonian as $H_{\rm CI}$, and the details can be found in Appendix~\ref{app: model detail}.

For each symmetry class, we numerically diagonalize $2\times 10^5$ samples of 
Hamiltonians %
with the parameters provided in Appendix~\ref{app: model detail}. %
For the models in classes BDI and CI, a subextensive number of eigenvalues are real or purely 
imaginary
[Figs.~\ref{figs: AZ_real model}\,(a) and (b)]. 
The probability of $z_{\min}$ being real or purely imaginary coincides with that of non-Hermitian random matrices in the same symmetry classes (see Table~\ref{tab: model AZ_real}).
For the model in class \AIIp, by contrast, all eigenvalues are complex, also consistent with the random-matrix behavior [Fig.~\ref{figs: AZ_real model}\,(c)]. 
Furthermore, the distributions of $z_{\min}$, including the radial distributions of real, purely imaginary, and complex $z_{\min}$, and the angular distributions of complex $z_{\min}$, in all these models are well described by non-Hermitian random matrices [Figs.~\ref{figs: AZ_real model}\,(d)-(f)].
In Table~\ref{tab: model AZ_real}, we compare 
$\langle |z_{\min}|^2 \rangle$
from non-Hermitian random matrices and 
that from the non-Hermitian Hamiltonians. The values are consistent among different systems in the same symmetry class.

\begin{table}[bt]
  \caption{Mean value $\langle  |z_{\min}| \rangle$ and probability of $z_{\min}$ being real or purely imaginary for non-Hermitian random matrices, the Hamiltonians in %
  Sec.~\ref{subsec: AZ_real free model}, %
  and 
  the SYK Lindbladians in %
  Sec.~\ref{subsec: SYK Lindbladian}. %
  Here, $z_{\min}$ denotes the complex eigenvalue with the smallest modulus.}
  \begin{tabular}{c|ccc}
  \hline \hline
  \makecell[c]{Symmetry class \\ and systems} & $\langle |z_{\rm min}|^2 \rangle$ & Pr($z_{\min} \in \mathbb{R}$) &  Pr($\ii z_{\min} \in \mathbb{R}$) \\ \hline
Class BDI & & \\  \hline
Random matrices &  $1.4488(6)$ & $0.5373(2)$ & $0.1960(2)$ \\
      $h_{\rm eff}^{\rm BDI}$ [Eq.~(\ref{eq: h_BDI})] & $1.447(4)$ & $0.536(2)$ &  $0.196(1)$ \\
     SYK Lindbladian & $1.448(6)$ & $0.537(2)$ &  $0.198(2)$   \\ 
\hline
Class CI & & \\  \hline
Random matrices   & $1.2223(4)$ &$0.4247(2)$ & $0.2732(2)$  \\
 $H_{\rm CI}$ [Eq.~(\ref{eq: AZ_real model CI})] & $1.223(3)$ & $0.424(2)$ &  $0.277(2)$ \\
SYK Lindbladian & $1.226(4)$ & $0.425(2)$ &  $0.272(2)$   \\ 
\hline
Class \AIIp & & \\  \hline
Random matrices & $1.0623(2)$ & $ 0 $ & $ 0 $ \\
 $H_{\rm eff}^{{\rm AII} + \mathcal{S}_+}$ [Eq.~(\ref{eq: h_AII_p})] & $1.062(2)$ & $ 0 $ &  $ 0 $ \\
\hline  
\end{tabular}
  \label{tab: model AZ_real}
  \end{table}

\subsection{Many-body Lindbladians}
\label{subsec: SYK Lindbladian}

\begin{figure*}[htb]
  \centering
  \includegraphics[width=1\linewidth]{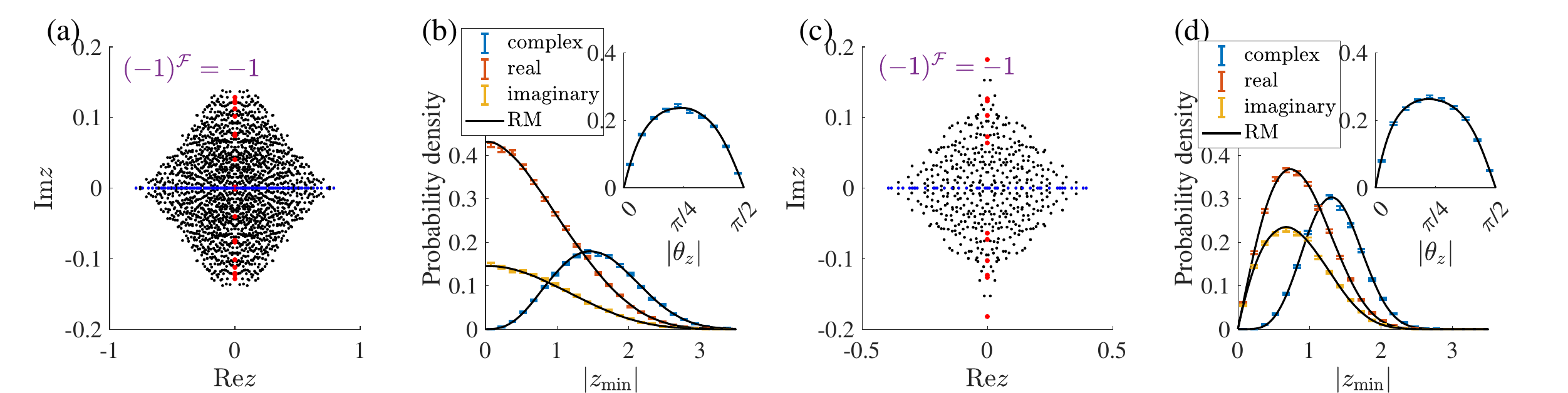}
  \caption{Complex eigenvalues in the symmetry sector $(-1)^{\mF} = -1$ of single realizations of the many-body Lindbladians 
  in (a)~class BDI ($N=12$)
  and 
  (b)~class CI ($N=10$).
  Complex, real, and purely imaginary eigenvalues are denoted by the black, blue, and red points, respectively.
 The distributions of the complex eigenvalue $z_{\min}$ with the smallest modulus for the 
 shifted many-body Lindbladians $\mathcal{L} - {\rm Tr} \mathcal{L}/{\rm Tr} 1$ in (b)~class BDI and (d)~class CI.
  The blue, orange, and yellow points with the error bars represent the radial distributions of $z_{\min}$ when it is complex, real, and purely imaginary, respectively. 
  Inset:~angular distribution of complex $z_{\min}$. 
  The solid curves correspond to non-Hermitian random matrices (RM).}
  \label{figs: AZ_real model many body}
\end{figure*} 

As a prototypical physical model with strongly correlated fermions coupled to the environment, we study the Lindbladian $\mathcal{L}$ for the SYK model~\cite{sa2022SYK, kulkarni2022SYK, kawabata2023, sa2023SYK}.
The Hamiltonian is given by the original SYK model consisting of the $q$-body 
random interactions,
\begin{equation} 
    \label{eq: SYK_q}
  H = \ii^{\frac{q}{2}} \sum_{1 \leq i_1  < \cdots < i_q \leq N}
  J_{i_1i_2\cdots i_q } \gamma_{i_1 } \cdots \gamma_{i_q}\, ,  
\end{equation}
where $\gamma_i$'s are Majorana operators, $N$ is the number of Majorana fermions, and $J_{i_1i_2\cdots i_q}$'s are independent real Gaussian variables with zero means and $\langle J_{i_1i_2\cdots i_q}^2 \rangle = ((q-1)!/N^q)\,J^2$. 
Then, we add the following linear dissipators $L_m$, 
\begin{equation} \label{eq: SYK_L}
  L_m = \sum_{1 \leq i \leq N} K_{m;i} \gamma_{i } \quad (m = 1, 2, \cdots, M),
\end{equation}  
where $K_{m;i}$'s are independent complex Gaussian variables with zero means and $\left\langle |K_{m;i}|^2  \right\rangle = ((q-1)!/N^q)\,K^2$.

As discussed in Sec.~\ref{subsec: Lindblad}, after the vectorization of the density matrix, the Lindbladian $\mathcal{L}$ becomes a non-Hermitian operator in the double Hilbert space.
It can be represented with the basis of adjoint fermions~\cite{costa2023} (see also Appendix~\ref{app: SYK Lindbladian} for details), facilitating analytic analyses.
The SYK Lindbladian $\mathcal{L}$ respects a unitary symmetry
\begin{equation}
    \left( -1 \right)^{\mF} \mathcal{L} \left( -1 \right)^{\mF} = \mathcal{L},
\end{equation}
where $(-1)^{\mF}$ is the total fermion parity operator 
in the double Hilbert space.
For $q = 6$, the traceless Lindbladian $\mathcal{L} - {\rm Tr} \mathcal{L}/{\rm Tr} 1$ belongs to symmetry class BDI for $N \in 4 \mathbb{Z}$ and class CI 
for $N \in 4 \mathbb{Z} + 2$ in each sector of 
total fermion parity $(-1)^{\mF} = \pm 1$ (see Ref.~\cite{kawabata2023} 
and Appendix~\ref{app: SYK Lindbladian} for derivations).

We numerically diagonalize $10^5$ realizations of the traceless 
Lindbladians with the following two sets of parameters: 
(i)~Class BDI.---$N = M = 12$, $J = 1$, and $K = 0.4$;
(ii)~Class CI.---$N = M = 10$, $J = 1$, and $K = 0.3$.  The SYK Lindbladians exhibit distinct distributions of $z_{\min}$ for different $N$ [Figs.~\ref{figs: AZ_real model many body}\,(b), (d)], consistent with non-Hermitian random matrices in the corresponding symmetry classes (see Table~\ref{tab: model AZ_real}). 
As also discussed in Sec.~\ref{subsec: quadratic L}, the hard-edge statistics should be relevant to the short-time dynamics of the SYK Lindbladians.

\section{Conclusions and Discussions}
    \label{sec: conclusion}

In 28 out of 38 symmetry classes of non-Hermitian random matrices, we find that the spectral origin in the complex plane respects higher symmetry compared to other points.
We comprehensively investigate the hard-edge statistics in these 28 symmetry classes.
This work, combined with the prior works on the threefold universal statistics in the bulk~\cite{hamazaki20} and the sevenfold universal statistics on the real axis~\cite{xiao22}, forms a foundational comprehension of 
non-Hermitian random matrix theory across all the 38 symmetry classes.
In the seven symmetry classes of the AZ$_0$ classification, 
we introduce complex level ratios around the spectral origin and 
analyze the universality classes of the hard-edge statistics. 
We numerically obtain the seven universal ratio distributions, with both 
radial and angular components exhibiting the characteristic behavior.
In 21 symmetry classes beyond the AZ$_0$ classification, we utilize the distributions 
of complex eigenvalues $z_{\min}$ with the smallest modulus to characterize the hard-edge 
statistics. In certain symmetry classes, $z_{\min}$ can be real or purely imaginary 
with the universal probability. This probability serves as practically useful indicators for determining universality classes. 
Additionally, we perform analytic calculations for small random matrices and elucidate the essential features of our numerical results.

The results of this paper also serve as a diagnostic tool of non-integrability or its absence in open quantum systems with symmetry, as well as the transitions between them. 
We investigate diverse physical models in the seven AZ$_0$ symmetry classes and 
the three representative symmetry classes beyond the AZ$_0$ classification. 
In contrast to dense random matrices in the Gaussian ensembles, ensembles of physical Hamiltonians or Lindbladians are regarded as ensembles of sparse matrices because of the locality constraint.
Nevertheless, the random-matrix hard-edge statistics emerge in both physical Hamiltonians and Lindbladians, showcasing their universality. 
Notably, in the presence of symmetry, the spectral statistics around the origin can exhibit distinct behavior compared to that in the bulk.
In such systems, chaos and integrability cannot be diagnosed solely by the bulk statistics, and thus both bulk and hard-edge statistics are significant.
Prime examples in closed quantum systems include two-dimensional disordered superconductors, in which only zero modes can be free from the Anderson localization and exhibit quantum chaotic behavior~\cite{Mildenberger07,Evers-review}.
Examining the spectral statistics in both the bulk and around the origin also provides insights into fundamental questions regarding the impact of symmetry on the universality classes of chaos-integrability transitions.

Symmetry of ensembles of physical %
systems
is generally lower than %
symmetry of the Gaussian random-matrix ensembles.
However, the hard-edge statistics of physical systems conform to the random-matrix behavior, which implies 
the emergent spectral U(1) or D$_4$ symmetry of the ensemble-average distributions of the complex eigenvalues around the spectral origin.
It is worth mentioning that the ensemble-average distributions of real eigenvalues of chaotic Hermitian physical systems in the standard symmetry classes (i.e., classes A, AI, and AII) should also be symmetric about the origin %
within several mean level spacings, %
provided that 
the origin is away from the edge of the spectra. 
Since generic Hamiltonian ensembles 
of Hermitian systems
do not have %
such spectral inversion symmetry, this %
implies emergent $\mathbb{Z}_2$ symmetry.
Nonetheless, the spectral origin %
of Hermitian systems in the standard 
symmetry classes %
does not %
possess
any higher symmetry than generic eigenvalues, 
and the $\mathbb{Z}_2$ symmetry arises because the density of eigenvalues 
should be almost constant within the range of several mean level spacings.
By contrast, the emergent symmetry in the non-Hermitian systems studied in this work is distinct and more intricate than the Hermitian case.
The spectral origin of these non-Hermitian systems
exhibits higher symmetry, and the density around the origin displays a non-constant 
but D$_4$-symmetric (or U(1)-symmetric) distribution within 
the range of several mean level spacings.

It is interesting to explore alternative characterizations of hard-edge statistics and their applicability to physical systems. 
For example, in Appendix~\ref{app: sec alternative}, for non-Hermitian random matrices in classes BDI, CI, and \AIIp, 
we consider the distributions of complex, real, and purely imaginary eigenvalues, separately. 
Despite slower convergence with respect to the matrix size, these distributions also appear to 
be universal and align with physical Hamiltonians. 
Other potential characterizations, 
including %
the level number variance within a fixed range around the spectral origin %
and $n$-point correlation functions~\cite{mehta2004}, 
are also left for future study.

The symmetry classification of open quantum systems is not limited to Hamiltonians or Lindbladians but also applies to dynamical generators of the non-Markovian 
dynamics~\cite{deVega-RMP17}. 
Investigating and comparing their hard-edge statistics with our results can provide 
insights into these less-explored, but physically relevant, open quantum systems. 
The applications of non-Hermitian random matrix theory in 
classical systems, including complex networks with symmetry, 
also merit future studies.

\begin{acknowledgments}
We thank Tomi Ohtsuki for helpful discussions and valuable comments on the manuscript.
Z.X. thanks Lingxian Kong for helpful comments.
Z.X. and R.S. are supported by the National Basic Research Programs of China (No.~2019YFA0308401) and by the National Natural Science Foundation of China (No.~11674011 and No.~12074008).
K.K. is supported by MEXT KAKENHI Grant-in-Aid for Transformative
Research Areas A ``Extreme Universe" No.~24H00945.
\end{acknowledgments}

\appendix

\section{Form of non-Hermitian random matrices in each symmetry class}
\label{app: form RM}

We provide explicit forms of non-Hermitian random matrices in chiral and BdG symmetry classes in the AZ$_0$ classification, whose spectral statistics are investigated in Sec.~\ref{sec: RM AZ0}.
Below, $\tau_{\mu}$ and $\sigma_{\mu}$ $(\mu = 0,x,y,z)$ represent Pauli matrices.
A generic non-Hermitian random matrix $H$ in class D with $\mC_- = \sigma_0$ is an anti-symmetric complex matrix, satisfying 
\begin{equation} 
 H = -H^{\rm T} \, .
\end{equation} 
A generic non-Hermitian random matrix $H$ in class C with $\mC_- = \sigma_y$ is given as 
\begin{equation} 
 H = \begin{pmatrix} 
   A & B \\ C & -A^{\rm T} \\
 \end{pmatrix}  \text{ with } B = B^{\rm T}, \, C = C^{\rm T} \, .
\end{equation}
A generic non-Hermitian random matrix $H$ in class \DIIIz\ is given as
\begin{equation} 
  H = \begin{pmatrix} 
    0 & A \\ B & 0
  \end{pmatrix} \text{ with } A =- A^{\rm T}, \, B = -B^{\rm T} \, ,
\end{equation}
where the symmetry operators are chosen as $\mS = \sigma_z$ and 
$\mC_+%
= \sigma_y$.
A generic non-Hermitian random matrix $H$ in class \CIz\ is given as
\begin{equation} 
  H = \begin{pmatrix} 
    0 & A \\ B & 0
  \end{pmatrix} \text{ with } A = A^{\rm T}, \, B = B^{\rm T} \, ,
\end{equation}
with the symmetry operators $\mS = \sigma_z$ and 
$\mC_+%
= \sigma_x$.
A generic non-Hermitian random matrix $H$ in class \AIIIz\ is given as
\begin{equation} 
  H =  \begin{pmatrix} 
    0 & A \\ B & 0
  \end{pmatrix} \, ,
\end{equation}
with $\mS = \sigma_z$.
A generic non-Hermitian random matrix $H$ in class \BDIz\ is given as
 \begin{equation} 
  H = \begin{pmatrix} 
    0 & A \\ A^{\rm T} & 0
  \end{pmatrix} \, ,
\end{equation}
with $\mS = \sigma_z$ and 
$\mC_+%
= \sigma_0$.
A generic non-Hermitian random matrix $H$ in class \CIIz\ is given as
\begin{equation} 
  H = \begin{pmatrix} 
     0 & 0 & A & B \\
     0 & 0 & C & D \\
     D^{\rm T} & -B^{\rm T} & 0 & 0  \\
     -C^{\rm T} & A^{\rm T} & 0 & 0 & 
  \end{pmatrix}  \text{ with } A_{ij},B_{ij},C_{ij},D_{ij} \in \mathbb{C} \, ,
\end{equation} 
where the symmetry operators are chosen as $\mS = \tau_z \otimes \sigma_0$ and $\mC_+ %
= \tau_0 \otimes \sigma_y$.

Additionally, we give explicit forms of non-Hermitian random 
matrices in classes BDI, CI, and \AIIp, whose 
spectral statistics are investigated in Sec.~\ref{sec: RM AZ_real}. 
A generic random matrix $H$ in class BDI is given as 
\begin{align}
      H = \begin{pmatrix} 
     A & B \\
    B^{\rm T} & C \\
  \end{pmatrix} \, &\text{with } A = -A^{\rm T}, \, C = -C^{\rm T} , \nonumber \\
  &\quad \text{and } A_{ij},B_{ij},C_{ij} \in \mathbb{R} \, ,
\end{align} 
where the symmetry operators are chosen as $\mT_+ = \tau_0$ and $\mC_- = \tau_z$.
A generic non-Hermitian random matrix $H$ in class CI is given as 
\begin{align} 
    \label{eq: CI def}
    H = \begin{pmatrix} 
    A & B \\
    C & -A^{\rm T}\\
    \end{pmatrix}  \, & \text{with } B = B^{\rm T}, \, C = C^{\rm T} ,\, \nonumber \\
    & \text{and } A_{ij},B_{ij}, C_{ij} \in \mathbb{R} \, ,
\end{align}
with $\mT_+ = \tau_0$ and $\mC_- = \tau_y$.
A generic non-Hermitian random matrix $H$ in class \AIIp\ is given as 
\begin{equation} 
  H = \begin{pmatrix} 
     0 & 0 & A & B \\
     0 & 0 & -B^* & A^* \\
     C & D & 0 & 0  \\
     -D^* & C^* & 0 & 0 & 
  \end{pmatrix}  \text{ with } A_{ij},B_{ij},C_{ij},D_{ij} \in \mathbb{C} \, ,
\end{equation} 
with $\mT_+ = \tau_0 \otimes \sigma_y$ and $\mS = \tau_z \otimes \sigma_0$.

\section{Analytic results}
\label{app: analytic}
\subsection{Level-ratio distributions around the spectral origin}

\subsubsection{Class \AIIIz}

We consider $4 \times 4$ non-Hermitian random matrices in the Gaussian ensemble in class \AIIIz. 
Let $\pm z_1, \pm z_2$ be its eigenvalues. 
We also introduce $y_i \equiv z_i^2$ ($i = 1,2$).
The joint probability density function %
$p_{\rm joint}(y_1,y_2)$ %
of $y_1,y_2$ is given as~\cite{osborn2004} 
\begin{equation} 
  p_{\rm joint}(y_1,y_2)  \propto K_0(|y_1|) K_0(|y_2|) |y_1 - y_2|^2 \, ,
\end{equation}
where $K_0(\cdot)$ is the modified Bessel function of the second kind. 
With $y_i \equiv |y_i| e^{\ii \phi_i}$, the probability density function for $|y_i|,\phi_i$ is given as 
\begin{align} 
   & p_{\rm joint}(|y_1|,\phi_1,|y_2|,\phi_2 ) \propto K_0(|y_1|) K_0(|y_2|) |y_1||y_2| \nonumber \\
   &\qquad\quad \times \left[|y_1|^2 + |y_2|^2 - 2 |y_1||y_2| \cos(\phi_1 - \phi_2)  \right] \, . 
  \label{app eq: jpdf AIII0}
\end{align}
Integrating $p_{\rm joint}(|y_1|,\phi_1,|y_2|,\phi_2 )$ over $\phi_1,\phi_2$, we have
\begin{equation} 
  p_{\rm joint}(|y_1|,|y_2| ) \propto K_0(|y_1|) K_0(|y_2|) |y_1||y_2| \left(|y_1|^2 + |y_2|^2 \right)  \, .
\end{equation}
Let us introduce $|y_2| = R |y_1|$. 
Using $ d|y_1| d|y_2| = |y_1| d|y_1| dR$, we further have
\begin{equation} 
  p_{\rm joint}(|y_1|,R ) \propto K_0(|y_1|) K_0(R|y_1| ) 
  |y_1|^{4} %
  R \left( 1 + R^2\right)\, .
\end{equation}
We then integrate $p_{\rm joint}(|y_1|,R )$ over $|y_1|$, %
resulting in the distribution function of $R$ %
\begin{equation} 
  p_R(R) = \frac{4 R \left(R^2+1\right) \left(-3 R^4+2 \left(R^4+4 R^2+1\right) \ln R+3\right)}{\left(R^2-1\right)^5} \, .
\end{equation}
The distribution function of the level ratio $r \equiv |z_1/z_2 | = \sqrt{R}$ is given as
\begin{align}
      & p_r^{N = 4}(r) \nonumber \\
       =&  \frac{16 r^3 \left(r^4+1\right) \left(-3 r^8+4 \left(r^8+4 r^4+1\right) \ln r+3\right)}{\left(r^4-1\right)^5} \, . 
\end{align}
Here, an extra factor of two is due to the assumption $|z_1| \leq |z_2|$.
For $r \ll 1$, we have $\rho_r(r) \propto -r^3 \ln r$, which is consistent with the numerical results from large random matrices (see Table~\ref{tab: AZ0 classification}).  

We can also integrate $p_{\rm joint}(|y_1|,\phi_1,|y_2|,\phi_2 )$ in Eq.~(\ref{app eq: jpdf AIII0}) over $ |y_1|,|y_2|$, leading to
\begin{equation} 
  p_{\rm joint}(\phi_1,\phi_2 ) \propto 8 - \frac{\pi^2}{2} \cos(\phi_1 - \phi_2) \,.
\end{equation} 
Since the phase $\theta$ of the level ratio $z_1/z_2$ equals $ (\phi_1 - \phi_2)/2$, its distribution function is obtained as
\begin{equation}
    p_{\theta}^{N = 4} = \frac{16 - \pi^2 \cos 2\theta}{16\pi} \,.
\end{equation}
The angular distribution $p_{\theta}^{N = 4}$ reaches its maximum and minimum at $\theta = \pm \pi/2$ and $\theta = 0$, respectively, consistent with the numerical results of large random matrices [see Fig.~\ref{figs: rho_ratio chiral_AZ0}\,(c)].

\subsubsection{Class D}
A $4 \times 4$ non-Hermitian random matrix $H$ in class D, satisfying $H = - H^{\rm T}$, can be given as 
\begin{equation}
    H =  \sum_{\mu = 0,x,z} \left(a_{\mu} \tau_{\mu} \sy + b_{\mu} \ty \sigma_{\mu}\right) \quad 
    \label{app eq: RM_D_4}
\end{equation}
with $a_{\mu},b_{\mu} \in \mathbb{C}$.
In the Gaussian ensemble with the probability density $p(H) \propto e^{-\beta\,{\rm Tr}\,(H^{\dagger}H)}$, $a_{\mu}$ and $b_{\mu}$ are independent complex Gaussian variables with zero means.  
The four eigenvalues of $H$ are 
\begin{equation}    
    \pm z_1 = \pm \left( w_1 - w_2 \right),\quad
    \pm z_2 = \pm \left(  w_1 + w_2 \right)
\end{equation}
with
\begin{equation}
    w_1 \equiv \sqrt{b_0^2 + a_x^2 + a_z^2},\quad w_2 \equiv \sqrt{a_0^2 + b_x^2 + b_z^2} \, .
\end{equation} 
Notably, $w_1 $ and $w_2 $ are independent complex random variables with the identical probability distribution in the complex plane~\cite{hamazaki20},
\begin{equation}
    p_w(w) = \frac{1}{\pi} |w|^2 e^{-|w|^2}  \, .
\end{equation} 
The radial distribution of $w$ is given as
\begin{align}
    p_w^{(r)}(|w| = s) &= \int 
    p_w(w) \delta( |w| - s) dw dw^* \nonumber \\
    &= 2 |s|^3 e^{-|s|^2} \, .
\end{align}
The radial distribution of $t \equiv w_1/w_2$ is given as
\begin{align}
    p_t^{(r)}(|t| = x) &= \int p_{w}^{(r)} (|w| = s_1) p_{w}^{(r)} (|w| = s_2) \nonumber \\
    &\qquad\qquad\qquad \times \delta( s_1/s_2 - x ) ds_1 ds_2 \nonumber \\ 
    & = \frac{12 x^3}{(1+x^2)^4} \, .
\end{align}
Thus, the probability density of $t$ in the complex plane is 
\begin{equation}
  p_t (t) = \frac{6 |t|^2}{\pi (1+|t|^2)^4}  \, .
\end{equation} 
The complex level ratio $v \equiv r e^{\ii \theta} \equiv z_1/z_2 = (1 + t)/(1 - t)$ is a meromorphic function of $t$.
Then, the probability density of $v$ in the complex plane is given as
\begin{align}
  p_v (v) &= \frac{6 |t|^2}{\pi (1+|t|^2)^4}   \left \lvert \frac{d t}{d v} \right \rvert^2 \nonumber \\
         & =  \frac{24 }{\pi} \frac{\left( 1 + r^2 + 2r \cos \theta \right)\left( 1 + r^2 - 2r \cos \theta \right)}{  \left( 2 + 2r^2 \right)^4} \, .
\end{align}
Integrating $ p_v (v)$ over $\theta$ or $r$, we obtain the radial and angular distributions of
the complex level ratio $v$ as
\begin{align} 
  p_r^{N = 4}(r) & = \frac{6(1 + r^4)r}{(1+r^2)^4} \, , \\
  p_{\theta}^{N = 4}(\theta) & =  \frac{2 - \cos 2 \theta}{4 \pi} \, ,
\end{align} 
respectively.
For $r \ll 1$, we have $p_r^{N = 4}(r) \propto r$, and $p_{\theta}^{N = 4}$ reaches its maximum and minimum at $\theta = \pm \pi/2$ and $\theta = 0$, respectively, consistent with numerical results from large random matrices [see Table~\ref{tab: AZ0 classification} and Fig.~\ref{figs: rho_ratio chiral_AZ0} (c)].

\subsection{Density of complex eigenvalues}

\subsubsection{Classes D and \BDIz}

A generic $2 \times 2$ non-Hermitian matrix $H$ in class D, which 
respects PHS $H^{\rm T} = - H$, is given as
\begin{equation}
H = a_y \sigma_y \quad
\left( a_y \in \mathbb{C} \right).
    \label{eq: 2*2-D}
\end{equation}
The two complex eigenvalues of $H$ are obtained as $\pm z = \pm a_y$.
Notably, $\left| z \right| = \left| a_y \right|$ is
essentially
equivalent to the level spacing between the two eigenvalues of $2 \times 2$ Hermitian matrices in class AI. 
Consequently, the probability distribution function $\rho \left( \left| z \right| \right)$ is the same as the level-spacing distribution for $2 \times 2$ Hermitian random matrices in class AI. 

Suppose that $H$ obeys the Gaussian probability distribution $\propto e^{-\beta\,\mathrm{Tr}\,( H^{\dag} H )/2} = e^{-\beta\,( y_{\rm r}^2 + y_{\rm i}^2 )}$ with a constant $\beta > 0$ and $y_{\rm r} \equiv \mathrm{Re}\,a_y$, $y_{\rm i} \equiv \mathrm{Im}\,a_y$. 
Then, the probability distribution function $\rho \left( \left| z \right| \right)$ reads
\begin{align}
\rho \left( \left| z \right| \right) 
&= \frac{\beta}{\pi} \int_{-\infty}^{\infty} dy_{\rm r} 
dy_{\rm i}~\delta \left( \left| z \right| - \sqrt{y_{\rm r}^2 + y_{\rm i}^2} \right) e^{-\beta\,( y_{\rm r}^2 + y_{\rm i}^2 )}.
\end{align}
Introducing the polar coordinate $r \equiv \sqrt{y_{\rm r}^2 + y_{\rm i}^2}$, we have
\begin{align}
\rho \left( \left| z \right| \right)
&= 2\beta \int_{0}^{\infty} dr~r~\delta \left( \left| z \right| - r \right) e^{-\beta r^2} \nonumber \\
&= 2\beta \left| z \right| e^{-\beta \left| z \right|^2}. 
    \label{eq: p-D}
\end{align}
To further impose the normalization condition $\int_0^{\infty} \left| z \right| \rho \left( \left| z \right| \right) d\left| z \right| = 1$, we should choose $\beta$ as $\beta = \pi/4$.
Notably, $\rho \left( \left| z \right| \right)$ linearly vanishes at the spectral origin, 
\begin{equation}
\rho \left( \left| z \right| \right) \simeq 2\beta \left| z \right| 
\end{equation}
for small $\left| z \right| \ll 1$.
This linear decay of $\rho \left( \left| z \right| \right)$ does not necessarily mean the level repulsion around the origin since it just arises from the two-dimensional integral measure along the angle direction.
A similar result was obtained in Ref.~\cite{garcia22}.

A generic $2 \times 2$ non-Hermitian matrix $H$ in class \BDIz~(or equivalently, class D + $\mS_{+}$), which 
respects SLS $\sigma_z H \sigma_z = - H$ and PHS $H^{\rm T} = - H$, is again given as Eq.~(\ref{eq: 2*2-D}).
Consequently, the probability distribution function $\rho \left( \left| z \right| \right)$ is the same as $\rho \left( \left| z \right| \right)$ in class D [i.e., Eq.~(\ref{eq: p-D})].
However, this is specific to small non-Hermitian matrices.
For larger non-Hermitian matrices, $\rho \left( \left| z \right| \right)$'s are different between classes D and \BDIz 
owing to the many-level effect, as shown in Fig.~\ref{figs: AZ0_RM_DoS_modulus}.

\subsubsection{Class C}

A generic $2 \times 2$ non-Hermitian matrix $H$ in class C, which  
respects PHS $\tau_y H^{\rm T} \tau_y = - H$, is given as
\begin{equation}
H = a_x \tau_x + a_y \tau_y + a_z \tau_z \quad
\left( a_x, a_y, a_z \in \mathbb{C} \right).
\end{equation}
The probability distribution function $\rho \left( \left| z \right| \right)$ of $\left| z \right| = \left| \sqrt{a_x^2 + a_y^2 + a_z^2 }\right|$ is equivalent to the level-spacing distribution of $2 \times 2$ non-Hermitian random matrices in class A, which is analytically obtained as~\cite{hamazaki20}
\begin{align}
    \rho \left( \left| z \right| \right) = 2 C_3^4 \left| z \right|^3 e^{-C_3^2 \left| z \right|^2}
\end{align}
with the normalization constant $C_3 \equiv 3\sqrt{\pi}/4 = 1.32934 \cdots$.
Here, $\rho \left( \left| z \right| \right)$ is normalized by $\int_0^{\infty} \rho \left( \left| z \right| \right) d\left| z \right| = \int_0^{\infty} \left| z \right| \rho \left( \left| z \right| \right) d\left| z \right| = 1$.
For small $\left| z \right| \ll 1$, we have 
\begin{align}
    \rho \left( \left| z \right| \right) \simeq 2 C_3^4 \left| z \right|^3.
\end{align}
The cubic decay of $\rho \left( \left| z \right| \right)$ means the level repulsion around the origin.
A similar result was obtained in Ref.~\cite{garcia22}.

\subsubsection{Class \CIIz}

No generic $2 \times 2$ non-Hermitian matrix $H$ is present in class \CIIz~(or equivalently, class C + $\mathcal{S}_{+}$).
A generic $4 \times 4$ non-Hermitian matrix $H$ in class \CIIz, which respects TRS$^{\dag}$ $\sigma_y H^{\rm T} \sigma_y = H$ and PHS $\tau_y H^{\rm T} \tau_y = - H$, is given as
\begin{equation}
H = a_x \tau_x + \left( b_x \sigma_x + b_y \sigma_y + b_z \sigma_z \right) \tau_y + a_z \tau_z
\end{equation}
with $a_x, a_z, b_x, b_y, b_z \in \mathbb{C}$.
The eigenvalues are obtained as
\begin{equation}
    \pm z = \pm \sqrt{a_x^2 + a_z^2 + b_x^2 + b_y^2 + b_z^2},
\end{equation}
and exhibit the two-fold degeneracy due to TRS$^{\dag}$.
Thus, the probability distribution function $\rho \left( \left| z \right| \right)$ is equivalent to the level-spacing distribution of $2 \times 2$ non-Hermitian random matrices in class AII$^{\dag}$, which is analytically obtained as~\cite{hamazaki20}
\begin{align}
    \rho \left( \left| z \right| \right) = \frac{2C_5^4}{3} \left| z \right|^3 \left( 1 + C_5^2 \left| z \right|^2 \right) e^{-C_5^2 \left| z \right|^2}
\end{align}
with the normalization constant $C_5 \equiv 7\sqrt{\pi}/8 = 1.5509 \cdots$.
Here, $\rho \left( \left| z \right| \right)$ is normalized by $\int_0^{\infty} \rho \left( \left| z \right| \right) d\left| z \right| = \int_0^{\infty} \left| z \right| \rho \left( \left| z \right| \right) d\left| z \right| = 1$.
For small $\left| z \right| \ll 1$, we have 
\begin{align}
    \rho \left( \left| z \right| \right) \simeq \frac{2C_5^4}{3} \left| z \right|^3.
\end{align}
The cubic decay of $\rho \left( \left| z \right| \right)$ means the level repulsion around the origin.

\subsubsection{Classes \AIIIz, \DIIIz, and \CIz}

A generic $2 \times 2$ non-Hermitian matrix $H$ in class \AIIIz, which 
respects SLS $\sigma_z H \sigma_z = - H$, is given as
\begin{equation}
H = a_x \sigma_x + a_y \sigma_y \quad
\left( a_x, a_y \in \mathbb{C} \right).
\end{equation}
The two eigenvalues of $H$ are obtained as
\begin{equation}
\pm z = \pm \sqrt{a_x^2 + a_y^2}.
\end{equation}
Notably, the density of complex eigenvalues, $\rho \left( \left| z \right| \right)$, is equivalent to the level-spacing statistics of $2 \times 2$ non-Hermitian matrices in class AI$^{\dag}$, which are analytically obtained as~\cite{hamazaki20}
\begin{align}
    \rho \left( \left| z \right| \right) = 2C_2^4 \left| z \right|^3 K_0\,( C_2^2 \left| z \right|^2 ),
        \label{eq: p-AIII}
\end{align}
where $C_2 \equiv \left[ \Gamma \left( 1/4 \right)\right]^2/8\sqrt{2} = 1.16187 \cdots$ is a normalization constant, and $K_0$ is the modified Bessel function of the second kind.
Here, $\rho \left( \left| z \right| \right)$ is normalized by $\int_0^{\infty} \rho\left( \left| z \right| \right) d\left| z \right| = \int_0^{\infty} \left| z \right| \rho\left( \left| z \right| \right) d\left| z \right| =  1$.
For small $\left| z \right| \ll 1$, we have $K_0\,( C_2^2 \left| z \right|^2 ) \simeq - \ln\,( C_2^2 \left| z \right|^2 )$ and hence
\begin{align}
    \rho \left( \left| z \right| \right) \simeq - 4 C_2^4 \left| z \right|^3 \ln \left( C_2 \left| z \right| \right).
\end{align}
A similar result was obtained in Refs.~\cite{splittorff04, garcia22}. 
In the previous analyses~\cite{hamazaki20, xiao22}, the logarithmic correction arises only in the presence of TRS$^{\dag}$ with the positive sign $+1$.
By contrast, the logarithmic correction appears in the hard-edge statistics with SLS, even in the absence of TRS$^{\dag}$.

No generic $2 \times 2$ non-Hermitian matrix $H$ is present in class \DIIIz~(or equivalently, class D + $\mathcal{S}_{-}$).
A generic $4 \times 4$ non-Hermitian matrix $H$ in class \DIIIz, which respects TRS$^{\dag}$ $\sigma_y H^{\rm T} \sigma_y = H$ and SLS $\left( \sigma_z \tau_z \right) H \left( \sigma_z \tau_z \right) = - H$, is given as
\begin{align}
    H = a_x \tau_x + a_y \sigma_z \tau_y\quad
\left( a_x, a_y \in \mathbb{C} \right).
\end{align}
The modulus of the two eigenvalues is $\left|z \right| = \left| \sqrt{a_x^2 + a_y^2} \right|$, and hence the probability distribution function $\rho \left( \left| z \right| \right)$ is equivalent to $\rho \left( \left| z \right| \right)$ in class \AIIIz\ [i.e., Eq.~(\ref{eq: p-AIII})].

A generic $2 \times 2$ non-Hermitian matrix $H$ in class \CIz~(or equivalently, class C + $\mS_{-}$), which respects PHS $\tau_y H^{\rm T} \tau_y = - H$ and TRS$^{\dag}$ $H^{\rm T} = H$, is given as
\begin{equation}
H = a_x \tau_x + a_z \tau_z \quad
\left( a_x, a_z \in \mathbb{C} \right).
\end{equation}
Hence, the probability distribution function $\rho \left( \left| z \right| \right)$ of $\left| z \right| = \left| \sqrt{a_x^2 + a_z^2 }\right|$ is equivalent to $\rho \left( \left| z \right| \right)$ in class \AIIIz~[i.e., Eq.~(\ref{eq: p-AIII})].

\subsection{Distributions of the eigenvalue with the smallest modulus}
\subsubsection{Class BDI}
The minimal dimension of generic non-Hermitian random matrices in class BDI is four.
A $4 \times 4$ non-Hermitian random matrix $H$ in class BDI with $\mT_+ = \tz$ and $\mC_- = 1$ is expressed as Eq.~(\ref{app eq: RM_D_4}) with $a_x,b_0,b_x,b_z, \ii a_0, \ii a_z \in \mathbb{R}$. 
In the Gaussian ensemble, $a_x,b_0,b_x,b_z, \ii a_0, \ii a_z$ are independent real Gaussian variables with zero means. 
The four eigenvalues of $H$ are 
\begin{equation}    
    \pm z_1 = \pm \left( w_1 - w_2 \right),\quad
    \pm z_2 = \pm \left(  w_1 + w_2 \right)
\end{equation}
with
\begin{equation}
    w_1 \equiv \sqrt{b_0^2 + a_x^2 + a_z^2},\quad w_2 \equiv \sqrt{a_0^2 + b_x^2 + b_z^2} \, .
\end{equation} 
Therefore, $w_1$ and $w_2$ are independent random variables with the same probability density function.
Here, $w_{1}$ and $w_2$ are either real or purely imaginary, and we require $\Re\,w_1 \geq 0$ and $\Im\,w_2 \geq 0$.
The probability density $p_w(w_1 = \alpha + \ii \beta)$ of $w_1$ in the complex plane is given as~\cite{xiao22}   
\begin{equation} \label{eq: pdf BDI w}
  p_w(w_1 = \alpha + \ii \beta) = \frac{\alpha e^{-\alpha^2/2}}{\sqrt{2}}  \delta(\beta) +  \frac{\beta e^{\beta^2/2}{\rm erfc}(\beta)}{\sqrt{2}}  \delta(\alpha) \, .
\end{equation}
According to whether $w_{1}$ and $w_2$ are real or purely imaginary, we have three possible cases, as follows.

(i)~{\it Both $w_1$ and $w_2$ are real}.---In this case, all the four eigenvalues of $H$ are real. 
Without loss of generality, we can assume $w_1 \geq w_2 \geq 0$. 
Then, we have $z_2 = w_1 + w_2  > z_1 =  w_1 - w_2 \geq 0$, and hence $z_1$ is the eigenvalue with the smallest modulus. 
The probability density $p_{\min;r}(x)$ of $z_1 = x  \geq 0$ is given as 
\begin{align} 
   &p_{\rm min;r}^{N = 4}(x) = \int_0^{\infty} d\alpha~\alpha \left( \alpha + x \right) e^{- (\alpha^2 + (\alpha+x)^2)/2} \nonumber \\
   &= \frac{e^{-x^2/2}}{8} \left(2 x-\sqrt{\pi } e^{x^2/4} \left(x^2-2\right) \text{erfc}\left(x/2\right)\right)\, ,
\end{align}
satisfying $p_{\min;r}(0) = \sqrt{\pi}/4 \approx 0.443$.

(ii)~{\it Both $w_1$ and $w_2$ are purely imaginary}.---In this case, the four eigenvalues of $H$ are purely imaginary.
Without loss of generality, we can assume $|w_1| \geq |w_2| \geq 0$.
The probability density $p_{\min;i}(|z_1|)$ of the eigenvalue $z_1 = \ii y$ with the smallest modulus is given as
\begin{equation}  
     \quad p_{\rm min;i}^{N = 4}(y) = \int_0^{\infty} d\beta~\beta \left( \beta + y \right) {\rm erfc} \left( \beta \right) {\rm erfc} \left( \beta+y \right),
\end{equation}
satisfying $p_{\rm min; i}(|z_1| = 0) = \left( 1 -\ln 2 \right)/\left( 2 \sqrt{\pi } \right) \approx 0.087$. 
For generic $y$, this integral cannot be expressed by elementary functions.

(iii)~{\it One of $w_1$ and $w_2$ is real, and the other is purely imaginary}.---In this case, $z_1$ and $z_2$ are both complex with $|z_1| = |z_2|$. 
Without loss of generality, we can assume $w_1 \in \mathbb{R}$ and $\ii w_2 \in \mathbb{R}$.
The probability density $p_{\rm min;c}(z_2 = x + \ii y )$ of $z_{\min} = |z_{\min}| e^{\ii \theta_z}$ in the complex plane is given as
\begin{equation} 
  p_{\rm min;c}^{N = 4}(x,y)= x y e^{(-x^2+y^2)/2} {\rm erfc} \left( y \right).
\end{equation}
For $|z_{\rm min}| \ll 1$, we have $  p_{\rm min;c}(x,y) \propto |z_{\rm min}|^2$. 
Thus, the radial distribution $p^{(r);N=4}_{\rm min;c}(|z_{\rm min}|)$ of $|z_{\min}|$ is proportional to $|z_{\min}|^3$ for $|z_{\min}| \ll 1$. 
We can also verify that the angular distribution satisfies $p^{(\theta_z);N=4}_{\rm min;c}(\theta_z) \propto |\theta_z|$ for $|\theta_z| \ll 1$ and $p^{(\theta_z);N=4}_{\rm min;c}(\theta_z) \propto |\pi/2 - \theta_z|$ for $|\pi/2 - \theta_z| \ll 1$.
Thus, the distributions of the eigenvalues with the smallest modulus obtained by the $4 \times 4$ non-Hermitian random matrix are qualitatively the same as those of large random matrices [see Figs.~\ref{figs: pdf_z_min_AZ_real}~(a) and (g)]. 

\subsubsection{Class \AIIp}
\label{app: AIIp}
The minimal dimension of generic non-Hermitian random matrices in class \AIIp\ is four, whose eigenvalues appear in quartets as $\left( z, -z, z^{*}, - z^{*} \right)$ with the same modulus. 
The probability distribution $p_{\min}^{N = 4}(z) $
is analytically obtained as~\cite{akemann2005}
\begin{equation}
    p_{\min}^{N = 4}(z)  =  \frac{1}{128 \pi}  \left \lvert z^2 - (z^{*})^{2} \right \rvert ^2   \lvert z \rvert^2 K_0\left(4  \lvert z \rvert^2 \right) \, .
\end{equation}
In class \AIIp, the probability of an eigenvalue being real or purely imaginary is zero, leading to
\begin{equation}
p_{\min;c}^{N = 4}(z) 
= 
p_{\min}^{N = 4}(z) 
\end{equation}
Thus, the radial distribution $p_{\min;c}^{(r);N = 4}(|z_{\min}|) $ of $z_{\min} = |z_{\min}| e^{\ii \theta_z}$ with $\theta_z \in [0,\pi)$ is given as 
\begin{equation}
    p_{\min;c}^{(r);N = 4}(|z_{\min}|) = \frac{1}{32} |z_{\min}|^7 K_0(|z_{\min}|^2)  \, ,
\end{equation}
and the angular distribution 
\begin{equation}
    p_{\min;c}^{(\theta_z);N = 4} = \frac{4}{\pi} \cos^2 2 \theta_z.
\end{equation}
Specifically, we have 
\begin{equation}
    p_{\min;c}^{(r);N = 4}(|z_{\min}|) \propto -|z_{\min}|^7\ln|z_{\min}| \quad \left( |z_{\min}| \ll 1  \right)
\end{equation}
and
\begin{equation}
    p^{(\theta_z);N=4}_{\rm min;c}(\theta_z) \propto \begin{cases}
        |\theta_z|^2 & \left( |\theta_z| \ll 1 \right); \\
        |\pi/2-\theta_z|^2 & \left( |\pi/2 - \theta_z| \ll 1 \right).
    \end{cases}
\end{equation}
These behaviors for $N=4$ are both consistent with the behaviors 
for large $N$
[see Figs.~\ref{figs: pdf_z_min_AZ_real}\,(c) and (g)].

\begin{figure*}[h]
    \centering
    \includegraphics[width=0.85\linewidth]{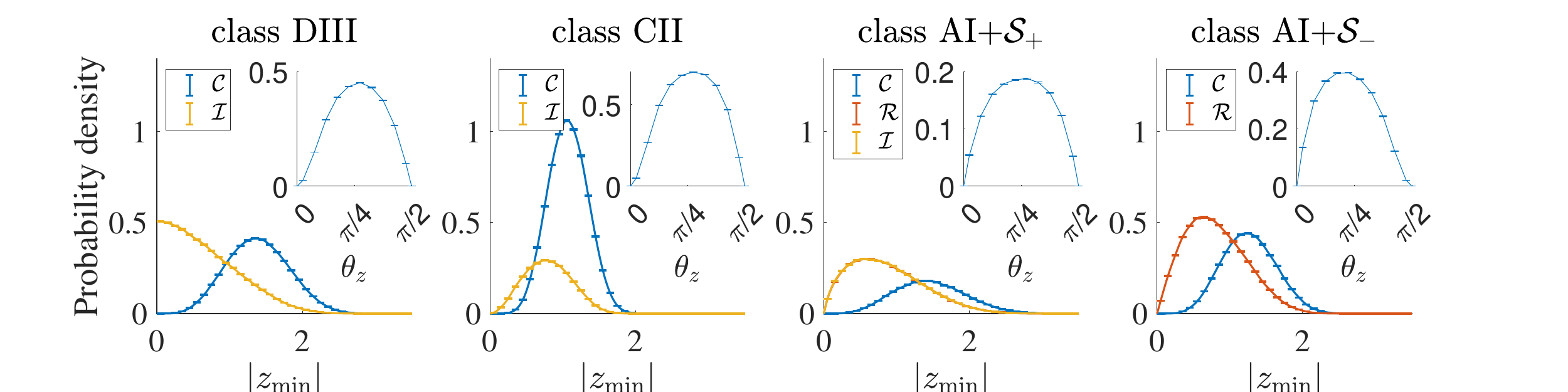}

    \includegraphics[width=0.85\linewidth]{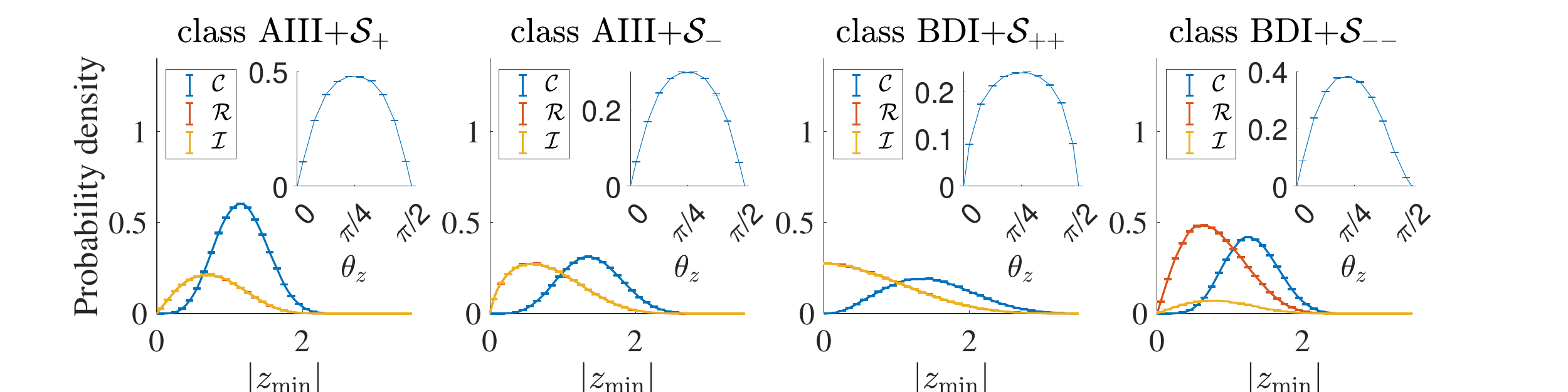}

    \includegraphics[width=0.85\linewidth]{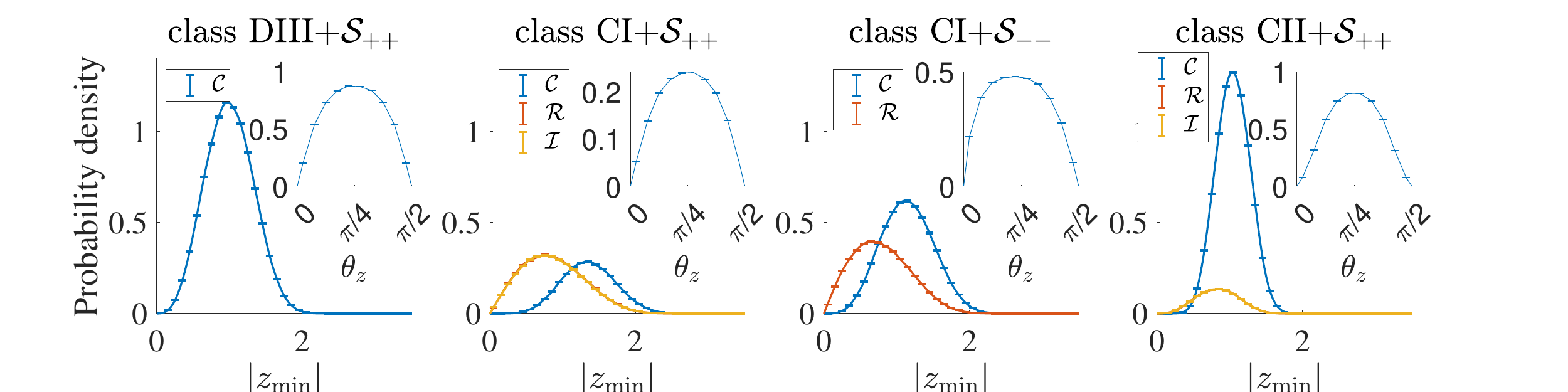}

    \includegraphics[width=0.85\linewidth]{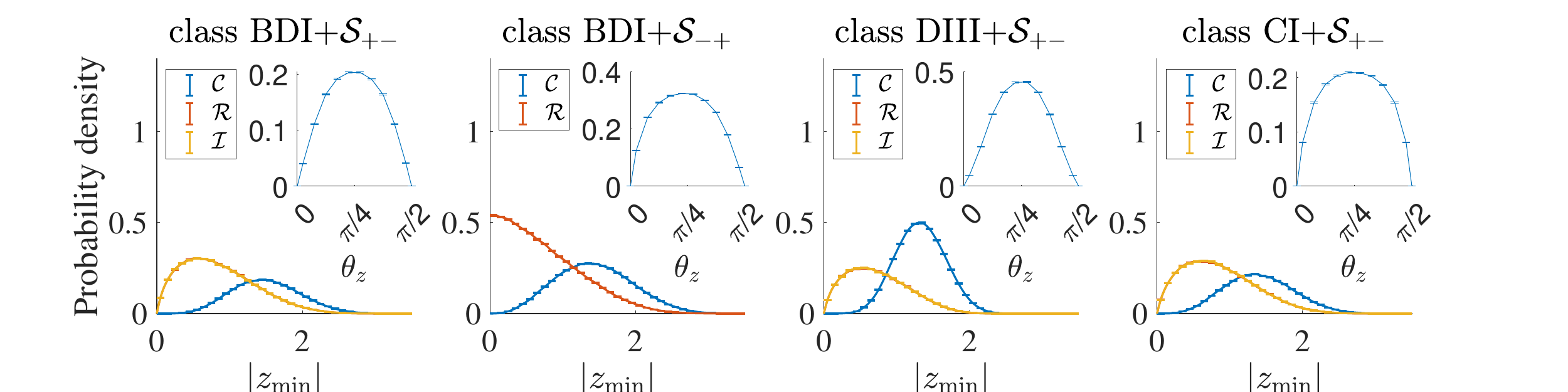}

    \includegraphics[width=0.85\linewidth]{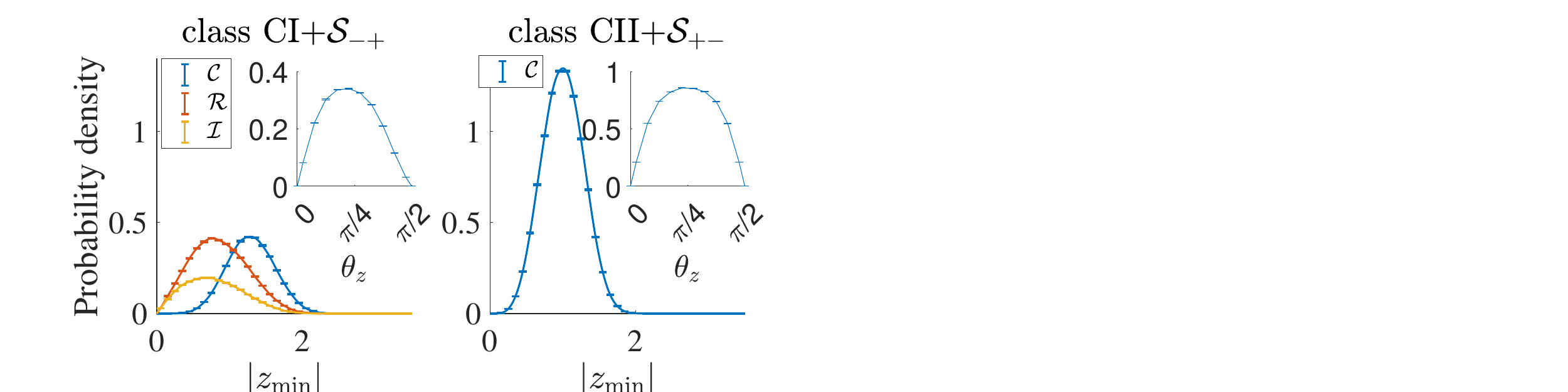}
    \caption{Distributions of the eigenvalue $z_{\min}$ with the smallest modulus of non-Hermitian random matrices in 18 out of 21 symmetry classes beyond the AZ$^{\dagger}$ and AZ$_0$ classification.
    The distributions in the three symmetry classes (i.e., classes BDI, CI, and \AIIp) not shown here are found in Fig.~\ref{figs: pdf_z_min_AZ_real}.
    The blue, orange, and yellow curves represent the radial distributions of $z_{\min}$ when it is complex ($\mathcal{C}$), real ($\mathcal{R}$), and purely imaginary ($\mathcal{I}$), respectively.
    In some symmetry classes (e.g., class AI + $\mathcal{S}_+$), the orange and yellow curves overlap almost perfectly, and the orange curve is hardly seen. 
    Inset:~angular distribution of complex $z_{\min}$.
    The distributions are obtained by diagonalizing $10^6$ samples of $256 \times 256$ random matrices in each symmetry class.}
    \label{fig: z_min AZ all}
\end{figure*}

\section{Hard-edge statistics beyond the AZ$^{\dagger}$ and AZ$_0$ classification}
\label{app: sec alternative}

\subsection{Distributions of $z_{\min}$ in general symmetry classes}
\label{app: hard-edge supp}

In Sec.~\ref{sec: RM AZ_real - BDI, CI, AIIp}, we investigate the hard-edge statistics of non-Hermitian random matrices in classes BDI, CI, and \AIIp\ in detail.
Here, we present the distributions of the eigenvalue $z_{\min}$ with the smallest modulus in all the remaining 18 symmetry classes beyond the AZ$^{\dagger}$ and AZ$_0$ classification (see Fig.~\ref{fig: z_min AZ all}).

\subsection{Additional characterization of hard-edge statistics}

\begin{figure}[b]    
  \centering
      \includegraphics[width=1\linewidth]{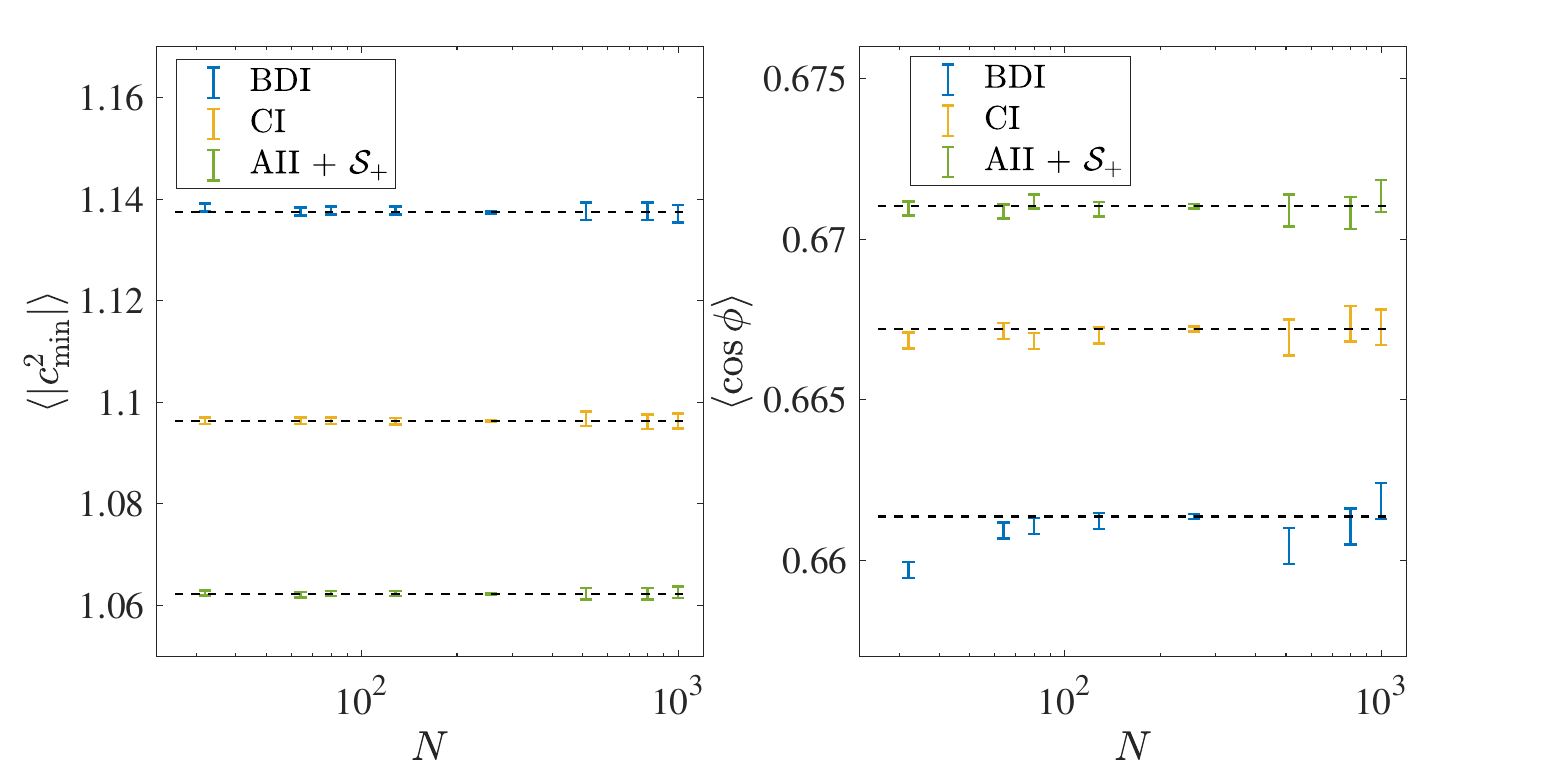}
      \caption{Mean values $\left\langle |c_{\min}|^2 \right\rangle$ and $\left\langle \cos \phi \right\rangle$ 
      as functions of the matrix size $N$, where $c_{\min} = |c_{\min}| e^{\ii \phi}$ is the complex eigenvalue with the smallest modulus.
      For each symmetry class and each matrix size, $N_{\rm sample}$ samples are diagonalized, with $N_{\rm sample} = 10^7,10^6,2\times 10^5$ for $N = 256$, $N<256$, $N > 256$, respectively. 
      The dashed lines denote the mean values obtained by matrices with $N = 256$.}
      \label{figs: mean_r_cos_AZ_real}
\end{figure}

\begin{figure}[htb]
  \centering
      \includegraphics[width=1\linewidth]{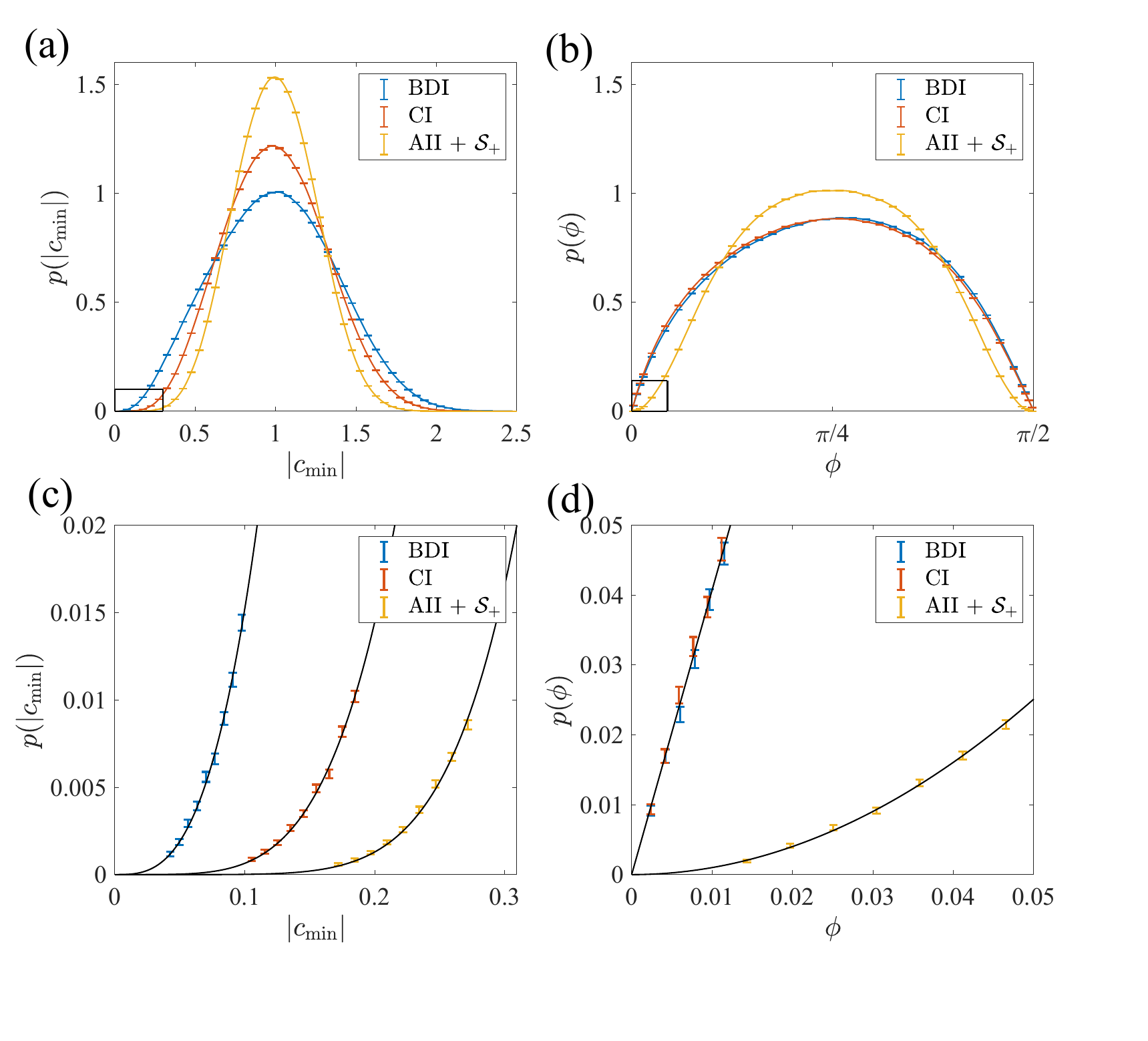}
\caption{(a)~Radial and (b) angular distributions of the complex eigenvalues with the smallest modulus for $256\times 256$ non-Hermitian random matrices in classes BDI, CI, and \AIIp. 
(c)~Zoom-in of subfigure~(a) around $\left| c_{\rm min} \right| = 0$. 
The three black curves from left to right are the fitting curves $p (\left| c_{\rm min} \right|) \propto \left| c_{\rm min} \right|^3$, $p (\left| c_{\rm min} \right|) \propto -\left| c_{\rm min} \right|^5 \ln \left| c_{\rm min} \right|$, and $p (\left| c_{\rm min} \right|) \propto -\left| c_{\rm min} \right|^7 \ln \left| c_{\rm min} \right|$, respectively. 
(d)~Zoom-in of subfigure~(b) around $\left| c_{\rm min} \right| = 0$.
The two black curves from left to right are the fitting curves $p (\left| c_{\rm min} \right|) \propto \left| c_{\rm min} \right|$ and $p (\left| c_{\rm min} \right|) \propto \left| c_{\rm min} \right|^2$, respectively.}
\label{figs: pdf_complex_r_angle_AZ_real}
\end{figure} 

\begin{figure}[hbt]
  \centering
      \includegraphics[width=1\linewidth]{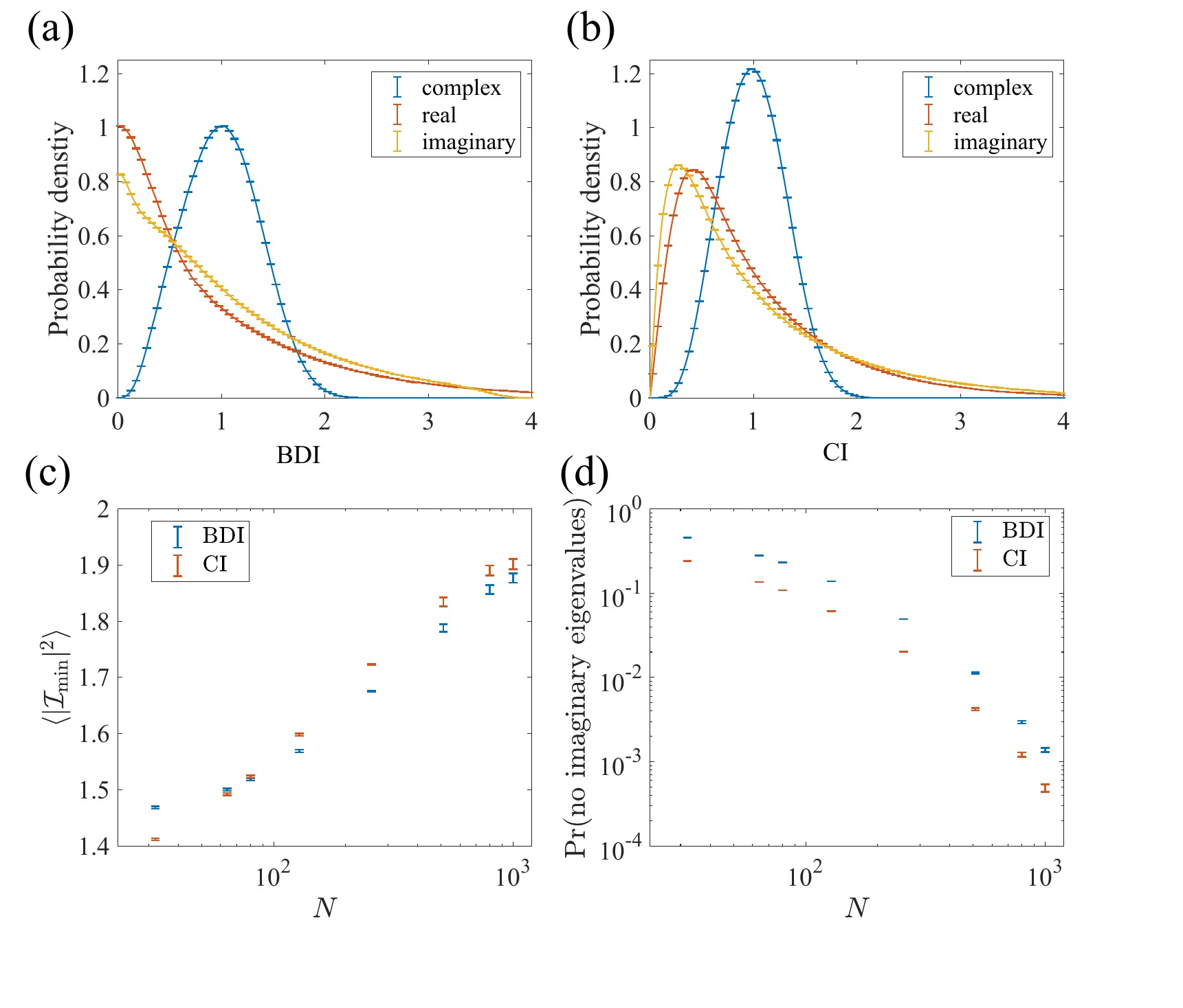}
\caption{Distributions of the complex, real, and purely imaginary eigenvalues with the smallest modulus for $256\times 256$ non-Hermitian random matrices in (a)~class BDI and (b)~class CI. 
For a small proportion $p$ of random matrices, real eigenvalues ($p \approx 0.02\%$) or purely imaginary eigenvalues ($ p \approx 3.5\% $) are absent. 
We exclude such matrices in the statistics of real or imaginary eigenvalues. 
(c)~Mean values $\left\langle\mathcal{I}_{\min}^2\right\rangle$ and (d)~probability that no purely imaginary eigenvalues exist as functions of the matrix size $N$.}
    \label{figs: pdf_ric_AZ_real}
\end{figure} 

\begin{figure}[tbh]
  \centering
      \includegraphics[width=1\linewidth]{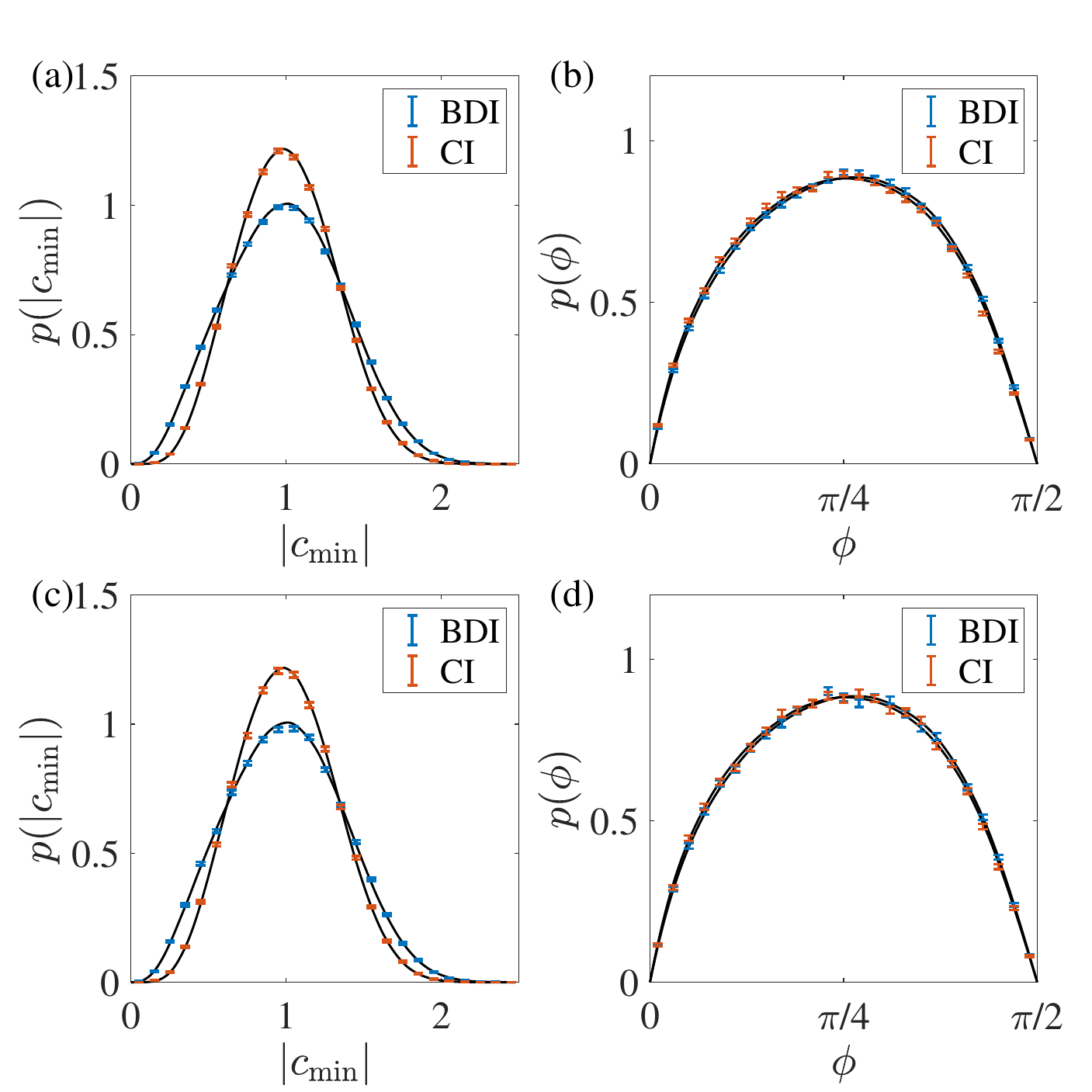}
\caption{(a), (c)~Radial and (b), (d)~angular distributions of the complex eigenvalue $c_{\min}$ with the smallest modulus for the non-Hermitian Hamiltonians and many-body Lindbladians. 
The points with the error bars correspond to the quantum systems, and the solid curves are obtained from %
the random matrices in the same symmetry classes. 
(a), (b)~Effective non-Hermitian Hamiltonian in Eq.~(\ref{eq: h_BDI}) for class BDI and non-Hermitian superconductor in Eq.~(\ref{eq: AZ_real model CI}) for class CI. 
(c), (d)~Many-body Lindbladians of the SYK model in Eqs.~(\ref{eq: SYK_q}) and (\ref{eq: SYK_L}) for classes BDI ($N = 12$) and CI ($N = 10$), where we use the eigenvalues in the subsector of $(-1)^{\mF} = -1$ to obtain the statistics.}
\label{fig: model_p_c_min}
\end{figure} 

For non-Hermitian random matrices in some symmetry classes beyond the AZ$^{\dagger}$ and AZ$_0$ classification (see Table~\ref{tab: AZ_real}), a 
subextensive number of eigenvalues are real 
or purely imaginary. 
We also find that in 
these symmetry classes,
the level correlations of complex eigenvalues are qualitatively 
different from those of the real or purely imaginary eigenvalues 
(see Fig.~\ref{figs: pdf_z_min_AZ_real}). To characterize the level 
statistics in these symmetry classes, besides studying the 
eigenvalue $z_{\min}$ with the smallest modulus (see Sec.~\ref{sec: RM AZ_real} for details), 
we here study complex, real, and purely imaginary eigenvalues, separately. 
Let $c_{\min}$, $\mathcal{R}_{\min}$, and $\mathcal{I}_{\min}$ be the complex, real, 
and purely imaginary eigenvalues with the smallest modulus, respectively. 
Owing to the defining symmetries, we can require $\Re\,c_{\min}, \Im\,c_{\min} >0 $, $\mathcal{R}_{\min}\geq 0$, and $\mathcal{I}_{\min}\geq 0$ without loss of generality. 
As in Sec.~\ref{sec: RM AZ_real - BDI, CI, AIIp}, we here focus on the three representative symmetry classes, i.e., classes BDI, CI, and \AIIp.

We first investigate the distribution of $c_{\min} = |c_{\min}| e^{\ii \phi}$ for non-Hermitian random matrices in classes BDI, CI, and \AIIp.
We normalize $c_{\min}$ such that $\left\langle |c_{\rm min}|\right\rangle = 1$.
In class \AIIp, all eigenvalues are complex, and hence we have $c_{\min} = z_{\min}$. 
For $N \geq 256$, $\left\langle |c_{\min}|^2 \right\rangle$ and $\left\langle \cos\phi \right\rangle$ converge to the characteristic values in each symmetry class (see Fig.~\ref{figs: mean_r_cos_AZ_real}), implying the convergence of the distributions of $z_{\min}$. 
In each symmetry class, the distribution of $c_{\min}$ is 
similar to that of complex $z_{\min}$. 
The radial distributions $p(|c_{\min}|)$ exhibit the characteristic small-$|c_{\min}|$ behavior [see Figs.~\ref{figs: pdf_complex_r_angle_AZ_real}\,(a), (c)], 
\begin{equation} \label{eq: small c_min}
  p(|c_{\min} |) \propto \begin{cases}
    |c_{\min} |^3 & \text{(class BDI)} \, ; \\
    -|c_{\min} |^5\ln |c_{\min}| & \text{(class CI)} \, ; \\
    -|c_{\min} |^7\ln |c_{\min}| & \text{(class AII} + \mathcal{S}_+) \, .\\
  \end{cases}
\end{equation}
This is the same as the small-$|z_{\min}|$ behavior of $p_{\rm min;c}(|z_{\min}|)$ in the same symmetry classes (Fig.~\ref{figs: pdf_z_min_AZ_real}).
The angular distributions $p(\phi)$ of $c_{\min}$ are also similar to those of complex $z_{\min}$ [compare Figs.~\ref{figs: pdf_complex_r_angle_AZ_real}\,(b), (d) with Fig.~\ref{figs: AZ_real model}].

Then, we study the distributions of $\mathcal{R}_{\min}$ and $\mathcal{I}_{\min}$ in classes BDI and CI.
In each symmetry class, the distributions $p(\mathcal{R}_{\min})$ and $p(\mathcal{I}_{\min})$ share the same 
small-$\mathcal{R}_{\min}$ or small-$\mathcal{I}_{\min}$
behavior as $p_{\min:{\rm r}/{\rm i}}(|z_{\min}|)$.  
However, $p(\mathcal{R}_{\min})$ and $p(\mathcal{I}_{\min})$ are long-tailed in contrast to $p_{\min:{\rm r}/{\rm i}}(|z_{\min}|)$. 
The statistics of $\mathcal{I}_{\min}$ converge much more slowly than the statistics of $c_{\min}$ and do not converge even for $N = 1000$ [see Figs.~\ref{figs: pdf_ric_AZ_real}\,(c) and (d)].
Thus, $p(|c_{\min}|)$ or $p(\mathcal{R}_{\min})$ 
should be a more computationally efficient measure of the level statistics than $p(\mathcal{I}_{\min})$.

Furthermore, we compare $p(|c_{\min}|)$ and $p(\phi)$ of open quantum systems studied in
Sec.~\ref{subsec: AZ_real free model}
with those of non-Hermitian random matrices.
We observe that $p(|c_{\min}|)$ of open quantum systems in different symmetry classes exhibit distinct behavior consistent with those of non-Hermitian random matrices (Fig.~\ref{fig: model_p_c_min}).
The angular distributions $p(\phi)$ also match well with those of non-Hermitian random matrices (Fig.~\ref{fig: model_p_c_min}). 
These results further demonstrate that 
the level statistics of random matrices well describe universal properties of quantum chaotic systems with symmetry.
\section{Details on models and numerical simulation}
\label{app: model detail}

\subsection{Effective non-Hermitian Hamiltonians of quadratic Lindbladians}

In Sec.~\ref{subsec: quadratic L}, we consider Hermitian Hamiltonians $H_{\rm AIII}$, $H_{\rm BDI}$, and $H_{\rm CII}$ in the form $  \sum_{\left\langle \iii,\jjj \right\rangle} c_{\iii}^{\dagger} t_{\iii,\jjj}  c_{\jjj} + \text{H.c.}$ [Eq.~(\ref{eq: H_chiral_class})] and their dissipators. %
Their effective non-Hermitian Hamiltonians in Eq.~(\ref{eq: h_eff}) are given as
\begin{align} 
h_{\rm eff}^{\rm AIII^{\dagger}}  & = H_{\rm AIII} + \gamma \sum_{\iii} \left(  c_{\iii + e_z}^{\dagger} c_{\iii } -  c_{\iii}^{\dagger} c_{\iii + \e_z}  \right) \, , \label{eq: H_AIIId}\\
h_{\rm eff}^{\rm BDI_0}  & = H_{\rm BDI} - \ii \gamma \sum_{\iii} \left(  c_{\iii + e_z}^{\dagger} c_{\iii } +  c_{\iii}^{\dagger} c_{\iii + \e_z}  \right) \, , \label{eq: H_BDI0} \\
h_{\rm eff}^{\rm CII_0}  & = H_{\rm CII} %
-\ii
\gamma \sum_{\iii} \left( c_{\iii + e_z}^{\dagger} \sigma_0 c_{\iii } + %
c_{\iii}^{\dagger} \sigma_0 c_{\iii + \e_z}  \right) \label{eq: H_CII0}\, ,
\end{align}
respectively.
These effective non-Hermitian Hamiltonians still respect SLS and TRS$^{\dagger}$ with the same symmetry operators.
In fact, we have
\begin{equation}
    \left( h_{\rm eff}^{\rm BDI_0} \right)^{\rm T} = h_{\rm eff}^{\rm BDI_0}, \quad \sigma_y \left( h_{\rm eff}^{\rm CII_0} \right)^{\rm T} \sigma_y = h_{\rm eff}^{\rm CII_0}.
\end{equation}
Thus, $h_{\rm eff}^{\rm AIII^{\dagger}}$, $h_{\rm eff}^{\rm BDI_0}$, and $h_{\rm eff}^{\rm CII_0}$ belong to symmetry classes \AIIIz, \BDIz, and \CIIz, respectively.

For each class, we diagonalize $2 \times 10^5$ samples in the disorder ensemble, with the parameters chosen as follows:

\begin{itemize}
    \item[(i)] { Class \AIIIz}~($h_{\rm eff}^{\rm AIII^{\dagger}}$).---$t_{\iii, \jjj} = e^{\ii \theta_{\iii, \jjj}}$ with $\theta_{\iii, \jjj} \in [0,2 \pi)$, $\gamma = 0.1$, and $L = 18$;
    
    \item[(ii)] Class \BDIz~($h_{\rm eff}^{\rm BDI_0}$).---$ t_{\iii, \jjj} \in [t-W/2,t+W/2]$ with $t = 1$, $W= 1.6$, $\gamma = 0.1$, and $L = 18$;

    \item[(iii)] Class \CIIz~($h_{\rm eff}^{\rm CII_0}$).---$t_{\iii, \jjj}$'s are random SU(2) matrices; $\gamma = 0.4$ and $L = 12$.
\end{itemize}

In Sec.~\ref{subsec: AZ_real free model}, we consider Hamiltonians $H_{\rm BDI}$ and $H_{\rm CII}$ with their dissipators in Eqs.~(\ref{eq: L BDI}) and (\ref{eq: L AIIp}). 
The effective non-Hermitian Hamiltonians of these quadratic Lindbladians in Eq.~(\ref{eq: h_eff}) are
given as
\begin{align} 
  h_{\rm eff}^{\rm BDI}   & = H_{\rm BDI} +  \gamma \sum_{\iii} \left(  -c_{\iii + 2\e_z}^{\dagger} c_{\iii } +  c_{\iii}^{\dagger} c_{\iii + 2\e_z}  \right) \, , \label{eq: h_BDI} \\
  h_{\rm eff}^{{\rm AII} + \mathcal{S}_+}   & = H_{\rm CII} +  \gamma \sum_{\iii} \left(  -c_{\iii + \e_z}^{\dagger} c_{\iii } +  c_{\iii}^{\dagger} c_{\iii + \e_z}  \right) \, , \label{eq: h_AII_p} 
\end{align}
respectively. 
The Hamiltonian $h_{\rm eff}^{\rm BDI}$ satisfies 
\begin{equation}
    (h_{\rm eff}^{\rm BDI})^* = h_{\rm eff}^{\rm BDI}, \quad \mS (h_{\rm eff}^{\rm BDI})^{\dagger} \mS^{-1} = - h_{\rm eff}^{\rm BDI}
\end{equation}
with $\mS = \delta_{\iii,\jjj} (-1)^{(\iii_x + \iii_y + \iii_z)}$ and hence belongs to class BDI. 
On the other hand, the Hamiltonian $h_{\rm eff}^{{\rm AII} + \mathcal{S}_+}$ satisfies
\begin{equation}
    \sigma_y(h_{\rm eff}^{{\rm AII} + \mathcal{S}_+})^*\sigma_y = h_{\rm eff}^{{\rm AII}}, \quad \mS (h_{\rm eff}^{{\rm AII} + \mathcal{S}_+}) \mS^{-1} = - h_{\rm eff}^{{\rm AII} + \mathcal{S}_+},
\end{equation}
and hence belongs to class \AIIp. 
For each class, we diagonalize $2 \times 10^5$ disorder realizations, with the parameters chosen as follows: 
\begin{itemize}
\item[(i)] Class BDI ($h_{\rm eff}^{\rm BDI}$).---$t = 1, W = 1.6, \gamma = 0.15$, and $L = 16$;
\item[(ii)] Class \AIIp~($h_{\rm eff}^{{\rm AII} + \mathcal{S}_+}$).---$t_{\iii, \jjj}$'s are random SU(2) matrices; $\gamma = 0.4$ and $L = 12$.
\end{itemize}

\subsection{Non-Hermitian superconductors}

In Sec.~\ref{subsec: model BdG}, we consider non-Hermitian superconductors in classes D, C, CI$_{0}$, and DIII$_0$ in the AZ$_0$ classification. 
Here, we provide some details of these models. %

(i)~For symmetry class D, the phase $e^{\ii\phi}$ of the hopping term in the Hamiltonian [Eq.~(\ref{eq: H_D})] can originate from a spatial modulation of the superconducting phase. 
In fact, with a gauge transformation 
$ c_{\iii}  \rightarrow e^{\ii  \iii_x \phi} c_{\iii}$, the hopping phase can 
be eliminated~\cite{pientka13}, and the paring terms change as $\ii \Delta c_{\iii + \e_x} c_{\iii } \rightarrow \ii \Delta e^{\ii  (2\iii_x \phi + \phi)}  c_{\iii  + \e_x} c_{\iii}$ and $  \Delta c_{\iii  + \e_y} c_{\iii}   \rightarrow \Delta e^{\ii  (2\iii_x \phi)}  c_{\iii  + \e_y} c_{\iii}$. 
In the Nambu basis $\Psi_{\iii} = (c_{\iii}, c_{\iii}^{\dagger})^{\rm T}$, the Hamiltonian is written as 
\begin{align} 
\label{eq: AZ0 model D} 
   H_{\rm D} & = \sum_{\iii} \left\{ \left( \epsilon^r_{\iii} + \ii \epsilon^i_{\iii} -\mu\right)  \Psi_{\iii}^{\dagger} \tau_z \Psi_{\iii} \right. \nonumber \\
   &\left. + \left[ \Psi_{\iii+\e_x}^{\dagger} ( t \cos \phi \tau_z + \ii t \sin \phi \tau_0  + \ii \Delta \tau_x) \Psi_{\iii} \right. \right. \nonumber \\
    & \left. \left. + \Psi_{\iii+\e_y}^{\dagger} ( t  \tau_z  + \ii \Delta \tau_y   ) \Psi_{\iii}+ {\rm H.c.} \right]\right\} \, .
\end{align}

(ii)~%
For symmetry class C, the phase $\phi$ in $H_{\rm C}$ [Eq.~(\ref{eq: AZ0 model C})] can originate from a spatial modulation of the superconducting order parameter, as discussed for the Hamiltonian $H_{\rm D}$. %

(iii)~For symmetry class CI$_0$, %
the Hamiltonian $H_{\rm CI_0}$ in the Nambu basis is given as %
\begin{align}   
    \label{eq: AZ0 model CI0}
  H_{\rm CI_0} = & \sum_{\iii} \left\{ \Psi_{\iii}^{\dagger} \left[ (  \epsilon_{\iii}-\mu) \tau_z + \ii \Delta^s_{\iii} \tau_y  \right] \Psi_{\iii} \right. \nonumber \\ + 
  & \left. \sum_{\mu = x,y,z} \left[ \Psi_{\iii + \e_{\mu}}^{\dagger} \left(t \tau_z + \Delta_{\mu} \tau_y\right) \Psi_{\iii} + {\rm H.c.} \right]   \right\} \, ,
\end{align}
where $\epsilon_{\iii}$ is a random chemical potential distributed uniformly in $[-W_1/2,W_1/2]$, $\Delta_{\mu} \in \mathbb{R}$ ($\mu = x,y,z$) is a Hermitian pairing potential, and $\Delta^s_{\iii}$ is a random imbalanced pairing potential distributed uniformly in $[-W_2/2,W_2/2]$.
The Hamiltonian $H_{\rm CI_0}$ satisfies 
\begin{equation}
    \tau_y H_{\rm CI_0}^{\rm T} \tau_y = - H_{\rm CI_0},\quad \tau_z H_{\rm CI_0}^{\rm T} \tau_z = H_{\rm CI_0}, 
\end{equation}
and hence belongs to class \CIz.

(iv)~For symmetry class DIII$_0$, %
the normal-state Bloch Hamiltonian reads
\begin{equation}
h \left( {\bm k} \right) = \sum_{\mu = x,y,z} \left( \left( 2t \cos k_{\mu} -\mu \right) \sigma_0 - \left( 2 \lambda \sin k_{\mu} \right) \sigma_{\mu} \right), 
\end{equation}
and the spin-triplet paring term reads, 
\begin{equation}
\Delta \sum_{\mu = x,y,z} ( \psi^{\dagger}_{\iii + \e_{\mu}} )^{\rm T} \left( \ii  \sigma_{\mu} \right) \left( \ii \sigma_y \right) \psi_{\iii}^{\dagger},
\end{equation}
where $\sigma_{\mu}$'s are Pauli matrices acting on spin space. 
In terms of the Nambu basis $\Psi_{\iii} = ( c_{\uparrow},c_{\downarrow},c_{\uparrow}^{\dagger},c_{\downarrow}^{\dagger} )^{\rm T}$ and $\tau_{\mu}$'s that act on its  particle-hole space, the Hamiltonian reads,  
\begin{align} 
\label{eq: AZ0 model DIII0}
   &H_{\rm DIII_0} = \sum_{\iii} \left\{ \Psi_{\iii}^{\dagger} \left[ (  \epsilon_{\iii}^r  + \ii \epsilon_{\iii}^i -\mu) \tau_z \sigma_0\right] \Psi_{\iii} \right. \nonumber \\
   &+ \left. \left[ \Psi_{\iii + \e_{x}}^{\dagger} \left(t \tau_z \sigma_0 + \ii \lambda \tau_0 \sigma_x + \ii \Delta \tau_x \sigma_z \right) \Psi_{\iii} \right. \right. \nonumber \\
   &+ \left. \Psi_{\iii + \e_{y}}^{\dagger} \left(t \tau_z \sigma_0 + \ii \lambda \tau_z \sigma_y  -\ii \Delta \tau_y \sigma_0 \right) \Psi_{\iii}  \right. \nonumber \\
   &+ \left. \left.  \Psi_{\iii + \e_{z}}^{\dagger} \left(t \tau_z \sigma_0 + \ii \lambda \tau_0 \sigma_z - \ii \Delta \tau_x \sigma_x \right) \Psi_{\iii}  + {\rm H.c.} \right]
   \right\} \, ,
\end{align}
where $\epsilon_{\iii}^r$ and $\epsilon_{\iii}^i$ are a 
random chemical potential and random 
energy gain and loss that are distributed uniformly in $[-W_1/2,W_1/2]$ and $[-W_2/2,W_2/2]$, respectively. This Hamiltonian respects
\begin{equation}
    \tau_x H_{\rm DIII_0}^{\rm T} \tau_x = - H_{\rm DIII_0}, \quad \sigma_y H_{\rm DIII_0}^{\rm T} \sigma_y = H_{\rm DIII_0},
\end{equation}
and hence belongs to class \DIIIz.

For each Hamiltonian, we diagonalize $2 \times 10^5$ disorder realizations, with the parameters chosen as follows: 
\begin{itemize}
\item[(i)] Class D ($H_{\rm D}$).---$\mu = 0$, $t = 2$, $\Delta = 1$, $\phi = \pi/3$, $W_1 = 1$, $W_2 = 2$, and $L = 40$; 
\item[(ii)] Class C ($H_{\rm C}$).---$\mu = 0$, $t = 2$, $\Delta = 1$, $\phi = \pi/3$, $W_1 = 6$, $W_2 = 3$, and $L = 12$;
\item[(iii)] Class \CIz~($H_{\rm CI_0}$).---$\mu = 0$, $t = 2$, $\Delta_{x/y} = 1$, $\Delta_z = 1/2$, $W_1 = 6$, $W_2 = 3$, and $L = 12$;
\item[(iv)] Class \DIIIz~($H_{\rm DIII_0}$).---$\mu = 0$, $t = 2$, $\lambda=1$, 
$\Delta = 1$, $W_1 = W_2 = 9$, and $L = 10$. 
\end{itemize}
Here, the lattice is either square lattice or cubic lattice with the linear dimension $L$. 

In Sec.~\ref{subsec: AZ_real free model}, for symmetry class CI, we consider a three-dimensional non-Hermitian superconductor with imbalanced pairing potentials.
In the Nambu basis, the Hamiltonian $H_{\rm CI}$ reads
\begin{align}
     H_{\rm CI} &  = \sum_{\iii} \left\{ \Psi_{\iii}^{\dagger} \left[ (  \epsilon_{\iii}-\mu) \tau_z + \ii \Delta^s_{\iii} \tau_x  \right] \Psi_{\iii} \right. \nonumber \\ 
    & \left. + \sum_{\mu = x,y,z} \left[ \Psi_{\iii + \e_{\mu}}^{\dagger} \left(t \tau_z + \Delta_{\mu} \tau_y\right) \Psi_{\iii} + {\rm H.c.} \right]
    \right\} \, , 
  \label{eq: AZ_real model CI}
\end{align}
where the random chemical potential $\epsilon_{\iii}$ is distributed uniformly in $[-W_1/2,W_1/2]$, and the random imbalanced paring potential $\Delta^s_{\iii}$ is distributed uniformly in $[-W_2/2,W_2/2]$.
This Hamiltonian $H_{\rm CI}$ satisfies 
\begin{equation}
    \tau_z H_{\rm CI}^* \tau_z= H_{\rm CI}, \quad \tau_x H_{\rm CI}^{\dagger} \tau_x = - H_{\rm CI}
\end{equation}
and hence belongs to class CI. We diagonalize $2\times 10^5$ realizations with the parameters: $t = 2, \mu = 0, \Delta_x = \Delta_y =1, \Delta_z = 1/2$, $W_1 = 6, W_2 = 3$, and $L = 12$.

\section{Symmetry of SYK Lindbladian}
\label{app: SYK Lindbladian}

We analyze symmetry of the SYK Lindbladian~\cite{sa2022SYK, kulkarni2022SYK, kawabata2023} in Sec.~\ref{subsec: SYK Lindbladian}. 
Through the vectorization $\rho = \rho_{ij}\ket{i}\bra{j} \rightarrow \ket{\rho} = \rho_{ij}\ket{i}\ket{j}$, the Lindbladian becomes a non-Hermitian operator in the double Hilbert space, given by Eq.~(\ref{eq: Lindbladian}). 
For simplicity, we assume that the number $N$ of Majorana fermions $\gamma_i$'s is even.
To be specific, we choose a basis of $\gamma_i$'s such that $\gamma_{2i-1} = \gamma_{2i-1}^*$ and $\gamma_{2i} = -\gamma_{2i}^*$.
We define adjoint Majorana operators $\tg$ in the double Hilbert space by
\begin{equation} \label{eq: adjoint fermion}
 \begin{aligned} 
  \tg & \equiv (\gamma^+_1 , \cdots \gamma^+_N, \gamma^-_1 , \cdots \gamma^+_N)   \\
  &\qquad \text{with } \gamma^+_i \equiv \gamma_1 \otimes P , \,  \gamma^-_i \equiv 1 \otimes \ii P \gamma_i^* \, , 
 \end{aligned}  
\end{equation}
with the fermion parity operator
\begin{equation} 
P =   \prod_{i = 1}^{N/2} 2 \ii  \gamma_{2i-1} \gamma_{2i}.
\end{equation}
With this construction, we can verify $\{ \tg_i , \tg_j \} = \delta_{i,j}$ and $\tg_i = \tg_i^{\dagger}$ ($i,j = 1,\cdots, 2N$), 
ensuring that
$\tg$ consists of the Majorana operators in the double Hilbert space.
Additionally, we have $\gamma_{2i-1}^+ = (\gamma_{2i-1}^+)^*$ and $\gamma_{2i}^+ = -(\gamma_{2i}^+)^*$, as well as $\gamma_{2i-1}^- = -(\gamma_{2i-1}^-)^*$ and $\gamma_{2i} = (\gamma_{2i}^-)^*$.
Notably, the basis of the adjoint fermions in Eq.~(\ref{eq: adjoint fermion}) differs from that employed in Ref.~\cite{kawabata2023}, and the following expressions of symmetry operations may not be exactly identical 
to those in Ref.~\cite{kawabata2023}.
However, 
the algebra of symmetry remains consistent between the two formulations.

On the basis of $\tg$, we formulate the Lindbladian $\mathcal{L}$ in Eq.~(\ref{eq: Lindbladian}) with the $q$-body ($q \in 2 \mathbb{Z}$) SYK Hamiltonian $H$ in Eq.~(\ref{eq: SYK_q}) and the linear dissipators $L_m$ in Eq.~(\ref{eq: SYK_L}).
The Lindbladian $\mathcal{L}$ can be decomposed as $\mathcal{L}_H + \mathcal{L_D}$.
Here, $\mathcal{L}_H$ arises from the Hamiltonian, 
\begin{align} 
   \mathcal{L}_H & = -\ii \left(H \otimes 1 - 1 \otimes H^* \right) \nonumber \\
   & = -\ii  \sum_{1 \leq i_1  < \cdots < i_q \leq N}  J_{i_1i_2\cdots i_q } \ii^{q/2} \left( \gamma_{i_1} \cdots \gamma_{i_q} \otimes 1 \right. \nonumber \\
   &\qquad\qquad\qquad\qquad \left. - (-1)^q  \otimes \gamma_{i_1}^* \cdots \gamma_{i_q}^* \right) \nonumber \\
   & = - \ii \sum_{1 \leq i_1  < \cdots < i_q \leq N}  J_{i_1i_2\cdots i_q } \ii^{q/2} \left(\gamma_{i_1}^+ \cdots \gamma_{i_q}^+  \right. \nonumber \\
  &\qquad\qquad\qquad\qquad  \left. - (-1)^{q/2} \gamma_{i_1}^- \cdots \gamma_{i_q}^- \right) \, .
\end{align}
On the other hand, $\mathcal{L_D}$ arises from the dissipators,
\begin{align} 
  \mathcal{L_D}  & = \sum_{m} \left[ L_{m} \otimes L_{m}^* - \frac{1}{2} \left(L_{m}^{\dagger}L_{m} \otimes 1 \right)  - \frac{1}{2} \left(1 \otimes L_{m}^{\rm T}L_{m}^* \right) \right] \nonumber \\
  & =  \sum_{m,i,j}   K_{m;i} K_{m;j}^* \left(  \gamma_{i} \otimes \gamma_{j}^* -  \frac{1}{2} \gamma_j \gamma_i \otimes 1 - \frac{1}{2} \otimes \gamma_i^*\gamma_j^*
  \right) \nonumber \\
  & =  \sum_{m,i,j}   K_{m;i} K_{m;j}^* \left(  -\ii \gamma_{i}^+ \gamma_{j}^- -  \frac{1}{2} \gamma_j^+ \gamma_i^+ - \frac{1}{2} \gamma_i^-\gamma_j^-
  \right) \nonumber \\
  & = \sum_{m,i,j}   K_{m;i} K_{m;j}^* \left(  -\ii \gamma_{i}^+ \gamma_{j}^- 
  \right) + \nonumber \\
  & \sum_{m, i \neq j} K_{m;i} K_{m;j}^* \left(  -  \frac{1}{2} \gamma_j^+ \gamma_i^+ - \frac{1}{2} \gamma_i^-\gamma_j^-
  \right) - \frac{1}{2}\sum_{m,i} |K_{m;i}|^2 \, .
\end{align}
We have ${\rm Tr}\,\mathcal{L}_H = 0$ and ${\rm Tr}  \mathcal{L_D}/{\rm Tr}1 = -\frac{1}{2}\sum_{m,i} |K_{m;i}|^2$.

We study symmetry of the shifted Lindbladian $\mathcal{L} - {\rm Tr}\mathcal{L}/{\rm Tr}1$ with $q = 6$. 
The fermion parity operator $(-1)^{\mF}$ in the double Hilbert space is given as
\begin{equation} 
  (-1)^{\mF} = \prod_{i = 1}^{N} 2 \ii  \tg_{2i-1} \tg_{2i} =   \prod_{i = 1}^{N/2} ( -4 \gamma_{2i-1}^+ \gamma_{2i}^+   \gamma_{2i-1}^- \gamma_{2i}^-) \, ,
\end{equation}
respecting $(-1)^{\mF} \mathcal{L} (-1)^{\mF} = \mathcal{L}$ and $\left( (-1)^{\mF} \right)^2 = 1$.
We construct an anti-unitary operator 
\begin{equation} 
  \mT_+ \mK  \equiv \prod_{i = 1}^{N/2} \left[ \frac{1}{2}\left( 1 -  2 \gamma_{2i-1}^+\gamma_{2i-1}^- \right)\left(  1 +  2\gamma_{2i}^+\gamma_{2i}^- \right)  \right] \mK \, ,
\end{equation}
where $\mK$ denotes complex conjugation, satisfying $(\mT_+ \mK)^2 = 1$.
This anti-unitary operation switches $\gamma_i^{+}$ and $\gamma_i^{-}$, i.e.,  
\begin{equation} 
 \left( \mT_+ \mK \right) \gamma_i^{\pm} \left( \mT_+ \mK \right)^{-1} =  \gamma_i^{\mp} \, ,
\end{equation} 
and respects $(\mT_+ \mK)  \mathcal{L} \left( \mT_+ \mK \right)^{-1} =\mathcal{L} $.
This anti-unitary symmetry, known as modular conjugation symmetry, originates solely from the Hermiticity-preserving nature and hence is respected by arbitrary Lindbladians.
We construct another anti-unitary operator
\begin{equation} 
  \mC_- \mK \equiv  \prod_{i = 1}^{N/2} \left( 2 \ii \gamma_{2i-1}^{+}  \gamma_{2i}^- \right)\mK \, , 
\end{equation}
which satisfies
\begin{equation} 
 (\mC_- \mK)^2 = (-1)^{N/2} \, .
\end{equation}
It acts on $\gamma_i^{\pm}$ as 
\begin{equation} 
  (\mC_- \mK  ) \gamma_i^{\pm} (\mC_- \mK  )^{-1}= - \gamma_i^{\pm} \, ,
\end{equation}
and we have
\begin{equation} 
  (\mC_- \mK  )  ( \mathcal{L} - {\rm Tr}\mathcal{L}/{\rm Tr}1 )^{\dagger} (\mC_- \mK  )^{-1} = \mathcal{L} - {\rm Tr}\mathcal{L}/{\rm Tr}1\,.
\end{equation}

We consider the algebra between $(-1)^{\mF}$, $ \mT_+ \mK$, and $\mC_- \mK$. 
From straightforward calculations, we find
\begin{align} 
  (-1)^{\mF} \mT_+ \mK (-1)^{\mF} = \mT_+ \mK \, , \\
  (-1)^{\mF} \mC_- \mK (-1)^{\mF} = \mC_- \mK \, .
\end{align}
Thus, in each symmetry sector of fermion parity $(-1)^{\mF} = \pm 1$, the shifted Lindbladian $\mathcal{L} - {\rm Tr}\mathcal{L}/{\rm Tr}1$ respects both $ \mT_+ \mK$ and $\mC_- \mK$.
For $N \in 4 \mathbb{Z}$, we have $(\mC_- \mK  )^2 = + 1$, and the shifted Lindbladian belongs to class BDI;
for $N \in 4 \mathbb{Z} + 2$, we have $(\mC_- \mK  )^2 = -1$, and it belongs to class CI.

\bibliography{ref}

\begin{thebibliography}{159}%
\makeatletter
\providecommand \@ifxundefined [1]{%
 \@ifx{#1\undefined}
}%
\providecommand \@ifnum [1]{%
 \ifnum #1\expandafter \@firstoftwo
 \else \expandafter \@secondoftwo
 \fi
}%
\providecommand \@ifx [1]{%
 \ifx #1\expandafter \@firstoftwo
 \else \expandafter \@secondoftwo
 \fi
}%
\providecommand \natexlab [1]{#1}%
\providecommand \enquote  [1]{``#1''}%
\providecommand \bibnamefont  [1]{#1}%
\providecommand \bibfnamefont [1]{#1}%
\providecommand \citenamefont [1]{#1}%
\providecommand \href@noop [0]{\@secondoftwo}%
\providecommand \href [0]{\begingroup \@sanitize@url \@href}%
\providecommand \@href[1]{\@@startlink{#1}\@@href}%
\providecommand \@@href[1]{\endgroup#1\@@endlink}%
\providecommand \@sanitize@url [0]{\catcode `\\12\catcode `\$12\catcode
  `\&12\catcode `\#12\catcode `\^12\catcode `\_12\catcode `\%12\relax}%
\providecommand \@@startlink[1]{}%
\providecommand \@@endlink[0]{}%
\providecommand \url  [0]{\begingroup\@sanitize@url \@url }%
\providecommand \@url [1]{\endgroup\@href {#1}{\urlprefix }}%
\providecommand \urlprefix  [0]{URL }%
\providecommand \Eprint [0]{\href }%
\providecommand \doibase [0]{https://doi.org/}%
\providecommand \selectlanguage [0]{\@gobble}%
\providecommand \bibinfo  [0]{\@secondoftwo}%
\providecommand \bibfield  [0]{\@secondoftwo}%
\providecommand \translation [1]{[#1]}%
\providecommand \BibitemOpen [0]{}%
\providecommand \bibitemStop [0]{}%
\providecommand \bibitemNoStop [0]{.\EOS\space}%
\providecommand \EOS [0]{\spacefactor3000\relax}%
\providecommand \BibitemShut  [1]{\csname bibitem#1\endcsname}%
\let\auto@bib@innerbib\@empty
%</preamble>
\bibitem [{\citenamefont {Haake}\ \emph {et~al.}(2018)\citenamefont {Haake},
  \citenamefont {Gnutzmann},\ and\ \citenamefont {Ku\'{s}}}]{Haake-textbook}%
  \BibitemOpen
  \bibfield  {author} {\bibinfo {author} {\bibfnamefont {F.}~\bibnamefont
  {Haake}}, \bibinfo {author} {\bibfnamefont {S.}~\bibnamefont {Gnutzmann}},\
  and\ \bibinfo {author} {\bibfnamefont {M.}~\bibnamefont {Ku\'{s}}},\
  }\href@noop {} {\emph {\bibinfo {title} {{Quantum Signatures of Chaos}}}}\
  (\bibinfo  {publisher} {Springer},\ \bibinfo {address} {Cham},\ \bibinfo
  {year} {2018})\BibitemShut {NoStop}%
\bibitem [{\citenamefont {Dyson}(1962)}]{Dyson-62}%
  \BibitemOpen
  \bibfield  {author} {\bibinfo {author} {\bibfnamefont {F.~J.}\ \bibnamefont
  {Dyson}},\ }\bibfield  {title} {\bibinfo {title} {{The Threefold Way.
  Algebraic Structure of Symmetry Groups and Ensembles in Quantum Mechanics}},\
  }\href {https://doi.org/10.1063/1.1703863} {\bibfield  {journal} {\bibinfo
  {journal} {J. Math. Phys.}\ }\textbf {\bibinfo {volume} {3}},\ \bibinfo
  {pages} {1199} (\bibinfo {year} {1962})}\BibitemShut {NoStop}%
\bibitem [{\citenamefont {Wigner}(1951)}]{Wigner-51}%
  \BibitemOpen
  \bibfield  {author} {\bibinfo {author} {\bibfnamefont {E.~P.}\ \bibnamefont
  {Wigner}},\ }\bibfield  {title} {\bibinfo {title} {{On the statistical
  distribution of the widths and spacings of nuclear resonance levels}},\
  }\href {https://doi.org/10.1017/S0305004100027237} {\bibfield  {journal}
  {\bibinfo  {journal} {Math. Proc. Cambridge Philos. Soc.}\ }\textbf {\bibinfo
  {volume} {47}},\ \bibinfo {pages} {790} (\bibinfo {year} {1951})}\BibitemShut
  {NoStop}%
\bibitem [{\citenamefont {Wigner}(1958)}]{Wigner-58}%
  \BibitemOpen
  \bibfield  {author} {\bibinfo {author} {\bibfnamefont {E.~P.}\ \bibnamefont
  {Wigner}},\ }\bibfield  {title} {\bibinfo {title} {{On the Distribution of
  the Roots of Certain Symmetric Matrices}},\ }\href
  {https://doi.org/10.2307/1970008} {\bibfield  {journal} {\bibinfo  {journal}
  {Ann. Math.}\ }\textbf {\bibinfo {volume} {67}},\ \bibinfo {pages} {325}
  (\bibinfo {year} {1958})}\BibitemShut {NoStop}%
\bibitem [{\citenamefont {Bohigas}\ \emph {et~al.}(1984)\citenamefont
  {Bohigas}, \citenamefont {Giannoni},\ and\ \citenamefont
  {Schmit}}]{bohigas1984}%
  \BibitemOpen
  \bibfield  {author} {\bibinfo {author} {\bibfnamefont {O.}~\bibnamefont
  {Bohigas}}, \bibinfo {author} {\bibfnamefont {M.~J.}\ \bibnamefont
  {Giannoni}},\ and\ \bibinfo {author} {\bibfnamefont {C.}~\bibnamefont
  {Schmit}},\ }\bibfield  {title} {\bibinfo {title} {Characterization of
  {{Chaotic Quantum Spectra}} and {{Universality}} of {{Level Fluctuation
  Laws}}},\ }\href {https://doi.org/10.1103/PhysRevLett.52.1} {\bibfield
  {journal} {\bibinfo  {journal} {Phys. Rev. Lett.}\ }\textbf {\bibinfo
  {volume} {52}},\ \bibinfo {pages} {1} (\bibinfo {year} {1984})}\BibitemShut
  {NoStop}%
\bibitem [{\citenamefont {Friedrich}\ and\ \citenamefont
  {Wintgen}(1989)}]{Friedrich1987}%
  \BibitemOpen
  \bibfield  {author} {\bibinfo {author} {\bibfnamefont {H.}~\bibnamefont
  {Friedrich}}\ and\ \bibinfo {author} {\bibfnamefont {H.}~\bibnamefont
  {Wintgen}},\ }\bibfield  {title} {\bibinfo {title} {The hydrogen atom in a
  uniform magnetic field {\textemdash} {{An}} example of chaos},\ }\href
  {https://doi.org/10.1016/0370-1573(89)90121-X} {\bibfield  {journal}
  {\bibinfo  {journal} {Phys. Rep.}\ }\textbf {\bibinfo {volume} {183}},\
  \bibinfo {pages} {37} (\bibinfo {year} {1989})}\BibitemShut {NoStop}%
\bibitem [{\citenamefont {Poilblanc}\ \emph {et~al.}(1993)\citenamefont
  {Poilblanc}, \citenamefont {Ziman}, \citenamefont {Bellissard}, \citenamefont
  {Mila},\ and\ \citenamefont {Montambaux}}]{poilblanc1993}%
  \BibitemOpen
  \bibfield  {author} {\bibinfo {author} {\bibfnamefont {D.}~\bibnamefont
  {Poilblanc}}, \bibinfo {author} {\bibfnamefont {T.}~\bibnamefont {Ziman}},
  \bibinfo {author} {\bibfnamefont {J.}~\bibnamefont {Bellissard}}, \bibinfo
  {author} {\bibfnamefont {F.}~\bibnamefont {Mila}},\ and\ \bibinfo {author}
  {\bibfnamefont {G.}~\bibnamefont {Montambaux}},\ }\bibfield  {title}
  {\bibinfo {title} {Poisson vs. {{GOE Statistics}} in {{Integrable}} and
  {{Non-Integrable Quantum Hamiltonians}}},\ }\href
  {https://doi.org/10.1209/0295-5075/22/7/010} {\bibfield  {journal} {\bibinfo
  {journal} {Europhys. Lett.}\ }\textbf {\bibinfo {volume} {22}},\ \bibinfo
  {pages} {537} (\bibinfo {year} {1993})}\BibitemShut {NoStop}%
\bibitem [{\citenamefont {Mirlin}(2000)}]{mirlin00}%
  \BibitemOpen
  \bibfield  {author} {\bibinfo {author} {\bibfnamefont {A.~D.}\ \bibnamefont
  {Mirlin}},\ }\bibfield  {title} {\bibinfo {title} {{Statistics of energy
  levels and eigenfunctions in disordered systems}},\ }\href
  {https://doi.org/10.1016/S0370-1573(99)00091-5} {\bibfield  {journal}
  {\bibinfo  {journal} {Phys. Rep.}\ }\textbf {\bibinfo {volume} {326}},\
  \bibinfo {pages} {259} (\bibinfo {year} {2000})}\BibitemShut {NoStop}%
\bibitem [{\citenamefont {Serbyn}\ and\ \citenamefont
  {Moore}(2016)}]{Serbyn2016}%
  \BibitemOpen
  \bibfield  {author} {\bibinfo {author} {\bibfnamefont {M.}~\bibnamefont
  {Serbyn}}\ and\ \bibinfo {author} {\bibfnamefont {J.~E.}\ \bibnamefont
  {Moore}},\ }\bibfield  {title} {\bibinfo {title} {Spectral statistics across
  the many-body localization transition},\ }\href
  {https://doi.org/10.1103/PhysRevB.93.041424} {\bibfield  {journal} {\bibinfo
  {journal} {Phys. Rev. B}\ }\textbf {\bibinfo {volume} {93}},\ \bibinfo
  {pages} {041424} (\bibinfo {year} {2016})}\BibitemShut {NoStop}%
\bibitem [{\citenamefont {Verbaarschot}\ and\ \citenamefont
  {Zahed}(1993)}]{Verbaarschot-93}%
  \BibitemOpen
  \bibfield  {author} {\bibinfo {author} {\bibfnamefont {J.~J.~M.}\
  \bibnamefont {Verbaarschot}}\ and\ \bibinfo {author} {\bibfnamefont
  {I.}~\bibnamefont {Zahed}},\ }\bibfield  {title} {\bibinfo {title} {{Spectral
  density of the QCD Dirac operator near zero virtuality}},\ }\href
  {https://doi.org/10.1103/PhysRevLett.70.3852} {\bibfield  {journal} {\bibinfo
   {journal} {Phys. Rev. Lett.}\ }\textbf {\bibinfo {volume} {70}},\ \bibinfo
  {pages} {3852} (\bibinfo {year} {1993})}\BibitemShut {NoStop}%
\bibitem [{\citenamefont {Verbaarschot}(1994)}]{Verbaarschot-94}%
  \BibitemOpen
  \bibfield  {author} {\bibinfo {author} {\bibfnamefont {J.}~\bibnamefont
  {Verbaarschot}},\ }\bibfield  {title} {\bibinfo {title} {{Spectrum of the QCD
  Dirac operator and chiral random matrix theory}},\ }\href
  {https://doi.org/10.1103/PhysRevLett.72.2531} {\bibfield  {journal} {\bibinfo
   {journal} {Phys. Rev. Lett.}\ }\textbf {\bibinfo {volume} {72}},\ \bibinfo
  {pages} {2531} (\bibinfo {year} {1994})}\BibitemShut {NoStop}%
\bibitem [{\citenamefont {Altland}\ and\ \citenamefont
  {Zirnbauer}(1997)}]{altland1997}%
  \BibitemOpen
  \bibfield  {author} {\bibinfo {author} {\bibfnamefont {A.}~\bibnamefont
  {Altland}}\ and\ \bibinfo {author} {\bibfnamefont {M.~R.}\ \bibnamefont
  {Zirnbauer}},\ }\bibfield  {title} {\bibinfo {title} {Nonstandard symmetry
  classes in mesoscopic normal-superconducting hybrid structures},\ }\href
  {https://doi.org/10.1103/PhysRevB.55.1142} {\bibfield  {journal} {\bibinfo
  {journal} {Phys. Rev. B}\ }\textbf {\bibinfo {volume} {55}},\ \bibinfo
  {pages} {1142} (\bibinfo {year} {1997})}\BibitemShut {NoStop}%
\bibitem [{\citenamefont {Beenakker}(1997)}]{beenakker1997}%
  \BibitemOpen
  \bibfield  {author} {\bibinfo {author} {\bibfnamefont {C.~W.~J.}\
  \bibnamefont {Beenakker}},\ }\bibfield  {title} {\bibinfo {title}
  {Random-matrix theory of quantum transport},\ }\href
  {https://doi.org/10.1103/RevModPhys.69.731} {\bibfield  {journal} {\bibinfo
  {journal} {Rev. Mod. Phys.}\ }\textbf {\bibinfo {volume} {69}},\ \bibinfo
  {pages} {731} (\bibinfo {year} {1997})}\BibitemShut {NoStop}%
\bibitem [{\citenamefont {Beenakker}(2015)}]{beenakkerRMT15}%
  \BibitemOpen
  \bibfield  {author} {\bibinfo {author} {\bibfnamefont {C.~W.~J.}\
  \bibnamefont {Beenakker}},\ }\bibfield  {title} {\bibinfo {title}
  {Random-matrix theory of {{Majorana}} fermions and topological
  superconductors},\ }\href {https://doi.org/10.1103/RevModPhys.87.1037}
  {\bibfield  {journal} {\bibinfo  {journal} {Rev. Mod. Phys.}\ }\textbf
  {\bibinfo {volume} {87}},\ \bibinfo {pages} {1037} (\bibinfo {year}
  {2015})}\BibitemShut {NoStop}%
\bibitem [{\citenamefont {Evers}\ and\ \citenamefont
  {Mirlin}(2008)}]{Evers-review}%
  \BibitemOpen
  \bibfield  {author} {\bibinfo {author} {\bibfnamefont {F.}~\bibnamefont
  {Evers}}\ and\ \bibinfo {author} {\bibfnamefont {A.~D.}\ \bibnamefont
  {Mirlin}},\ }\bibfield  {title} {\bibinfo {title} {{Anderson transitions}},\
  }\href {https://doi.org/10.1103/RevModPhys.80.1355} {\bibfield  {journal}
  {\bibinfo  {journal} {Rev. Mod. Phys.}\ }\textbf {\bibinfo {volume} {80}},\
  \bibinfo {pages} {1355} (\bibinfo {year} {2008})}\BibitemShut {NoStop}%
\bibitem [{\citenamefont {Nandkishore}\ and\ \citenamefont
  {Huse}(2015)}]{Huse-review}%
  \BibitemOpen
  \bibfield  {author} {\bibinfo {author} {\bibfnamefont {R.}~\bibnamefont
  {Nandkishore}}\ and\ \bibinfo {author} {\bibfnamefont {D.~A.}\ \bibnamefont
  {Huse}},\ }\bibfield  {title} {\bibinfo {title} {{Many-Body Localization and
  Thermalization in Quantum Statistical Mechanics}},\ }\href
  {https://doi.org/10.1146/annurev-conmatphys-031214-014726} {\bibfield
  {journal} {\bibinfo  {journal} {Annu. Rev. Condens. Matter Phys.}\ }\textbf
  {\bibinfo {volume} {6}},\ \bibinfo {pages} {15} (\bibinfo {year}
  {2015})}\BibitemShut {NoStop}%
\bibitem [{\citenamefont {Abanin}\ \emph {et~al.}(2019)\citenamefont {Abanin},
  \citenamefont {Altman}, \citenamefont {Bloch},\ and\ \citenamefont
  {Serbyn}}]{Abanin19}%
  \BibitemOpen
  \bibfield  {author} {\bibinfo {author} {\bibfnamefont {D.~A.}\ \bibnamefont
  {Abanin}}, \bibinfo {author} {\bibfnamefont {E.}~\bibnamefont {Altman}},
  \bibinfo {author} {\bibfnamefont {I.}~\bibnamefont {Bloch}},\ and\ \bibinfo
  {author} {\bibfnamefont {M.}~\bibnamefont {Serbyn}},\ }\bibfield  {title}
  {\bibinfo {title} {Colloquium: Many-body localization, thermalization, and
  entanglement},\ }\href {https://doi.org/10.1103/RevModPhys.91.021001}
  {\bibfield  {journal} {\bibinfo  {journal} {Rev. Mod. Phys.}\ }\textbf
  {\bibinfo {volume} {91}},\ \bibinfo {pages} {021001} (\bibinfo {year}
  {2019})}\BibitemShut {NoStop}%
\bibitem [{\citenamefont {Verbaarschot}\ and\ \citenamefont
  {Wettig}(2000)}]{verbaarschot2000}%
  \BibitemOpen
  \bibfield  {author} {\bibinfo {author} {\bibfnamefont {J.}~\bibnamefont
  {Verbaarschot}}\ and\ \bibinfo {author} {\bibfnamefont {T.}~\bibnamefont
  {Wettig}},\ }\bibfield  {title} {\bibinfo {title} {Random {{Matrix Theory}}
  and {{Chiral Symmetry}} in {{QCD}}},\ }\href
  {https://doi.org/10.1146/annurev.nucl.50.1.343} {\bibfield  {journal}
  {\bibinfo  {journal} {Annu. Rev. Nucl. Part. Sci.}\ }\textbf {\bibinfo
  {volume} {50}},\ \bibinfo {pages} {343} (\bibinfo {year} {2000})}\BibitemShut
  {NoStop}%
\bibitem [{\citenamefont {Osborn}(2004)}]{osborn2004}%
  \BibitemOpen
  \bibfield  {author} {\bibinfo {author} {\bibfnamefont {J.~C.}\ \bibnamefont
  {Osborn}},\ }\bibfield  {title} {\bibinfo {title} {Universal {{Results}} from
  an {{Alternate Random-Matrix Model}} for {{QCD}} with a {{Baryon Chemical
  Potential}}},\ }\href {https://doi.org/10.1103/PhysRevLett.93.222001}
  {\bibfield  {journal} {\bibinfo  {journal} {Phys. Rev. Lett.}\ }\textbf
  {\bibinfo {volume} {93}},\ \bibinfo {pages} {222001} (\bibinfo {year}
  {2004})}\BibitemShut {NoStop}%
\bibitem [{\citenamefont {Akemann}\ and\ \citenamefont
  {Wettig}(2004)}]{akemann2004}%
  \BibitemOpen
  \bibfield  {author} {\bibinfo {author} {\bibfnamefont {G.}~\bibnamefont
  {Akemann}}\ and\ \bibinfo {author} {\bibfnamefont {T.}~\bibnamefont
  {Wettig}},\ }\bibfield  {title} {\bibinfo {title} {{{QCD Dirac Operator}} at
  {{Nonzero Chemical Potential}}: {{Lattice Data}} and {{Matrix Model}}},\
  }\href {https://doi.org/10.1103/PhysRevLett.92.102002} {\bibfield  {journal}
  {\bibinfo  {journal} {Phys. Rev. Lett.}\ }\textbf {\bibinfo {volume} {92}},\
  \bibinfo {pages} {102002} (\bibinfo {year} {2004})}\BibitemShut {NoStop}%
\bibitem [{\citenamefont {Kanazawa}\ and\ \citenamefont
  {Wettig}(2021)}]{kanazawa2021}%
  \BibitemOpen
  \bibfield  {author} {\bibinfo {author} {\bibfnamefont {T.}~\bibnamefont
  {Kanazawa}}\ and\ \bibinfo {author} {\bibfnamefont {T.}~\bibnamefont
  {Wettig}},\ }\bibfield  {title} {\bibinfo {title} {New universality classes
  of the non-{{Hermitian Dirac}} operator in {{QCD-like}} theories},\ }\href
  {https://doi.org/10.1103/PhysRevD.104.014509} {\bibfield  {journal} {\bibinfo
   {journal} {Phys. Rev. D}\ }\textbf {\bibinfo {volume} {104}},\ \bibinfo
  {pages} {014509} (\bibinfo {year} {2021})}\BibitemShut {NoStop}%
\bibitem [{\citenamefont {May}(1976)}]{may1976}%
  \BibitemOpen
  \bibfield  {author} {\bibinfo {author} {\bibfnamefont {R.~M.}\ \bibnamefont
  {May}},\ }\bibfield  {title} {\bibinfo {title} {Simple mathematical models
  with very complicated dynamics},\ }\href {https://doi.org/10.1038/261459a0}
  {\bibfield  {journal} {\bibinfo  {journal} {Nature}\ }\textbf {\bibinfo
  {volume} {261}},\ \bibinfo {pages} {459} (\bibinfo {year}
  {1976})}\BibitemShut {NoStop}%
\bibitem [{\citenamefont {Sommers}\ \emph
  {et~al.}(1988{\natexlab{a}})\citenamefont {Sommers}, \citenamefont
  {Crisanti}, \citenamefont {Sompolinsky},\ and\ \citenamefont
  {Stein}}]{sommers88}%
  \BibitemOpen
  \bibfield  {author} {\bibinfo {author} {\bibfnamefont {H.~J.}\ \bibnamefont
  {Sommers}}, \bibinfo {author} {\bibfnamefont {A.}~\bibnamefont {Crisanti}},
  \bibinfo {author} {\bibfnamefont {H.}~\bibnamefont {Sompolinsky}},\ and\
  \bibinfo {author} {\bibfnamefont {Y.}~\bibnamefont {Stein}},\ }\bibfield
  {title} {\bibinfo {title} {{Spectrum of Large Random Asymmetric Matrices}},\
  }\href {https://doi.org/10.1103/PhysRevLett.60.1895} {\bibfield  {journal}
  {\bibinfo  {journal} {Phys. Rev. Lett.}\ }\textbf {\bibinfo {volume} {60}},\
  \bibinfo {pages} {1895} (\bibinfo {year} {1988}{\natexlab{a}})}\BibitemShut
  {NoStop}%
\bibitem [{\citenamefont {Sompolinsky}\ \emph {et~al.}(1988)\citenamefont
  {Sompolinsky}, \citenamefont {Crisanti},\ and\ \citenamefont
  {Sommers}}]{sompolinsky88}%
  \BibitemOpen
  \bibfield  {author} {\bibinfo {author} {\bibfnamefont {H.}~\bibnamefont
  {Sompolinsky}}, \bibinfo {author} {\bibfnamefont {A.}~\bibnamefont
  {Crisanti}},\ and\ \bibinfo {author} {\bibfnamefont {H.~J.}\ \bibnamefont
  {Sommers}},\ }\bibfield  {title} {\bibinfo {title} {{Chaos in Random Neural
  Networks}},\ }\href {https://doi.org/10.1103/PhysRevLett.61.259} {\bibfield
  {journal} {\bibinfo  {journal} {Phys. Rev. Lett.}\ }\textbf {\bibinfo
  {volume} {61}},\ \bibinfo {pages} {259} (\bibinfo {year} {1988})}\BibitemShut
  {NoStop}%
\bibitem [{\citenamefont {Ott}\ \emph {et~al.}(1990)\citenamefont {Ott},
  \citenamefont {Grebogi},\ and\ \citenamefont {Yorke}}]{ott1990}%
  \BibitemOpen
  \bibfield  {author} {\bibinfo {author} {\bibfnamefont {E.}~\bibnamefont
  {Ott}}, \bibinfo {author} {\bibfnamefont {C.}~\bibnamefont {Grebogi}},\ and\
  \bibinfo {author} {\bibfnamefont {J.~A.}\ \bibnamefont {Yorke}},\ }\bibfield
  {title} {\bibinfo {title} {Controlling chaos},\ }\href
  {https://doi.org/10.1103/PhysRevLett.64.1196} {\bibfield  {journal} {\bibinfo
   {journal} {Phys. Rev. Lett.}\ }\textbf {\bibinfo {volume} {64}},\ \bibinfo
  {pages} {1196} (\bibinfo {year} {1990})}\BibitemShut {NoStop}%
\bibitem [{\citenamefont {Nelson}\ and\ \citenamefont
  {Shnerb}(1998)}]{nelson98}%
  \BibitemOpen
  \bibfield  {author} {\bibinfo {author} {\bibfnamefont {D.~R.}\ \bibnamefont
  {Nelson}}\ and\ \bibinfo {author} {\bibfnamefont {N.~M.}\ \bibnamefont
  {Shnerb}},\ }\bibfield  {title} {\bibinfo {title} {{Non-Hermitian
  localization and population biology}},\ }\href
  {https://doi.org/10.1103/PhysRevE.58.1383} {\bibfield  {journal} {\bibinfo
  {journal} {Phys. Rev. E}\ }\textbf {\bibinfo {volume} {58}},\ \bibinfo
  {pages} {1383} (\bibinfo {year} {1998})}\BibitemShut {NoStop}%
\bibitem [{\citenamefont {Amir}\ \emph {et~al.}(2016)\citenamefont {Amir},
  \citenamefont {Hatano},\ and\ \citenamefont {Nelson}}]{amir2016}%
  \BibitemOpen
  \bibfield  {author} {\bibinfo {author} {\bibfnamefont {A.}~\bibnamefont
  {Amir}}, \bibinfo {author} {\bibfnamefont {N.}~\bibnamefont {Hatano}},\ and\
  \bibinfo {author} {\bibfnamefont {D.~R.}\ \bibnamefont {Nelson}},\ }\bibfield
   {title} {\bibinfo {title} {Non-{{Hermitian}} localization in biological
  networks},\ }\href {https://doi.org/10.1103/PhysRevE.93.042310} {\bibfield
  {journal} {\bibinfo  {journal} {Phys. Rev. E}\ }\textbf {\bibinfo {volume}
  {93}},\ \bibinfo {pages} {042310} (\bibinfo {year} {2016})}\BibitemShut
  {NoStop}%
\bibitem [{\citenamefont {Murugan}\ and\ \citenamefont
  {Vaikuntanathan}(2017)}]{murugan2017}%
  \BibitemOpen
  \bibfield  {author} {\bibinfo {author} {\bibfnamefont {A.}~\bibnamefont
  {Murugan}}\ and\ \bibinfo {author} {\bibfnamefont {S.}~\bibnamefont
  {Vaikuntanathan}},\ }\bibfield  {title} {\bibinfo {title} {Topologically
  protected modes in non-equilibrium stochastic systems},\ }\href
  {https://doi.org/10.1038/ncomms13881} {\bibfield  {journal} {\bibinfo
  {journal} {Nat. Commun.}\ }\textbf {\bibinfo {volume} {8}},\ \bibinfo {pages}
  {13881} (\bibinfo {year} {2017})}\BibitemShut {NoStop}%
\bibitem [{\citenamefont {Zhang}\ and\ \citenamefont
  {Nelson}(2019)}]{zhang2019a}%
  \BibitemOpen
  \bibfield  {author} {\bibinfo {author} {\bibfnamefont {G.~H.}\ \bibnamefont
  {Zhang}}\ and\ \bibinfo {author} {\bibfnamefont {D.~R.}\ \bibnamefont
  {Nelson}},\ }\bibfield  {title} {\bibinfo {title} {Eigenvalue repulsion and
  eigenvector localization in sparse non-{{Hermitian}} random matrices},\
  }\href {https://doi.org/10.1103/PhysRevE.100.052315} {\bibfield  {journal}
  {\bibinfo  {journal} {Phys. Rev. E}\ }\textbf {\bibinfo {volume} {100}},\
  \bibinfo {pages} {052315} (\bibinfo {year} {2019})}\BibitemShut {NoStop}%
\bibitem [{\citenamefont {Breuer}\ and\ \citenamefont
  {Petruccione}(2007)}]{breuer2002theory}%
  \BibitemOpen
  \bibfield  {author} {\bibinfo {author} {\bibfnamefont {H.-P.}\ \bibnamefont
  {Breuer}}\ and\ \bibinfo {author} {\bibfnamefont {F.}~\bibnamefont
  {Petruccione}},\ }\href@noop {} {\emph {\bibinfo {title} {{The Theory of Open
  Quantum Systems}}}}\ (\bibinfo  {publisher} {Oxford University Press},\
  \bibinfo {address} {Oxford},\ \bibinfo {year} {2007})\BibitemShut {NoStop}%
\bibitem [{\citenamefont {Gorini}\ \emph {et~al.}(1976)\citenamefont {Gorini},
  \citenamefont {Kossakowski},\ and\ \citenamefont {Sudarshan}}]{GKS-76}%
  \BibitemOpen
  \bibfield  {author} {\bibinfo {author} {\bibfnamefont {V.}~\bibnamefont
  {Gorini}}, \bibinfo {author} {\bibfnamefont {A.}~\bibnamefont
  {Kossakowski}},\ and\ \bibinfo {author} {\bibfnamefont {E.~C.~G.}\
  \bibnamefont {Sudarshan}},\ }\bibfield  {title} {\bibinfo {title}
  {{Completely positive dynamical semigroups of $N$‐level systems}},\ }\href
  {https://doi.org/10.1063/1.522979} {\bibfield  {journal} {\bibinfo  {journal}
  {J. Math. Phys.}\ }\textbf {\bibinfo {volume} {17}},\ \bibinfo {pages} {821}
  (\bibinfo {year} {1976})}\BibitemShut {NoStop}%
\bibitem [{\citenamefont {Lindblad}(1976)}]{Lindblad-76}%
  \BibitemOpen
  \bibfield  {author} {\bibinfo {author} {\bibfnamefont {G.}~\bibnamefont
  {Lindblad}},\ }\bibfield  {title} {\bibinfo {title} {{On the generators of
  quantum dynamical semigroups}},\ }\href {https://doi.org/10.1007/BF01608499}
  {\bibfield  {journal} {\bibinfo  {journal} {Commun. Math. Phys.}\ }\textbf
  {\bibinfo {volume} {48}},\ \bibinfo {pages} {119} (\bibinfo {year}
  {1976})}\BibitemShut {NoStop}%
\bibitem [{\citenamefont {Makris}\ \emph {et~al.}(2008)\citenamefont {Makris},
  \citenamefont {El-Ganainy}, \citenamefont {Christodoulides},\ and\
  \citenamefont {Musslimani}}]{Makris08}%
  \BibitemOpen
  \bibfield  {author} {\bibinfo {author} {\bibfnamefont {K.~G.}\ \bibnamefont
  {Makris}}, \bibinfo {author} {\bibfnamefont {R.}~\bibnamefont {El-Ganainy}},
  \bibinfo {author} {\bibfnamefont {D.~N.}\ \bibnamefont {Christodoulides}},\
  and\ \bibinfo {author} {\bibfnamefont {Z.~H.}\ \bibnamefont {Musslimani}},\
  }\bibfield  {title} {\bibinfo {title} {{Beam Dynamics in
  $\mathcal{P}\mathcal{T}$ Symmetric Optical Lattices}},\ }\href
  {https://doi.org/10.1103/PhysRevLett.100.103904} {\bibfield  {journal}
  {\bibinfo  {journal} {Phys. Rev. Lett.}\ }\textbf {\bibinfo {volume} {100}},\
  \bibinfo {pages} {103904} (\bibinfo {year} {2008})}\BibitemShut {NoStop}%
\bibitem [{\citenamefont {Guo}\ \emph {et~al.}(2009)\citenamefont {Guo},
  \citenamefont {Salamo}, \citenamefont {Duchesne}, \citenamefont {Morandotti},
  \citenamefont {Volatier-Ravat}, \citenamefont {Aimez}, \citenamefont
  {Siviloglou},\ and\ \citenamefont {Christodoulides}}]{guo2009}%
  \BibitemOpen
  \bibfield  {author} {\bibinfo {author} {\bibfnamefont {A.}~\bibnamefont
  {Guo}}, \bibinfo {author} {\bibfnamefont {G.~J.}\ \bibnamefont {Salamo}},
  \bibinfo {author} {\bibfnamefont {D.}~\bibnamefont {Duchesne}}, \bibinfo
  {author} {\bibfnamefont {R.}~\bibnamefont {Morandotti}}, \bibinfo {author}
  {\bibfnamefont {M.}~\bibnamefont {Volatier-Ravat}}, \bibinfo {author}
  {\bibfnamefont {V.}~\bibnamefont {Aimez}}, \bibinfo {author} {\bibfnamefont
  {G.~A.}\ \bibnamefont {Siviloglou}},\ and\ \bibinfo {author} {\bibfnamefont
  {D.~N.}\ \bibnamefont {Christodoulides}},\ }\bibfield  {title} {\bibinfo
  {title} {{Observation of $\mathcal{P}\mathcal{T}$-Symmetry Breaking in
  Complex Optical Potentials}},\ }\href
  {https://doi.org/10.1103/PhysRevLett.103.093902} {\bibfield  {journal}
  {\bibinfo  {journal} {Phys. Rev. Lett.}\ }\textbf {\bibinfo {volume} {103}},\
  \bibinfo {pages} {093902} (\bibinfo {year} {2009})}\BibitemShut {NoStop}%
\bibitem [{\citenamefont {Zeuner}\ \emph {et~al.}(2015)\citenamefont {Zeuner},
  \citenamefont {Rechtsman}, \citenamefont {Plotnik}, \citenamefont {Lumer},
  \citenamefont {Nolte}, \citenamefont {Rudner}, \citenamefont {Segev},\ and\
  \citenamefont {Szameit}}]{zeuner2015}%
  \BibitemOpen
  \bibfield  {author} {\bibinfo {author} {\bibfnamefont {J.~M.}\ \bibnamefont
  {Zeuner}}, \bibinfo {author} {\bibfnamefont {M.~C.}\ \bibnamefont
  {Rechtsman}}, \bibinfo {author} {\bibfnamefont {Y.}~\bibnamefont {Plotnik}},
  \bibinfo {author} {\bibfnamefont {Y.}~\bibnamefont {Lumer}}, \bibinfo
  {author} {\bibfnamefont {S.}~\bibnamefont {Nolte}}, \bibinfo {author}
  {\bibfnamefont {M.~S.}\ \bibnamefont {Rudner}}, \bibinfo {author}
  {\bibfnamefont {M.}~\bibnamefont {Segev}},\ and\ \bibinfo {author}
  {\bibfnamefont {A.}~\bibnamefont {Szameit}},\ }\bibfield  {title} {\bibinfo
  {title} {Observation of a {{Topological Transition}} in the {{Bulk}} of a
  {{Non-Hermitian System}}},\ }\href
  {https://doi.org/10.1103/PhysRevLett.115.040402} {\bibfield  {journal}
  {\bibinfo  {journal} {Phys. Rev. Lett.}\ }\textbf {\bibinfo {volume} {115}},\
  \bibinfo {pages} {040402} (\bibinfo {year} {2015})}\BibitemShut {NoStop}%
\bibitem [{\citenamefont {Tzortzakakis}\ \emph {et~al.}(2020)\citenamefont
  {Tzortzakakis}, \citenamefont {Makris},\ and\ \citenamefont
  {Economou}}]{tzortzakakis2020}%
  \BibitemOpen
  \bibfield  {author} {\bibinfo {author} {\bibfnamefont {A.~F.}\ \bibnamefont
  {Tzortzakakis}}, \bibinfo {author} {\bibfnamefont {K.~G.}\ \bibnamefont
  {Makris}},\ and\ \bibinfo {author} {\bibfnamefont {E.~N.}\ \bibnamefont
  {Economou}},\ }\bibfield  {title} {\bibinfo {title} {Non-{{Hermitian}}
  disorder in two-dimensional optical lattices},\ }\href
  {https://doi.org/10.1103/PhysRevB.101.014202} {\bibfield  {journal} {\bibinfo
   {journal} {Phys. Rev. B}\ }\textbf {\bibinfo {volume} {101}},\ \bibinfo
  {pages} {014202} (\bibinfo {year} {2020})}\BibitemShut {NoStop}%
\bibitem [{\citenamefont {Lee}\ and\ \citenamefont {Chan}(2014)}]{lee2014}%
  \BibitemOpen
  \bibfield  {author} {\bibinfo {author} {\bibfnamefont {T.~E.}\ \bibnamefont
  {Lee}}\ and\ \bibinfo {author} {\bibfnamefont {C.-K.}\ \bibnamefont {Chan}},\
  }\bibfield  {title} {\bibinfo {title} {Heralded {{Magnetism}} in
  {{Non-Hermitian Atomic Systems}}},\ }\href
  {https://doi.org/10.1103/PhysRevX.4.041001} {\bibfield  {journal} {\bibinfo
  {journal} {Phys. Rev. X}\ }\textbf {\bibinfo {volume} {4}},\ \bibinfo {pages}
  {041001} (\bibinfo {year} {2014})}\BibitemShut {NoStop}%
\bibitem [{\citenamefont {Li}\ \emph {et~al.}(2020)\citenamefont {Li},
  \citenamefont {Lee},\ and\ \citenamefont {Gong}}]{li2020}%
  \BibitemOpen
  \bibfield  {author} {\bibinfo {author} {\bibfnamefont {L.}~\bibnamefont
  {Li}}, \bibinfo {author} {\bibfnamefont {C.~H.}\ \bibnamefont {Lee}},\ and\
  \bibinfo {author} {\bibfnamefont {J.}~\bibnamefont {Gong}},\ }\bibfield
  {title} {\bibinfo {title} {Topological {{Switch}} for {{Non-Hermitian Skin
  Effect}} in {{Cold-Atom Systems}} with {{Loss}}},\ }\href
  {https://doi.org/10.1103/PhysRevLett.124.250402} {\bibfield  {journal}
  {\bibinfo  {journal} {Phys. Rev. Lett.}\ }\textbf {\bibinfo {volume} {124}},\
  \bibinfo {pages} {250402} (\bibinfo {year} {2020})}\BibitemShut {NoStop}%
\bibitem [{\citenamefont {Liang}\ \emph {et~al.}(2022)\citenamefont {Liang},
  \citenamefont {Xie}, \citenamefont {Dong}, \citenamefont {Li}, \citenamefont
  {Li}, \citenamefont {Gadway}, \citenamefont {Yi},\ and\ \citenamefont
  {Yan}}]{liang2022}%
  \BibitemOpen
  \bibfield  {author} {\bibinfo {author} {\bibfnamefont {Q.}~\bibnamefont
  {Liang}}, \bibinfo {author} {\bibfnamefont {D.}~\bibnamefont {Xie}}, \bibinfo
  {author} {\bibfnamefont {Z.}~\bibnamefont {Dong}}, \bibinfo {author}
  {\bibfnamefont {H.}~\bibnamefont {Li}}, \bibinfo {author} {\bibfnamefont
  {H.}~\bibnamefont {Li}}, \bibinfo {author} {\bibfnamefont {B.}~\bibnamefont
  {Gadway}}, \bibinfo {author} {\bibfnamefont {W.}~\bibnamefont {Yi}},\ and\
  \bibinfo {author} {\bibfnamefont {B.}~\bibnamefont {Yan}},\ }\bibfield
  {title} {\bibinfo {title} {Dynamic {{Signatures}} of {{Non-Hermitian Skin
  Effect}} and {{Topology}} in {{Ultracold Atoms}}},\ }\href
  {https://doi.org/10.1103/PhysRevLett.129.070401} {\bibfield  {journal}
  {\bibinfo  {journal} {Phys. Rev. Lett.}\ }\textbf {\bibinfo {volume} {129}},\
  \bibinfo {pages} {070401} (\bibinfo {year} {2022})}\BibitemShut {NoStop}%
\bibitem [{\citenamefont {Daley}(2014)}]{Daley-review}%
  \BibitemOpen
  \bibfield  {author} {\bibinfo {author} {\bibfnamefont {A.~J.}\ \bibnamefont
  {Daley}},\ }\bibfield  {title} {\bibinfo {title} {{Quantum trajectories and
  open many-body quantum systems}},\ }\href
  {https://doi.org/10.1080/00018732.2014.933502} {\bibfield  {journal}
  {\bibinfo  {journal} {Adv. Phys.}\ }\textbf {\bibinfo {volume} {63}},\
  \bibinfo {pages} {77} (\bibinfo {year} {2014})}\BibitemShut {NoStop}%
\bibitem [{\citenamefont {Bender}(2007)}]{Bender-review}%
  \BibitemOpen
  \bibfield  {author} {\bibinfo {author} {\bibfnamefont {C.~M.}\ \bibnamefont
  {Bender}},\ }\bibfield  {title} {\bibinfo {title} {{Making sense of
  non-Hermitian Hamiltonians}},\ }\href
  {https://doi.org/10.1088/0034-4885/70/6/R03} {\bibfield  {journal} {\bibinfo
  {journal} {Rep. Prog. Phys.}\ }\textbf {\bibinfo {volume} {70}},\ \bibinfo
  {pages} {947} (\bibinfo {year} {2007})}\BibitemShut {NoStop}%
\bibitem [{\citenamefont {Konotop}\ \emph {et~al.}(2016)\citenamefont
  {Konotop}, \citenamefont {Yang},\ and\ \citenamefont
  {Zezyulin}}]{Konotop-review}%
  \BibitemOpen
  \bibfield  {author} {\bibinfo {author} {\bibfnamefont {V.~V.}\ \bibnamefont
  {Konotop}}, \bibinfo {author} {\bibfnamefont {J.}~\bibnamefont {Yang}},\ and\
  \bibinfo {author} {\bibfnamefont {D.~A.}\ \bibnamefont {Zezyulin}},\
  }\bibfield  {title} {\bibinfo {title} {{Nonlinear waves in
  $\mathcal{PT}$-symmetric systems}},\ }\href
  {https://doi.org/10.1103/RevModPhys.88.035002} {\bibfield  {journal}
  {\bibinfo  {journal} {Rev. Mod. Phys.}\ }\textbf {\bibinfo {volume} {88}},\
  \bibinfo {pages} {035002} (\bibinfo {year} {2016})}\BibitemShut {NoStop}%
\bibitem [{\citenamefont {El-Ganainy}\ \emph {et~al.}(2018)\citenamefont
  {El-Ganainy}, \citenamefont {Makris}, \citenamefont {Khajavikhan},
  \citenamefont {Musslimani}, \citenamefont {Rotter},\ and\ \citenamefont
  {Christodoulides}}]{Christodoulides-review}%
  \BibitemOpen
  \bibfield  {author} {\bibinfo {author} {\bibfnamefont {R.}~\bibnamefont
  {El-Ganainy}}, \bibinfo {author} {\bibfnamefont {K.~G.}\ \bibnamefont
  {Makris}}, \bibinfo {author} {\bibfnamefont {M.}~\bibnamefont {Khajavikhan}},
  \bibinfo {author} {\bibfnamefont {Z.~H.}\ \bibnamefont {Musslimani}},
  \bibinfo {author} {\bibfnamefont {S.}~\bibnamefont {Rotter}},\ and\ \bibinfo
  {author} {\bibfnamefont {D.~N.}\ \bibnamefont {Christodoulides}},\ }\bibfield
   {title} {\bibinfo {title} {{Non-Hermitian physics and PT symmetry}},\ }\href
  {https://doi.org/10.1038/nphys4323} {\bibfield  {journal} {\bibinfo
  {journal} {Nat. Phys.}\ }\textbf {\bibinfo {volume} {14}},\ \bibinfo {pages}
  {11} (\bibinfo {year} {2018})}\BibitemShut {NoStop}%
\bibitem [{\citenamefont {Bergholtz}\ \emph {et~al.}(2021)\citenamefont
  {Bergholtz}, \citenamefont {Budich},\ and\ \citenamefont
  {Kunst}}]{bergholtz2021}%
  \BibitemOpen
  \bibfield  {author} {\bibinfo {author} {\bibfnamefont {E.~J.}\ \bibnamefont
  {Bergholtz}}, \bibinfo {author} {\bibfnamefont {J.~C.}\ \bibnamefont
  {Budich}},\ and\ \bibinfo {author} {\bibfnamefont {F.~K.}\ \bibnamefont
  {Kunst}},\ }\bibfield  {title} {\bibinfo {title} {Exceptional topology of
  non-{{Hermitian}} systems},\ }\href
  {https://doi.org/10.1103/RevModPhys.93.015005} {\bibfield  {journal}
  {\bibinfo  {journal} {Rev. Mod. Phys.}\ }\textbf {\bibinfo {volume} {93}},\
  \bibinfo {pages} {015005} (\bibinfo {year} {2021})}\BibitemShut {NoStop}%
\bibitem [{\citenamefont {Bernard}\ and\ \citenamefont {LeClair}(2002)}]{BL02}%
  \BibitemOpen
  \bibfield  {author} {\bibinfo {author} {\bibfnamefont {D.}~\bibnamefont
  {Bernard}}\ and\ \bibinfo {author} {\bibfnamefont {A.}~\bibnamefont
  {LeClair}},\ }\bibfield  {title} {\bibinfo {title} {{A Classification of
  Non-Hermitian Random Matrices}},\ }in\ \href
  {https://doi.org/10.1007/978-94-010-0514-2_19} {\emph {\bibinfo {booktitle}
  {Statistical Field Theories}}},\ \bibinfo {series} {NATO Science Series},
  Vol.~\bibinfo {volume} {73},\ \bibinfo {editor} {edited by\ \bibinfo {editor}
  {\bibfnamefont {A.}~\bibnamefont {Cappelli}}\ and\ \bibinfo {editor}
  {\bibfnamefont {G.}~\bibnamefont {Mussardo}}}\ (\bibinfo {year} {2002})\ pp.\
  \bibinfo {pages} {207--214}\BibitemShut {NoStop}%
\bibitem [{\citenamefont {Kawabata}\ \emph
  {et~al.}(2019{\natexlab{a}})\citenamefont {Kawabata}, \citenamefont
  {Shiozaki}, \citenamefont {Ueda},\ and\ \citenamefont {Sato}}]{kawabata19}%
  \BibitemOpen
  \bibfield  {author} {\bibinfo {author} {\bibfnamefont {K.}~\bibnamefont
  {Kawabata}}, \bibinfo {author} {\bibfnamefont {K.}~\bibnamefont {Shiozaki}},
  \bibinfo {author} {\bibfnamefont {M.}~\bibnamefont {Ueda}},\ and\ \bibinfo
  {author} {\bibfnamefont {M.}~\bibnamefont {Sato}},\ }\bibfield  {title}
  {\bibinfo {title} {Symmetry and {{Topology}} in {{Non-Hermitian Physics}}},\
  }\href {https://doi.org/10.1103/PhysRevX.9.041015} {\bibfield  {journal}
  {\bibinfo  {journal} {Phys. Rev. X}\ }\textbf {\bibinfo {volume} {9}},\
  \bibinfo {pages} {041015} (\bibinfo {year} {2019}{\natexlab{a}})}\BibitemShut
  {NoStop}%
\bibitem [{\citenamefont {Xu}\ \emph {et~al.}(2017)\citenamefont {Xu},
  \citenamefont {Wang},\ and\ \citenamefont {Duan}}]{Xu17}%
  \BibitemOpen
  \bibfield  {author} {\bibinfo {author} {\bibfnamefont {Y.}~\bibnamefont
  {Xu}}, \bibinfo {author} {\bibfnamefont {S.-T.}\ \bibnamefont {Wang}},\ and\
  \bibinfo {author} {\bibfnamefont {L.-M.}\ \bibnamefont {Duan}},\ }\bibfield
  {title} {\bibinfo {title} {{Weyl Exceptional Rings in a Three-Dimensional
  Dissipative Cold Atomic Gas}},\ }\href
  {https://doi.org/10.1103/PhysRevLett.118.045701} {\bibfield  {journal}
  {\bibinfo  {journal} {Phys. Rev. Lett.}\ }\textbf {\bibinfo {volume} {118}},\
  \bibinfo {pages} {045701} (\bibinfo {year} {2017})}\BibitemShut {NoStop}%
\bibitem [{\citenamefont {Gong}\ \emph {et~al.}(2018)\citenamefont {Gong},
  \citenamefont {Ashida}, \citenamefont {Kawabata}, \citenamefont {Takasan},
  \citenamefont {Higashikawa},\ and\ \citenamefont {Ueda}}]{gong2018}%
  \BibitemOpen
  \bibfield  {author} {\bibinfo {author} {\bibfnamefont {Z.}~\bibnamefont
  {Gong}}, \bibinfo {author} {\bibfnamefont {Y.}~\bibnamefont {Ashida}},
  \bibinfo {author} {\bibfnamefont {K.}~\bibnamefont {Kawabata}}, \bibinfo
  {author} {\bibfnamefont {K.}~\bibnamefont {Takasan}}, \bibinfo {author}
  {\bibfnamefont {S.}~\bibnamefont {Higashikawa}},\ and\ \bibinfo {author}
  {\bibfnamefont {M.}~\bibnamefont {Ueda}},\ }\bibfield  {title} {\bibinfo
  {title} {Topological {{Phases}} of {{Non-Hermitian Systems}}},\ }\href
  {https://doi.org/10.1103/PhysRevX.8.031079} {\bibfield  {journal} {\bibinfo
  {journal} {Phys. Rev. X}\ }\textbf {\bibinfo {volume} {8}},\ \bibinfo {pages}
  {031079} (\bibinfo {year} {2018})}\BibitemShut {NoStop}%
\bibitem [{\citenamefont {Yao}\ and\ \citenamefont {Wang}(2018)}]{Yao18}%
  \BibitemOpen
  \bibfield  {author} {\bibinfo {author} {\bibfnamefont {S.}~\bibnamefont
  {Yao}}\ and\ \bibinfo {author} {\bibfnamefont {Z.}~\bibnamefont {Wang}},\
  }\bibfield  {title} {\bibinfo {title} {{Edge States and Topological
  Invariants of Non-Hermitian Systems}},\ }\href
  {https://doi.org/10.1103/PhysRevLett.121.086803} {\bibfield  {journal}
  {\bibinfo  {journal} {Phys. Rev. Lett.}\ }\textbf {\bibinfo {volume} {121}},\
  \bibinfo {pages} {086803} (\bibinfo {year} {2018})}\BibitemShut {NoStop}%
\bibitem [{\citenamefont {Yao}\ \emph {et~al.}(2018)\citenamefont {Yao},
  \citenamefont {Song},\ and\ \citenamefont {Wang}}]{Yao18_2}%
  \BibitemOpen
  \bibfield  {author} {\bibinfo {author} {\bibfnamefont {S.}~\bibnamefont
  {Yao}}, \bibinfo {author} {\bibfnamefont {F.}~\bibnamefont {Song}},\ and\
  \bibinfo {author} {\bibfnamefont {Z.}~\bibnamefont {Wang}},\ }\bibfield
  {title} {\bibinfo {title} {{Non-Hermitian Chern Bands}},\ }\href
  {https://doi.org/10.1103/PhysRevLett.121.136802} {\bibfield  {journal}
  {\bibinfo  {journal} {Phys. Rev. Lett.}\ }\textbf {\bibinfo {volume} {121}},\
  \bibinfo {pages} {136802} (\bibinfo {year} {2018})}\BibitemShut {NoStop}%
\bibitem [{\citenamefont {Kawabata}\ \emph
  {et~al.}(2019{\natexlab{b}})\citenamefont {Kawabata}, \citenamefont
  {Higashikawa}, \citenamefont {Gong}, \citenamefont {Ashida},\ and\
  \citenamefont {Ueda}}]{kawabata2019a}%
  \BibitemOpen
  \bibfield  {author} {\bibinfo {author} {\bibfnamefont {K.}~\bibnamefont
  {Kawabata}}, \bibinfo {author} {\bibfnamefont {S.}~\bibnamefont
  {Higashikawa}}, \bibinfo {author} {\bibfnamefont {Z.}~\bibnamefont {Gong}},
  \bibinfo {author} {\bibfnamefont {Y.}~\bibnamefont {Ashida}},\ and\ \bibinfo
  {author} {\bibfnamefont {M.}~\bibnamefont {Ueda}},\ }\bibfield  {title}
  {\bibinfo {title} {Topological unification of time-reversal and particle-hole
  symmetries in non-{{Hermitian}} physics},\ }\href
  {https://doi.org/10.1038/s41467-018-08254-y} {\bibfield  {journal} {\bibinfo
  {journal} {Nat. Commun.}\ }\textbf {\bibinfo {volume} {10}},\ \bibinfo
  {pages} {297} (\bibinfo {year} {2019}{\natexlab{b}})}\BibitemShut {NoStop}%
\bibitem [{\citenamefont {Zhou}\ and\ \citenamefont {Lee}(2019)}]{Zhou19}%
  \BibitemOpen
  \bibfield  {author} {\bibinfo {author} {\bibfnamefont {H.}~\bibnamefont
  {Zhou}}\ and\ \bibinfo {author} {\bibfnamefont {J.~Y.}\ \bibnamefont {Lee}},\
  }\bibfield  {title} {\bibinfo {title} {{Periodic table for topological bands
  with non-Hermitian symmetries}},\ }\href
  {https://doi.org/10.1103/PhysRevB.99.235112} {\bibfield  {journal} {\bibinfo
  {journal} {Phys. Rev. B}\ }\textbf {\bibinfo {volume} {99}},\ \bibinfo
  {pages} {235112} (\bibinfo {year} {2019})}\BibitemShut {NoStop}%
\bibitem [{\citenamefont {Hatano}\ and\ \citenamefont
  {Nelson}(1996)}]{hatano1996localization}%
  \BibitemOpen
  \bibfield  {author} {\bibinfo {author} {\bibfnamefont {N.}~\bibnamefont
  {Hatano}}\ and\ \bibinfo {author} {\bibfnamefont {D.~R.}\ \bibnamefont
  {Nelson}},\ }\bibfield  {title} {\bibinfo {title} {{Localization Transitions
  in Non-Hermitian Quantum Mechanics}},\ }\href
  {https://doi.org/10.1103/PhysRevLett.77.570} {\bibfield  {journal} {\bibinfo
  {journal} {Phys. Rev. Lett.}\ }\textbf {\bibinfo {volume} {77}},\ \bibinfo
  {pages} {570} (\bibinfo {year} {1996})}\BibitemShut {NoStop}%
\bibitem [{\citenamefont {Hatano}\ and\ \citenamefont
  {Nelson}(1997)}]{HatanoNelson97PRB}%
  \BibitemOpen
  \bibfield  {author} {\bibinfo {author} {\bibfnamefont {N.}~\bibnamefont
  {Hatano}}\ and\ \bibinfo {author} {\bibfnamefont {D.~R.}\ \bibnamefont
  {Nelson}},\ }\bibfield  {title} {\bibinfo {title} {{Vortex pinning and
  non-Hermitian quantum mechanics}},\ }\href
  {https://doi.org/10.1103/PhysRevB.56.8651} {\bibfield  {journal} {\bibinfo
  {journal} {Phys. Rev. B}\ }\textbf {\bibinfo {volume} {56}},\ \bibinfo
  {pages} {8651} (\bibinfo {year} {1997})}\BibitemShut {NoStop}%
\bibitem [{\citenamefont {Efetov}(1997)}]{Efetov97}%
  \BibitemOpen
  \bibfield  {author} {\bibinfo {author} {\bibfnamefont {K.~B.}\ \bibnamefont
  {Efetov}},\ }\bibfield  {title} {\bibinfo {title} {Quantum disordered systems
  with a direction},\ }\href {https://doi.org/10.1103/PhysRevB.56.9630}
  {\bibfield  {journal} {\bibinfo  {journal} {Phys. Rev. B}\ }\textbf {\bibinfo
  {volume} {56}},\ \bibinfo {pages} {9630} (\bibinfo {year}
  {1997})}\BibitemShut {NoStop}%
\bibitem [{\citenamefont {Xu}\ \emph {et~al.}(2016)\citenamefont {Xu},
  \citenamefont {Ohtsuki},\ and\ \citenamefont {Shindou}}]{Xu16}%
  \BibitemOpen
  \bibfield  {author} {\bibinfo {author} {\bibfnamefont {B.}~\bibnamefont
  {Xu}}, \bibinfo {author} {\bibfnamefont {T.}~\bibnamefont {Ohtsuki}},\ and\
  \bibinfo {author} {\bibfnamefont {R.}~\bibnamefont {Shindou}},\ }\bibfield
  {title} {\bibinfo {title} {{Integer quantum magnon Hall plateau-plateau
  transition in a spin-ice model}},\ }\href
  {https://doi.org/10.1103/PhysRevB.94.220403} {\bibfield  {journal} {\bibinfo
  {journal} {Phys. Rev. B}\ }\textbf {\bibinfo {volume} {94}},\ \bibinfo
  {pages} {220403(R)} (\bibinfo {year} {2016})}\BibitemShut {NoStop}%
\bibitem [{\citenamefont {Longhi}(2019)}]{longhi2019}%
  \BibitemOpen
  \bibfield  {author} {\bibinfo {author} {\bibfnamefont {S.}~\bibnamefont
  {Longhi}},\ }\bibfield  {title} {\bibinfo {title} {Topological {{Phase
  Transition}} in non-{{Hermitian Quasicrystals}}},\ }\href
  {https://doi.org/10.1103/PhysRevLett.122.237601} {\bibfield  {journal}
  {\bibinfo  {journal} {Phys. Rev. Lett.}\ }\textbf {\bibinfo {volume} {122}},\
  \bibinfo {pages} {237601} (\bibinfo {year} {2019})}\BibitemShut {NoStop}%
\bibitem [{\citenamefont {Zeng}\ and\ \citenamefont {Xu}(2020)}]{Zeng20}%
  \BibitemOpen
  \bibfield  {author} {\bibinfo {author} {\bibfnamefont {Q.-B.}\ \bibnamefont
  {Zeng}}\ and\ \bibinfo {author} {\bibfnamefont {Y.}~\bibnamefont {Xu}},\
  }\bibfield  {title} {\bibinfo {title} {{Winding numbers and generalized
  mobility edges in non-Hermitian systems}},\ }\href
  {https://doi.org/10.1103/PhysRevResearch.2.033052} {\bibfield  {journal}
  {\bibinfo  {journal} {Phys. Rev. Research}\ }\textbf {\bibinfo {volume}
  {2}},\ \bibinfo {pages} {033052} (\bibinfo {year} {2020})}\BibitemShut
  {NoStop}%
\bibitem [{\citenamefont {Wang}\ and\ \citenamefont {Wang}(2020)}]{Wang20}%
  \BibitemOpen
  \bibfield  {author} {\bibinfo {author} {\bibfnamefont {C.}~\bibnamefont
  {Wang}}\ and\ \bibinfo {author} {\bibfnamefont {X.~R.}\ \bibnamefont
  {Wang}},\ }\bibfield  {title} {\bibinfo {title} {{Level statistics of
  extended states in random non-Hermitian Hamiltonians}},\ }\href
  {https://doi.org/10.1103/PhysRevB.101.165114} {\bibfield  {journal} {\bibinfo
   {journal} {Phys. Rev. B}\ }\textbf {\bibinfo {volume} {101}},\ \bibinfo
  {pages} {165114} (\bibinfo {year} {2020})}\BibitemShut {NoStop}%
\bibitem [{\citenamefont {Huang}\ and\ \citenamefont
  {Shklovskii}(2020{\natexlab{a}})}]{huang2020}%
  \BibitemOpen
  \bibfield  {author} {\bibinfo {author} {\bibfnamefont {Y.}~\bibnamefont
  {Huang}}\ and\ \bibinfo {author} {\bibfnamefont {B.~I.}\ \bibnamefont
  {Shklovskii}},\ }\bibfield  {title} {\bibinfo {title} {Anderson transition in
  three-dimensional systems with non-{{Hermitian}} disorder},\ }\href
  {https://doi.org/10.1103/PhysRevB.101.014204} {\bibfield  {journal} {\bibinfo
   {journal} {Phys. Rev. B}\ }\textbf {\bibinfo {volume} {101}},\ \bibinfo
  {pages} {014204} (\bibinfo {year} {2020}{\natexlab{a}})}\BibitemShut
  {NoStop}%
\bibitem [{\citenamefont {Huang}\ and\ \citenamefont
  {Shklovskii}(2020{\natexlab{b}})}]{Huang20SR}%
  \BibitemOpen
  \bibfield  {author} {\bibinfo {author} {\bibfnamefont {Y.}~\bibnamefont
  {Huang}}\ and\ \bibinfo {author} {\bibfnamefont {B.~I.}\ \bibnamefont
  {Shklovskii}},\ }\bibfield  {title} {\bibinfo {title} {{Spectral rigidity of
  non-Hermitian symmetric random matrices near the Anderson transition}},\
  }\href {https://doi.org/10.1103/PhysRevB.102.064212} {\bibfield  {journal}
  {\bibinfo  {journal} {Phys. Rev. B}\ }\textbf {\bibinfo {volume} {102}},\
  \bibinfo {pages} {064212} (\bibinfo {year} {2020}{\natexlab{b}})}\BibitemShut
  {NoStop}%
\bibitem [{\citenamefont {Kawabata}\ and\ \citenamefont
  {Ryu}(2021)}]{kawabata20}%
  \BibitemOpen
  \bibfield  {author} {\bibinfo {author} {\bibfnamefont {K.}~\bibnamefont
  {Kawabata}}\ and\ \bibinfo {author} {\bibfnamefont {S.}~\bibnamefont {Ryu}},\
  }\bibfield  {title} {\bibinfo {title} {{Nonunitary Scaling Theory of
  Non-Hermitian Localization}},\ }\href
  {https://doi.org/10.1103/PhysRevLett.126.166801} {\bibfield  {journal}
  {\bibinfo  {journal} {Phys. Rev. Lett.}\ }\textbf {\bibinfo {volume} {126}},\
  \bibinfo {pages} {166801} (\bibinfo {year} {2021})}\BibitemShut {NoStop}%
\bibitem [{\citenamefont {Luo}\ \emph {et~al.}(2021{\natexlab{a}})\citenamefont
  {Luo}, \citenamefont {Ohtsuki},\ and\ \citenamefont {Shindou}}]{Luo21}%
  \BibitemOpen
  \bibfield  {author} {\bibinfo {author} {\bibfnamefont {X.}~\bibnamefont
  {Luo}}, \bibinfo {author} {\bibfnamefont {T.}~\bibnamefont {Ohtsuki}},\ and\
  \bibinfo {author} {\bibfnamefont {R.}~\bibnamefont {Shindou}},\ }\bibfield
  {title} {\bibinfo {title} {{Universality Classes of the Anderson Transitions
  Driven by Non-Hermitian Disorder}},\ }\href
  {https://doi.org/10.1103/PhysRevLett.126.090402} {\bibfield  {journal}
  {\bibinfo  {journal} {Phys. Rev. Lett.}\ }\textbf {\bibinfo {volume} {126}},\
  \bibinfo {pages} {090402} (\bibinfo {year} {2021}{\natexlab{a}})}\BibitemShut
  {NoStop}%
\bibitem [{\citenamefont {Luo}\ \emph {et~al.}(2021{\natexlab{b}})\citenamefont
  {Luo}, \citenamefont {Ohtsuki},\ and\ \citenamefont {Shindou}}]{Luo21TM}%
  \BibitemOpen
  \bibfield  {author} {\bibinfo {author} {\bibfnamefont {X.}~\bibnamefont
  {Luo}}, \bibinfo {author} {\bibfnamefont {T.}~\bibnamefont {Ohtsuki}},\ and\
  \bibinfo {author} {\bibfnamefont {R.}~\bibnamefont {Shindou}},\ }\bibfield
  {title} {\bibinfo {title} {{Transfer matrix study of the Anderson transition
  in non-Hermitian systems}},\ }\href
  {https://doi.org/10.1103/PhysRevB.104.104203} {\bibfield  {journal} {\bibinfo
   {journal} {Phys. Rev. B}\ }\textbf {\bibinfo {volume} {104}},\ \bibinfo
  {pages} {104203} (\bibinfo {year} {2021}{\natexlab{b}})}\BibitemShut
  {NoStop}%
\bibitem [{\citenamefont {Luo}\ \emph {et~al.}(2022)\citenamefont {Luo},
  \citenamefont {Xiao}, \citenamefont {Kawabata}, \citenamefont {Ohtsuki},\
  and\ \citenamefont {Shindou}}]{luo2021unifying}%
  \BibitemOpen
  \bibfield  {author} {\bibinfo {author} {\bibfnamefont {X.}~\bibnamefont
  {Luo}}, \bibinfo {author} {\bibfnamefont {Z.}~\bibnamefont {Xiao}}, \bibinfo
  {author} {\bibfnamefont {K.}~\bibnamefont {Kawabata}}, \bibinfo {author}
  {\bibfnamefont {T.}~\bibnamefont {Ohtsuki}},\ and\ \bibinfo {author}
  {\bibfnamefont {R.}~\bibnamefont {Shindou}},\ }\bibfield  {title} {\bibinfo
  {title} {{Unifying the Anderson transitions in Hermitian and non-Hermitian
  systems}},\ }\href {https://doi.org/10.1103/PhysRevResearch.4.L022035}
  {\bibfield  {journal} {\bibinfo  {journal} {Phys. Rev. Research}\ }\textbf
  {\bibinfo {volume} {4}},\ \bibinfo {pages} {L022035} (\bibinfo {year}
  {2022})}\BibitemShut {NoStop}%
\bibitem [{\citenamefont {Xiao}\ \emph {et~al.}(2023)\citenamefont {Xiao},
  \citenamefont {Kawabata}, \citenamefont {Luo}, \citenamefont {Ohtsuki},\ and\
  \citenamefont {Shindou}}]{xiao2023}%
  \BibitemOpen
  \bibfield  {author} {\bibinfo {author} {\bibfnamefont {Z.}~\bibnamefont
  {Xiao}}, \bibinfo {author} {\bibfnamefont {K.}~\bibnamefont {Kawabata}},
  \bibinfo {author} {\bibfnamefont {X.}~\bibnamefont {Luo}}, \bibinfo {author}
  {\bibfnamefont {T.}~\bibnamefont {Ohtsuki}},\ and\ \bibinfo {author}
  {\bibfnamefont {R.}~\bibnamefont {Shindou}},\ }\bibfield  {title} {\bibinfo
  {title} {Anisotropic {{Topological Anderson Transitions}} in {{Chiral
  Symmetry Classes}}},\ }\href {https://doi.org/10.1103/PhysRevLett.131.056301}
  {\bibfield  {journal} {\bibinfo  {journal} {Phys. Rev. Lett.}\ }\textbf
  {\bibinfo {volume} {131}},\ \bibinfo {pages} {056301} (\bibinfo {year}
  {2023})}\BibitemShut {NoStop}%
\bibitem [{\citenamefont {Hamazaki}\ \emph {et~al.}(2019)\citenamefont
  {Hamazaki}, \citenamefont {Kawabata},\ and\ \citenamefont
  {Ueda}}]{hamazaki2019non}%
  \BibitemOpen
  \bibfield  {author} {\bibinfo {author} {\bibfnamefont {R.}~\bibnamefont
  {Hamazaki}}, \bibinfo {author} {\bibfnamefont {K.}~\bibnamefont {Kawabata}},\
  and\ \bibinfo {author} {\bibfnamefont {M.}~\bibnamefont {Ueda}},\ }\bibfield
  {title} {\bibinfo {title} {{Non-Hermitian Many-Body Localization}},\ }\href
  {https://doi.org/10.1103/PhysRevLett.123.090603} {\bibfield  {journal}
  {\bibinfo  {journal} {Phys. Rev. Lett.}\ }\textbf {\bibinfo {volume} {123}},\
  \bibinfo {pages} {090603} (\bibinfo {year} {2019})}\BibitemShut {NoStop}%
\bibitem [{\citenamefont {Suthar}\ \emph {et~al.}(2022)\citenamefont {Suthar},
  \citenamefont {Wang}, \citenamefont {Huang}, \citenamefont {Jen},\ and\
  \citenamefont {You}}]{suthar2022}%
  \BibitemOpen
  \bibfield  {author} {\bibinfo {author} {\bibfnamefont {K.}~\bibnamefont
  {Suthar}}, \bibinfo {author} {\bibfnamefont {Y.-C.}\ \bibnamefont {Wang}},
  \bibinfo {author} {\bibfnamefont {Y.-P.}\ \bibnamefont {Huang}}, \bibinfo
  {author} {\bibfnamefont {H.~H.}\ \bibnamefont {Jen}},\ and\ \bibinfo {author}
  {\bibfnamefont {J.-S.}\ \bibnamefont {You}},\ }\bibfield  {title} {\bibinfo
  {title} {Non-{{Hermitian}} many-body localization with open boundaries},\
  }\href {https://doi.org/10.1103/PhysRevB.106.064208} {\bibfield  {journal}
  {\bibinfo  {journal} {Phys. Rev. B}\ }\textbf {\bibinfo {volume} {106}},\
  \bibinfo {pages} {064208} (\bibinfo {year} {2022})}\BibitemShut {NoStop}%
\bibitem [{\citenamefont {Zhai}\ \emph {et~al.}(2020)\citenamefont {Zhai},
  \citenamefont {Yin},\ and\ \citenamefont {Huang}}]{zhai2020}%
  \BibitemOpen
  \bibfield  {author} {\bibinfo {author} {\bibfnamefont {L.-J.}\ \bibnamefont
  {Zhai}}, \bibinfo {author} {\bibfnamefont {S.}~\bibnamefont {Yin}},\ and\
  \bibinfo {author} {\bibfnamefont {G.-Y.}\ \bibnamefont {Huang}},\ }\bibfield
  {title} {\bibinfo {title} {Many-body localization in a non-{{Hermitian}}
  quasiperiodic system},\ }\href {https://doi.org/10.1103/PhysRevB.102.064206}
  {\bibfield  {journal} {\bibinfo  {journal} {Phys. Rev. B}\ }\textbf {\bibinfo
  {volume} {102}},\ \bibinfo {pages} {064206} (\bibinfo {year}
  {2020})}\BibitemShut {NoStop}%
\bibitem [{\citenamefont {Bender}\ and\ \citenamefont
  {Boettcher}(1998)}]{Bender98}%
  \BibitemOpen
  \bibfield  {author} {\bibinfo {author} {\bibfnamefont {C.~M.}\ \bibnamefont
  {Bender}}\ and\ \bibinfo {author} {\bibfnamefont {S.}~\bibnamefont
  {Boettcher}},\ }\bibfield  {title} {\bibinfo {title} {{Real Spectra in
  Non-Hermitian Hamiltonians Having $\mathcal{PT}$ Symmetry}},\ }\href
  {https://doi.org/10.1103/PhysRevLett.80.5243} {\bibfield  {journal} {\bibinfo
   {journal} {Phys. Rev. Lett.}\ }\textbf {\bibinfo {volume} {80}},\ \bibinfo
  {pages} {5243} (\bibinfo {year} {1998})}\BibitemShut {NoStop}%
\bibitem [{\citenamefont {Grobe}\ \emph {et~al.}(1988)\citenamefont {Grobe},
  \citenamefont {Haake},\ and\ \citenamefont {Sommers}}]{grobe88}%
  \BibitemOpen
  \bibfield  {author} {\bibinfo {author} {\bibfnamefont {R.}~\bibnamefont
  {Grobe}}, \bibinfo {author} {\bibfnamefont {F.}~\bibnamefont {Haake}},\ and\
  \bibinfo {author} {\bibfnamefont {H.-J.}\ \bibnamefont {Sommers}},\
  }\bibfield  {title} {\bibinfo {title} {{Quantum Distinction of Regular and
  Chaotic Dissipative Motion}},\ }\href
  {https://doi.org/10.1103/PhysRevLett.61.1899} {\bibfield  {journal} {\bibinfo
   {journal} {Phys. Rev. Lett.}\ }\textbf {\bibinfo {volume} {61}},\ \bibinfo
  {pages} {1899} (\bibinfo {year} {1988})}\BibitemShut {NoStop}%
\bibitem [{\citenamefont {Grobe}\ and\ \citenamefont {Haake}(1989)}]{grobe89}%
  \BibitemOpen
  \bibfield  {author} {\bibinfo {author} {\bibfnamefont {R.}~\bibnamefont
  {Grobe}}\ and\ \bibinfo {author} {\bibfnamefont {F.}~\bibnamefont {Haake}},\
  }\bibfield  {title} {\bibinfo {title} {{Universality of cubic-level repulsion
  for dissipative quantum chaos}},\ }\href
  {https://doi.org/10.1103/PhysRevLett.62.2893} {\bibfield  {journal} {\bibinfo
   {journal} {Phys. Rev. Lett.}\ }\textbf {\bibinfo {volume} {62}},\ \bibinfo
  {pages} {2893} (\bibinfo {year} {1989})}\BibitemShut {NoStop}%
\bibitem [{\citenamefont {Xu}\ \emph {et~al.}(2019)\citenamefont {Xu},
  \citenamefont {Garc\'{\i}a-Pintos}, \citenamefont {Chenu},\ and\
  \citenamefont {del Campo}}]{Xu-19}%
  \BibitemOpen
  \bibfield  {author} {\bibinfo {author} {\bibfnamefont {Z.}~\bibnamefont
  {Xu}}, \bibinfo {author} {\bibfnamefont {L.~P.}\ \bibnamefont
  {Garc\'{\i}a-Pintos}}, \bibinfo {author} {\bibfnamefont {A.}~\bibnamefont
  {Chenu}},\ and\ \bibinfo {author} {\bibfnamefont {A.}~\bibnamefont {del
  Campo}},\ }\bibfield  {title} {\bibinfo {title} {{Extreme Decoherence and
  Quantum Chaos}},\ }\href {https://doi.org/10.1103/PhysRevLett.122.014103}
  {\bibfield  {journal} {\bibinfo  {journal} {Phys. Rev. Lett.}\ }\textbf
  {\bibinfo {volume} {122}},\ \bibinfo {pages} {014103} (\bibinfo {year}
  {2019})}\BibitemShut {NoStop}%
\bibitem [{\citenamefont {Denisov}\ \emph {et~al.}(2019)\citenamefont
  {Denisov}, \citenamefont {Laptyeva}, \citenamefont {Tarnowski}, \citenamefont
  {Chru\ifmmode \acute{s}\else \'{s}\fi{}ci\ifmmode~\acute{n}\else
  \'{n}\fi{}ski},\ and\ \citenamefont {\ifmmode~\dot{Z}\else
  \.{Z}\fi{}yczkowski}}]{Denisov-19}%
  \BibitemOpen
  \bibfield  {author} {\bibinfo {author} {\bibfnamefont {S.}~\bibnamefont
  {Denisov}}, \bibinfo {author} {\bibfnamefont {T.}~\bibnamefont {Laptyeva}},
  \bibinfo {author} {\bibfnamefont {W.}~\bibnamefont {Tarnowski}}, \bibinfo
  {author} {\bibfnamefont {D.}~\bibnamefont {Chru\ifmmode \acute{s}\else
  \'{s}\fi{}ci\ifmmode~\acute{n}\else \'{n}\fi{}ski}},\ and\ \bibinfo {author}
  {\bibfnamefont {K.}~\bibnamefont {\ifmmode~\dot{Z}\else
  \.{Z}\fi{}yczkowski}},\ }\bibfield  {title} {\bibinfo {title} {{Universal
  Spectra of Random Lindblad Operators}},\ }\href
  {https://doi.org/10.1103/PhysRevLett.123.140403} {\bibfield  {journal}
  {\bibinfo  {journal} {Phys. Rev. Lett.}\ }\textbf {\bibinfo {volume} {123}},\
  \bibinfo {pages} {140403} (\bibinfo {year} {2019})}\BibitemShut {NoStop}%
\bibitem [{\citenamefont {Can}\ \emph {et~al.}(2019)\citenamefont {Can},
  \citenamefont {Oganesyan}, \citenamefont {Orgad},\ and\ \citenamefont
  {Gopalakrishnan}}]{Can-19PRL}%
  \BibitemOpen
  \bibfield  {author} {\bibinfo {author} {\bibfnamefont {T.}~\bibnamefont
  {Can}}, \bibinfo {author} {\bibfnamefont {V.}~\bibnamefont {Oganesyan}},
  \bibinfo {author} {\bibfnamefont {D.}~\bibnamefont {Orgad}},\ and\ \bibinfo
  {author} {\bibfnamefont {S.}~\bibnamefont {Gopalakrishnan}},\ }\bibfield
  {title} {\bibinfo {title} {{Spectral Gaps and Midgap States in Random Quantum
  Master Equations}},\ }\href {https://doi.org/10.1103/PhysRevLett.123.234103}
  {\bibfield  {journal} {\bibinfo  {journal} {Phys. Rev. Lett.}\ }\textbf
  {\bibinfo {volume} {123}},\ \bibinfo {pages} {234103} (\bibinfo {year}
  {2019})}\BibitemShut {NoStop}%
\bibitem [{\citenamefont {Can}(2019)}]{Can-19JPhysA}%
  \BibitemOpen
  \bibfield  {author} {\bibinfo {author} {\bibfnamefont {T.}~\bibnamefont
  {Can}},\ }\bibfield  {title} {\bibinfo {title} {{Random Lindblad dynamics}},\
  }\href {https://doi.org/10.1088/1751-8121/ab4d26} {\bibfield  {journal}
  {\bibinfo  {journal} {J. Phys. A}\ }\textbf {\bibinfo {volume} {52}},\
  \bibinfo {pages} {485302} (\bibinfo {year} {2019})}\BibitemShut {NoStop}%
\bibitem [{\citenamefont {Hamazaki}\ \emph {et~al.}(2020)\citenamefont
  {Hamazaki}, \citenamefont {Kawabata}, \citenamefont {Kura},\ and\
  \citenamefont {Ueda}}]{hamazaki20}%
  \BibitemOpen
  \bibfield  {author} {\bibinfo {author} {\bibfnamefont {R.}~\bibnamefont
  {Hamazaki}}, \bibinfo {author} {\bibfnamefont {K.}~\bibnamefont {Kawabata}},
  \bibinfo {author} {\bibfnamefont {N.}~\bibnamefont {Kura}},\ and\ \bibinfo
  {author} {\bibfnamefont {M.}~\bibnamefont {Ueda}},\ }\bibfield  {title}
  {\bibinfo {title} {Universality classes of non-{{Hermitian}} random
  matrices},\ }\href {https://doi.org/10.1103/PhysRevResearch.2.023286}
  {\bibfield  {journal} {\bibinfo  {journal} {Phys. Rev. Research}\ }\textbf
  {\bibinfo {volume} {2}},\ \bibinfo {pages} {023286} (\bibinfo {year}
  {2020})}\BibitemShut {NoStop}%
\bibitem [{\citenamefont {Akemann}\ \emph {et~al.}(2019)\citenamefont
  {Akemann}, \citenamefont {Kieburg}, \citenamefont {Mielke},\ and\
  \citenamefont {Prosen}}]{akemann19}%
  \BibitemOpen
  \bibfield  {author} {\bibinfo {author} {\bibfnamefont {G.}~\bibnamefont
  {Akemann}}, \bibinfo {author} {\bibfnamefont {M.}~\bibnamefont {Kieburg}},
  \bibinfo {author} {\bibfnamefont {A.}~\bibnamefont {Mielke}},\ and\ \bibinfo
  {author} {\bibfnamefont {T.}~\bibnamefont {Prosen}},\ }\bibfield  {title}
  {\bibinfo {title} {Universal {{Signature}} from {{Integrability}} to
  {{Chaos}} in {{Dissipative Open Quantum Systems}}},\ }\href
  {https://doi.org/10.1103/PhysRevLett.123.254101} {\bibfield  {journal}
  {\bibinfo  {journal} {Phys. Rev. Lett.}\ }\textbf {\bibinfo {volume} {123}},\
  \bibinfo {pages} {254101} (\bibinfo {year} {2019})}\BibitemShut {NoStop}%
\bibitem [{\citenamefont {S{\'a}}\ \emph {et~al.}(2020)\citenamefont {S{\'a}},
  \citenamefont {Ribeiro},\ and\ \citenamefont {Prosen}}]{sa20}%
  \BibitemOpen
  \bibfield  {author} {\bibinfo {author} {\bibfnamefont {L.}~\bibnamefont
  {S{\'a}}}, \bibinfo {author} {\bibfnamefont {P.}~\bibnamefont {Ribeiro}},\
  and\ \bibinfo {author} {\bibfnamefont {T.}~\bibnamefont {Prosen}},\
  }\bibfield  {title} {\bibinfo {title} {Complex {{Spacing Ratios}}: {{A
  Signature}} of {{Dissipative Quantum Chaos}}},\ }\href
  {https://doi.org/10.1103/PhysRevX.10.021019} {\bibfield  {journal} {\bibinfo
  {journal} {Phys. Rev. X}\ }\textbf {\bibinfo {volume} {10}},\ \bibinfo
  {pages} {021019} (\bibinfo {year} {2020})}\BibitemShut {NoStop}%
\bibitem [{\citenamefont {Wang}\ \emph {et~al.}(2020)\citenamefont {Wang},
  \citenamefont {Piazza},\ and\ \citenamefont {Luitz}}]{Wang-20}%
  \BibitemOpen
  \bibfield  {author} {\bibinfo {author} {\bibfnamefont {K.}~\bibnamefont
  {Wang}}, \bibinfo {author} {\bibfnamefont {F.}~\bibnamefont {Piazza}},\ and\
  \bibinfo {author} {\bibfnamefont {D.~J.}\ \bibnamefont {Luitz}},\ }\bibfield
  {title} {\bibinfo {title} {{Hierarchy of Relaxation Timescales in Local
  Random Liouvillians}},\ }\href
  {https://doi.org/10.1103/PhysRevLett.124.100604} {\bibfield  {journal}
  {\bibinfo  {journal} {Phys. Rev. Lett.}\ }\textbf {\bibinfo {volume} {124}},\
  \bibinfo {pages} {100604} (\bibinfo {year} {2020})}\BibitemShut {NoStop}%
\bibitem [{\citenamefont {Xu}\ \emph {et~al.}(2021)\citenamefont {Xu},
  \citenamefont {Chenu}, \citenamefont {Prosen},\ and\ \citenamefont {del
  Campo}}]{Xu-21}%
  \BibitemOpen
  \bibfield  {author} {\bibinfo {author} {\bibfnamefont {Z.}~\bibnamefont
  {Xu}}, \bibinfo {author} {\bibfnamefont {A.}~\bibnamefont {Chenu}}, \bibinfo
  {author} {\bibfnamefont {T.}~\bibnamefont {Prosen}},\ and\ \bibinfo {author}
  {\bibfnamefont {A.}~\bibnamefont {del Campo}},\ }\bibfield  {title} {\bibinfo
  {title} {{Thermofield dynamics: Quantum chaos versus decoherence}},\ }\href
  {https://doi.org/10.1103/PhysRevB.103.064309} {\bibfield  {journal} {\bibinfo
   {journal} {Phys. Rev. B}\ }\textbf {\bibinfo {volume} {103}},\ \bibinfo
  {pages} {064309} (\bibinfo {year} {2021})}\BibitemShut {NoStop}%
\bibitem [{\citenamefont {Li}\ \emph {et~al.}(2021)\citenamefont {Li},
  \citenamefont {Prosen},\ and\ \citenamefont {Chan}}]{li2021a}%
  \BibitemOpen
  \bibfield  {author} {\bibinfo {author} {\bibfnamefont {J.}~\bibnamefont
  {Li}}, \bibinfo {author} {\bibfnamefont {T.}~\bibnamefont {Prosen}},\ and\
  \bibinfo {author} {\bibfnamefont {A.}~\bibnamefont {Chan}},\ }\bibfield
  {title} {\bibinfo {title} {Spectral {{Statistics}} of {{Non-Hermitian
  Matrices}} and {{Dissipative Quantum Chaos}}},\ }\href
  {https://doi.org/10.1103/PhysRevLett.127.170602} {\bibfield  {journal}
  {\bibinfo  {journal} {Phys. Rev. Lett.}\ }\textbf {\bibinfo {volume} {127}},\
  \bibinfo {pages} {170602} (\bibinfo {year} {2021})}\BibitemShut {NoStop}%
\bibitem [{\citenamefont {{Garc{\'i}a-Garc{\'i}a}}\ \emph
  {et~al.}(2023)\citenamefont {{Garc{\'i}a-Garc{\'i}a}}, \citenamefont
  {S{\'a}},\ and\ \citenamefont {Verbaarschot}}]{garcia2023}%
  \BibitemOpen
  \bibfield  {author} {\bibinfo {author} {\bibfnamefont {A.~M.}\ \bibnamefont
  {{Garc{\'i}a-Garc{\'i}a}}}, \bibinfo {author} {\bibfnamefont
  {L.}~\bibnamefont {S{\'a}}},\ and\ \bibinfo {author} {\bibfnamefont
  {J.~J.~M.}\ \bibnamefont {Verbaarschot}},\ }\bibfield  {title} {\bibinfo
  {title} {Universality and its limits in non-{{Hermitian}} many-body quantum
  chaos using the {{Sachdev-Ye-Kitaev}} model},\ }\href
  {https://doi.org/10.1103/PhysRevD.107.066007} {\bibfield  {journal} {\bibinfo
   {journal} {Phys. Rev. D}\ }\textbf {\bibinfo {volume} {107}},\ \bibinfo
  {pages} {066007} (\bibinfo {year} {2023})}\BibitemShut {NoStop}%
\bibitem [{\citenamefont {Costa}\ \emph {et~al.}(2023)\citenamefont {Costa},
  \citenamefont {Ribeiro}, \citenamefont {De~Luca}, \citenamefont {Prosen},\
  and\ \citenamefont {S{\'a}}}]{costa2023}%
  \BibitemOpen
  \bibfield  {author} {\bibinfo {author} {\bibfnamefont {J.}~\bibnamefont
  {Costa}}, \bibinfo {author} {\bibfnamefont {P.}~\bibnamefont {Ribeiro}},
  \bibinfo {author} {\bibfnamefont {A.}~\bibnamefont {De~Luca}}, \bibinfo
  {author} {\bibfnamefont {T.}~\bibnamefont {Prosen}},\ and\ \bibinfo {author}
  {\bibfnamefont {L.}~\bibnamefont {S{\'a}}},\ }\bibfield  {title} {\bibinfo
  {title} {Spectral and steady-state properties of fermionic random quadratic
  {{Liouvillians}}},\ }\href {https://doi.org/10.21468/SciPostPhys.15.4.145}
  {\bibfield  {journal} {\bibinfo  {journal} {SciPost Phys.}\ }\textbf
  {\bibinfo {volume} {15}},\ \bibinfo {pages} {145} (\bibinfo {year}
  {2023})}\BibitemShut {NoStop}%
\bibitem [{\citenamefont {Kawabata}\ \emph
  {et~al.}(2023{\natexlab{a}})\citenamefont {Kawabata}, \citenamefont {Xiao},
  \citenamefont {Ohtsuki},\ and\ \citenamefont
  {Shindou}}]{kawabata2023singular}%
  \BibitemOpen
  \bibfield  {author} {\bibinfo {author} {\bibfnamefont {K.}~\bibnamefont
  {Kawabata}}, \bibinfo {author} {\bibfnamefont {Z.}~\bibnamefont {Xiao}},
  \bibinfo {author} {\bibfnamefont {T.}~\bibnamefont {Ohtsuki}},\ and\ \bibinfo
  {author} {\bibfnamefont {R.}~\bibnamefont {Shindou}},\ }\bibfield  {title}
  {\bibinfo {title} {Singular-{{Value Statistics}} of {{Non-Hermitian Random
  Matrices}} and {{Open Quantum Systems}}},\ }\href
  {https://doi.org/10.1103/PRXQuantum.4.040312} {\bibfield  {journal} {\bibinfo
   {journal} {PRX Quantum}\ }\textbf {\bibinfo {volume} {4}},\ \bibinfo {pages}
  {040312} (\bibinfo {year} {2023}{\natexlab{a}})}\BibitemShut {NoStop}%
\bibitem [{\citenamefont {Roccati}\ \emph {et~al.}(2024)\citenamefont
  {Roccati}, \citenamefont {Balducci}, \citenamefont {Shir},\ and\
  \citenamefont {Chenu}}]{roccati2023}%
  \BibitemOpen
  \bibfield  {author} {\bibinfo {author} {\bibfnamefont {F.}~\bibnamefont
  {Roccati}}, \bibinfo {author} {\bibfnamefont {F.}~\bibnamefont {Balducci}},
  \bibinfo {author} {\bibfnamefont {R.}~\bibnamefont {Shir}},\ and\ \bibinfo
  {author} {\bibfnamefont {A.}~\bibnamefont {Chenu}},\ }\bibfield  {title}
  {\bibinfo {title} {{Diagnosing non-Hermitian many-body localization and
  quantum chaos via singular value decomposition}},\ }\href
  {https://doi.org/10.1103/PhysRevB.109.L140201} {\bibfield  {journal}
  {\bibinfo  {journal} {Phys. Rev. B}\ }\textbf {\bibinfo {volume} {109}},\
  \bibinfo {pages} {L140201} (\bibinfo {year} {2024})}\BibitemShut {NoStop}%
\bibitem [{\citenamefont {Sachdev}\ and\ \citenamefont
  {Ye}(1993)}]{Sachdev-Ye-93}%
  \BibitemOpen
  \bibfield  {author} {\bibinfo {author} {\bibfnamefont {S.}~\bibnamefont
  {Sachdev}}\ and\ \bibinfo {author} {\bibfnamefont {J.}~\bibnamefont {Ye}},\
  }\bibfield  {title} {\bibinfo {title} {{Gapless spin-fluid ground state in a
  random quantum Heisenberg magnet}},\ }\href
  {https://doi.org/10.1103/PhysRevLett.70.3339} {\bibfield  {journal} {\bibinfo
   {journal} {Phys. Rev. Lett.}\ }\textbf {\bibinfo {volume} {70}},\ \bibinfo
  {pages} {3339} (\bibinfo {year} {1993})}\BibitemShut {NoStop}%
\bibitem [{\citenamefont {Kitaev}(2015)}]{kitaev15}%
  \BibitemOpen
  \bibfield  {author} {\bibinfo {author} {\bibfnamefont {A.}~\bibnamefont
  {Kitaev}},\ }\bibfield  {title} {\bibinfo {title} {{A simple model of quantum
  holography}}} (\bibinfo {year} {2015}),\ \bibinfo {note} {{KITP Program:
  Entanglement in Strongly-Correlated Quantum Matter}}\BibitemShut {NoStop}%
\bibitem [{\citenamefont {Sachdev}(2015)}]{Sachdev-15}%
  \BibitemOpen
  \bibfield  {author} {\bibinfo {author} {\bibfnamefont {S.}~\bibnamefont
  {Sachdev}},\ }\bibfield  {title} {\bibinfo {title} {{Bekenstein-Hawking
  Entropy and Strange Metals}},\ }\href
  {https://doi.org/10.1103/PhysRevX.5.041025} {\bibfield  {journal} {\bibinfo
  {journal} {Phys. Rev. X}\ }\textbf {\bibinfo {volume} {5}},\ \bibinfo {pages}
  {041025} (\bibinfo {year} {2015})}\BibitemShut {NoStop}%
\bibitem [{\citenamefont {Polchinski}\ and\ \citenamefont
  {Rosenhaus}(2016)}]{Polchinski-Rosenhaus-16}%
  \BibitemOpen
  \bibfield  {author} {\bibinfo {author} {\bibfnamefont {J.}~\bibnamefont
  {Polchinski}}\ and\ \bibinfo {author} {\bibfnamefont {V.}~\bibnamefont
  {Rosenhaus}},\ }\bibfield  {title} {\bibinfo {title} {{The spectrum in the
  Sachdev-Ye-Kitaev model}},\ }\href {https://doi.org/10.1007/JHEP04(2016)001}
  {\bibfield  {journal} {\bibinfo  {journal} {J. High Energ. Phys.}\ }\textbf
  {\bibinfo {volume} {2016}}\bibinfo  {number} { (4)},\ \bibinfo {pages}
  {1}}\BibitemShut {NoStop}%
\bibitem [{\citenamefont {Maldacena}\ and\ \citenamefont
  {Stanford}(2016)}]{Maldacena-Stanford-16}%
  \BibitemOpen
\bibfield  {number} {  }\bibfield  {author} {\bibinfo {author} {\bibfnamefont
  {J.}~\bibnamefont {Maldacena}}\ and\ \bibinfo {author} {\bibfnamefont
  {D.}~\bibnamefont {Stanford}},\ }\bibfield  {title} {\bibinfo {title}
  {{Remarks on the Sachdev-Ye-Kitaev model}},\ }\href
  {https://doi.org/10.1103/PhysRevD.94.106002} {\bibfield  {journal} {\bibinfo
  {journal} {Phys. Rev. D}\ }\textbf {\bibinfo {volume} {94}},\ \bibinfo
  {pages} {106002} (\bibinfo {year} {2016})}\BibitemShut {NoStop}%
\bibitem [{\citenamefont {Rosenhaus}(2019)}]{Rosenhaus-review}%
  \BibitemOpen
  \bibfield  {author} {\bibinfo {author} {\bibfnamefont {V.}~\bibnamefont
  {Rosenhaus}},\ }\bibfield  {title} {\bibinfo {title} {{An introduction to the
  SYK model}},\ }\href {https://doi.org/10.1088/1751-8121/ab2ce1} {\bibfield
  {journal} {\bibinfo  {journal} {J. Phys. A}\ }\textbf {\bibinfo {volume}
  {52}},\ \bibinfo {pages} {323001} (\bibinfo {year} {2019})}\BibitemShut
  {NoStop}%
\bibitem [{\citenamefont {Chowdhury}\ \emph {et~al.}(2022)\citenamefont
  {Chowdhury}, \citenamefont {Georges}, \citenamefont {Parcollet},\ and\
  \citenamefont {Sachdev}}]{Sachdev-review}%
  \BibitemOpen
  \bibfield  {author} {\bibinfo {author} {\bibfnamefont {D.}~\bibnamefont
  {Chowdhury}}, \bibinfo {author} {\bibfnamefont {A.}~\bibnamefont {Georges}},
  \bibinfo {author} {\bibfnamefont {O.}~\bibnamefont {Parcollet}},\ and\
  \bibinfo {author} {\bibfnamefont {S.}~\bibnamefont {Sachdev}},\ }\bibfield
  {title} {\bibinfo {title} {{Sachdev-Ye-Kitaev models and beyond: Window into
  non-Fermi liquids}},\ }\href {https://doi.org/10.1103/RevModPhys.94.035004}
  {\bibfield  {journal} {\bibinfo  {journal} {Rev. Mod. Phys.}\ }\textbf
  {\bibinfo {volume} {94}},\ \bibinfo {pages} {035004} (\bibinfo {year}
  {2022})}\BibitemShut {NoStop}%
\bibitem [{\citenamefont {{Garc{\'i}a-Garc{\'i}a}}\ \emph
  {et~al.}(2022{\natexlab{a}})\citenamefont {{Garc{\'i}a-Garc{\'i}a}},
  \citenamefont {S{\'a}},\ and\ \citenamefont {Verbaarschot}}]{garcia22}%
  \BibitemOpen
  \bibfield  {author} {\bibinfo {author} {\bibfnamefont {A.~M.}\ \bibnamefont
  {{Garc{\'i}a-Garc{\'i}a}}}, \bibinfo {author} {\bibfnamefont
  {L.}~\bibnamefont {S{\'a}}},\ and\ \bibinfo {author} {\bibfnamefont
  {J.~J.~M.}\ \bibnamefont {Verbaarschot}},\ }\bibfield  {title} {\bibinfo
  {title} {Symmetry {{Classification}} and {{Universality}} in {{Non-Hermitian
  Many-Body Quantum Chaos}} by the {{Sachdev-Ye-Kitaev Model}}},\ }\href
  {https://doi.org/10.1103/PhysRevX.12.021040} {\bibfield  {journal} {\bibinfo
  {journal} {Phys. Rev. X}\ }\textbf {\bibinfo {volume} {12}},\ \bibinfo
  {pages} {021040} (\bibinfo {year} {2022}{\natexlab{a}})}\BibitemShut
  {NoStop}%
\bibitem [{\citenamefont {Lieu}\ \emph {et~al.}(2020)\citenamefont {Lieu},
  \citenamefont {McGinley},\ and\ \citenamefont {Cooper}}]{lieu20a}%
  \BibitemOpen
  \bibfield  {author} {\bibinfo {author} {\bibfnamefont {S.}~\bibnamefont
  {Lieu}}, \bibinfo {author} {\bibfnamefont {M.}~\bibnamefont {McGinley}},\
  and\ \bibinfo {author} {\bibfnamefont {N.~R.}\ \bibnamefont {Cooper}},\
  }\bibfield  {title} {\bibinfo {title} {Tenfold {{Way}} for {{Quadratic
  Lindbladians}}},\ }\href {https://doi.org/10.1103/PhysRevLett.124.040401}
  {\bibfield  {journal} {\bibinfo  {journal} {Phys. Rev. Lett.}\ }\textbf
  {\bibinfo {volume} {124}},\ \bibinfo {pages} {040401} (\bibinfo {year}
  {2020})}\BibitemShut {NoStop}%
\bibitem [{\citenamefont {Altland}\ \emph {et~al.}(2021)\citenamefont
  {Altland}, \citenamefont {Fleischhauer},\ and\ \citenamefont
  {Diehl}}]{altland2021}%
  \BibitemOpen
  \bibfield  {author} {\bibinfo {author} {\bibfnamefont {A.}~\bibnamefont
  {Altland}}, \bibinfo {author} {\bibfnamefont {M.}~\bibnamefont
  {Fleischhauer}},\ and\ \bibinfo {author} {\bibfnamefont {S.}~\bibnamefont
  {Diehl}},\ }\bibfield  {title} {\bibinfo {title} {Symmetry {{Classes}} of
  {{Open Fermionic Quantum Matter}}},\ }\href
  {https://doi.org/10.1103/PhysRevX.11.021037} {\bibfield  {journal} {\bibinfo
  {journal} {Phys. Rev. X}\ }\textbf {\bibinfo {volume} {11}},\ \bibinfo
  {pages} {021037} (\bibinfo {year} {2021})}\BibitemShut {NoStop}%
\bibitem [{\citenamefont {S{\'a}}\ \emph {et~al.}(2023)\citenamefont {S{\'a}},
  \citenamefont {Ribeiro},\ and\ \citenamefont {Prosen}}]{sa2023}%
  \BibitemOpen
  \bibfield  {author} {\bibinfo {author} {\bibfnamefont {L.}~\bibnamefont
  {S{\'a}}}, \bibinfo {author} {\bibfnamefont {P.}~\bibnamefont {Ribeiro}},\
  and\ \bibinfo {author} {\bibfnamefont {T.}~\bibnamefont {Prosen}},\
  }\bibfield  {title} {\bibinfo {title} {Symmetry {{Classification}} of
  {{Many-Body Lindbladians}}: {{Tenfold Way}} and {{Beyond}}},\ }\href
  {https://doi.org/10.1103/PhysRevX.13.031019} {\bibfield  {journal} {\bibinfo
  {journal} {Phys. Rev. X}\ }\textbf {\bibinfo {volume} {13}},\ \bibinfo
  {pages} {031019} (\bibinfo {year} {2023})}\BibitemShut {NoStop}%
\bibitem [{\citenamefont {Kawabata}\ \emph
  {et~al.}(2023{\natexlab{b}})\citenamefont {Kawabata}, \citenamefont
  {Kulkarni}, \citenamefont {Li}, \citenamefont {Numasawa},\ and\ \citenamefont
  {Ryu}}]{kawabata2023}%
  \BibitemOpen
  \bibfield  {author} {\bibinfo {author} {\bibfnamefont {K.}~\bibnamefont
  {Kawabata}}, \bibinfo {author} {\bibfnamefont {A.}~\bibnamefont {Kulkarni}},
  \bibinfo {author} {\bibfnamefont {J.}~\bibnamefont {Li}}, \bibinfo {author}
  {\bibfnamefont {T.}~\bibnamefont {Numasawa}},\ and\ \bibinfo {author}
  {\bibfnamefont {S.}~\bibnamefont {Ryu}},\ }\bibfield  {title} {\bibinfo
  {title} {Symmetry of {{Open Quantum Systems}}: {{Classification}} of
  {{Dissipative Quantum Chaos}}},\ }\href
  {https://doi.org/10.1103/PRXQuantum.4.030328} {\bibfield  {journal} {\bibinfo
   {journal} {PRX Quantum}\ }\textbf {\bibinfo {volume} {4}},\ \bibinfo {pages}
  {030328} (\bibinfo {year} {2023}{\natexlab{b}})}\BibitemShut {NoStop}%
\bibitem [{\citenamefont {Pikulin}\ and\ \citenamefont
  {Nazarov}(2012)}]{pikulin12}%
  \BibitemOpen
  \bibfield  {author} {\bibinfo {author} {\bibfnamefont {D.~I.}\ \bibnamefont
  {Pikulin}}\ and\ \bibinfo {author} {\bibfnamefont {Y.~V.}\ \bibnamefont
  {Nazarov}},\ }\bibfield  {title} {\bibinfo {title} {{Topological properties
  of superconducting junction}},\ }\href
  {https://doi.org/10.1134/S0021364011210090} {\bibfield  {journal} {\bibinfo
  {journal} {JETP Lett.}\ }\textbf {\bibinfo {volume} {94}},\ \bibinfo {pages}
  {693} (\bibinfo {year} {2012})}\BibitemShut {NoStop}%
\bibitem [{\citenamefont {San-Jose}\ \emph {et~al.}(2016)\citenamefont
  {San-Jose}, \citenamefont {Cayao}, \citenamefont {Prada},\ and\ \citenamefont
  {Aguado}}]{sanjose16}%
  \BibitemOpen
  \bibfield  {author} {\bibinfo {author} {\bibfnamefont {P.}~\bibnamefont
  {San-Jose}}, \bibinfo {author} {\bibfnamefont {J.}~\bibnamefont {Cayao}},
  \bibinfo {author} {\bibfnamefont {E.}~\bibnamefont {Prada}},\ and\ \bibinfo
  {author} {\bibfnamefont {R.}~\bibnamefont {Aguado}},\ }\bibfield  {title}
  {\bibinfo {title} {{Majorana bound states from exceptional points in
  non-topological superconductors}},\ }\href
  {https://doi.org/10.1038/srep21427} {\bibfield  {journal} {\bibinfo
  {journal} {Sci. Rep.}\ }\textbf {\bibinfo {volume} {6}},\ \bibinfo {pages}
  {21427} (\bibinfo {year} {2016})}\BibitemShut {NoStop}%
\bibitem [{\citenamefont {Ghatak}\ and\ \citenamefont
  {Das}(2018)}]{ghatak2018}%
  \BibitemOpen
  \bibfield  {author} {\bibinfo {author} {\bibfnamefont {A.}~\bibnamefont
  {Ghatak}}\ and\ \bibinfo {author} {\bibfnamefont {T.}~\bibnamefont {Das}},\
  }\bibfield  {title} {\bibinfo {title} {Theory of superconductivity with
  non-{{Hermitian}} and parity-time reversal symmetric {{Cooper}} pairing
  symmetry},\ }\href {https://doi.org/10.1103/PhysRevB.97.014512} {\bibfield
  {journal} {\bibinfo  {journal} {Phys. Rev. B}\ }\textbf {\bibinfo {volume}
  {97}},\ \bibinfo {pages} {014512} (\bibinfo {year} {2018})}\BibitemShut
  {NoStop}%
\bibitem [{\citenamefont {Okuma}\ \emph {et~al.}(2020)\citenamefont {Okuma},
  \citenamefont {Kawabata}, \citenamefont {Shiozaki},\ and\ \citenamefont
  {Sato}}]{okuma2020}%
  \BibitemOpen
  \bibfield  {author} {\bibinfo {author} {\bibfnamefont {N.}~\bibnamefont
  {Okuma}}, \bibinfo {author} {\bibfnamefont {K.}~\bibnamefont {Kawabata}},
  \bibinfo {author} {\bibfnamefont {K.}~\bibnamefont {Shiozaki}},\ and\
  \bibinfo {author} {\bibfnamefont {M.}~\bibnamefont {Sato}},\ }\bibfield
  {title} {\bibinfo {title} {Topological {{Origin}} of {{Non-Hermitian Skin
  Effects}}},\ }\href {https://doi.org/10.1103/PhysRevLett.124.086801}
  {\bibfield  {journal} {\bibinfo  {journal} {Phys. Rev. Lett.}\ }\textbf
  {\bibinfo {volume} {124}},\ \bibinfo {pages} {086801} (\bibinfo {year}
  {2020})}\BibitemShut {NoStop}%
\bibitem [{\citenamefont {Cayao}\ and\ \citenamefont {Sato}()}]{cayao23}%
  \BibitemOpen
  \bibfield  {author} {\bibinfo {author} {\bibfnamefont {J.}~\bibnamefont
  {Cayao}}\ and\ \bibinfo {author} {\bibfnamefont {M.}~\bibnamefont {Sato}},\
  }\bibfield  {title} {\bibinfo {title} {{Non-Hermitian phase-biased Josephson
  junctions}},\ }\Eprint {https://arxiv.org/abs/arXiv:2307.15472}
  {arXiv:2307.15472} \BibitemShut {NoStop}%
\bibitem [{\citenamefont {Esaki}\ \emph {et~al.}(2011)\citenamefont {Esaki},
  \citenamefont {Sato}, \citenamefont {Hasebe},\ and\ \citenamefont
  {Kohmoto}}]{Esaki11}%
  \BibitemOpen
  \bibfield  {author} {\bibinfo {author} {\bibfnamefont {K.}~\bibnamefont
  {Esaki}}, \bibinfo {author} {\bibfnamefont {M.}~\bibnamefont {Sato}},
  \bibinfo {author} {\bibfnamefont {K.}~\bibnamefont {Hasebe}},\ and\ \bibinfo
  {author} {\bibfnamefont {M.}~\bibnamefont {Kohmoto}},\ }\bibfield  {title}
  {\bibinfo {title} {{Edge states and topological phases in non-Hermitian
  systems}},\ }\href {https://doi.org/10.1103/PhysRevB.84.205128} {\bibfield
  {journal} {\bibinfo  {journal} {Phys. Rev. B}\ }\textbf {\bibinfo {volume}
  {84}},\ \bibinfo {pages} {205128} (\bibinfo {year} {2011})}\BibitemShut
  {NoStop}%
\bibitem [{\citenamefont {Lieu}(2018)}]{lieu18}%
  \BibitemOpen
  \bibfield  {author} {\bibinfo {author} {\bibfnamefont {S.}~\bibnamefont
  {Lieu}},\ }\bibfield  {title} {\bibinfo {title} {{Topological phases in the
  non-Hermitian Su-Schrieffer-Heeger model}},\ }\href
  {https://doi.org/10.1103/PhysRevB.97.045106} {\bibfield  {journal} {\bibinfo
  {journal} {Phys. Rev. B}\ }\textbf {\bibinfo {volume} {97}},\ \bibinfo
  {pages} {045106} (\bibinfo {year} {2018})}\BibitemShut {NoStop}%
\bibitem [{\citenamefont {Su}\ \emph {et~al.}(1980)\citenamefont {Su},
  \citenamefont {Schrieffer},\ and\ \citenamefont {Heeger}}]{SSH1980}%
  \BibitemOpen
  \bibfield  {author} {\bibinfo {author} {\bibfnamefont {W.~P.}\ \bibnamefont
  {Su}}, \bibinfo {author} {\bibfnamefont {J.~R.}\ \bibnamefont {Schrieffer}},\
  and\ \bibinfo {author} {\bibfnamefont {A.~J.}\ \bibnamefont {Heeger}},\
  }\bibfield  {title} {\bibinfo {title} {Soliton excitations in
  polyacetylene},\ }\href {https://doi.org/10.1103/PhysRevB.22.2099} {\bibfield
   {journal} {\bibinfo  {journal} {Phys. Rev. B}\ }\textbf {\bibinfo {volume}
  {22}},\ \bibinfo {pages} {2099} (\bibinfo {year} {1980})}\BibitemShut
  {NoStop}%
\bibitem [{\citenamefont {Ginibre}(1965)}]{ginibre1965statistical}%
  \BibitemOpen
  \bibfield  {author} {\bibinfo {author} {\bibfnamefont {J.}~\bibnamefont
  {Ginibre}},\ }\bibfield  {title} {\bibinfo {title} {{Statistical Ensembles of
  Complex, Quaternion, and Real Matrices}},\ }\href
  {https://doi.org/10.1063/1.1704292} {\bibfield  {journal} {\bibinfo
  {journal} {J. Math. Phys.}\ }\textbf {\bibinfo {volume} {6}},\ \bibinfo
  {pages} {440} (\bibinfo {year} {1965})}\BibitemShut {NoStop}%
\bibitem [{\citenamefont {Edelman}\ and\ \citenamefont
  {Kostlan}(1995)}]{edelman1995}%
  \BibitemOpen
  \bibfield  {author} {\bibinfo {author} {\bibfnamefont {A.}~\bibnamefont
  {Edelman}}\ and\ \bibinfo {author} {\bibfnamefont {E.}~\bibnamefont
  {Kostlan}},\ }\bibfield  {title} {\bibinfo {title} {How many zeros of a
  random polynomial are real?},\ }\href
  {https://doi.org/https://doi.org/10.1090/S0273-0979-1995-00571-9} {\bibfield
  {journal} {\bibinfo  {journal} {Bull. Amer. Math. Soc.}\ }\textbf {\bibinfo
  {volume} {32}},\ \bibinfo {pages} {1} (\bibinfo {year} {1995})}\BibitemShut
  {NoStop}%
\bibitem [{\citenamefont {Sommers}\ \emph
  {et~al.}(1988{\natexlab{b}})\citenamefont {Sommers}, \citenamefont
  {Crisanti}, \citenamefont {Sompolinsky},\ and\ \citenamefont
  {Stein}}]{Sommers98}%
  \BibitemOpen
  \bibfield  {author} {\bibinfo {author} {\bibfnamefont {H.~J.}\ \bibnamefont
  {Sommers}}, \bibinfo {author} {\bibfnamefont {A.}~\bibnamefont {Crisanti}},
  \bibinfo {author} {\bibfnamefont {H.}~\bibnamefont {Sompolinsky}},\ and\
  \bibinfo {author} {\bibfnamefont {Y.}~\bibnamefont {Stein}},\ }\bibfield
  {title} {\bibinfo {title} {{Spectrum of Large Random Asymmetric Matrices}},\
  }\href {https://doi.org/10.1103/PhysRevLett.60.1895} {\bibfield  {journal}
  {\bibinfo  {journal} {Phys. Rev. Lett.}\ }\textbf {\bibinfo {volume} {60}},\
  \bibinfo {pages} {1895} (\bibinfo {year} {1988}{\natexlab{b}})}\BibitemShut
  {NoStop}%
\bibitem [{\citenamefont {Kanzieper}\ and\ \citenamefont
  {Akemann}(2005)}]{kanzieper2005}%
  \BibitemOpen
  \bibfield  {author} {\bibinfo {author} {\bibfnamefont {E.}~\bibnamefont
  {Kanzieper}}\ and\ \bibinfo {author} {\bibfnamefont {G.}~\bibnamefont
  {Akemann}},\ }\bibfield  {title} {\bibinfo {title} {Statistics of {{Real
  Eigenvalues}} in {{Ginibre}}'s {{Ensemble}} of {{Random Real Matrices}}},\
  }\href {https://doi.org/10.1103/PhysRevLett.95.230201} {\bibfield  {journal}
  {\bibinfo  {journal} {Phys. Rev. Lett.}\ }\textbf {\bibinfo {volume} {95}},\
  \bibinfo {pages} {230201} (\bibinfo {year} {2005})}\BibitemShut {NoStop}%
\bibitem [{\citenamefont {Forrester}\ and\ \citenamefont
  {Nagao}(2007)}]{Forrester07}%
  \BibitemOpen
  \bibfield  {author} {\bibinfo {author} {\bibfnamefont {P.~J.}\ \bibnamefont
  {Forrester}}\ and\ \bibinfo {author} {\bibfnamefont {T.}~\bibnamefont
  {Nagao}},\ }\bibfield  {title} {\bibinfo {title} {{Eigenvalue Statistics of
  the Real Ginibre Ensemble}},\ }\href
  {https://doi.org/10.1103/PhysRevLett.99.050603} {\bibfield  {journal}
  {\bibinfo  {journal} {Phys. Rev. Lett.}\ }\textbf {\bibinfo {volume} {99}},\
  \bibinfo {pages} {050603} (\bibinfo {year} {2007})}\BibitemShut {NoStop}%
\bibitem [{\citenamefont {Xiao}\ \emph {et~al.}(2022)\citenamefont {Xiao},
  \citenamefont {Kawabata}, \citenamefont {Luo}, \citenamefont {Ohtsuki},\ and\
  \citenamefont {Shindou}}]{xiao22}%
  \BibitemOpen
  \bibfield  {author} {\bibinfo {author} {\bibfnamefont {Z.}~\bibnamefont
  {Xiao}}, \bibinfo {author} {\bibfnamefont {K.}~\bibnamefont {Kawabata}},
  \bibinfo {author} {\bibfnamefont {X.}~\bibnamefont {Luo}}, \bibinfo {author}
  {\bibfnamefont {T.}~\bibnamefont {Ohtsuki}},\ and\ \bibinfo {author}
  {\bibfnamefont {R.}~\bibnamefont {Shindou}},\ }\bibfield  {title} {\bibinfo
  {title} {Level statistics of real eigenvalues in non-{{Hermitian}} systems},\
  }\href {https://doi.org/10.1103/PhysRevResearch.4.043196} {\bibfield
  {journal} {\bibinfo  {journal} {Phys. Rev. Research}\ }\textbf {\bibinfo
  {volume} {4}},\ \bibinfo {pages} {043196} (\bibinfo {year}
  {2022})}\BibitemShut {NoStop}%
\bibitem [{\citenamefont {Splittorff}\ and\ \citenamefont
  {Verbaarschot}(2004)}]{splittorff04}%
  \BibitemOpen
  \bibfield  {author} {\bibinfo {author} {\bibfnamefont {K.}~\bibnamefont
  {Splittorff}}\ and\ \bibinfo {author} {\bibfnamefont {J.}~\bibnamefont
  {Verbaarschot}},\ }\bibfield  {title} {\bibinfo {title} {{Factorization of
  correlation functions and the replica limit of the Toda lattice equation}},\
  }\href {https://doi.org/10.1016/j.nuclphysb.2004.01.031} {\bibfield
  {journal} {\bibinfo  {journal} {Nucl. Phys. B}\ }\textbf {\bibinfo {volume}
  {683}},\ \bibinfo {pages} {467} (\bibinfo {year} {2004})}\BibitemShut
  {NoStop}%
\bibitem [{\citenamefont {Akemann}\ \emph {et~al.}(2009)\citenamefont
  {Akemann}, \citenamefont {Bittner}, \citenamefont {Phillips},\ and\
  \citenamefont {Shifrin}}]{akemann09}%
  \BibitemOpen
  \bibfield  {author} {\bibinfo {author} {\bibfnamefont {G.}~\bibnamefont
  {Akemann}}, \bibinfo {author} {\bibfnamefont {E.}~\bibnamefont {Bittner}},
  \bibinfo {author} {\bibfnamefont {M.~J.}\ \bibnamefont {Phillips}},\ and\
  \bibinfo {author} {\bibfnamefont {L.}~\bibnamefont {Shifrin}},\ }\bibfield
  {title} {\bibinfo {title} {{Wigner surmise for Hermitian and non-Hermitian
  chiral random matrices}},\ }\href
  {https://doi.org/10.1103/PhysRevE.80.065201} {\bibfield  {journal} {\bibinfo
  {journal} {Phys. Rev. E}\ }\textbf {\bibinfo {volume} {80}},\ \bibinfo
  {pages} {065201} (\bibinfo {year} {2009})}\BibitemShut {NoStop}%
\bibitem [{foo()}]{footnote1}%
  \BibitemOpen
  \href@noop {} {}\bibinfo {note} {We numerically evaluate the proportionality
  coefficients $c$'s of $\rho(|z|) \propto f(|z|)$ ($|z| \ll 1$) with $f(z) =
  |z|$, $-|z|^3 \ln|z|$, or $|z|^3$ for different symmetry classes. While $c$
  is given as $ c = \lim_{|z| \to 0} f(|z|)^{-1}\rho(|z|)$ by definition, we
  numerically evaluate $c$ by $ \frac{\int_0^{\Delta z} \rho(|z|)
  d|z|}{\int_0^{\Delta z} f(|z|) d|z|}$ with $\Delta z = 0.1$. Here,
  $\int_0^{\Delta z} \rho(|z|) d|z|$ is evaluated by $N_{[0,\Delta z]}/N_{\rm
  sample}$, where $N_{[0,\Delta z]}$ is the total number of eigenvalues whose
  moduli are smaller than $\Delta z$, and $N_{\rm sample}$ is the number of
  samples. According to the definition of $c$, we should choose small enough
  $\Delta z$, meaning small $N_{[0,\Delta z]}$ and a large relative statistical
  error of $N_{[0,\Delta z]}$. Thus, the evaluated proportionality coefficients
  $c$'s might have relatively large statistical errors.}\BibitemShut {Stop}%
\bibitem [{\citenamefont {Press}\ \emph {et~al.}(2007)\citenamefont {Press},
  \citenamefont {Teukolsky}, \citenamefont {Vetterling},\ and\ \citenamefont
  {Flannery}}]{press07}%
  \BibitemOpen
  \bibfield  {author} {\bibinfo {author} {\bibfnamefont {W.~H.}\ \bibnamefont
  {Press}}, \bibinfo {author} {\bibfnamefont {S.~A.}\ \bibnamefont
  {Teukolsky}}, \bibinfo {author} {\bibfnamefont {W.~T.}\ \bibnamefont
  {Vetterling}},\ and\ \bibinfo {author} {\bibfnamefont {B.~P.}\ \bibnamefont
  {Flannery}},\ }\href@noop {} {\emph {\bibinfo {title} {Numerical Recipes: The
  Art of Scientific Computing}}}\ (\bibinfo  {publisher} {Cambridge University
  Press},\ \bibinfo {year} {2007})\BibitemShut {NoStop}%
\bibitem [{\citenamefont {Br{\'e}zin}\ \emph {et~al.}(1999)\citenamefont
  {Br{\'e}zin}, \citenamefont {Hikami},\ and\ \citenamefont
  {Larkin}}]{brezin1999}%
  \BibitemOpen
  \bibfield  {author} {\bibinfo {author} {\bibfnamefont {E.}~\bibnamefont
  {Br{\'e}zin}}, \bibinfo {author} {\bibfnamefont {S.}~\bibnamefont {Hikami}},\
  and\ \bibinfo {author} {\bibfnamefont {A.~I.}\ \bibnamefont {Larkin}},\
  }\bibfield  {title} {\bibinfo {title} {Level statistics inside the vortex of
  a superconductor and symplectic random-matrix theory in an external source},\
  }\href {https://doi.org/10.1103/PhysRevB.60.3589} {\bibfield  {journal}
  {\bibinfo  {journal} {Phys. Rev. B}\ }\textbf {\bibinfo {volume} {60}},\
  \bibinfo {pages} {3589} (\bibinfo {year} {1999})}\BibitemShut {NoStop}%
\bibitem [{\citenamefont {Ivanov}(2002)}]{ivanov2002}%
  \BibitemOpen
  \bibfield  {author} {\bibinfo {author} {\bibfnamefont {D.~A.}\ \bibnamefont
  {Ivanov}},\ }\bibfield  {title} {\bibinfo {title} {Random-{{Matrix
  Ensembles}} in $p$-{{Wave Vortices}}},\ }in\ \href
  {https://doi.org/10.1007/978-3-662-04665-4_15} {\emph {\bibinfo {booktitle}
  {Vortices in {{Unconventional Superconductors}} and {{Superfluids}}}}},\
  \bibinfo {series and number} {Springer {{Series}} in {{Solid-State
  Sciences}}},\ \bibinfo {editor} {edited by\ \bibinfo {editor} {\bibfnamefont
  {R.~P.}\ \bibnamefont {Huebener}}, \bibinfo {editor} {\bibfnamefont
  {N.}~\bibnamefont {Schopohl}},\ and\ \bibinfo {editor} {\bibfnamefont
  {G.~E.}\ \bibnamefont {Volovik}}}\ (\bibinfo  {publisher} {{Springer}},\
  \bibinfo {address} {{Berlin, Heidelberg}},\ \bibinfo {year} {2002})\ pp.\
  \bibinfo {pages} {253--265}\BibitemShut {NoStop}%
\bibitem [{\citenamefont {Garc{\'i}a-Garc{\'i}a}\ \emph {et~al.}()\citenamefont
  {Garc{\'i}a-Garc{\'i}a}, \citenamefont {S{\'a}}, \citenamefont
  {Verbaarschot},\ and\ \citenamefont {Yin}}]{garcia2023topo}%
  \BibitemOpen
  \bibfield  {author} {\bibinfo {author} {\bibfnamefont {A.~M.}\ \bibnamefont
  {Garc{\'i}a-Garc{\'i}a}}, \bibinfo {author} {\bibfnamefont {L.}~\bibnamefont
  {S{\'a}}}, \bibinfo {author} {\bibfnamefont {J.~J.~M.}\ \bibnamefont
  {Verbaarschot}},\ and\ \bibinfo {author} {\bibfnamefont {C.}~\bibnamefont
  {Yin}},\ }\bibfield  {title} {\bibinfo {title} {{Emergent Topology in
  Many-Body Dissipative Quantum Chaos}},\ }\Eprint
  {https://arxiv.org/abs/arXiv:2311.14640} {arXiv:2311.14640} \BibitemShut
  {NoStop}%
\bibitem [{\citenamefont {Oganesyan}\ and\ \citenamefont
  {Huse}(2007)}]{Oganesyan07}%
  \BibitemOpen
  \bibfield  {author} {\bibinfo {author} {\bibfnamefont {V.}~\bibnamefont
  {Oganesyan}}\ and\ \bibinfo {author} {\bibfnamefont {D.~A.}\ \bibnamefont
  {Huse}},\ }\bibfield  {title} {\bibinfo {title} {Localization of interacting
  fermions at high temperature},\ }\href
  {https://doi.org/10.1103/PhysRevB.75.155111} {\bibfield  {journal} {\bibinfo
  {journal} {Phys. Rev. B}\ }\textbf {\bibinfo {volume} {75}},\ \bibinfo
  {pages} {155111} (\bibinfo {year} {2007})}\BibitemShut {NoStop}%
\bibitem [{\citenamefont {Atas}\ \emph {et~al.}(2013)\citenamefont {Atas},
  \citenamefont {Bogomolny}, \citenamefont {Giraud},\ and\ \citenamefont
  {Roux}}]{atas13}%
  \BibitemOpen
  \bibfield  {author} {\bibinfo {author} {\bibfnamefont {Y.~Y.}\ \bibnamefont
  {Atas}}, \bibinfo {author} {\bibfnamefont {E.}~\bibnamefont {Bogomolny}},
  \bibinfo {author} {\bibfnamefont {O.}~\bibnamefont {Giraud}},\ and\ \bibinfo
  {author} {\bibfnamefont {G.}~\bibnamefont {Roux}},\ }\bibfield  {title}
  {\bibinfo {title} {Distribution of the {{Ratio}} of {{Consecutive Level
  Spacings}} in {{Random Matrix Ensembles}}},\ }\href
  {https://doi.org/10.1103/PhysRevLett.110.084101} {\bibfield  {journal}
  {\bibinfo  {journal} {Phys. Rev. Lett.}\ }\textbf {\bibinfo {volume} {110}},\
  \bibinfo {pages} {084101} (\bibinfo {year} {2013})}\BibitemShut {NoStop}%
\bibitem [{\citenamefont {Sun}\ and\ \citenamefont {Ye}(2020)}]{sun20}%
  \BibitemOpen
  \bibfield  {author} {\bibinfo {author} {\bibfnamefont {F.}~\bibnamefont
  {Sun}}\ and\ \bibinfo {author} {\bibfnamefont {J.}~\bibnamefont {Ye}},\
  }\bibfield  {title} {\bibinfo {title} {Periodic {{Table}} of the {{Ordinary}}
  and {{Supersymmetric Sachdev-Ye-Kitaev Models}}},\ }\href
  {https://doi.org/10.1103/PhysRevLett.124.244101} {\bibfield  {journal}
  {\bibinfo  {journal} {Phys. Rev. Lett.}\ }\textbf {\bibinfo {volume} {124}},\
  \bibinfo {pages} {244101} (\bibinfo {year} {2020})}\BibitemShut {NoStop}%
\bibitem [{\citenamefont {Akemann}(2005)}]{akemann2005}%
  \BibitemOpen
  \bibfield  {author} {\bibinfo {author} {\bibfnamefont {G.}~\bibnamefont
  {Akemann}},\ }\bibfield  {title} {\bibinfo {title} {The complex {{Laguerre}}
  symplectic ensemble of non-{{Hermitian}} matrices},\ }\href
  {https://doi.org/10.1016/j.nuclphysb.2005.09.039} {\bibfield  {journal}
  {\bibinfo  {journal} {Nucl. Phys. B}\ }\textbf {\bibinfo {volume} {730}},\
  \bibinfo {pages} {253} (\bibinfo {year} {2005})}\BibitemShut {NoStop}%
\bibitem [{\citenamefont {Prosen}(2008)}]{prosen2008a}%
  \BibitemOpen
  \bibfield  {author} {\bibinfo {author} {\bibfnamefont {T.}~\bibnamefont
  {Prosen}},\ }\bibfield  {title} {\bibinfo {title} {Third quantization: A
  general method to solve master equations for quadratic open {{Fermi}}
  systems},\ }\href {https://doi.org/10.1088/1367-2630/10/4/043026} {\bibfield
  {journal} {\bibinfo  {journal} {New J. Phys.}\ }\textbf {\bibinfo {volume}
  {10}},\ \bibinfo {pages} {043026} (\bibinfo {year} {2008})}\BibitemShut
  {NoStop}%
\bibitem [{\citenamefont {Prosen}(2010)}]{prosen2010}%
  \BibitemOpen
  \bibfield  {author} {\bibinfo {author} {\bibfnamefont {T.}~\bibnamefont
  {Prosen}},\ }\bibfield  {title} {\bibinfo {title} {Spectral theorem for the
  {{Lindblad}} equation for quadratic open fermionic systems},\ }\href
  {https://doi.org/10.1088/1742-5468/2010/07/P07020} {\bibfield  {journal}
  {\bibinfo  {journal} {J. Stat. Mech.}\ }\textbf {\bibinfo {volume} {2010}},\
  \bibinfo {pages} {P07020} (\bibinfo {year} {2010})}\BibitemShut {NoStop}%
\bibitem [{\citenamefont {Prosen}(2012)}]{Prosen-12}%
  \BibitemOpen
  \bibfield  {author} {\bibinfo {author} {\bibfnamefont {T.}~\bibnamefont
  {Prosen}},\ }\bibfield  {title} {\bibinfo {title}
  {{$\mathbb{P}\mathbb{T}$-Symmetric Quantum Liouvillean Dynamics}},\ }\href
  {https://doi.org/10.1103/PhysRevLett.109.090404} {\bibfield  {journal}
  {\bibinfo  {journal} {Phys. Rev. Lett.}\ }\textbf {\bibinfo {volume} {109}},\
  \bibinfo {pages} {090404} (\bibinfo {year} {2012})}\BibitemShut {NoStop}%
\bibitem [{\citenamefont {Kawasaki}\ \emph {et~al.}(2022)\citenamefont
  {Kawasaki}, \citenamefont {Mochizuki},\ and\ \citenamefont
  {Obuse}}]{kawasaki2022}%
  \BibitemOpen
  \bibfield  {author} {\bibinfo {author} {\bibfnamefont {M.}~\bibnamefont
  {Kawasaki}}, \bibinfo {author} {\bibfnamefont {K.}~\bibnamefont
  {Mochizuki}},\ and\ \bibinfo {author} {\bibfnamefont {H.}~\bibnamefont
  {Obuse}},\ }\bibfield  {title} {\bibinfo {title} {Topological phases potected
  by shifted sublattice symmetry in dissipative quantum systems},\ }\href
  {https://doi.org/10.1103/PhysRevB.106.035408} {\bibfield  {journal} {\bibinfo
   {journal} {Phys. Rev. B}\ }\textbf {\bibinfo {volume} {106}},\ \bibinfo
  {pages} {035408} (\bibinfo {year} {2022})}\BibitemShut {NoStop}%
\bibitem [{\citenamefont {Lee}\ and\ \citenamefont {Fisher}(1981)}]{lee1981}%
  \BibitemOpen
  \bibfield  {author} {\bibinfo {author} {\bibfnamefont {P.~A.}\ \bibnamefont
  {Lee}}\ and\ \bibinfo {author} {\bibfnamefont {D.~S.}\ \bibnamefont
  {Fisher}},\ }\bibfield  {title} {\bibinfo {title} {Anderson {{Localization}}
  in {{Two Dimensions}}},\ }\href {https://doi.org/10.1103/PhysRevLett.47.882}
  {\bibfield  {journal} {\bibinfo  {journal} {Phys. Rev. Lett.}\ }\textbf
  {\bibinfo {volume} {47}},\ \bibinfo {pages} {882} (\bibinfo {year}
  {1981})}\BibitemShut {NoStop}%
\bibitem [{\citenamefont {Furusaki}(1999)}]{furusaki1999}%
  \BibitemOpen
  \bibfield  {author} {\bibinfo {author} {\bibfnamefont {A.}~\bibnamefont
  {Furusaki}},\ }\bibfield  {title} {\bibinfo {title} {Anderson
  {{Localization}} due to a {{Random Magnetic Field}} in {{Two Dimensions}}},\
  }\href {https://doi.org/10.1103/PhysRevLett.82.604} {\bibfield  {journal}
  {\bibinfo  {journal} {Phys. Rev. Lett.}\ }\textbf {\bibinfo {volume} {82}},\
  \bibinfo {pages} {604} (\bibinfo {year} {1999})}\BibitemShut {NoStop}%
\bibitem [{\citenamefont {Asada}\ \emph {et~al.}(2002)\citenamefont {Asada},
  \citenamefont {Slevin},\ and\ \citenamefont {Ohtsuki}}]{asada2002}%
  \BibitemOpen
  \bibfield  {author} {\bibinfo {author} {\bibfnamefont {Y.}~\bibnamefont
  {Asada}}, \bibinfo {author} {\bibfnamefont {K.}~\bibnamefont {Slevin}},\ and\
  \bibinfo {author} {\bibfnamefont {T.}~\bibnamefont {Ohtsuki}},\ }\bibfield
  {title} {\bibinfo {title} {Anderson {{Transition}} in {{Two-Dimensional
  Systems}} with {{Spin-Orbit Coupling}}},\ }\href
  {https://doi.org/10.1103/PhysRevLett.89.256601} {\bibfield  {journal}
  {\bibinfo  {journal} {Phys. Rev. Lett.}\ }\textbf {\bibinfo {volume} {89}},\
  \bibinfo {pages} {256601} (\bibinfo {year} {2002})}\BibitemShut {NoStop}%
\bibitem [{\citenamefont {Kondov}\ \emph {et~al.}(2011)\citenamefont {Kondov},
  \citenamefont {McGehee}, \citenamefont {Zirbel},\ and\ \citenamefont
  {DeMarco}}]{kondov2011}%
  \BibitemOpen
  \bibfield  {author} {\bibinfo {author} {\bibfnamefont {S.~S.}\ \bibnamefont
  {Kondov}}, \bibinfo {author} {\bibfnamefont {W.~R.}\ \bibnamefont {McGehee}},
  \bibinfo {author} {\bibfnamefont {J.~J.}\ \bibnamefont {Zirbel}},\ and\
  \bibinfo {author} {\bibfnamefont {B.}~\bibnamefont {DeMarco}},\ }\bibfield
  {title} {\bibinfo {title} {Three-{{Dimensional Anderson Localization}} of
  {{Ultracold Matter}}},\ }\href {https://doi.org/10.1126/science.1209019}
  {\bibfield  {journal} {\bibinfo  {journal} {Science}\ }\textbf {\bibinfo
  {volume} {334}},\ \bibinfo {pages} {66} (\bibinfo {year} {2011})}\BibitemShut
  {NoStop}%
\bibitem [{\citenamefont {Meier}\ \emph {et~al.}(2018)\citenamefont {Meier},
  \citenamefont {An}, \citenamefont {Dauphin}, \citenamefont {Maffei},
  \citenamefont {Massignan}, \citenamefont {Hughes},\ and\ \citenamefont
  {Gadway}}]{meier2018}%
  \BibitemOpen
  \bibfield  {author} {\bibinfo {author} {\bibfnamefont {E.~J.}\ \bibnamefont
  {Meier}}, \bibinfo {author} {\bibfnamefont {F.~A.}\ \bibnamefont {An}},
  \bibinfo {author} {\bibfnamefont {A.}~\bibnamefont {Dauphin}}, \bibinfo
  {author} {\bibfnamefont {M.}~\bibnamefont {Maffei}}, \bibinfo {author}
  {\bibfnamefont {P.}~\bibnamefont {Massignan}}, \bibinfo {author}
  {\bibfnamefont {T.~L.}\ \bibnamefont {Hughes}},\ and\ \bibinfo {author}
  {\bibfnamefont {B.}~\bibnamefont {Gadway}},\ }\bibfield  {title} {\bibinfo
  {title} {Observation of the topological {{Anderson}} insulator in disordered
  atomic wires},\ }\href {https://doi.org/10.1126/science.aat3406} {\bibfield
  {journal} {\bibinfo  {journal} {Science}\ }\textbf {\bibinfo {volume}
  {362}},\ \bibinfo {pages} {929} (\bibinfo {year} {2018})}\BibitemShut
  {NoStop}%
\bibitem [{\citenamefont {St{\"u}tzer}\ \emph {et~al.}(2018)\citenamefont
  {St{\"u}tzer}, \citenamefont {Plotnik}, \citenamefont {Lumer}, \citenamefont
  {Titum}, \citenamefont {Lindner}, \citenamefont {Segev}, \citenamefont
  {Rechtsman},\ and\ \citenamefont {Szameit}}]{stutzer2018}%
  \BibitemOpen
  \bibfield  {author} {\bibinfo {author} {\bibfnamefont {S.}~\bibnamefont
  {St{\"u}tzer}}, \bibinfo {author} {\bibfnamefont {Y.}~\bibnamefont
  {Plotnik}}, \bibinfo {author} {\bibfnamefont {Y.}~\bibnamefont {Lumer}},
  \bibinfo {author} {\bibfnamefont {P.}~\bibnamefont {Titum}}, \bibinfo
  {author} {\bibfnamefont {N.~H.}\ \bibnamefont {Lindner}}, \bibinfo {author}
  {\bibfnamefont {M.}~\bibnamefont {Segev}}, \bibinfo {author} {\bibfnamefont
  {M.~C.}\ \bibnamefont {Rechtsman}},\ and\ \bibinfo {author} {\bibfnamefont
  {A.}~\bibnamefont {Szameit}},\ }\bibfield  {title} {\bibinfo {title}
  {Photonic topological {{Anderson}} insulators},\ }\href
  {https://doi.org/10.1038/s41586-018-0418-2} {\bibfield  {journal} {\bibinfo
  {journal} {Nature}\ }\textbf {\bibinfo {volume} {560}},\ \bibinfo {pages}
  {461} (\bibinfo {year} {2018})}\BibitemShut {NoStop}%
\bibitem [{\citenamefont {Liu}\ \emph {et~al.}(2020)\citenamefont {Liu},
  \citenamefont {Yang}, \citenamefont {Ren}, \citenamefont {Xue}, \citenamefont
  {Lin}, \citenamefont {Hu}, \citenamefont {Sun}, \citenamefont {Peng},
  \citenamefont {Zhou}, \citenamefont {Chong},\ and\ \citenamefont
  {Zhang}}]{liu2020}%
  \BibitemOpen
  \bibfield  {author} {\bibinfo {author} {\bibfnamefont {G.-G.}\ \bibnamefont
  {Liu}}, \bibinfo {author} {\bibfnamefont {Y.}~\bibnamefont {Yang}}, \bibinfo
  {author} {\bibfnamefont {X.}~\bibnamefont {Ren}}, \bibinfo {author}
  {\bibfnamefont {H.}~\bibnamefont {Xue}}, \bibinfo {author} {\bibfnamefont
  {X.}~\bibnamefont {Lin}}, \bibinfo {author} {\bibfnamefont {Y.-H.}\
  \bibnamefont {Hu}}, \bibinfo {author} {\bibfnamefont {H.-x.}\ \bibnamefont
  {Sun}}, \bibinfo {author} {\bibfnamefont {B.}~\bibnamefont {Peng}}, \bibinfo
  {author} {\bibfnamefont {P.}~\bibnamefont {Zhou}}, \bibinfo {author}
  {\bibfnamefont {Y.}~\bibnamefont {Chong}},\ and\ \bibinfo {author}
  {\bibfnamefont {B.}~\bibnamefont {Zhang}},\ }\bibfield  {title} {\bibinfo
  {title} {Topological {{Anderson Insulator}} in {{Disordered Photonic
  Crystals}}},\ }\href {https://doi.org/10.1103/PhysRevLett.125.133603}
  {\bibfield  {journal} {\bibinfo  {journal} {Phys. Rev. Lett.}\ }\textbf
  {\bibinfo {volume} {125}},\ \bibinfo {pages} {133603} (\bibinfo {year}
  {2020})}\BibitemShut {NoStop}%
\bibitem [{\citenamefont {Livi}\ \emph {et~al.}(2016)\citenamefont {Livi},
  \citenamefont {Cappellini}, \citenamefont {Diem}, \citenamefont {Franchi},
  \citenamefont {Clivati}, \citenamefont {Frittelli}, \citenamefont {Levi},
  \citenamefont {Calonico}, \citenamefont {Catani}, \citenamefont {Inguscio},\
  and\ \citenamefont {Fallani}}]{livi2016}%
  \BibitemOpen
  \bibfield  {author} {\bibinfo {author} {\bibfnamefont {L.~F.}\ \bibnamefont
  {Livi}}, \bibinfo {author} {\bibfnamefont {G.}~\bibnamefont {Cappellini}},
  \bibinfo {author} {\bibfnamefont {M.}~\bibnamefont {Diem}}, \bibinfo {author}
  {\bibfnamefont {L.}~\bibnamefont {Franchi}}, \bibinfo {author} {\bibfnamefont
  {C.}~\bibnamefont {Clivati}}, \bibinfo {author} {\bibfnamefont
  {M.}~\bibnamefont {Frittelli}}, \bibinfo {author} {\bibfnamefont
  {F.}~\bibnamefont {Levi}}, \bibinfo {author} {\bibfnamefont {D.}~\bibnamefont
  {Calonico}}, \bibinfo {author} {\bibfnamefont {J.}~\bibnamefont {Catani}},
  \bibinfo {author} {\bibfnamefont {M.}~\bibnamefont {Inguscio}},\ and\
  \bibinfo {author} {\bibfnamefont {L.}~\bibnamefont {Fallani}},\ }\bibfield
  {title} {\bibinfo {title} {Synthetic {{Dimensions}} and {{Spin-Orbit
  Coupling}} with an {{Optical Clock Transition}}},\ }\href
  {https://doi.org/10.1103/PhysRevLett.117.220401} {\bibfield  {journal}
  {\bibinfo  {journal} {Phys. Rev. Lett.}\ }\textbf {\bibinfo {volume} {117}},\
  \bibinfo {pages} {220401} (\bibinfo {year} {2016})}\BibitemShut {NoStop}%
\bibitem [{\citenamefont {Ozawa}\ and\ \citenamefont
  {Price}(2019)}]{ozawa2019}%
  \BibitemOpen
  \bibfield  {author} {\bibinfo {author} {\bibfnamefont {T.}~\bibnamefont
  {Ozawa}}\ and\ \bibinfo {author} {\bibfnamefont {H.~M.}\ \bibnamefont
  {Price}},\ }\bibfield  {title} {\bibinfo {title} {Topological quantum matter
  in synthetic dimensions},\ }\href {https://doi.org/10.1038/s42254-019-0045-3}
  {\bibfield  {journal} {\bibinfo  {journal} {Nat. Rev. Phys.}\ }\textbf
  {\bibinfo {volume} {1}},\ \bibinfo {pages} {349} (\bibinfo {year}
  {2019})}\BibitemShut {NoStop}%
\bibitem [{\citenamefont {Diehl}\ \emph {et~al.}(2011)\citenamefont {Diehl},
  \citenamefont {Rico}, \citenamefont {Baranov},\ and\ \citenamefont
  {Zoller}}]{diehl2011}%
  \BibitemOpen
  \bibfield  {author} {\bibinfo {author} {\bibfnamefont {S.}~\bibnamefont
  {Diehl}}, \bibinfo {author} {\bibfnamefont {E.}~\bibnamefont {Rico}},
  \bibinfo {author} {\bibfnamefont {M.~A.}\ \bibnamefont {Baranov}},\ and\
  \bibinfo {author} {\bibfnamefont {P.}~\bibnamefont {Zoller}},\ }\bibfield
  {title} {\bibinfo {title} {Topology by dissipation in atomic quantum wires},\
  }\href {https://doi.org/10.1038/nphys2106} {\bibfield  {journal} {\bibinfo
  {journal} {Nat. Phys}\ }\textbf {\bibinfo {volume} {7}},\ \bibinfo {pages}
  {971} (\bibinfo {year} {2011})}\BibitemShut {NoStop}%
\bibitem [{\citenamefont {Song}\ \emph {et~al.}(2019)\citenamefont {Song},
  \citenamefont {Yao},\ and\ \citenamefont {Wang}}]{song2019a}%
  \BibitemOpen
  \bibfield  {author} {\bibinfo {author} {\bibfnamefont {F.}~\bibnamefont
  {Song}}, \bibinfo {author} {\bibfnamefont {S.}~\bibnamefont {Yao}},\ and\
  \bibinfo {author} {\bibfnamefont {Z.}~\bibnamefont {Wang}},\ }\bibfield
  {title} {\bibinfo {title} {Non-{{Hermitian Skin Effect}} and {{Chiral
  Damping}} in {{Open Quantum Systems}}},\ }\href
  {https://doi.org/10.1103/PhysRevLett.123.170401} {\bibfield  {journal}
  {\bibinfo  {journal} {Phys. Rev. Lett.}\ }\textbf {\bibinfo {volume} {123}},\
  \bibinfo {pages} {170401} (\bibinfo {year} {2019})}\BibitemShut {NoStop}%
\bibitem [{\citenamefont {Girko}(1985)}]{Girko-85}%
  \BibitemOpen
  \bibfield  {author} {\bibinfo {author} {\bibfnamefont {V.~L.}\ \bibnamefont
  {Girko}},\ }\bibfield  {title} {\bibinfo {title} {{Circular Law}},\ }\href
  {https://doi.org/10.1137/1129095} {\bibfield  {journal} {\bibinfo  {journal}
  {Theory Probab. Appl.}\ }\textbf {\bibinfo {volume} {29}},\ \bibinfo {pages}
  {694} (\bibinfo {year} {1985})}\BibitemShut {NoStop}%
\bibitem [{\citenamefont {Edwards}\ and\ \citenamefont
  {Thouless}(1972)}]{JTEdwards1972}%
  \BibitemOpen
  \bibfield  {author} {\bibinfo {author} {\bibfnamefont {J.~T.}\ \bibnamefont
  {Edwards}}\ and\ \bibinfo {author} {\bibfnamefont {D.~J.}\ \bibnamefont
  {Thouless}},\ }\bibfield  {title} {\bibinfo {title} {Numerical studies of
  localization in disordered systems},\ }\href
  {https://doi.org/10.1088/0022-3719/5/8/007} {\bibfield  {journal} {\bibinfo
  {journal} {J. Phys. C}\ }\textbf {\bibinfo {volume} {5}},\ \bibinfo {pages}
  {807} (\bibinfo {year} {1972})}\BibitemShut {NoStop}%
\bibitem [{\citenamefont {Altshuler}\ and\ \citenamefont
  {Shklovskii}(1986)}]{altshuler1986repulsion}%
  \BibitemOpen
  \bibfield  {author} {\bibinfo {author} {\bibfnamefont {B.}~\bibnamefont
  {Altshuler}}\ and\ \bibinfo {author} {\bibfnamefont {B.}~\bibnamefont
  {Shklovskii}},\ }\bibfield  {title} {\bibinfo {title} {Repulsion of energy
  levels and conductivity of small metal samples},\ }\href@noop {} {\bibfield
  {journal} {\bibinfo  {journal} {J. Exp. Theor. Phys.}\ }\textbf {\bibinfo
  {volume} {64}},\ \bibinfo {pages} {127} (\bibinfo {year} {1986})}\BibitemShut
  {NoStop}%
\bibitem [{\citenamefont {Pientka}\ \emph {et~al.}(2013)\citenamefont
  {Pientka}, \citenamefont {Glazman},\ and\ \citenamefont {von
  Oppen}}]{pientka13}%
  \BibitemOpen
  \bibfield  {author} {\bibinfo {author} {\bibfnamefont {F.}~\bibnamefont
  {Pientka}}, \bibinfo {author} {\bibfnamefont {L.~I.}\ \bibnamefont
  {Glazman}},\ and\ \bibinfo {author} {\bibfnamefont {F.}~\bibnamefont {von
  Oppen}},\ }\bibfield  {title} {\bibinfo {title} {Topological superconducting
  phase in helical {{Shiba}} chains},\ }\href
  {https://doi.org/10.1103/PhysRevB.88.155420} {\bibfield  {journal} {\bibinfo
  {journal} {Phys. Rev. B}\ }\textbf {\bibinfo {volume} {88}},\ \bibinfo
  {pages} {155420} (\bibinfo {year} {2013})}\BibitemShut {NoStop}%
\bibitem [{\citenamefont {Li}\ \emph {et~al.}(2018)\citenamefont {Li},
  \citenamefont {Zhang}, \citenamefont {Zhang},\ and\ \citenamefont
  {Song}}]{li2018}%
  \BibitemOpen
  \bibfield  {author} {\bibinfo {author} {\bibfnamefont {C.}~\bibnamefont
  {Li}}, \bibinfo {author} {\bibfnamefont {X.~Z.}\ \bibnamefont {Zhang}},
  \bibinfo {author} {\bibfnamefont {G.}~\bibnamefont {Zhang}},\ and\ \bibinfo
  {author} {\bibfnamefont {Z.}~\bibnamefont {Song}},\ }\bibfield  {title}
  {\bibinfo {title} {Topological phases in a {{Kitaev}} chain with imbalanced
  pairing},\ }\href {https://doi.org/10.1103/PhysRevB.97.115436} {\bibfield
  {journal} {\bibinfo  {journal} {Phys. Rev. B}\ }\textbf {\bibinfo {volume}
  {97}},\ \bibinfo {pages} {115436} (\bibinfo {year} {2018})}\BibitemShut
  {NoStop}%
\bibitem [{\citenamefont {Kornich}\ and\ \citenamefont
  {Trauzettel}(2022{\natexlab{a}})}]{kornich2022}%
  \BibitemOpen
  \bibfield  {author} {\bibinfo {author} {\bibfnamefont {V.}~\bibnamefont
  {Kornich}}\ and\ \bibinfo {author} {\bibfnamefont {B.}~\bibnamefont
  {Trauzettel}},\ }\bibfield  {title} {\bibinfo {title} {{Signature of
  $\mathcal{P}\mathcal{T}$-symmetric non-Hermitian superconductivity in
  angle-resolved photoelectron fluctuation spectroscopy}},\ }\href
  {https://doi.org/10.1103/PhysRevResearch.4.L022018} {\bibfield  {journal}
  {\bibinfo  {journal} {Phys. Rev. Research}\ }\textbf {\bibinfo {volume}
  {4}},\ \bibinfo {pages} {L022018} (\bibinfo {year}
  {2022}{\natexlab{a}})}\BibitemShut {NoStop}%
\bibitem [{\citenamefont {Kornich}\ and\ \citenamefont
  {Trauzettel}(2022{\natexlab{b}})}]{kornich2022a}%
  \BibitemOpen
  \bibfield  {author} {\bibinfo {author} {\bibfnamefont {V.}~\bibnamefont
  {Kornich}}\ and\ \bibinfo {author} {\bibfnamefont {B.}~\bibnamefont
  {Trauzettel}},\ }\bibfield  {title} {\bibinfo {title} {{Andreev bound states
  in junctions formed by conventional and $\mathcal{PT}$-symmetric
  non-Hermitian superconductors}},\ }\href
  {https://doi.org/10.1103/PhysRevResearch.4.033201} {\bibfield  {journal}
  {\bibinfo  {journal} {Phys. Rev. Research}\ }\textbf {\bibinfo {volume}
  {4}},\ \bibinfo {pages} {033201} (\bibinfo {year}
  {2022}{\natexlab{b}})}\BibitemShut {NoStop}%
\bibitem [{\citenamefont {Kornich}(2023)}]{kornich23}%
  \BibitemOpen
  \bibfield  {author} {\bibinfo {author} {\bibfnamefont {V.}~\bibnamefont
  {Kornich}},\ }\bibfield  {title} {\bibinfo {title} {{Current-Voltage
  Characteristics of the Normal Metal-Insulator-PT-Symmetric Non-Hermitian
  Superconductor Junction as a Probe of Non-Hermitian Formalisms}},\ }\href
  {https://doi.org/10.1103/PhysRevLett.131.116001} {\bibfield  {journal}
  {\bibinfo  {journal} {Phys. Rev. Lett.}\ }\textbf {\bibinfo {volume} {131}},\
  \bibinfo {pages} {116001} (\bibinfo {year} {2023})}\BibitemShut {NoStop}%
\bibitem [{\citenamefont {Zhang}\ and\ \citenamefont {Sheng}(2023)}]{zhang23a}%
  \BibitemOpen
  \bibfield  {author} {\bibinfo {author} {\bibfnamefont {C.}~\bibnamefont
  {Zhang}}\ and\ \bibinfo {author} {\bibfnamefont {L.}~\bibnamefont {Sheng}},\
  }\bibfield  {title} {\bibinfo {title} {{Fate of magnetic impurity induced
  states in a non-Hermitian $s$-wave superconductor}},\ }\href
  {https://doi.org/10.1103/PhysRevB.107.184507} {\bibfield  {journal} {\bibinfo
   {journal} {Phys. Rev. B}\ }\textbf {\bibinfo {volume} {107}},\ \bibinfo
  {pages} {184507} (\bibinfo {year} {2023})}\BibitemShut {NoStop}%
\bibitem [{\citenamefont {Schnyder}\ \emph {et~al.}(2008)\citenamefont
  {Schnyder}, \citenamefont {Ryu}, \citenamefont {Furusaki},\ and\
  \citenamefont {Ludwig}}]{schnyder08a}%
  \BibitemOpen
  \bibfield  {author} {\bibinfo {author} {\bibfnamefont {A.~P.}\ \bibnamefont
  {Schnyder}}, \bibinfo {author} {\bibfnamefont {S.}~\bibnamefont {Ryu}},
  \bibinfo {author} {\bibfnamefont {A.}~\bibnamefont {Furusaki}},\ and\
  \bibinfo {author} {\bibfnamefont {A.~W.~W.}\ \bibnamefont {Ludwig}},\
  }\bibfield  {title} {\bibinfo {title} {Classification of topological
  insulators and superconductors in three spatial dimensions},\ }\href
  {https://doi.org/10.1103/PhysRevB.78.195125} {\bibfield  {journal} {\bibinfo
  {journal} {Phys. Rev. B}\ }\textbf {\bibinfo {volume} {78}},\ \bibinfo
  {pages} {195125} (\bibinfo {year} {2008})}\BibitemShut {NoStop}%
\bibitem [{\citenamefont {Schnyder}\ and\ \citenamefont
  {Ryu}(2011)}]{schnyder11}%
  \BibitemOpen
  \bibfield  {author} {\bibinfo {author} {\bibfnamefont {A.~P.}\ \bibnamefont
  {Schnyder}}\ and\ \bibinfo {author} {\bibfnamefont {S.}~\bibnamefont {Ryu}},\
  }\bibfield  {title} {\bibinfo {title} {Topological phases and surface flat
  bands in superconductors without inversion symmetry},\ }\href
  {https://doi.org/10.1103/PhysRevB.84.060504} {\bibfield  {journal} {\bibinfo
  {journal} {Phys. Rev. B}\ }\textbf {\bibinfo {volume} {84}},\ \bibinfo
  {pages} {060504} (\bibinfo {year} {2011})}\BibitemShut {NoStop}%
\bibitem [{\citenamefont {Mildenberger}\ \emph {et~al.}(2007)\citenamefont
  {Mildenberger}, \citenamefont {Evers}, \citenamefont {Mirlin},\ and\
  \citenamefont {Chalker}}]{Mildenberger07}%
  \BibitemOpen
  \bibfield  {author} {\bibinfo {author} {\bibfnamefont {A.}~\bibnamefont
  {Mildenberger}}, \bibinfo {author} {\bibfnamefont {F.}~\bibnamefont {Evers}},
  \bibinfo {author} {\bibfnamefont {A.~D.}\ \bibnamefont {Mirlin}},\ and\
  \bibinfo {author} {\bibfnamefont {J.~T.}\ \bibnamefont {Chalker}},\
  }\bibfield  {title} {\bibinfo {title} {{Density of quasiparticle states for a
  two-dimensional disordered system: Metallic, insulating, and critical
  behavior in the class-D thermal quantum Hall effect}},\ }\href
  {https://doi.org/10.1103/PhysRevB.75.245321} {\bibfield  {journal} {\bibinfo
  {journal} {Phys. Rev. B}\ }\textbf {\bibinfo {volume} {75}},\ \bibinfo
  {pages} {245321} (\bibinfo {year} {2007})}\BibitemShut {NoStop}%
\bibitem [{\citenamefont {Liu}\ \emph {et~al.}(2021)\citenamefont {Liu},
  \citenamefont {Zhang},\ and\ \citenamefont {Chen}}]{liu2021}%
  \BibitemOpen
  \bibfield  {author} {\bibinfo {author} {\bibfnamefont {C.}~\bibnamefont
  {Liu}}, \bibinfo {author} {\bibfnamefont {P.}~\bibnamefont {Zhang}},\ and\
  \bibinfo {author} {\bibfnamefont {X.}~\bibnamefont {Chen}},\ }\bibfield
  {title} {\bibinfo {title} {Non-unitary dynamics of {{Sachdev-Ye-Kitaev}}
  chain},\ }\href {https://doi.org/10.21468/SciPostPhys.10.2.048} {\bibfield
  {journal} {\bibinfo  {journal} {SciPost Phys.}\ }\textbf {\bibinfo {volume}
  {10}},\ \bibinfo {pages} {048} (\bibinfo {year} {2021})}\BibitemShut
  {NoStop}%
\bibitem [{\citenamefont {{Garc{\'i}a-Garc{\'i}a}}\ \emph
  {et~al.}(2022{\natexlab{b}})\citenamefont {{Garc{\'i}a-Garc{\'i}a}},
  \citenamefont {Jia}, \citenamefont {Rosa},\ and\ \citenamefont
  {Verbaarschot}}]{garcia22prl}%
  \BibitemOpen
  \bibfield  {author} {\bibinfo {author} {\bibfnamefont {A.~M.}\ \bibnamefont
  {{Garc{\'i}a-Garc{\'i}a}}}, \bibinfo {author} {\bibfnamefont
  {Y.}~\bibnamefont {Jia}}, \bibinfo {author} {\bibfnamefont {D.}~\bibnamefont
  {Rosa}},\ and\ \bibinfo {author} {\bibfnamefont {J.~J.~M.}\ \bibnamefont
  {Verbaarschot}},\ }\bibfield  {title} {\bibinfo {title} {{Dominance of
  Replica Off-Diagonal Configurations and Phase Transitions in a $PT$ Symmetric
  Sachdev-Ye-Kitaev Model}},\ }\href
  {https://doi.org/10.1103/PhysRevLett.128.081601} {\bibfield  {journal}
  {\bibinfo  {journal} {Phys. Rev. Lett.}\ }\textbf {\bibinfo {volume} {128}},\
  \bibinfo {pages} {081601} (\bibinfo {year} {2022}{\natexlab{b}})}\BibitemShut
  {NoStop}%
\bibitem [{\citenamefont {Zhang}\ \emph {et~al.}(2021)\citenamefont {Zhang},
  \citenamefont {Jian}, \citenamefont {Liu},\ and\ \citenamefont
  {Chen}}]{zhang2021}%
  \BibitemOpen
  \bibfield  {author} {\bibinfo {author} {\bibfnamefont {P.}~\bibnamefont
  {Zhang}}, \bibinfo {author} {\bibfnamefont {S.-K.}\ \bibnamefont {Jian}},
  \bibinfo {author} {\bibfnamefont {C.}~\bibnamefont {Liu}},\ and\ \bibinfo
  {author} {\bibfnamefont {X.}~\bibnamefont {Chen}},\ }\bibfield  {title}
  {\bibinfo {title} {Emergent {{Replica Conformal Symmetry}} in {{Non-Hermitian
  SYK$_2$}} {{Chains}}},\ }\href {https://doi.org/10.22331/q-2021-11-16-579}
  {\bibfield  {journal} {\bibinfo  {journal} {Quantum}\ }\textbf {\bibinfo
  {volume} {5}},\ \bibinfo {pages} {579} (\bibinfo {year} {2021})}\BibitemShut
  {NoStop}%
\bibitem [{\citenamefont {Jian}\ \emph {et~al.}(2021)\citenamefont {Jian},
  \citenamefont {Liu}, \citenamefont {Chen}, \citenamefont {Swingle},\ and\
  \citenamefont {Zhang}}]{jian2021}%
  \BibitemOpen
  \bibfield  {author} {\bibinfo {author} {\bibfnamefont {S.-K.}\ \bibnamefont
  {Jian}}, \bibinfo {author} {\bibfnamefont {C.}~\bibnamefont {Liu}}, \bibinfo
  {author} {\bibfnamefont {X.}~\bibnamefont {Chen}}, \bibinfo {author}
  {\bibfnamefont {B.}~\bibnamefont {Swingle}},\ and\ \bibinfo {author}
  {\bibfnamefont {P.}~\bibnamefont {Zhang}},\ }\bibfield  {title} {\bibinfo
  {title} {Measurement-{{Induced Phase Transition}} in the {{Monitored
  Sachdev-Ye-Kitaev Model}}},\ }\href
  {https://doi.org/10.1103/PhysRevLett.127.140601} {\bibfield  {journal}
  {\bibinfo  {journal} {Phys. Rev. Lett.}\ }\textbf {\bibinfo {volume} {127}},\
  \bibinfo {pages} {140601} (\bibinfo {year} {2021})}\BibitemShut {NoStop}%
\bibitem [{\citenamefont {S\'a}\ \emph {et~al.}(2022)\citenamefont {S\'a},
  \citenamefont {Ribeiro},\ and\ \citenamefont {Prosen}}]{sa2022SYK}%
  \BibitemOpen
  \bibfield  {author} {\bibinfo {author} {\bibfnamefont {L.}~\bibnamefont
  {S\'a}}, \bibinfo {author} {\bibfnamefont {P.}~\bibnamefont {Ribeiro}},\ and\
  \bibinfo {author} {\bibfnamefont {T.}~\bibnamefont {Prosen}},\ }\bibfield
  {title} {\bibinfo {title} {{Lindbladian dissipation of strongly-correlated
  quantum matter}},\ }\href {https://doi.org/10.1103/PhysRevResearch.4.L022068}
  {\bibfield  {journal} {\bibinfo  {journal} {Phys. Rev. Research}\ }\textbf
  {\bibinfo {volume} {4}},\ \bibinfo {pages} {L022068} (\bibinfo {year}
  {2022})}\BibitemShut {NoStop}%
\bibitem [{\citenamefont {Kulkarni}\ \emph {et~al.}(2022)\citenamefont
  {Kulkarni}, \citenamefont {Numasawa},\ and\ \citenamefont
  {Ryu}}]{kulkarni2022SYK}%
  \BibitemOpen
  \bibfield  {author} {\bibinfo {author} {\bibfnamefont {A.}~\bibnamefont
  {Kulkarni}}, \bibinfo {author} {\bibfnamefont {T.}~\bibnamefont {Numasawa}},\
  and\ \bibinfo {author} {\bibfnamefont {S.}~\bibnamefont {Ryu}},\ }\bibfield
  {title} {\bibinfo {title} {{Lindbladian dynamics of the Sachdev-Ye-Kitaev
  model}},\ }\href {https://doi.org/10.1103/PhysRevB.106.075138} {\bibfield
  {journal} {\bibinfo  {journal} {Phys. Rev. B}\ }\textbf {\bibinfo {volume}
  {106}},\ \bibinfo {pages} {075138} (\bibinfo {year} {2022})}\BibitemShut
  {NoStop}%
\bibitem [{\citenamefont {Garc\'{\i}a-Garc\'{\i}a}\ \emph
  {et~al.}()\citenamefont {Garc\'{\i}a-Garc\'{\i}a}, \citenamefont {S\'a},
  \citenamefont {Verbaarschot},\ and\ \citenamefont {Yin}}]{sa2023SYK}%
  \BibitemOpen
  \bibfield  {author} {\bibinfo {author} {\bibfnamefont {A.~M.}\ \bibnamefont
  {Garc\'{\i}a-Garc\'{\i}a}}, \bibinfo {author} {\bibfnamefont
  {L.}~\bibnamefont {S\'a}}, \bibinfo {author} {\bibfnamefont {J.~J.~M.}\
  \bibnamefont {Verbaarschot}},\ and\ \bibinfo {author} {\bibfnamefont
  {C.}~\bibnamefont {Yin}},\ }\bibfield  {title} {\bibinfo {title} {{Towards a
  classification of PT-symmetric quantum systems: from dissipative dynamics to
  topology and wormholes}},\ }\Eprint {https://arxiv.org/abs/arXiv:2311.15677}
  {arXiv:2311.15677} \BibitemShut {NoStop}%
\bibitem [{\citenamefont {Mehta}(2004)}]{mehta2004}%
  \BibitemOpen
  \bibfield  {author} {\bibinfo {author} {\bibfnamefont {M.~L.}\ \bibnamefont
  {Mehta}},\ }\href@noop {} {\emph {\bibinfo {title} {{Random Matrices}}}}\
  (\bibinfo  {publisher} {Elsevier},\ \bibinfo {address} {Amsterdam},\ \bibinfo
  {year} {2004})\BibitemShut {NoStop}%
\bibitem [{\citenamefont {de~Vega}\ and\ \citenamefont
  {Alonso}(2017)}]{deVega-RMP17}%
  \BibitemOpen
  \bibfield  {author} {\bibinfo {author} {\bibfnamefont {I.}~\bibnamefont
  {de~Vega}}\ and\ \bibinfo {author} {\bibfnamefont {D.}~\bibnamefont
  {Alonso}},\ }\bibfield  {title} {\bibinfo {title} {{Dynamics of non-Markovian
  open quantum systems}},\ }\href
  {https://doi.org/10.1103/RevModPhys.89.015001} {\bibfield  {journal}
  {\bibinfo  {journal} {Rev. Mod. Phys.}\ }\textbf {\bibinfo {volume} {89}},\
  \bibinfo {pages} {015001} (\bibinfo {year} {2017})}\BibitemShut {NoStop}%
\end{thebibliography}%
\end{document}